%% file: main.tex
\documentclass[nobib,xcolor=table]{tufte-book}

\hypersetup{colorlinks} %

\usepackage{ulem}
\usepackage{microtype} %

\usepackage{lipsum} %

\usepackage{graphicx} %

\usepackage{tikz} %

\graphicspath{{graphics/}} %
\setkeys{Gin}{width=\linewidth,totalheight=\textheight,keepaspectratio} %

\usepackage{fancyvrb} %
\fvset{fontsize=\normalsize} %

\usepackage{xspace} %

\usepackage{units} %

\usepackage{makeidx} %
\makeindex %

\usepackage{subcaption} %
\usepackage{booktabs} %
\usepackage{multirow} %
\usepackage{enumitem} %
\usepackage{wrapfig}

\usepackage[most]{tcolorbox} %
\usepackage{epigraph}
\usepackage{lettrine}

\usepackage{multicol}
\usepackage{amsmath,amssymb,amsfonts}
\usepackage{listings}
\usepackage{url}
\usepackage{tabularx}
\usepackage{rotating}
\usepackage{pifont}

\usepackage{caption}
\usepackage{adjustbox} %
\usepackage{latexsym}
\usepackage{paralist}

\input{tufte-book-style/macro}

\title{Towards a Science of \\ Causal Interpretability in \\ Deep Learning for \\ Software Engineering
} %

\author[]{David N. Palacio} %

\begin{document}

\frontmatter

\maketitle %

\newpage
\begin{fullwidth}
~\vfill
\thispagestyle{empty}
\setlength{\parindent}{0pt}
\setlength{\parskip}{\baselineskip}
Copyright \copyright\ \the\year\\ \thanklessauthor

\par A Dissertation presented to the Graduate Faculty of the College of William and Mary in Virginia in Candidacy for the Degree of Doctor of Philosophy

\par\textit{First printing, \monthyear}
\end{fullwidth}

\tableofcontents %

\listoffigures %

\listoftables %

\input{chapters/chap_00_prolog/0-dedication}

\input{chapters/chap_00_prolog/0-abstract-tufte}

\input{chapters/chap_00_prolog/0-acknowledgments-tufte}

\mainmatter

\input{chapters/part_00_chap_01/1-introduction-wm}
\part{Preliminaries}
\input{chapters/part_00_chap_02/preliminaries} %
\input{chapters/part_00_chap_03/decomposition} %
\part{Associational Interpretability}
\input{chapters/part_01_chap_01/bayes} %
\input{chapters/part_01_chap_02/nature} %
\input{chapters/part_01_chap_03/explanations} %
\part{Interventional Interpretability}
\input{chapters/part_02_chap_01/do_code} %
\input{chapters/part_02_chap_02/case} %
\input{chapters/part_02_chap_03/benchmarking} %
\part{Counterfactual Interpretability}
\input{chapters/part_03_chap_01/autopoietic}

\part{Consolidation}
\input{chapters/part_04_chap_01/conclusions} %

\appendix
\input{appendixA}

\input{appendixB}

\backmatter

\bibliography{
references
} %

\bibliographystyle{alpha} %

\printindex %

\end{document}

%% file: tufte-book-style/macro.tex
\definecolor{MidnightBlue}{HTML}{006895}

\newcommand{\monthyear}{\ifcase\month\or January\or February\or March\or April\or May\or June\or July\or August\or September\or October\or November\or December\fi\space\number\year} %

\newcommand{\hlred}[1]{\textcolor{Maroon}{#1}} %
\newcommand{\hangleft}[1]{\makebox[0pt][r]{#1}} %

\newcommand{\tuftebs}{\symbol{'134}} %

\providecommand{\XeLaTeX}{X\lower.5ex\hbox{\kern-0.15em\reflectbox{E}}\kern-0.1em\LaTeX}

\newcommand{\doccmddef}[2][]{\hlred{\texttt{\tuftebs#2}}\label{cmd:#2}\ifthenelse{\isempty{#1}} %
{ %
\index{#2 command@\protect\hangleft{\texttt{\tuftebs}}\texttt{#2}}%
}
{ %
\index{#2 command@\protect\hangleft{\texttt{\tuftebs}}\texttt{#2} (\texttt{#1} package)}%
\index{#1 package@\texttt{#1} package}\index{packages!#1@\texttt{#1}}%
}}

\newcommand{\doccmd}[2][]{%
\texttt{\tuftebs#2}%
\ifthenelse{\isempty{#1}}%
{%
\index{#2 command@\protect\hangleft{\texttt{\tuftebs}}\texttt{#2}}%
}
{%
\index{#2 command@\protect\hangleft{\texttt{\tuftebs}}\texttt{#2} (\texttt{#1} package)}%
\index{#1 package@\texttt{#1} package}\index{packages!#1@\texttt{#1}}%
}}

\makeatletter

\useunder{\uline}{\ul}{} %

\newcommand{\xmark}{\ding{55}}

\newsavebox\CBox
\newlength\CLength

\def\circledlong#1{\sbox\CBox{#1}%
  \ifdim\wd\CBox>\ht\CBox \CLength=\wd\CBox\else\CLength=\ht\CBox\fi
    \makebox[1.2\CLength]{\makebox(0,0.6\CLength){\put(0,0){\circle{1.3\CLength}}}%
    \makebox(0,0.6\CLength){\put(-.5\wd\CBox,0){#1}}}}

\newcommand{\scm}{SCM\xspace}
\newcommand{\scms}{SCMs\xspace}
\newcommand{\docalculus}{\textit{do-calculus}\xspace}

\newcommand{\ntp}{NTP\xspace}

\newcommand{\docode}{\textit{do$_{code}$}\xspace}

\newcommand{\nlms}{NCMs\xspace}
\newcommand{\nlm}{NCM\xspace}
\newcommand{\llms}{LLMs\xspace}
\newcommand{\llm}{LLM\xspace}
\newcommand{\llmc}{LLMc\xspace}

\newcommand{\dlse}{DL\textit{4}SE\xspace}

\newcommand{\datainterI}{\textit{ProgramRepair}\xspace}
\newcommand{\datainterII}{\textit{SemanticPreserving}\xspace}
\newcommand{\datainterIII}{\textit{UnCommenting}\xspace}
\newcommand{\datainterIV}{\textit{ASTNodeTypes}\xspace}

\newcommand{\modelinterI}{\textit{NumberLayers}\xspace}
\newcommand{\modelinterII}{\textit{NumberUnits}\xspace}

\newcommand{\control}{\textit{control}\xspace}
\newcommand{\Ta}{$T_1$\xspace}
\newcommand{\Tb}{$T_2$\xspace}

\newcommand{\assoJS}{JS Dist.\xspace}
\newcommand{\assoPR}{Pearson\xspace}

\newcommand{\ASTerrors}{\textit{n\_ast\_errors}\xspace}
\newcommand{\ASTlevels}{\textit{n\_ast\_levels}\xspace}
\newcommand{\ASTnodes}{\textit{n\_ast\_nodes}\xspace}
\newcommand{\whitespaces}{\textit{n\_whitespaces}\xspace}
\newcommand{\complexity}{\textit{complexity}\xspace}
\newcommand{\nloc}{\textit{nloc}\xspace}
\newcommand{\tokenCount}{\textit{token\_count}\xspace}
\newcommand{\identifiers}{\textit{n\_identifiers}\xspace}
\newcommand{\commitID}{\textit{commit\_id}\xspace}
\newcommand{\funName}{\textit{fun\_name}\xspace}
\newcommand{\commitMessage}{\textit{commit\_message}\xspace}
\newcommand{\docstring}{\textit{docstring}\xspace}
\newcommand{\promptSize}{\textit{prompt\_size}\xspace}

\newcommand{\jaccard}{\texttt{\small<jaccard>}\xspace}
\newcommand{\levenshtein}{\texttt{\small<levenshtein>}\xspace}
\newcommand{\soren}{\texttt{\small<sorensen-dice>}\xspace}

\newcommand{\rfi}{$\mathcal{R}_1$\xspace}

\newcommand{\training}{\textit{CodeSearchNet}\xspace}
\newcommand{\BuggyTB}{\textit{BuggyTB}\xspace}
\newcommand{\CommentsTB}{\textit{CommentsTB}\xspace}
\newcommand{\BigCloneIITB}{\textit{BigClone2TB}\xspace}
\newcommand{\BigCloneIIITB}{\textit{BigClone3TB}\xspace}
\newcommand{\BigCloneTB}{\textit{BigCloneTB}\xspace}
\newcommand{\Galeras}{\textit{Galeras}\xspace}
\newcommand{\galeras}{\textit{Galeras}\xspace}

\newcommand{\randomCut}{\textit{RandomCut}\xspace}
\newcommand{\withDocstring}{\textit{WithDocString}\xspace}
\newcommand{\fromDocstring}{\textit{FromDocString}\xspace}
\newcommand{\commitGen}{\textit{CommitGen}\xspace}
\newcommand{\summarizationGen}{\textit{SummarizationGen}\xspace}

\newcommand{\blocks}{\texttt{\small[blocks]}\xspace}
\newcommand{\tests}{\texttt{\small[tests]}\xspace}
\newcommand{\oop}{\texttt{\small[oop]}\xspace}
\newcommand{\declarations}{\texttt{\small[declarations]}\xspace}
\newcommand{\exceptions}{\texttt{\small[exceptions]}\xspace}
\newcommand{\datatype}{\texttt{\small[datatype]}\xspace}
\newcommand{\loops}{\texttt{\small[loops]}\xspace}
\newcommand{\operators}{\texttt{\small[operators]}\xspace}
\newcommand{\conditionals}{\texttt{\small[conditionals]}\xspace}
\newcommand{\extra}{\texttt{\small[extraTokens]}\xspace}

\newcommand{\secref}[1]{Sec.~\S~\ref{#1}\xspace}
\newcommand{\chapref}[1]{Chapter~\S~\ref{#1}\xspace}

\newcommand{\figref}[1]{Fig.~\ref{#1}\xspace}

\newcommand{\equaref}[1]{Equation~\ref{#1}\xspace}
\newcommand{\tabref}[1]{Table~\ref{#1}\xspace}

\newcommand{\rnn}{RNN$_{1,1024}$\xspace}
\newcommand{\gru}{GRU$_{1,1024}$\xspace}
\newcommand{\grui}{GRU$_{2,1024}$\xspace}
\newcommand{\gruii}{GRU$_{3,1024}$\xspace}
\newcommand{\gruiii}{GRU$_{1,512}$\xspace}
\newcommand{\gruiv}{GRU$_{1,2048}$\xspace}
\newcommand{\chatgpt}{ChatGPT\xspace}

\newcommand{\tf}{TF$_{6,12}$\xspace}
\newcommand{\tfi}{TF$_{12,12}$\xspace}
\newcommand{\tfii}{TF$_{24,12}$\xspace}

\newcommand{\bert}{BERT$_{12,12}$\xspace}

\newcommand{\Comet}{{\sc Comet}\xspace}
\newcommand{\Comets}{{\sc Comet's}\xspace}

\newcommand{\node}{\texttt{\small[node]}\xspace}
\newcommand{\forinclause}{\texttt{\small[for\_in\_clause]}\xspace}

\newcommand{\hairsp}{\hspace{1pt}} %
\newcommand{\ie}{\textit{i.\hairsp{}e.,}\xspace} %
\newcommand{\eg}{\textit{e.\hairsp{}g.,}\xspace} %
\newcommand{\etc}{\textit{etc.}\xspace}
\newcommand{\etal}{et al.\xspace}
\newcommand{\etals}{et al.'s\xspace}
\newcommand{\aka}{\textit{a.k.a.}\xspace}

\newtheorem{definition}{Definition}[chapter]
\newtheorem{exmp}{\underline{Example \textbf{. ---}} }[chapter]

\newcommand*\circled[1]{\tikz[baseline=(char.base)]{
            \node[shape=circle,draw,inner sep=0.5pt] (char) {#1};}}

\newboolean{showcomments} %
\setboolean{showcomments}{true} %
\ifthenelse{\boolean{showcomments}} %
  {\newcommand{\nb}[2]{
    \fbox{\bfseries\sffamily\scriptsize#1}
    {\sf\small$\blacktriangleright$\textit{#2}$\blacktriangleleft$}
   }
   
  }
  {\newcommand{\nb}[2]{}
   
  }

\newtcolorbox{boxK}{
    fontupper = \small,
    sharpish corners, %
    boxrule = 0pt,
    toprule = 4.5pt, %
    enhanced,
    fuzzy shadow = {0pt}{-2pt}{-0.5pt}{0.5pt}{black!35} %
}

\newcommand{\infotheory}{{\small\texttt{TraceXplainer}}\xspace}

\newcommand{\libest}{\texttt{libEST}\xspace}
\newcommand{\cisco}{\texttt{csc}\xspace}
\newcommand{\albergate}{\texttt{albergate}\xspace}
\newcommand{\ebt}{\texttt{EBT}\xspace}
\newcommand{\etour}{\texttt{etour}\xspace}
\newcommand{\itrust}{\texttt{itrust}\xspace}
\newcommand{\smos}{\texttt{SMOS}\xspace}

\newcommand{\bits}{\textit{$\mathcal{B}$}\xspace}

\newcommand{\wv}{\textit{word2vec}\xspace}
\newcommand{\dv}{\textit{doc2vec}\xspace}

\newcommand{\exbase}{$EX_0$\xspace}
\newcommand{\exfirst}{$EX_1$\xspace}
\newcommand{\exsecond}{$EX_2$\xspace}
\newcommand{\exthird}{$EX_3$\xspace}
\newcommand{\exfourth}{$EX_4$\xspace}

\newcommand{\wmd}{WMD\xspace}
\newcommand{\cosine}{COS\xspace}
\newcommand{\euc}{EUC\xspace}

\newcommand{\sm}{\textit{SM}\xspace}
\newcommand{\alife}{A\textit{Life}\xspace}
\newcommand{\sdlc}{\textit{SDLC}\xspace}
\newcommand{\autopo}{$\Psi$\textit{-Arch}\xspace}

\newtheorem{defn}{Definition}[section]

\newcommand{\lmc}{LM4Code\xspace}
\newcommand{\lmsc}{LM4Code\xspace}
\newcommand{\lms}{LMs\xspace}

\newcommand{\codeRational}{{{Code${\mathbb{Q}}$}}\xspace}
\newcommand{\ewash}{\textit{eWASH}\xspace}
\newcommand{\codeparrot}{\textit{codeparrot-small}\xspace}

\newcommand{\sgbd}{$TB_1$\xspace}
\newcommand{\dcsgbd}{$TB_2$\xspace}
\newcommand{\dcsg}{$TB_3$\xspace}
\newcommand{\dc}{$TB_4$\xspace}

\newcommand{\astrust}{AST\textit{rust}\xspace}
\newcommand{\asc}{\textit{SC}\xspace}
\newcommand{\ascs}{\textit{SCs}\xspace}

\newcommand{\dags}{\textit{DAGs}\xspace}

\newcommand{\gptI}{\textit{gpt-3 [125M]}\xspace}
\newcommand{\gptII}{\textit{gpt-3 [1.3B]}\xspace}
\newcommand{\gptIII}{\textit{gpt-3 [2.7B]}\xspace}

\newcommand{\codegenII}{\textit{codegen-nl [2B]}\xspace}

\newcommand{\multiI}{\textit{multi-lang [110M]}\xspace}

\newcommand{\multiIII}{\textit{multi-lang [2B]}\xspace}

\newcommand{\monoI}{\textit{mono-lang [110M]}\xspace}
\newcommand{\monoII}{\textit{mono-lang [1.5B]}\xspace}

\newcommand{\monoIIII}{\textit{mono-lang [2B]}\xspace}

\newcommand{\nlgpt}{\textit{NL GPT-3}\xspace}
\newcommand{\nlcodegen}{\textit{NL Codegen}\xspace}
\newcommand{\monolang}{\textit{Mono-Language-Type}\xspace}
\newcommand{\multilang}{\textit{Multi-Language-Type}\xspace}

\newcommand{\surveyControl}{\textit{$U_{CTR}$}\xspace}

\newcommand{\surveySequence}{\textit{$U_{SEQ}$}\xspace}
\newcommand{\surveyAstBased}{\textit{$U_{AST}$}\xspace}

\newcommand{\surveyAstPartial}{\textit{$U_{AST[p]}$}\xspace}
\newcommand{\surveyAstComplete}{\textit{$U_{AST[c]}$}\xspace}

%% file: chapters/chap_00_prolog/0-dedication.tex
\cleardoublepage
~\vfill
\begin{doublespace}
\noindent\fontsize{18}{22}\selectfont\itshape
\nohyphenation
To Myrin and Friends
\end{doublespace}
\vfill
\vfill

%% file: chapters/chap_00_prolog/0-abstract-tufte.tex
\cleardoublepage
\chapter{Abstract} %
\label{ch:abstract}

This dissertation addresses achieving causal interpretability in Deep Learning for Software Engineering (\dlse). Although Neural Code Models (\nlms) demonstrate promising performance in automating software engineering tasks, their lack of transparency on causal relationships between inputs and outputs hinders a complete understanding of their capabilities and limitations. Software researchers and, more generally, practitioners should develop the ability to \textit{explain} code predictions to increase the \textit{trust} and \textit{trustworthiness} of \nlms. However, traditional associational interpretability, which focuses on identifying correlations, is considered insufficient for tasks requiring interventions and understanding changes' impact. To overcome this limitation, this dissertation introduces \docode, a novel post hoc interpretability method specifically designed for \nlms. \docode leverages causal inference to provide programming language-oriented explanations of model predictions. It comprises a formal four-step pipeline: \textit{modeling} a causal problem using Structural Causal Models (SCMs), \textit{identifying} the causal estimand, \textit{estimating} causal effects using metrics such as Average Treatment Effect (ATE), and \textit{refuting} the effect estimates. The theoretical underpinnings of \docode are extensible, and a concrete instantiation is provided to mitigate the impact of spurious correlations by grounding explanations in properties of programming languages. A comprehensive case study on deep code generation across various interpretability scenarios and on the uses of different deep learning architectures demonstrates the practical benefits of \docode. The findings of the study reveal insights into the sensitivity of \nlms to changes in code syntax and \nlms' ability to learn certain concepts of programming languages with minimized confounding bias.
Next, the dissertation explores the role of associational interpretability as a foundation, examining the causal nature of software information and using information theory with tools such as COMET (a Hierarchical Bayesian Software Retrieval Model) and \infotheory to understand software traceability. Furthermore, the work emphasizes the importance of identifying code confounders for a more rigorous evaluation of \dlse models. Finally, the dissertation offers guidelines for applying causal interpretability to Neural Code Models, contributing a formal framework and practical considerations towards building more reliable and trustworthy AI in Software Engineering.

%% file: chapters/chap_00_prolog/0-acknowledgments-tufte.tex
\chapter{Acknowledgements} %
\label{ch:ack}

My time at William \& Mary has been the most gratifying and inspirational for me, a whole learning experience on a personal and academic level. It has shaped me not only into a determined scientist but also into a more conscious, adaptable, down-to-earth, and minimalist person. The perseverance, consistency, and assertiveness of my advisor, Denys Poshyvanyk, have tremendously contributed to my formation and, therefore, the success of every publication and chapter of this dissertation. I am grateful I have joined \textit{SEMERU} right after my master's to continue with my PhD in his research group. Denys has created a dynamic environment for his students where we can all contribute and work as a team, sharing ideas and supporting each other. I am immensely grateful to be part of the research software community.

My dissertation has benefited from email exchanges, feedback, research collaborations, and conversations with several remarkable professionals in the field: Kevin Moran, Daniel Rodriguez-Cardenas, Nathan Cooper, Dipin Khati, Antonio Mastropaolo, Alejandro Velasco, Gabriele Bavota, Cody Watson, Amit Seal Ami, Yanfu Yan, Yang Son, Martin White, Carlos Sierra, and Mario Linares-Vasquez. I especially thank the members of my dissertation committee for their support and feedback: professors Pieter Peers, Oscar Chaparro, Huajie Shao, and Collin McMillan. I also want to thank my former students and undergrad collaborators, Alvaro Rodriguez, Henry Burke, Daniel McCrystal, Avi Urbach, and Ali Yachnes. My particular thanks go to Carlos Bernal-Cardenas, Aya Garryyeva, and Michele Tufano for their invaluable friendship, academic, and professional support. I also thank my undergraduate advisor, Jonatan Gomez Perdomo, for introducing me to the wonderful world of artificial life/intelligence, complex systems, and computer science research. I thank Cisco Systems -with professor Chris Shenefiel and Jeff Johnson- and Microsoft Corporation -with Colin Clement and Neelakantan Sundaresan- for their support during my doctoral internships. The completion of this project could not have been accomplished without the support of the office management and systems at the Department of Computer Science in W\&M, Vanessa Godwin, Dale Hayes, and Joseph Hause. Special thanks to Sarah Glosson, the Graduate Center Director, for her invaluable support in helping me enhance my writing.

My deepest and warmest thanks extend to my mom for her unconditional love and support. I am grateful for my aunts (Mirella, Janneth, Belva, Ruth, and Eddy), grandma, cousins, and close friends (Ali, Miguel, Diego, Andrea, Laura, Karen, Tomos, Olga, and Erna)  who have always been there for me when starting and/or during this PhD adventure. I treasure their advice, guidance, long conversations, and incredible support over these years.

%% file: chapters/part_00_chap_01/1-introduction-wm.tex
\chapter{Introduction} %
\label{ch:intro}

\epigraph{\scriptsize (...) The ideal technology that causal inference strives to emulate resides within our own minds (...) All because we asked a simple question: Why? (...)}{\scriptsize\textit{Judea Pearl \citep{Pearl2018Causality}}}

\lettrine[lines=2]{\textbf{S}}{} \textit{cientific explanations} and \textit{causality} are two interconnected concepts. For most science philosophers, explaining a phenomenon is determining what exactly caused it. From Aristotle (322 a.c.), who claims that \textit{why-type} questions are the essence of scientific explanations, to David Hume (1775), who assesses that the only thing we experience is that some events are conjoined; a clear definition for \textit{causality} comprises an open philosophical and scientific problem today \citep{molak_causal_2023}. 

Contemporary authors such as Cartwright \citep{cartwright1989capacities,cartwright1999dappled,cartwright2007hunting} and Bunge \citep{bunge1959causality,bunge2003emergence,bunge2011philosophy} reject \textit{Humean} causation, which sees causality as just a regular sequence of events, in favor of emphasizing \textit{mechanisms over statistical correlations} to determine causal relationships\footnote{Bunge defines a \textit{``mechanism''} as a process in a concrete system whereby one state or event brings about another under laws of nature \citep{bunge2011philosophy}}. As an example, Cartwright allows for \textit{context-dependent causation} rather than strict deterministic laws. Bunge, nevertheless, insists on \textit{deterministic causality} but allows complex, multilayered interactions\footnote{According to Bunge, the world is organized in systems nested within systems, creating multiple levels. Thus, \textit{``multilayered interactions''} mean that events at one level \textit{affect} and \textit{are affected} by events at other levels \citep{bunge2011philosophy,bunge2003emergence,bunge1959causality}. For example, a causal chain spans multiple layers \eg gene mutation → brain chemistry → personality → economic decision.}. Remarkably, Judea Pearl introduced an alternative definition of causality from Computer Science. He coined the term \textbf{Causal Inference} to overcome the constraints of a purely statistical and probabilistic definition of causation \citep{Pearl2018Causality, Pearl2009Causality,pearl2009overview}. Pearl’s causal inference framework provides the mathematical analysis of causal queries for which Cartwright and Bunge have offered both criticism and acknowledgment of usefulness upon proper adaptation. 

\begin{marginfigure}
		\centering
		\includegraphics[width=\linewidth]{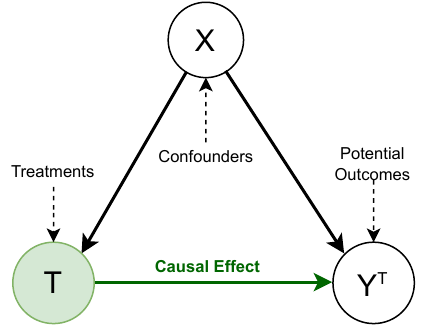}
		\caption{A generic \scm.}
        \label{fig:scm_0}
\end{marginfigure}

Pearl introduced a mathematical model of causality centered on \textbf{Structural Causal Models} (\scms) and a symbolic engine (\ie \docalculus), which predicts the effect of interventions controlling for \textit{confounding bias}. The aphorism \textit{``correlation is not causation''} represents what confounding bias measures: the disparity between observed associations and true causal effects. Formally, confounding bias refers to the distortion in the estimated effect of a treatment $T$ on an outcome $Y$ due to the presence of confounding variables $Z$ - variables that influence both treatments and outcomes (see \figref{fig:scm_0}). Under the \docalculus notation, the \textit{do-operator} denotes interventions \eg $p(y|do(t))$. Thus, confounding bias occurs when associations do not correspond to interventions $p(y|t) \neq p(y|do(t))$ by dint of the influence of confounders.

Furthermore, \scms enable AI systems to distinguish between three levels/rungs in a hierarchy of causal information:
\begin{itemize}
\item \textbf{L$_1$- Association (correlation):} ``People who eat ice cream are more likely to swim.''
\item \textbf{L$_2$- Intervention (causation):} ``If we make people eat ice cream, will they swim more?''
\item \textbf{L$_3$- Counterfactual reasoning:} ``If people had not eaten ice cream, would they have gone swimming?''
\end{itemize}

This classification of causal information responds to the type of questions each class can answer. The levels are tagged L$_1$-Association, L$_2$-Intervention, and L$_3$-Counterfactual. The first level invokes solely statistical relationships: $p(y|t)$. The second level simulates the intervention of the treatment using the do-operator: $p(y|do(t))$. The third and highest level enunciates complex retrospective reasoning: $p(y_t|t',y')$, which stands for the probability that event $Y=y$ would be observed had $T$ been $t$, given that we observed $T$ to be $t'$ and $Y$ to be $y'$. Counterfactuals are arranged at the top as they subsume questions about associations and interventions.  

Pearl’s \textit{Causal Inference} framework has revolutionized AI by providing a formal structure for systems to reason about cause and effect (\ie interventions and retrospection). The formal structure, or \scm, deploys three parts: 1) Graphical models, 2) Structural equations, and 3) Counterfactual and interventional logic. \textit{Graphical models}, or Directed Acyclic Graphs (\dags), serve as a data structure to represent assumptions about the environment, while \textit{counterfactuals and interventions} are semantic tools to articulate causal questions. \textit{Structural equations}, interestingly, connect both \dags and counterfactuals to determine analytically whether the probability of counterfactuals is estimable from experimental or observational data~\citep{Pearl_WSDM_2018}.   

One of the multiple applications of \textit{Causal Inference} entails removing the obstacle of \textit{explainability} in Machine Learning models (\ie ML models). An ML model, which remains a black box, disregards any reasons behind their predictions or recommendations, eroding users' trust. \scms are fundamental for producing \textit{scientific explanations} and allowing rigor in empirical studies involving ML models, such as those performed in Software Engineering research. 

In software research, for example, if practitioners are required to adopt a deep learning model that generates functionally correct code \ie Language Models for Code (\lmsc), they must have a degree of \textit{trustworthiness} in such a model \citep{palacio2024trustworthyinterpretablellmscode}. This degree of trustworthiness implies that the model must generate snippets free of errors \footnote{ The term ``errors'' refers to any program failure, product of an inadequate or incorrect code generation at the semantic or syntactic level.}. Although the software research community has not yet reached a consensus on the definitions of \textit{trust} and \textit{trustworthiness}, we have referred to recent systematic literature reviews and surveys approximating trust concepts in \lmsc~\citep{khati2025mappingtrustterrainllms,trustLLM}. Thus, we understand \textit{trust} as the willingness to rely on LLM outputs under uncertainty. In contrast, \textit{trustworthiness} is the intrinsic qualities of the system that ensure reliable, transparent, and ethical performance. 

Consequently, software researchers should develop the ability to \textit{explain} code predictions to control the amount of errors inserted by deep learning models. In explaining an erroneous output, researchers noticeably look for the cause of the error. In fact, the question \textit{why does a deep learning model fail in code generation?} is semantically the same as \textit{what are the causes of code errors generated by a deep learning model?} Therefore, both \textit{explanations} and \textit{causality} are the conceptual basis for establishing a \textit{formalism} whose purpose is to understand and interpret \textit{deep code generation}. This dissertation, broadly, concerns the phenomenon of \textit{interpreting} deep learning models to automate software engineering tasks, within a field known as \textit{Deep Learning for Software Engineering} (\dlse). 

Deep learning models are capable of automating software engineering tasks with, presumably, high accuracy. To verify this, \textit{Causal Inference} provides a mathematical foundation to articulate causal queries by quantifying the influence of a formulated treatment (\ie factors that influence models' predictions) on a potential outcome (\ie accuracy). While, in general, the performance of Neural Code Models (\nlms) and \lmsc appears promising \citep{watson2020dl4se,feng2020codebertpretrainedmodelprogramming,openai2023gpt4,GPTNeoX,codeparrot}, currently much is unknown about how such models arrive at decisions \citep{doshivelez2017rigorousscienceinterpretablemachine,lipton_mythos_2017,molnar2025}. This dissertation explores the intersection of these two fields:

\begin{center}
{ \centering \textbf{\dlse} $\cap$ \textbf{Causal Inference} $=$ \fbox{\textbf{?}} } 
\end{center}

The postulated intersection stems from the fact that an \textit{interpretability science} is indispensable to explain deep learning models trained on software data. Here, the \textit{formalism} employed for the proposed science is defined using a \scm that the reader will find recurrently throughout the dissertation (see \figref{fig:scm_0}). These \scms, as described before, are mathematical and probabilistic representations of causal assumptions (\ie the directionality of the variables) and observable data. Moreover, these graphical representations of causality provide researchers with algorithmic methods and probabilistic tools to estimate causal effects. Particularly, we adapt the calculus of causation (\ie \docalculus~\citep{pearl2009overview}) to investigate how to build \textit{code-based explanations} of deep learning models, which are used in software engineering research. 

The central focus of this work is the exploration and application of causal interpretability techniques within the domain of \dlse. This dissertation advocates for a shift from traditional \textit{canonical} interpretability to \textit{causal} interpretability~\citep{moraffah2020causalinterpretabilitymachinelearning}, highlighting the need to understand \textit{why} \dlse models (\ie \nlms) make certain predictions and enable interventions based on these causal explanations.

To this end, we introduce \docode, a post hoc interpretability method specific to \nlms capable of explaining model predictions. \docode is based on causal inference to enable programming language-oriented explanations. Although the theoretical underpinnings of \docode are extensible to explore different model properties, we provide a concrete instantiation that aims to mitigate the impact of \textit{spurious correlations} by grounding explanations of model behavior in properties of programming languages. To prove the practical benefit of \docode and illustrate the insights our framework can provide, we performed a case study on two popular deep learning architectures and ten \nlms. The results of this case study illustrate that the studied \nlms are sensitive to changes in code syntax. These insights demonstrate the potential of \docode as a useful method to detect and facilitate the elimination of confounding bias in \nlms.

Upon investigating how scientific explanations contribute to the interpretability of \dlse models, this dissertation resulted in several notable findings that illustrate the usefulness of adopting the causal inference framework. Our most relevant findings are as follows:

First, in associational interpretability, the dissertation explored software traceability. \Comet, a Hierarchical Bayesian Software Retrieval Model, was introduced and shown to improve traceability links' accuracy by combining Information Retrieval / Machine Learning (IR / ML) techniques, developer feedback, and transitive links.   

Second, the use of \infotheory demonstrated information-theoretic limits on the effectiveness of unsupervised traceability techniques. A key finding was that mutual information between tokens in source code and target documentation can be zero despite semantic similarity if they operate in different semantic spaces (natural language vs. code). The analysis highlighted potential information imbalances in the traceability datasets.

Third, the dissertation establishes the limitations of associational interpretability in \dlse, emphasizing that identifying correlations is insufficient to understand and leverage these \nlms for interventions. It argues for a crucial shift towards causal interpretability to understand the why behind model predictions and enable effective software engineering-based interventions with predictable outcomes.

Finally, the case study on deep code generation using \docode revealed several notable findings: \textcircled{1}~The presence or absence of buggy code did not appear to causally influence the prediction performance of the studied NCMs, even under high correlation measured, suggesting the presence of confounding bias. \textcircled{2}~The presence of Type II clones in training data impacted (or causally explained) the effectiveness of \nlms in terms of cross-entropy, indicating the influence of the characteristics of the training data on the performance of the model. \textcircled{3}~The intervention of masking random tokens had a more significant impact on code predictions than masking grammar-based categories for BERT-like models, suggesting that these models might not fully capture the structural information of programming languages. \textcircled{4}~The studied \nlms, except for the BERT-like model, statistically learned to predict tokens related to code blocks (\eg brackets, parentheses, semicolon) with less confounding bias compared with other programming language constructs. \textcircled{5}~The prompt engineering methods used in the \Galeras benchmark case study on ChatGPT showed a positive causal influence of prompt semantics on generative performance.

\textbf{Contributions.} Aforementioned findings suggest the need to integrate causal interpretability into post hoc analysis of \nlms. We provide researchers with an approach that integrates causal analysis into their research. To do this, in addition to our case study showing the potential that such an analysis can provide, we have created a checklist that summarizes the steps required to apply Causal Interpretability to \nlms (see \figref{fig:checklist}). In summary, this dissertation makes the following contributions. 

\underline{$Product_1$:} \textit{A Syntax (De)Composition Statistical Technique To Enable Code-based Explanations.} This document explores how syntactic (AST-based) and semantic (keyword-based) clustering of code predictions can be used within the \docode framework to provide more interpretable explanations of model behavior. More details in \chapref{ch:decomposition}.

\underline{$Product_2$:} \textit{A Hierarchi\textbf{C}al Pr\textbf{O}babilistic \textbf{M}odel for Softwar\textbf{E} \textbf{T}raceability.} \Comet is a tool to improve the effectiveness of software traceability automation. Furthermore, this technique demonstrates that software artifacts and their relationships can be represented as statistical distributions. More details in \chapref{ch:bayes}. 

\underline{$Product_3$:} \textit{A Method That Leverages Information Theory for Interpretability.} Information theory metrics can provide valuable formal analyses on information flow and relationships between software artifacts, helping to interpret the effectiveness of traceability models. See \infotheory in \chapref{ch:nature}.

\underline{$Product_4$:} \textit{A Patent for a Debugging Tool and an Interpretability Technique for Code-based Explanations.} We introduce \textit{code rationales} (\codeRational), a technique with rigorous mathematical and statistical foundation, to identify subsets of tokens that can explain individual code predictions. This technique was patented as a debugging tool \citep{clement2024debugging}. More details in \chapref{ch:rationales}. 

\underline{$Product_5$:} \textit{A Methodology for Causal Interpretability (\docode).} This dissertation emphasizes that simply identifying correlations between software artifacts or code features and model predictions is insufficient to truly understand and leverage models for intervention. \dlse To effectively use \dlse for tasks that require action and understanding of consequences, a change to causal interpretability is necessary. This involves identifying and quantifying cause-and-effect relationships. The \docode approach provides a structured methodology to achieve causal interpretability in \dlse, which involves the formulation of \scms, the definition of interventions, the estimation of causal effects, and the implementation of refutation tests. More details in \chapref{ch:docode}.

\underline{$Product_6$:} \textit{Several Empirical Intepretability Scenarios in Software Engineering.} We illustrate how \docode is actionable in Software Engineering. Seven scenarios were proposed within a permutation framework that can be extended as practitioners or researchers necessitate - more details in \chapref{ch:case}.  

\underline{$Product_7$:} \textit{A Benchmarking Strategy to Enable Causal Evaluations.} This work stresses the need to consider interventions that are meaningful within the software engineering domain (\eg code uncommenting, changing the number of layers) and to account for potential confounders (\ie common causes) that could lead to spurious correlations. The \Galeras benchmark is introduced as a tool for causal evaluation in code generation, emphasizing the need to go beyond traditional performance metrics and assess the causal effects of code features on model predictions. More details in \chapref{ch:benchmarking}. 

\underline{$Product_8$:} \textit{A Prospective Analysis on Counterfactual Applications in Software Engineering.} Drawing inspiration from Artificial Life (\textit{ALife}), we postulate a fundamental shift in the Software Development Life-Cycle (\textit{SDLC}) by introducing self-construction mechanisms that enable software to evolve and maintain autonomously. Autopoietic Architectures, specifically $\Psi$-\textit{Arch}, as a foundational framework for self-constructing software are proposed in \chapref{ch:autopoietic}.  

\underline{$Artifacts$:} \textit{Replication Packages and Repositories.} To encourage the use of our methodology in future work on developing and evaluating code generation models and for replicating our experiments: \docode~\citep{icodegen},\codeRational~\citep{CodeQ2025}, \astrust~\citep{RepoASTrust24}.

\textbf{How to read this dissertation.} Several ideas presented are communicated with a broad audience in mind. We assume that the reader has notions of probability theory and deep learning. Although this work was designed to be a primer in interpretability for software researchers with simple terms and mathematical definitions, the \textit{ten chapters} are organized into \textit{four parts} with increasing complexity. Then, to fully understand a chapter, the reader needs to cover prior ones, making the latter chapters more challenging and, therefore, the most interesting to read.

The \textit{first part}, with two chapters, highlights relevant concepts that the reader must fully understand before jumping to the approach sections. \chapref{ch:preliminaries} includes formal definitions of causal inference and a complete background in interpretability theory. It introduces \textbf{The Causal Interpretability Hypothesis}, which is, withal, the core of this research. \chapref{ch:decomposition} poses the concept of Syntax (De)Composition relevant to present potential outcomes in Structural Causal Models (\scms).

The \textit{second part}, with three chapters, focuses on the representation of code artifacts with Bayesian networks. \chapref{ch:bayes} poses a method to model the relationships of artifacts in a probabilistic fashion. While \chapref{ch:nature} and \chapref{ch:rationales} describe a way of presenting \textit{code-based explanations} and their properties based on information theory and \textit{Abstract Syntax Trees}, respectively.

The \textit{third part}, with three chapters, concentrates on deep code generation and the interpretability problem of estimating causal effects that help explain \nlms' outputs. \chapref{ch:docode} uses \textit{Pearl's Ladder of Causation} definition in the preliminary chapter and extends it to outline our interpretability approach \docode. This approach embodies four subsequent steps that can involve explanations of deep learning models: \textit{1. modeling a causal problem}, \textit{2. identifying causal estimand}, \textit{3. estimating causal effects}, and \textit{4. refuting effect estimates}. All four steps are required to compute causal effects. \chapref{ch:case} presents a practical case study in deep code generation for seven different interpretability scenarios. \chapref{ch:benchmarking} proposes a benchmarking pipeline to collect, filter, and analyze observable software data aligned to \scms. 

The fourth part, with one \chapref{ch:autopoietic}, presents a position of our counterfactual-based agent investigation in the software community. The last part, with one \chapref{ch:conclusion}, consolidates the most impactful findings, actionable takeaways, contributions, and a guide for practitioners to venture into causal analysis.

%% file: chapters/part_00_chap_02/preliminaries.tex
\chapter{Interpretable \dlse}
\label{ch:preliminaries}

\epigraph{\scriptsize (...) deep learning works for certain tasks, but it is the antithesis of transparency (...)}{\scriptsize\textit{Judea Pearl \citep{Pearl2018Causality}}}

\input{chapters/part_00_chap_02/sec_01_artificial_se} %
\input{chapters/part_00_chap_02/sec_02_problem} %
\input{chapters/part_00_chap_02/sec_03_interpretability} %
\input{chapters/part_00_chap_02/sec_04_deep_code_interpret} %

%% file: chapters/part_00_chap_02/sec_01_artificial_se.tex
\section{Deep Learning for Software Engineering}
\label{ch:preliminaries:sec_01}

Software engineering (SE) research investigates questions related to the design, development, maintenance, testing, and evolution of software systems. As software continues to pervade a wide range of industries, both open- and closed-source code repositories have grown to become unprecedentedly large and complex. This has increased the number of unstructured, unlabeled, but important data, including requirements, design documents, source code files, test cases, and defect reports. Previously, the software engineering community has applied canonical Artificial Intelligence (AI) techniques to identify patterns and unique relationships within these data to automate or enhance many tasks typically performed manually by developers. Unfortunately, the process of implementing canonical AI techniques can be a tedious exercise in careful feature engineering, where researchers experiment with identifying salient attributes of data that can be leveraged to help solve a given problem or automate a given task.

However, with recent improvements in computational power and the amount of memory available in modern computer architectures, an advancement to traditional AI approaches called Deep Learning (DL) has arisen. Deep learning represents a fundamental shift in the manner in which machines learn patterns from data by \textit{automatically} extracting salient features for a given computational task, rather than relying on human intuition. Deep Learning approaches are characterized by architectures comprised of several layers that perform mathematical transformations on data passing through them. These transformations are controlled by sets of learnable parameters that are adjusted using a variety of learning and optimization algorithms. These computational layers and parameters form models that can be trained for specific tasks by updating the parameters according to a model's performance on a set of training data. Given the immense amount of structured and unstructured data in software repositories that are likely to contain hidden patterns, DL techniques have introduced advances in a variety of tasks in software engineering research, including automatic program repair~\citep{Tufano2018}, code suggestion~\citep{Gu2018}, defect prediction~\citep{Wang2016}, malware detection \citep{Li2018}, feature location~\citep{Corley2015}, among many others~\citep{Ma2018, Wan2018, Liu2018, White2016, Xu2016, Guo:ICSE'17, Tian2018a, Liu2017}. A recent report from the 2019 NSF Workshop on Deep Learning \& Software Engineering has referred to this area of work as Deep Learning for Software Engineering (\dlse)~\citep{dlse19-report}. 

The applications of DL to improve and automate SE tasks point to a clear synergy between ongoing research in SE and DL. However, in order to effectively chart the most impactful path forward for research at the intersection of these two fields, researchers need a clear map of what has been done, what has been successful and what can be improved. %
In an effort to map and guide research at the intersection of DL and SE, we conducted a systematic literature review (SLR) to identify and systematically enumerate the synergies between the two research fields. As a result of the analysis performed in our SLR, we synthesize a detailed \textit{research roadmap} of previous work on DL techniques applied to SE tasks\footnote{It should be noted that another area, known as Software Engineering for Deep Learning (SE4DL), which explores improvements to engineering processes for DL-based systems, was also identified at the 2019 NSF workshop. However, the number of papers we identified on this topic was small, mostly centered around emerging testing techniques for DL models. Therefore, we reserve a survey on this line of research for future work.} (\ie \dlse), complete with identified open challenges and best practices for applying DL techniques to SE-related tasks and data. In addition, we analyze the impacts of these DL-based approaches and discuss some observed concerns related to the potential reproducibility and replicability of our studied body of literature. %

\textbf{Applications of \nlms in SE:} \nlms in SE have a rich history, originating from the seminal work of Hindle \etal who proposed the concept of \textit{ naturalness} of software using \textit{n}-gram models~\citep{hindle2012natural}. %
Then, with the rise of Deep Learning, researchers began exploring the possibility of adapting NLMs to code, as initially demonstrated by White \etal~\citep{White:MSR15}. Since then, \nlms have been used for a number of SE tasks such as code completion \citep{Nguyen2013ASS, Raychev2014CodeCW, tu2014local, Hellendoorn2017AreCode, Karampatsis2019,  Karampatsis2020Open-VocabularyAbstract, Ciniselli.TSE, Chen2021} and translation \citep{Chen2019, Roziere2020transcoder, Hu2018comment, Mastropaolo2021StudyingTasks, Tufano2018, Tufano:tosem2019, Tufano:icse2019}. 
Researchers have also investigated various representations of \nlms for code~\citep{White2016} as well as graph-based representations based on ASTs~\citep{allamanis2018learning} and program dependency graphs~\citep{Nguyen:ICSE15}. It has also been shown that using a \nlm for code representation and then fine-tuning for specific SE tasks can achieve \textit{state-of-the-art} performance across tasks \citep{Hussain2020DeepTL, feng2020codebertpretrainedmodelprogramming, guo2021graphcodebert}.

%% file: chapters/part_00_chap_02/sec_02_problem.tex
\section{The Problem of Causal Interpretability in \dlse}
\label{ch:preliminaries:sec_02}

The combination of large amounts of freely available code-related data, which can be mined from open source repositories, and ever-more sophisticated deep learning architectures for code, which we refer to as Neural Code Models (\nlms), have fueled the development of Software Engineering (SE) tools with increasing effectiveness. \nlms have (seemingly) illustrated promising performance across a range of different SE tasks~\citep{Watson:ICSE20,White:MSR15,Ciniselli.TSE,Mastropaolo2021StudyingTasks, Tufano.MSR.2018, SANER.2019, Tufano:icse2021}. In particular, \textit{code generation} has been an important area of SE research for decades, enabling tools for downstream tasks such as code completion~\citep{MSR-Completion}, program repair~\citep{Chen2019}, and test case generation~\citep{Watson:ICSE20}. In addition, industry interest in the use of large language models (LLMs), a scalable version of \nlms, has also grown, as evidenced by tools such as Microsoft's IntelliCode \citep{intellicode}, Tabnine \citep{tabnine}, OpenAI's Codex \citep{openai_codex} and GitHub's Copilot \citep{github_copilot}. Given the prior popularity of code completion engines within IDEs~\citep{murphy2006ide} and the investment in commercial tools, \nlms for code generation will almost certainly be used soon to help build production software systems if they are not used already. 

Recently, there has been an increase in interest in evaluating \nlms for code generation. A recent study by Chen \etal \citep{Chen2021} illustrates that certain issues, such as alignment failures and biases, exist on a large scale \nlms. Most of the conclusions from Chen~\etal's study were discovered by manual analysis, \eg by sourcing counterexamples, making it difficult to rigorously quantify or systematically apply such an analysis to research prototypes~\citep{wu2019errudite}. Given the increasing profile and role \nlms for code generation plays in SE and the current limitations of adopted evaluation techniques, it is clear that \textit{new methods are needed to provide deeper insight into \nlms performance}.

Much of the quality assessment work on \nlms has mainly concentrated on accuracy-based metrics (\eg accuracy, BLEU, METEOR, ROUGE) as opposed to multimetric evaluations (\eg robustness, fairness, bias, efficiency). Moreover, skepticism within the NLP research community is growing about the efficacy of current accuracy-based metrics, as these metrics tend to overestimate model performance ~\citep{ribeiro2020checklist, rei2020comet, kocmi2021ship}. Even benchmarks that span multiple tasks and metrics have been shown to lack robustness, leading to incorrect assumptions of model comparisons~\citep{dehghani2021benchmark}. Notable work has called for a more systematic approach that aims to understand a given model's behavior according to its linguistic capabilities~\citep{ribeiro2020checklist}, while others have suggested the need for holistic evaluations of Language Models~\citep{liang_holistic_2022}. 

In addition to limitations with current methods of model quality assessment, some of the most popular \nlms have been adapted from the field of Natural Language Processing (NLP), and thus may inherit the various limitations often associated with such models --- including biases, memorization, and issues with data inefficiency, to name a few~\citep{bender2021parrots}. However, perhaps the most problematic aspect of current neural language models is the fact that they are incapable of explaining their reasoning \citep{doshivelez2017rigorousscienceinterpretablemachine,Doshi-Velez2018ConsiderationsLearning,Molnar2020InterpretableChallenges,lipton_mythos_2017}. \nlms and, in general, \textit{deep learning architectures} seemingly trade effectiveness for transparency, as the same complexity that allows for impressive learning and generalization leads models to operate in a \textit{black-box} fashion. That is, we are uncertain how neural models --- including \nlms --- \textit{arrive at decisions}; a phenomenon described as \textit{incompleteness in problem formalization}~\citep{Doshi-Velez2018ConsiderationsLearning}. Such incompleteness manifests as an inability to explain models' predictions in human-understandable terms. If neural models fail at justifying their outputs, \textit{Can we trust these models? Will these models work in deployment? How brittle are neural models in practical software engineering settings?} Researchers and practitioners require neural models to be robust not only at making predictions but also in the \textbf{interpretability} of those predictions \citep{lipton_mythos_2017}. Despite the increasing popularity and apparent effectiveness of code generation tools based on \nlms, there is still much that is unknown about their behavior. 

\begin{marginfigure}
  \begin{center}
    \includegraphics[width=\linewidth]{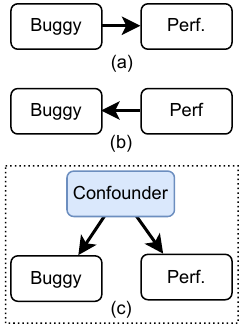}
  \end{center}
  \caption{Causal Interpretability of \nlms}
   \label{fig:example}
\end{marginfigure}

\textbf{A Motivating Example for  Interpretability.} Consider the scenario in which we \textit{observe} that a given \nlm is underperforming when predicting code from code inputs that are buggy. That is, we observe a correlation between \textit{buggy code} and \textit{worsened performance} after an initial exploratory analysis on this \nlm. However, a simple observation of a correlation between \textit{buggy code} and \textit{worsened performance} is insufficient evidence to explain the \nlms performance in this scenario, as different \textit{properties} of the buggy input code could be the \textit{cause} of the observation. To assert that \textit{input code being buggy} is an explanation or \textit{interpretation} for having a worsened output, we must first establish a \textbf{causal relationship} between these two random variables. Statistical dependencies or correlations are always induced by an underlying causal process \citep{Scholkopf2022}. In consequence, the correlation between \textit{buggy code} and \textit{worsened performance} has three possible causal explanations depicted as causal graphs in \figref{fig:example}: (a) \textit{worsened performance} is caused by \textit{buggy code}, (b) \textit{buggy code} is caused by \textit{worsened performance}; and (c) the reason for a correlation is that a confounding variable causally influences \textit{buggy code} and \textit{worsened performance}. Based on our \textit{domain knowledge} as an expert researcher, we may posit that option (c) is most likely to represent our causal assumptions, as many factors such as the confounding variable \textit{number of tokens or subwords} of a model's context window, can affect both variables. If a causal connection indeed exists between \textit{buggy code} and a model's predictions, then we can claim that the setting \textit{buggy code} is an interpretation for having a worsened output. However, in the above scenario, we would essentially be guessing, and as such the question remains: \textit{how can we quantify the causal effect of buggy code on \nlm's performance after controlling for the influence of hidden confounders?} The problem remains that, when attempting to understand the prediction performance of \nlms, no causal interpretability formalism is available to articulate or even answer the previous questions.

In this chapter, we cast this problem of achieving a more complete understanding of Neural Code Models as a \textit{Causal Interpretability} task and posit that we can leverage the \textit{theory of causation} as a mechanism to explain \nlms prediction performance. This mechanism includes a formalism to articulate and answer causal queries.  We hypothesize that this mechanism can serve as a useful verification tool to fulfill desiderata that we might want of \nlms. These desiderata, which are originally suggested in NLP literature, comprise notions such as facilitating debugging, detecting biases, providing recourse, and eventually, increasing the reliability and trust of \nlms \citep{molnar2025,doshivelez2017rigorousscienceinterpretablemachine,lo_trustworthy_2023}. As such, this dissertation introduces \docode, a novel global post hoc interpretability method specifically designed for understanding the effectiveness of \nlms using causal queries. Furthermore, \docode intends to establish a robust and adaptable methodology for \textit{interpreting predictions} of \nlms in contrast to simply \textit{measuring the accuracy} of these same \nlms.

\docode asks causal questions as to why prediction performance is affected by Software Engineering-based \textit{interventions}. {These interventions are generally data or model properties (\eg Bugginess of Code, Number of Inline Comments, Number of Model Layers) that a domain expert might think are affecting a given \nlm.} More specifically, \docode consists of four major conceptual steps, (1) \textbf{modeling} a \textit{structural causal graph}, (2) \textbf{identifying} causal estimand, (3) \textbf{estimating} causal effects, and (4) \textbf{refuting} the obtained effect estimate. Through the introduction of this interpretability method, we aim to help SE researchers and practitioners by allowing them to understand the potential limitations of a given model, work towards improving models and datasets based on these limitations, and ultimately make more informed decisions about how to build automated developer tools given a more holistic understanding of \textit{\textbf{what}} \nlms are predicting and \textit{\textbf{why}} the predictions are being made.

%% file: chapters/part_00_chap_02/sec_03_interpretability.tex
\section{From Canonical to Causal Interpretability}
\label{ch:preliminaries:sec_03}

\begin{marginfigure}
		\centering
		\includegraphics[width=\linewidth]{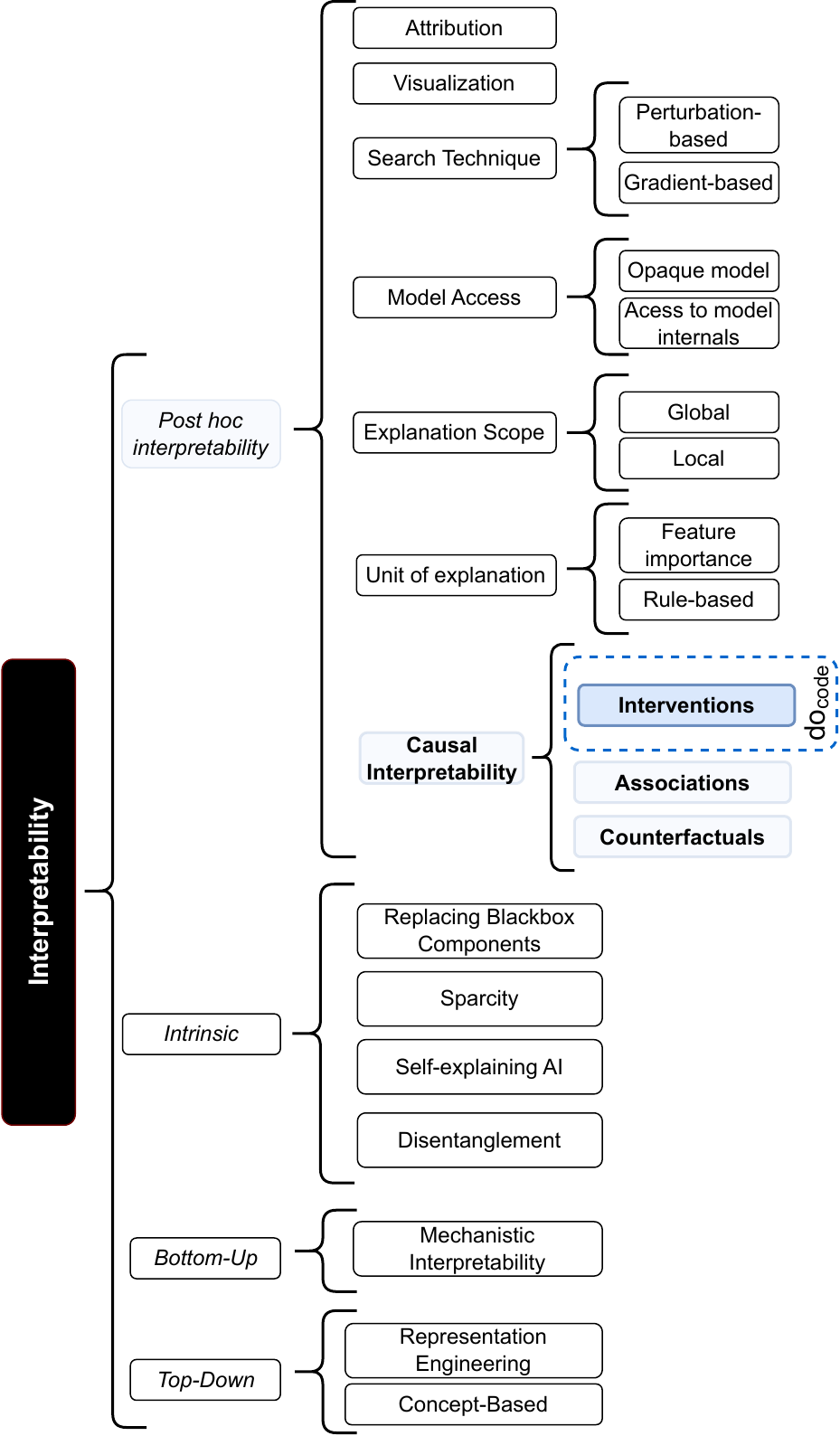}
		\caption{A classification of key methods in interpretability including \textit{Causal Interpretability}.}
        \label{fig:methods}
        \vspace{0.5cm}
\end{marginfigure}

\textit{Interpretable Machine Learning} is a research field aimed at understanding how opaque models give rise to predictions. This process is centered on human understanding. Although the research community has not reached a consensus on a precise definition of interpretability, researchers generally refer to this field as ``Interpretable Machine Learning'' or ``Explainable AI'' \citep{Molnar2020InterpretableChallenges,Doshi-Velez2018ConsiderationsLearning}. The main goal of the field is to create methods that explain the reasoning of the models and then verify whether this reasoning is valid \citep{doshivelez2017rigorousscienceinterpretablemachine}. These interpretability methods can be classified into different criteria depending on the concentration of the researcher. The most common criteria are post hoc vs. intrinsic, model-specific vs. model-agnostic, global vs. local explanation scope, feature importance vs. rule-based unit of explanation, and causal levels \citep{Molnar2020InterpretableChallenges,han_which_2022}. In \figref{fig:methods} we depict a summary of the most relevant methods published thus far. \figref{fig:methods} also positions our method \docode within the scope of \textit{Causal Interpretability}, which is based on the concept of Ladder of Causation by Pearl introduced in \secref{ch:nature:sec_04} \citep{Pearl2018Causality}.   

\begin{boxK}
\textit{Causal Interpretability} is a global post hoc approach by which Neural Code Models (\nlms) are interpreted or explained from a causal assumption encoded in a Structural Causal Model (SCM). Using the formalism of Pearl's Ladder of Causation, researchers can \textit{estimate} a quantifiable \textit{causal effect} of the proposed interventions.
\end{boxK}

\noindent\textbf{Interpretability of \nlms:} A growing body of research has investigated the potential of evaluating and understanding LMs using various techniques, which can be classified into two basic groups: (i) \textit{intrinsic techniques} such as probing for specific linguistic characteristics \citep{tenney2019bert} and examining neuron activations~\citep{dai2021knowledge}; and (ii) \textit{post hoc techniques} such as benchmarks \citep{wang2018glue, rajpurkar2018squad, husain2019codesearchnet, lu2021codexglue, Chen2021}, attention-based analysis \citep{liu2024reliability,mohammadkhani2023explain}, and intervention analyses \citep{khandelwal2018sharp, prabhakaran2019perturbation, ribeiro2020checklist, rabin2021generalizability}. Karpathy \etal were among the first to interpret NLMs for code using general next-token predictions~\citep{karpathy2015understand} as an interpretability method, as well as investigating individual neurons. Our work extends Karpathy \etal's interpretability analysis using causal inference and a more statistically rigorous methodology that grounds its evaluation in SE-specific features and settings. 

Complementary to our work, the work by Polyjuice \citep{wu-etal-2021-polyjuice} aims to automatically generate perturbations to language, to see how a given model performs in different scenarios. Our work is differentiated by the fact that we both define a mapping of source code tokens to categories and define a methodology for linking causal relationships of model changes to these various categories. As such, our work functions alongside techniques such as Polyjuice, providing deeper insights into why model performance varies across different perturbations. Our work is also complementary to the work by Cito \etal, \citep{cito_counterfactual_2022}  on Counterfactual explanations for models of source code. This work takes a significant step toward defining perturbations of code for counterfactual explanations of generative models. Our technique in conjunction with Cito et al.'s approach can provide deeper explanations as to \textit{why} a given \nlm's performance changed for a given perturbation.

Our method is different from the work on adversarial robustness via one key point, similar to the work of Cito \etal \textit{Our different testbeds are not intentionally designed to fool a model.} On the contrary, our different testbeds represent completely natural distributions of code tokens collected at scale. As such, our research has a completely different aim as compared to the work on adversarial robustness. {Other related works from Explainable Artificial Intelligence (XAI) have focused on the usage of causal theory applied to generate post hoc explanations about the logical structure of a neural network \citep{pmlr-v162-hu22b} or explore the causal relationships between the explanation and the prediction \citep{karimi_relationship_2023}. Both causal XAI methods differ from our work since we employ the notion of Pearl's Ladder of Causation directly on the inputs and outputs of \nlms.}

%% file: chapters/part_00_chap_02/sec_04_deep_code_interpret.tex
\section{Why do we need Causal Interpretability for \dlse?}
\label{ch:preliminaries:sec_04}

{Our research leverages Pearl's theory of Causal Inference and grounds it in interpreting Neural Code Models (\nlms). Given the wealth of metrics and benchmarks that exist for \nlms it is natural to ask \textit{why should we study causation in Deep Learning for Software Engineering?} Causation has two main goals in science (i) discovering causal variables and (ii) assessing counterfactual interventions \citep{Pearl2018Causality}. The field of \textit{Deep Learning for Software Engineering (DL4SE)} can take advantage of the latter when dealing with uncertainty and \textit{confounding bias} of \nlms. Estimating counterfactual interventions is a powerful tool to generate explanations of the model's performance. Our method, \docode, can be applied to a wide range of SE models for detecting and eliminating confounding bias.} {We do not intend to pose \docode as a tool to enhance prediction performance but as an adaptable approach to analyze why \nlms obtain their predictions}. However, quantifying the effects of interventions requires establishing a causal structure underlying the target data. 

Randomized controlled experiments were the general choice to explore causality before Pearl's definitions of graphical models. Nonetheless, it is not practical to force developers to perform interventions (such as removing comments from a training corpus) or even train hundreds of \nlms to test various treatments. Instead the $do-operator$ and causal graphs (\ie Structural Causal Models) are better tools for performing causal estimations from observational data. Reconstructing such graphical representations is challenging since it not only requires formalizing causation in the field of SE (\ie defining potential outcomes, common causes, and treatments) but also tracing and connecting software data to causal models (see the pipeline in \chapref{ch:docode}). In addition, formulating interventions is not an easy process. We must hypothesize feasible transformations or interventions that can occur in code to simulate real-world settings for \nlms, a concept that we synthesize as \textbf{The Causal Interpretability Hypothesis} (see \secref{sec:hypo}). To that end, we have proposed a pipeline to help aid in adapting the causal inference process to the interpretation of Neural Code Models (\nlms). This pipeline has been inspired by Pearl's notion of the \textit{Ladder of Causation}, introduced below, and the \textit{doWhy} library \citep{dowhy}. 

\subsection{Pearl's Ladder of Causation}

According to Pearl \& Mackenzie \citep{Pearl2018Causality}, Causal Inference (CI) seeks answers to questions of \textbf{association} (\textit{what is?}), counterfactual \textbf{interventions} (\textit{what if?}), and pure \textbf{counterfactuals} (\textit{why?}). The authors introduce the concept of \textit{Ladder of Causation} to match distinct levels of cognitive ability with concrete actions: seeing~(\textit{level/rung one}), doing~(\textit{level/rung two}), and imagining~(\textit{level/rung three}). Our proposed analysis is primarily concerned with levels one \& two. In particular, our method \docode is an extension of the intervention level.  

\begin{figure}[ht]
		\centering
		\includegraphics[width=0.9\textwidth]{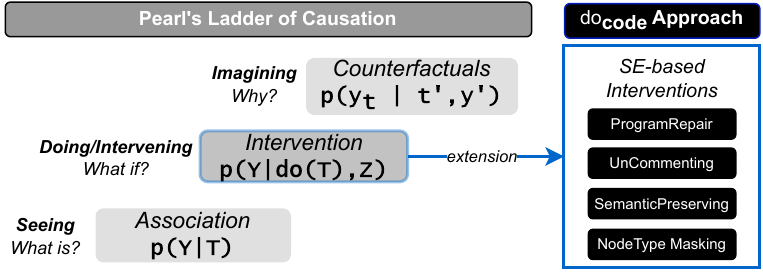}
		\caption{Ladder of Causation: \docode is an extension of the intervention level.}
        \label{fig:ladder}
        \vspace{1em}
\end{figure}

\textbf{Causal Association.} In level one causation, $p(Y|T)$ is estimated by using typical correlation methods (\eg Pearson, Spearman, or Covariance) in addition to functional associations such as $y=g(t)$, which can be predicted with regressions or ML methods. For binary treatments similar to those we used in $T_{[data]}$ (\eg Buggy/Fixed, Commented/Uncommented, Clone1/Clone2), we opt to employ Pearson's correlations $\rho_{YT}$ and Jensen-Shannon distance as association estimand for level 1 causation. 

\textbf{Causal Intervention.} Now, if we want to go beyond \textit{``what is''} type questions, we must go past simple correlations and associations. This requires the $do-operator$ found in level two, causation $p(y|do(t))$. It is relevant to identify cases of \textit{spurious correlations} (\ie \textit{Confounding Bias}) or cases where $p(Y|T)\neq p(Y|do(T))$. Typically, \textit{association is not causation} due to the influence of a common cause or confounding variable $Z$. This variable is the one that is being controlled or adjusted using the concept of the \textit{ adjustment formula} in Eq.~\ref{eqn:do-all-lines}. However, we can still compute the correlation $\rho_{YZ}$ to assess the variables that affect potential outcomes. 

\begin{marginfigure}
		\centering
		\includegraphics[width=\linewidth]{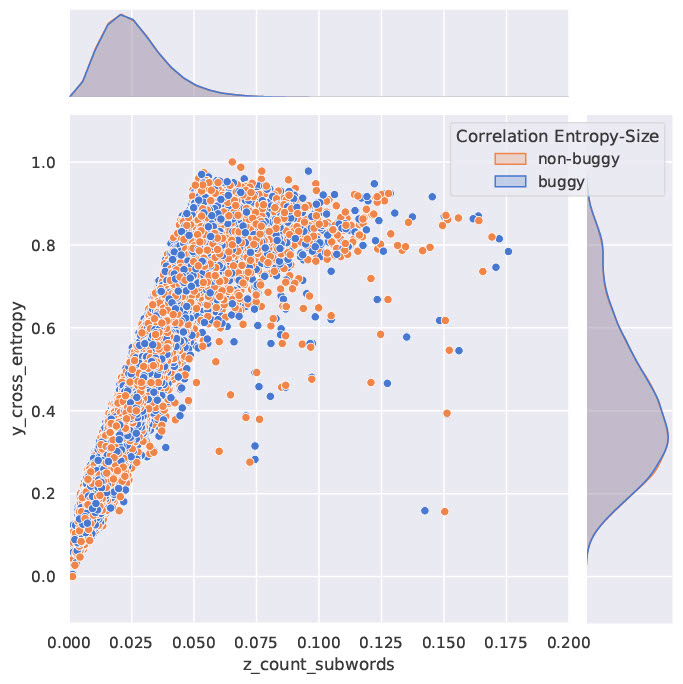}
		\caption{ \textit{Spurious Correlation} between the \textit{Number of Subwords} common cause and Cross-Entropy values ($p(Y|Z)\approx0.87$) for the \datainterI intervention generated from \tf. } 
        \vspace{0.2cm}
        \label{fig:covariate}
\end{marginfigure}

\textbf{An Example of Confounding Bias.} Consider an intervention that simulates the software engineering task of \datainterI where a treatment $T$ is predicting tokens following either Buggy/Fixed code, $Y$ is the cross-entropy of each method of a dataset of both buggy and fixed code (which we introduce in \chapref{ch:case}), and $Z$ is the \textit{Number of Subwords} for each method. Using \docode we find there exists a spurious correlation after estimating the association distribution $p(Y|T)\approx0.67$ with \textit{Jensen-Shannon Distance} (see Def.~\ref{def:js}) and intervention distribution $p(Y|do(T))\approx-2E-4$ with \textit{Average Treatment Effect} (ATE) (see Def.~\ref{def:ate}). One possible explanation is that the common cause $Z$ (the number of subwords) is confounding the relationship between the treatment and the outcome. Fig.~\ref{fig:covariate} depicts the influence of $Z$ on the potential outcome $Y$ for BuggyCode ($p(Y|Z,T=Buggy)\approx0.87$) and FixedCode ($p(Y|Z,T=Fixed)\approx0.86$). Blue and orange points in the plot are code snippets from the dataset. These points are equally distributed, which suggests that the \datainterI intervention has a negligible impact on model effectiveness as measured by cross-entropy. 

\subsection{Software Engineering-Based Interventions}

To enable our analysis based upon Pearl's Ladder of Causation, we need to design \textit{interventions}, or changes in the input data distribution that represent a meaningful concept (\ie commented vs. uncommented code), to attribute cause from the changes in the model's performance in predicting code tokens in different settings. That is, if we observe that a model is less effective at predicting operators after changing ``test'' data sequences to which the model is applied, we may be able to confirm that the change \textit{caused} this drop in effectiveness if we properly control for confounders. We design \docode's interventions based on the fact that \nlms are often not applied to the same types of code corpora upon which they are trained. For instance, if a model trained on a well-commented dataset is applied to predict segments of poorly commented code, this could potentially impact performance. As such, we define parallel code corpora that contain programming language-specific changes across the datasets, and specifically introduce four different initial interventions depicted on the right side of \figref{fig:ladder}. 

{We define \textit{SE-based Interventions} to understand model performance across different settings better. We formulate these settings as parallel code corpora with differing specified semantic properties. For instance, a testbed aimed at simulating a debugging environment may consist of two parallel corpora: the buggy code, and the (corresponding) fixed code. Therefore, these datasets describe some high-level SE properties, which we employ in \docode's causal analysis. We define four types of SE application settings, adapted from both our prior work and community datasets: (i) buggy/non-buggy~\citep{Tufano2019LearningBugFixes}, (ii) commented/non-commented~\citep{lu2021codexglue}, (iii) type II, and (iv) type III clone pairs~\citep{Svajlenko2015EvaluatingBigCloneBench}, and Abstract Syntax Trees node masking \citep{daniel23}. \docode is extensible, meaning that researchers can define their own code corpora interventions. In addition, we have also defined \textit{Model Hyper-parameter Interventions} to extend the causal analysis beyond data or code corpora perturbations (see \chapref{ch:case}).}

\subsection{The Causal Interpretability Hypothesis}
\label{sec:hypo}

{The following hypothesis aims to answer the question of why causal interpretability is necessary for the field of Deep Learning for Software Engineering (DL4SE). We claim that the prediction generated by any \nlm can be causally explained by employing the notion of the Ladder of Causation. Our proposed method \docode uses the process of causal inference to enable practitioners to answer their particular causal queries.} 

\begin{boxK}
\textbf{Hypothesis:} \docode is a causal interpretability method that aims to make DL4SE systems (\ie Neural Code Models) and their decision-making process understandable for researchers and practitioners. 
\end{boxK}

%% file: chapters/part_00_chap_03/decomposition.tex
\chapter{Syntax (De)Composition}
\label{ch:decomposition}

Trustworthiness and interpretability are concepts that are inextricably linked for \llms. The more interpretable \llm is, the more trustworthy it becomes. 
However, current techniques for interpreting \llms when applied to code-related tasks focus mainly on accuracy measurements, measures of how models react to change, or individual task performance instead of fine-grained explanations needed at prediction time for greater interpretability and hence trust. 
To improve on this status quo, this chapter introduces \astrust, an interpretability method for \llms of code that generates explanations grounded in the relationship between model confidence and syntactic structures of programming languages.
\astrust explains generated code in the context of \textit{syntax categories} based on Abstract Syntax Trees and helps practitioners understand model predictions at \textit{both} local (individual code snippets) and global (larger code data sets) levels.

By distributing and assigning model confidence scores to well-known syntactic structures that exist within ASTs, our approach moves beyond prior techniques that perform token-level confidence mapping by offering a view of model confidence that directly aligns with programming language concepts with which developers are familiar. 

To put \astrust into practice, we developed an automated visualization that illustrates the aggregated model confidence scores superimposed on sequence, heat map, and graph-based visuals of syntactic structures from ASTs. 
We examine both the \textit{practical benefit} that \astrust can provide through a data science study on 12 popular LLMs in a curated set of GitHub repos and the \textit{usefulness} of \astrust through a human study. 
Our findings illustrate that there is a \textit{causal} connection between learning error and an LLM's ability to predict different syntax categories according to \astrust~-- illustrating that our approach can be used to interpret model \textit{effectiveness} in the context of its syntactic categories. Finally, users generally found \astrust's visualizations useful in understanding the trustworthiness of model predictions.

\input{chapters/part_00_chap_03/sec_01_intro}
\input{chapters/part_00_chap_03/sec_02_background}
\input{chapters/part_00_chap_03/sec_03_approach}
\input{chapters/part_00_chap_03/sec_04_evaluation}
\input{chapters/part_00_chap_03/sec_05_results}
\input{chapters/part_00_chap_03/sec_06_discussion}
\input{chapters/part_00_chap_03/sec_07_conclusions}

%% file: chapters/part_00_chap_03/sec_01_intro.tex
\section{Introduction}
\label{ch:decomposition:sec_01}

The proliferation of open-source software projects and the rapid scaling of transformer-based Large Language Models (\llms) has catalyzed research leading to the increased effectiveness of automated Software Engineering (SE) tools. \llms have demonstrated considerable proficiency in a diverse array of generative SE tasks~\citep{Chen2021, watson2020dl4se}, including, but not limited to, code completion ~\citep{Raychev2014CodeCW,MSR-Completion}, program repair ~\citep{Chen2019,ahmad_unified_2021}, and test case generation ~\citep{Watson:ICSE20}. Current research in both designing \llms for code and applying them to programming tasks typically makes use of existing \textit{benchmarks} (\eg CodeSearchNet~\citep{husain2019codesearchnet}, or HumanEval~\citep{Chen2021}) and \textit{{canonical metrics}} (by \textit{canonical}, we refer to metrics that reflect an \textit{aggregate performance} across many model predictions, for example, percentage accuracy). These canonical metrics have been adapted from the field of Natural Language Processing (NLP) to evaluate the performance of deep code generation models.

Recent work has illustrated the limitations of benchmarks such as HumanEval~\citep{liu2023code} and there has been growing criticism of canonical metrics within the NLP community due to the lack of an \textit{interpretable context} that allows a deeper understanding of LLMs' predictions or outputs~\citep{molnar2025,kim_interpretability_2018,wan_what_2022,liu2024reliability,doshivelez2017rigorousscienceinterpretablemachine}. Although code-specific metrics such as CodeBLEU~\citep{Ren2020codebleu} may provide more robust aggregate pictures of model accuracy, they cannot provide the fine-grained context required to truly explain model predictions. The general lack of widely adopted interpretability or explainability tools is a barrier to the adoption of any deep learning model, and in particular LLMs of code, as practitioners are skeptical of models' trustworthiness~\citep{lo_trustworthy_2023}. This deficiency stems largely from the fact that such benchmarks and canonical metrics are often aimed at evaluating the functional correctness or standard performance of generated code \textit{at a glance}. That is, the evaluation is reduced to a single aggregate metric in which relevant information related to individual predictions is obfuscated~\citep{burnell_rethink_2023}.

\begin{figure}[ht]
		\centering
		\includegraphics[width=0.95\textwidth]{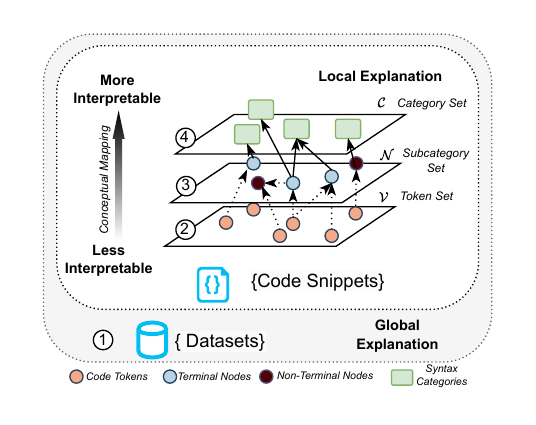}
		\caption{The Conceptual Framework of Syntax-Grounded Interpretability}
        \label{fig:overview}
\end{figure}

Methods for \textit{interpreting} and \textit{trusting} \llms for code are inextricably linked. A trustworthy \llm for code requires some degree of interpretability of its predictions, such that the behavior of the model can be understood at a fine-grained enough level to judge which parts of the output are correct or not and why. The more interpretable an LLM for code is, the higher the confidence and trust in the deployment and use of the model~\citep{Doshi-Velez2018ConsiderationsLearning, ji2024ai}. In particular, interpretability has been identified as an important component in improving trustworthiness in various studies ~\citep{lundberg2017unified,weller2019transparency,Liao_2020}. When evaluating trustworthiness, a clear understanding of how and why a model makes specific predictions is critical. This transparency not only addresses challenges related to uncertainty and the potential for bugs or vulnerabilities, but also plays a key role in the transformation of a model perceived as untrustworthy into one deemed reliable ~\citep{ribeiro2016why}. %

We claim that a \llm for code is \textit{interpretable}, and hence more trustworthy, if the reasoning behind its predictions is easy for a practitioner to understand. In other words, a useful interpretability technique must provide a \textit{conceptual mapping} between descriptions of a model's reasoning process and concepts inherently understood by programmers. In this chapter, we explore the possibility of using a model's \textit{confidence} in its predictions as a proxy for describing its reasoning process and develop a technique, which we call \textbf{\astrust} that automatically \textit{aligns} and \textit{clusters} model confidence measures with groups of tokens based on syntactic categories derived from Abstract Syntax Trees (ASTs) that we call \textit{Syntax Categories} (\ascs). This method enables a fine-grained understanding of the correctness of model predictions rooted in \textbf{ syntax-based explanations}. As illustrated by the overview of our approach in Fig.~\ref{fig:overview}, \astrust enables two different granularities of interpretability, \textit{local explanations} at the code snippet level and \textit{ global explanations} for large collections of code. 

\astrust also makes two main contributions: (i) a statistical technique for aligning and aggregating confidence scores to syntactic code structures of different granularities and (ii) an automated technique for generating visualizations of these aligned confidence scores. At the \textit{local} level these visualizations take the form of model confidence scores overlaid on both sequence- and graph-based illustrations of ASTs and different syntactic structures. At the global level, these take the form of a heat map with confidence values clustered around higher-level syntactic categories. An example of the type of explanation that a developer may derive from \astrust's visualizations is as follows, ``\textit{The model's prediction of the type of character} \texttt{\small character} \textit{parameter may be incorrect due to low confidence.}'' 

Grounding explanations of model confidence in code syntax provides an informative context for practitioners, allowing for interpretability. This is because the semantics of the code and the syntax are tightly coupled. That is, descriptions of code meaning, or semantics, are often \textit{grounded} in syntax. For instance, consider the following example of a developer describing program behavior in \texttt{\small numpy} in which the description of functionality is grounded in terms of data structures, \textit{``Convert an {array} representing the coefficients of a Legendre series,''}\footnote{\url{https://github.com/numpy/numpy/blob/main/numpy/polynomial/legendre.py\#L152C5-L152C72}} where the underlined word refers explicitly to the syntactic category of a data structure. One may ask \textit{``why not ground explanations in code semantics directly?''} However, such semantic-based grounding is difficult to achieve, as it requires reasoning among model confidence, input code, predicted code, and widely variable interpretations of code meaning -- leading to the potential for \textit{incorrect} explanations that would undermine a technique ultimately meant to build trust. However, as we illustrate in this chapter, it is possible to directly map the confidence measures of the model to different syntactic categories of code, providing a \textit{statistically sound} method of understanding the potential correctness of model predictions rooted in concepts that developers can easily understand.

We explore the \textit{practical benefit} of \astrust through a large-scale data science study that examines the relationship between model effectiveness and global explanations and evaluate the \textit{usefulness} of our method through a human study that focuses on local explanations of code snippets using different visualizations from \astrust. The context of our empirical evaluation includes 12 popular \llms for code and a curated set of code taken from recent commits of the 200 most popular Python projects on GitHub. Using a carefully crafted causal inference study, our analysis illustrates \textit{causal} connections between learning error and a model's ability to predict different syntax categories according to \astrust~-- showing that our approach can be used to interpret model \textit{effectiveness} in the context of its syntactic categories. Our human study included 27 participants who examined code snippets completed by GPT 3 and one of four of \astrust's visualization techniques for local explanations. Our results illustrate that developers generally found \astrust and its visualizations useful in understanding model predictions.

The results of our studies illustrate that mapping token-level predictions of \llms to segregated Syntax Categories is of considerable practical benefit to SE researchers and practitioners because it allows them to \textbf{interpret} and \textbf{trust} parts of generated code based on structural functionality, which contextualizes model predictions beyond canonical evaluation (\ie measuring intrinsic and extrinsic metrics). We hope that other researchers will build on our method to create new types of interpretability techniques in the future, and we provide an online appendix with the code for \astrust, and our data and experimental infrastructure to facilitate replication~\citep{RepoASTrust24}.

%% file: chapters/part_00_chap_03/sec_02_background.tex
\section{Background \& Related Work}
\label{ch:decomposition:sec_02}

In this section, we present the background on interpretability and trustworthiness as complementary terms to generate post hoc (\eg generated after training) syntax-based explanations for \llms of code. 

\textbf{Interpretability.} The brittleness of LLMs can be formulated as an \textit{incompleteness} in problem formalization ~\citep{doshivelez2017rigorousscienceinterpretablemachine}, which means that it is insufficient that models infer predictions only for certain tasks (the \textbf{what?}). The models must also explain how they arrive at such predictions (the \textbf{why?}). To mitigate such incompleteness in the formalization of problems, the field of \textit{interpretability} has grown to encompass techniques and methods that aim to solve the \textit{why} question. Although authors in this field generally use the terms \textit{explainability} and \textit{interpretability} interchangeably, these definitions are inconsistent in the literature~\citep{flora_comparing_2022}. We distinguish between the terms to avoid confusion with the purposes of our approach. We will use \textit{explainability} for methods whose goal is to understand how a \llm operates and makes a decision by exploring inner mechanisms or layers. In contrast, we will use \textit{interpretability} for methods that define \textit{conceptual mapping mechanisms} whose goal is to contextualize the predictions of models by associating them with an understandable concept, which in this chapter is the syntax of programming languages. 

\textbf{Related Work on Interpretability in NLP.} There are existing techniques in both natural language processing (NLP) and SE literature focused on interpretability, including LIME~\citep{ribeiro2016why}, Kernel SHAP~\citep{lundberg2017unified}, Integrated Gradient~\citep{sundararajan2017axiomatic} and Contextual Decomposition~\citep{murdoch2018beyond}. These techniques generally try to approximate an interpretable model that attempts to attribute meaning to hidden representations of neural networks or illustrate the relationship between input features and model performance. However, we argue that such techniques are difficult to make practical in the context of LLMs for code, given the lack of conceptual mappings explained earlier. However, the interpretability technique most closely related to \astrust, and one of the only ones to have adapted to LLMs of code, is that of \textit{ searching} which is a supervised analysis to determine which type of parameters (\eg input code snippets, tokenization process, number of hidden layers, and model size) influence the learning process in ML models \citep{troshin_probing_2022}. The purpose of probing is to assess whether hidden representations of \llms encode specific linguistic properties such as the syntactic structures of programming languages. Given our generated visualizations, there may be an inclination to characterize \astrust as a \textit{probing technique}.
However, it is essential to note that \astrust focuses on estimating the \textit{correctness} of predicted syntactic code elements rather than mapping meaning to internal model representations of data. 

\textbf{Related Work on Interpretability in SE.} In the realm of SE research, previous work has taken two main directions: (i) techniques for task-specific explanations~\citep{fu2023vul-explain,liu2022explainable,pornprasit2021explain}, and (ii) empirical interpretability studies using existing NLP techniques~\citep{liu2024reliability,tantithamthavorn2023explain,mohammadkhani2023explain}. Previous authors have proposed techniques for explaining specific tasks, including vulnerability explanation~\citep{fu2023vul-explain}, vulnerability prediction for Android~\citep{liu2022explainable}, and defect prediction models~\citep{pornprasit2021explain}. More recently, Liu \etal conducted a large empirical study using existing explainability techniques for global explanations of code to better understand generative language models of code~\citep{liu2024reliability}. Mohammadkhani \etal conducted a study using the LLM's attention mechanism to interpret their performance in code generation. Finally, a paper that proposed a code-specific interpretability technique is that of Cito \etal \citep{cito_counterfactual_2022} who formulated a method to generate explanations using counterfactual reasoning of models. Our work on \astrust complements this body of previous work by developing a \textit{new, generally applicable interpretability method} that can be applied to \textit{both} local and global code explanations, which no prior study or technique has done.

\textbf{Trustworthiness.} This research is inspired by definitions of \textit{trust} from automated systems, SE, and NLP. In automated systems, trust is defined as \textit{``the attitude that an agent will help achieve an individual's goal in a situation characterized by uncertainty and vulnerability''}~\citep{Lee_See_2004}. Bianco \etal define software trust as the degree of confidence when the software meets certain requirements~\citep{softtrust}. In NLP, Sun \etal argue that \llms must appropriately reflect truthfulness, safety, fairness, robustness, privacy, machine ethics, transparency, and accountability for them to be trustworthy~\citep{sun2024trustllmtrustworthinesslargelanguage}. We define trust as the confidence that practitioners and researchers have in \llms' code prediction, anticipating that these predictions will effectively align with their intended goals. Trustworthiness in \llms implies a sense of interpretability in a given \llm's performance, instilling confidence among practitioners in their abilities to perform code-related tasks. To the best of our knowledge, no paper proposes a concrete definition of trust based on interpretability within the SE research community. Yet, several researchers have called for the importance of trustworthiness in \llms for code~\citep{lo_trustworthy_2023,spiess2024quality}. In our work we present a concrete definition of trustworthiness, highlight its importance, and show how syntax-grounded explanations such as \astrust contribute to more trustworthy \llms.

%% file: chapters/part_00_chap_03/sec_03_approach.tex
\section{Syntax-Grounded Explanations}
\label{ch:decomposition:sec_03}

At a high level, \astrust queries a \llm for probabilities per token, estimates the median across tokens that are part of an AST node, and presents those averages as \textbf{confidence performance} values segregated by hand-assigned syntax categories. We also refer to this confidence performance as \textit{\astrust Interpretability Performance}.

\astrust consists of four steps as depicted in \figref{fig:overview}. In step \circled{1}, a code snippet for local or a testbed for global explanations is the starting point of the \textit{interpretability process}. Each sequence within the snippet or testbed is processed by a tokenizer (\eg Byte-Pair Encoding (BPE)). In step \circled{2}, the tokenizer sets a vocabulary that we named \textbf{ token set}. Once code sequences are preprocessed, an \llm under analysis generates \textbf{token-level predictions} (\ntp) for each position in a sequence. Next, in step \circled{3}, the token-level predictions generated are aligned with the associated Abstract Syntax Tree (AST) terminal nodes. Terminal nodes only store \ntp, while nonterminal nodes hierarchically store clustered and aggregated \ntp. The terminal and non-terminal nodes comprise the set of subcategories \textbf{}. For example, consider \fbox{\texttt{\small if\_}} BPE token from the \textit{token set}. This token is aligned with the\texttt{\small `if} terminal AST node while clustered in the \texttt{\small `if\_statement'} non-terminal node. Finally, in step \circled{4}, ten \textit{syntax categories} are proposed to summarize the predictions of a model. \textit{Syntax Categories} aim to group the subcategories into higher-level, more human-understandable categories. These syntax categories are a fixed \textbf{category set} that comprises more interpretable elements and include:

\begin{multicols}{3}
    \begin{itemize}
        \item {\small Decisions}
        \item {\small Data Structures}
        \item {\small Exceptions}
        \item {\small Iterations}
        \item {\small Functional Programming}
        \item {\small Operators}
        \item {\small Testing}
        \item {\small Scope}
        \item {\small Data Types}
        \item {\small Natural \\Language}
    \end{itemize}
\end{multicols}

For example, the subcategories \texttt{\small `if\_statement'} and  \texttt{\small `if'} are both clustered into one syntax category \textit{Decisions}. In the end, \astrust generates an averaged score per category for global explanations and an AST tree visualization with stored scores at each node for local explanations. In essence, we propose that the \textit{ syntax elements} contain semantic information that contextualizes the predicted probabilities. However, this semantic information varies between the granularity of these elements. We can claim, for example, that token-level elements carry less interpretable information than category-level elements.

\astrust produces \textit{post-hoc} local and global explanations of generated code snippets. A local explanation aims to interpret the generation of a code snippet by decomposing it into AST elements. In contrast, a global explanation uses a set of generated snippets (or an existing benchmark data set) to interpret a given model holistically into \textit{Syntax Categories} (\ascs). The following subsections introduce the building blocks of syntax-grounded explanations. \secref{sec:composable} defines the interpretable sets (\eg \textit{Token}, \textit{Subcategory}, and \textit{Category}) that contain the \textit{syntax elements} employed for the interpretability process. \secref{sec:formalism} formalizes two function interactions that communicate previously interpretable sets. This type of communication consists of aligning and clustering elements from code tokens to syntax categories. Finally, \secref{sec:explainations} shows the process of generating local and global explanations.

\subsection{Interpretable Syntax Sets}\label{sec:composable}
\textbf{Token Set $\mathcal{V}$.} Although \astrust was designed to be compatible with different types of \llms, this chapter concentrated on \textit{Decoder-Only} models due to their autoregressive capacity to generate code~\citep{xu_systematic_2022} by preserving \textit{long-range dependencies}~\citep{karpathy2015understand}. A decoder-only model can be used as a generative process, such as any token $w_i$ predicted by $\hat{w_i} \backsim P(w_i | w_{<i} ) = \sigma(y)_i = e^{y_{w_i}} / \Sigma_j e^{y_j}$. The term $y_j$ represents the \textit{non-normalized log-probabilities} for each output token $j$ (see \figref{fig:local_explaination}). We extracted and normalized these log-probabilities from the last layer of \llms to estimate \textit{Token-Level Predictions (\ntp)}. This estimation relies on the softmax function. The softmax $\sigma_i$ returns a distribution over predicted output classes, in this case, the classes are each token in the \textit{token set} $\mathcal{V}$. The predictions $\sigma_i$ are expected to be influenced by previous sequence inputs $w_{<i}$.

\textbf{Subcategory Set $\mathcal{N}$.} This set comprises elements of Context-Free Grammars (CFGs). Such elements are rules that contain the syntax and structural information of a programming language~\citep{10.5555/1196416}. CFGs define instructions that specify how different tokens (\ie Lexemes) are assembled to form valid statements for each language. Formally, a $CFG$ is defined as $\mathbb{G} = (\alpha, \lambda, \omega, \beta)$ where $\alpha$ denotes the finite set of non-terminal nodes, $\lambda$ the finite set of terminal nodes, $\omega$ the finite set of production rules and $\beta$ the start symbol. CFGs use the terminal and non-terminal nodes (or subcategories) to define the production rules $\omega$ for any statement (\eg conditional, assignation, operator). Furthermore, these terminal and non-terminal nodes retain different meanings. Note that these nodes are the elements of the subcategory set $\lambda,\alpha \in \mathcal{N}$.

\textbf{Category Set $\mathcal{C}$.} The steps three and four in \figref{fig:overview} illustrate the binding of $\alpha$ and $\lambda$ into a category $c \in \mathcal{C}$. We pose the term \textbf{Syntax Categories} (\ascs) as the elements within the Category Set $\mathcal{C}$. We propose ten different \ascs based on tree-sitter bindings~\citep{tree-sitter} for Python. \ascs are the semantic units to enable the syntax interpretability of \llms. As such, \astrust allows token-level predictions (\ntp) to be explained in a developer-centric way. In summary, each token in a sequence $s$ can be assigned to a category $c \in \mathcal{C}$. With \astrust, practitioners can easily associate the predictions of the code of \llms with specific structural attributes. For example, the nodes of \texttt{\small `identifier} and \texttt{\small `string'} nodes correspond to a common \textit{Natural Language} category in \figref{fig:global_explanation}. As such, we can group the nodes $\lambda$ and $\alpha$ into semantically meaningful \textit{categories} $\mathcal{C}$.

\subsection{Alignment and Clustering Formalism}\label{sec:formalism}

\begin{figure*}[ht]
    \includegraphics[width=\textwidth]{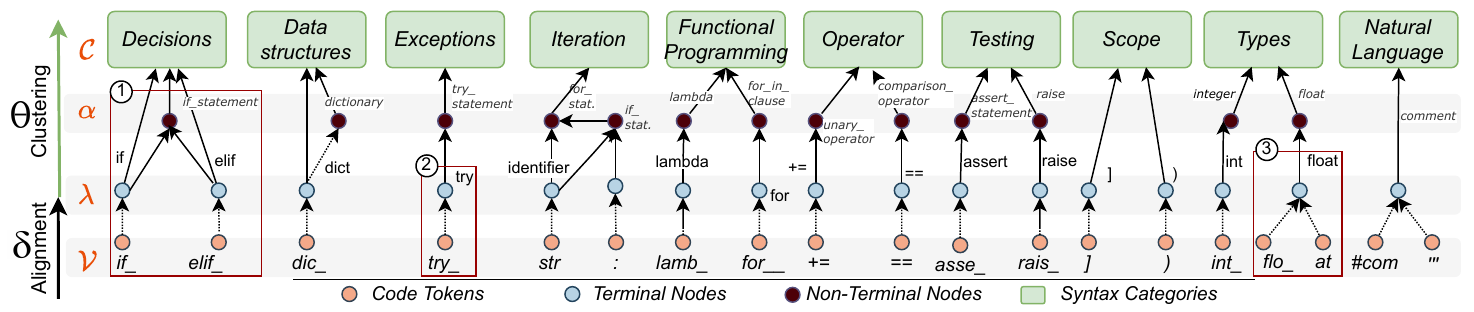}
		\caption{{\textit{Alignment \& Clustering Interactions}. The $\delta$ function aligns tokens $w_i$ to terminal nodes $\lambda$. Terminal and Non-terminal nodes $\lambda$, $\alpha$ $\in \mathcal{N}$ are clustered by Syntax Categories $\mathcal{C}$.}}
        \label{fig:runner_example}
\end{figure*}

The previous subsection describes the syntax elements for enabling \llms interpretability (\ie \textit{token-set}, $\alpha$ and $\lambda$ subcategories, and \textit{Syntax Categories} (\ascs)). This section elaborates on the interaction among these elements. Two interactions in the form of a function are defined. { First, the alignment function $\delta$ links code tokens  from the \textit{Token Set}  $\mathcal{V}$ to terminal nodes $\lambda$. Second, the clustering function $\theta$ groups the subcategories $\lambda$ (terminal nodes) and $\alpha$ (non-terminal nodes) by syntax categories (\ascs) from the \textit{Category Set}. \figref{fig:runner_example} showcases both function interactions $\delta$ and $\theta$ respectively.}  

{\textbf{Alignment Interaction.} \figref{fig:runner_example} illustrates the process of aligning the terminal nodes $\delta$ in the AST to their corresponding code tokens $w_i$. This alignment starts by decomposing an input snippet $s$ into tokens $w_{<=i}\in \mathcal{V}$. For instance, \figref{fig:runner_example}-\circled{2}  depicts the alignment of  \fbox{\texttt{\small try\_}} token to the terminal $\lambda$ \texttt{\small `try'} node. Note that the alignment ignores the character "\_" from \fbox{\texttt{\small try\_}}. A tokenizer may produce a sequence in which each token does not necessarily match one-to-one with a terminal $\lambda$ node, \eg \figref{fig:runner_example}-\circled{3} illustrates the tokens \fbox{\texttt{\small flo\_}} and \fbox{\texttt{\small at}} are aligned with the $\lambda$ node \texttt{\small `float'}. Formally, $\delta(flo\_,at) \to [float]$ in a many-to-one interaction. Consequently, the alignment between code tokens and terminal nodes is certainly many-to-one, including one-to-one, but never one-to-many or many-to-many. 
}

\begin{definition}
\label{def:delta}
\textit{Alignment ($\delta$).} The function $\delta: w_{<=i} \to \vec{\lambda}$ where $w_{<=i}$ corresponds to a code sub-sequence whose tokens are many-to-one associated to the corresponding terminal node vector $\vec{\lambda}$ of syntax subcategories.

\end{definition}
{
\textbf{Clustering Interaction.} A clustering function $\theta$ estimates the \textbf{confidence performance} of $\lambda$ and $\alpha$ nodes (subcategories) from an AST by hierarchically aggregating the Token-Level Predictions (\ntp) to a \textit{Category} $c\in\mathcal{C}$. Once the tokens are aligned with their corresponding nodes using $\delta$ from Def.\ref{def:delta}, \astrust clusters them into their respective category or non-terminal $\alpha$ node according to the AST representation. Some terminal $\lambda$ nodes can directly be aggregated into a category without considering intermediate non-terminal $\alpha$ nodes. A terminal $\lambda$ node can initiate a block sentence (\ie a category) and a block sequence parameters (\ie non-terminal if\_statement node). For instance, \figref{fig:runner_example}-\circled{1} depicts the terminal $\lambda$ \texttt{\small `if'} node aggregated into the \textit{Decisions} category and also starts the non-terminal $\alpha$ \texttt{\small `if\_statement'} node. To estimate the confidence performance, we traverse the entire AST and aggregate the \ntp probabilities of respective tokens. 

The $\theta$ function can adopt average, median, or max aggregations depending on the user configuration. \figref{fig:local_explaination} shows the clustering function applied to a concrete code generation sample. This application constitutes a local post hoc explanation: the parent node \texttt{\small `parameters'} has a $0.23$ associated confidence performance. This parent node average was aggregated with its terminal values: \texttt{\small `('} with $0.07$, \texttt{\small `identifier'} with $0.4$ and $0.1$, \texttt{\small `,'} with $0.5$, and \texttt{\small `)'} with $0.1$. Formally, $\theta( \vec{\lambda} = [0.07, 0.4, 0.1, 0.5, 0.1] ) \to [(parameters, 0.23)]$. If a sample snippet does not contain any particular syntax element (\ie token, subcategory, or category), such an element \textit{is therefore never considered for clustering}. An absent syntax element is reported as a \texttt{null} value to avoid biased syntax-grounded explanations. 
}

\begin{definition}
\label{def:theta}
\textit{Clustering ($\theta$).} The function $\theta: \vec{\lambda} \to median(\vec{n})$ where $\vec{\lambda}$ is the resulting vector of a sub-sequence and $\vec{n}$ is the vector of hierarchical associated non-terminal nodes for each terminal $\lambda$. The vector $\vec{n}$, therefore, contains the \ntp of non-terminal and the corresponding terminal nodes\footnote{In our study, we set the aggregation $\theta: N \to median(\hat{w}_{<=i})$ for a subset of tokens $w_{<=i}$.}. %
\end{definition}

\subsection{Post Hoc Local and Global Explanations}\label{sec:explainations}

\llms are more understandable when they \textit{reflect human knowledge}~\citep{kim_interpretability_2018}. One way of determining whether an \llm trained on code reflects human knowledge is testing it to see whether or not it operates \textit{similar to how a developer would estimate the prediction of a sequence}~\citep{palacio_toward_2023}. \astrust can adopt the form of a \textit{post-hoc} local or a global explanation to make code predictions humanly understandable.  

\begin{figure*}[ht]
		\centering
		\includegraphics[width=\textwidth]{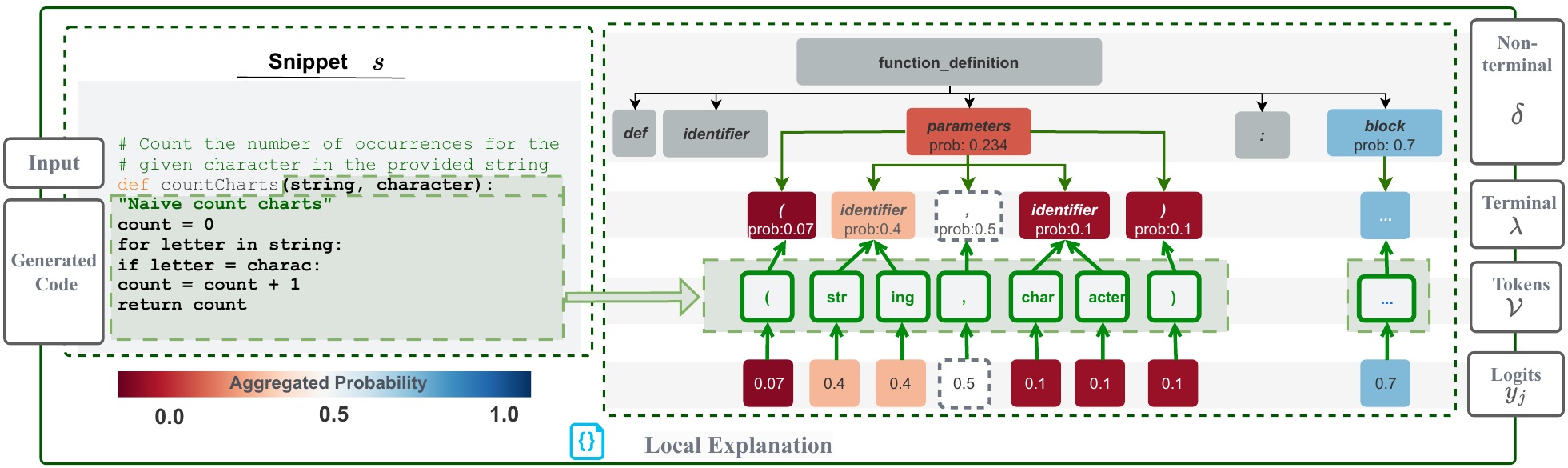}
		\caption{\textit{Post Hoc Local Explanation}.  A snippet is decomposed into code tokens. The highest annotated probabilities (\ie best predictions) are in blue.}
        \label{fig:local_explaination}
\end{figure*}

\textbf{\astrust for local interpretability} allows us to interpret a single snippet $s$ by generating a visual explanation based on an Abstract Syntax Tree (AST) as illustrated in \figref{fig:local_explaination}. A practitioner can explain the code predictions observing the probabilities associated with each element on the AST. In other words, we use $\theta$ from Def.~\ref{def:theta} to cluster around AST nodes across all levels (\ie AST probability annotations). Therefore, the syntax-grounded local explanation comprises a \textit{conceptual mapping} from the code prediction to a terminal and non-terminal node (or sub-categories). Fig.~\ref{fig:local_explaination} is a visual representation of the conceptual mapping using code predictions by \gptII model. The visualization displays a confidence value for each $\lambda$ and $\delta$ sub-categories after parsing the AST. The auto-completed snippet is processed with the clustering $\theta$ function.  

\begin{figure*}[ht]
\centering
\includegraphics[width=0.95\textwidth]{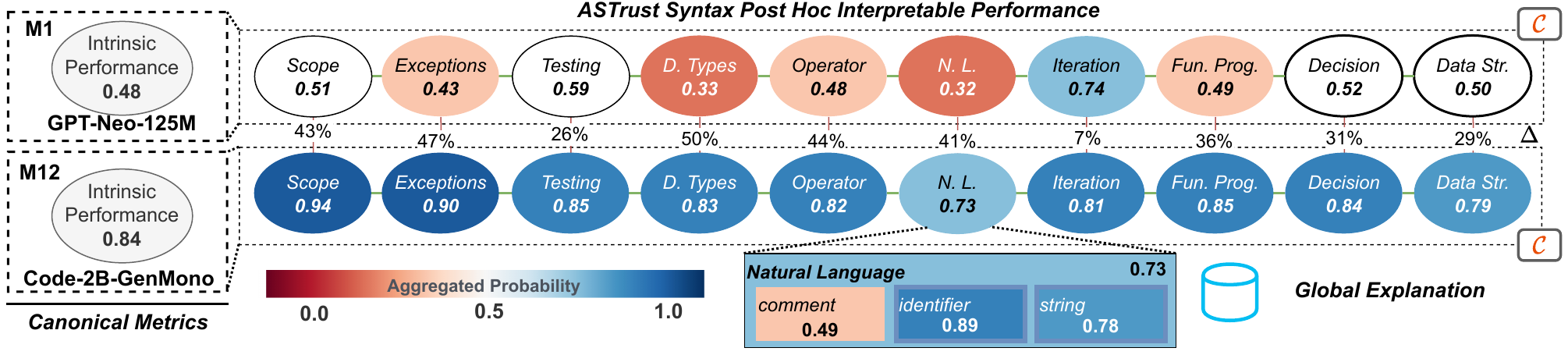}
\caption{\textit{Post Hoc Global Explanations} Segregated by Categories and Sub-Categories for \gptI and \monoIIII} 
\label{fig:global_explanation}
\end{figure*}

{ 
\textbf{\astrust for global interpretability} allows us to interpret a \llm by decomposing the canonical performance into segregated confidence performance. This segregated confidence is attached to Syntax Categories (\ascs). \ascs are tied to AST tree-sitter nodes~\citep{tree-sitter} and inspired by common object-oriented programming definitions. Although our approach is focused on the Python syntax, these categories may apply to multiple Programming Languages as they are being processed for similar AST elements.

Syntax-grounded global explanations comprise a \textit{conceptual mapping} from code predictions to ten syntax categories. These predictions are calculated for the entire testbed rather than a particular snippet following Def.~\ref{def:theta} about clustering function $\theta$. This clustering has an extra step in which a bootstrapping mechanism is used to estimate the \textit{confidence performance mean} of the category set.

Consequently, practitioners can compare Syntax Categories among models and explain the overall behavior of a given \llm by observing the segregated confidence performance. Note that previous analysis can be enriched with the information provided by canonical metrics. For example, \figref{fig:global_explanation} shows $M1$ with an intrinsic performance of $0.48$ while showing a \astrust interpretability performance of $0.74$ for the \textit{iteration} category. Details of this global categorization can be further explored and visualized in our online appendix~\citep{RepoASTrust24}}.

{It should be noted that the validity of global and local syntax-grounded explanations is dependent upon the \textit{calibration} of LLMs for code in terms of token probabilities and prediction correctness. That is, we assume that token probabilities are a reasonable proxy for the likelihood that a model is correct about a given token prediction. Recent work on calibration for LLMs of code has illustrated that, for code completion (which subsumes the experimental settings in this chapter), LLMs tend to be well calibrated for token probabilities~\citep{spiess2024quality}. We also further confirm this finding using causal inference in Section~\ref{sec:global_validity}.}

%% file: chapters/part_00_chap_03/sec_04_evaluation.tex
\section{Empirical Study Design}
\label{ch:decomposition:sec_04}

{We study the applicability of \astrust in interpreting code completion tasks. We conducted a \textit{human study} to investigate the \textit{usefulness} of local explanations in real-world settings. In contrast, we conducted a \textit{data science study} to showcase the \textit{effectiveness} of global explanations on a diverse set of LLMs. Finally, we carried out a \textit{causal inference study} to assess the \textit{validity} of the syntax-grounded explanations as they relate to the statistical learning error of the studied models. The following research questions were formulated:}

\begin{enumerate}[label= \textbf{RQ$_{\arabic*}$}, ref=RQ$_{\arabic*}$, wide,labelindent=5pt]\setlength{\itemsep}{0.2em}
    \item \label{rq:survey_usefulness} \textbf{[Usefulness]} \textit{How useful are local explanations in real-world settings?} We validate the extent to which AST probability annotations are useful in locally explaining code predictions. We measure usefulness in three key factors: complexity, readability, and \llms' reliability.
    \item \label{rq:performance} \textbf{[Effectiveness]} \textit{To what extent do \llms for code correctly predict different syntactic structures?} We interpret the performance of 12 \llms on each Syntax Category (\asc). The conceptual mapping allows us to obtain an interpretable and segregated confidence value per category, so we can detect categories that are easier or harder to predict -- moving beyond canonical aggregate metrics. 
    \item \label{rq:validity} \textbf{[Validity]} \textit{How do Syntax Concepts impact LLMs' statistical learning error?} We validate the causal connection between learning error and \llms' ability to predict different syntax categories using \astrust.
\end{enumerate}

\subsection{Experimental Context}

\textbf{Model Collection.} To perform our global analysis, we conducted an interpretability analysis of 12 open Decoder-only \llms, selected based on their popularity. The largest among these models boasts 2.7 billion parameters. \tabref{tab:models} categorizes these \llms into four distinct groups, each aligned with a specific fine-tuning strategy. The initial category comprises GPT-3-based models primarily trained on natural language, exemplified by Pile \citep{gao2020pile}. The second category encompasses models trained on natural language but constructed upon the \textit{codegen} architecture \citep{Nijkamp2022CodeGenAO}. Moving to the third category, we find models trained on multiple programming languages (PLs) using BigQuery \citep{GoogleBigQuery}, implemented on both the gpt-2 and codegen architectures. The final category consists of both \multilang models fine-tuned on BigPython \citep{Nijkamp2022CodeGenAO}, denoted as \monolang, and gpt-2 models such as codeparrot \citep{codeparrot}. All the datasets for training the \llms encompass repositories/files sourced from GitHub up to 2021.

\textbf{Evaluation Dataset} To ensure the integrity of our \astrust evaluation, it is imperative to avoid data contamination by excluding samples used in the training process of the \llms. We extended and used \galeras \citep{daniel23} to overcome this challenge. \galeras exclusively comprises code commits from the top 200 most popular Python GitHub repositories between 01/01/22 and 01/01/23. Notably, \galeras incorporates comprehensive data, including commit messages, method comments, the entire AST structure, node count, AST levels, AST errors, whitespace details, lines of code, cyclomatic complexity, and token counts.

\textbf{Machine Configuration.} We performed the experiments using 20.04 Ubuntu with an AMD EPYC 7532 32-Core CPU, A100 NVIDIA GPU with 40GB VRAM, and 1TB RAM. For the model inference process, we used HugginFace and Pytorch \citep{wolf2020transformers, pytorch}. All models were loaded into the GPU to boost the inference time.

\input{tables/chap_decomposition/tab1_design_1_models2}

\subsection{Human Study for \astrust Usefulness}
\label{sec:local_study_design}

{
This section presents a preliminary human study comprising a control/treatment experimental design to assess \astrust usefulness in practical settings. We followed a \textbf{purposive sampling approach} \citep{baltes_sampling_2021} since our primary goal was to gather preliminary data and insights from practitioners with expertise in ML and SE combined. We selected our subjects carefully rather than randomly to study the usefulness of \astrust (at local explanations). \astrust is designed to enhance the interpretation of model decisions in code completion tasks for practitioners with diverse backgrounds, including researchers, students, and data scientists. By targeting individuals with specific expertise, we ensured that the feedback received was relevant and informed, thereby enhancing the quality of our preliminary findings.}

\begin{figure}[ht]
		\centering
		\includegraphics[width=0.9\textwidth]{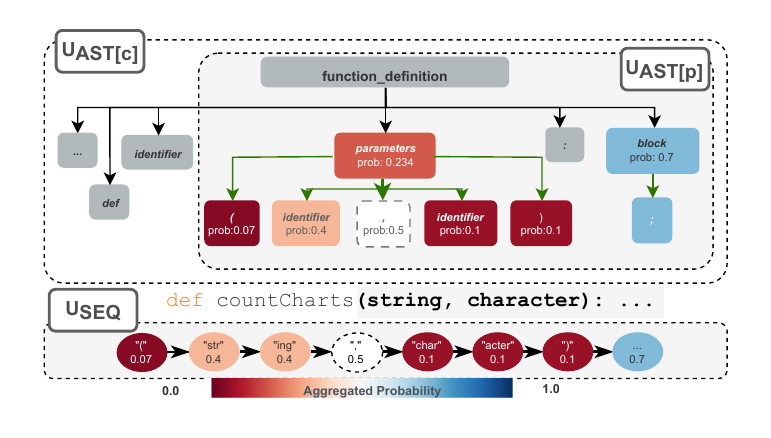}
		\caption{\astrust Local Explanation Treatments.}
        \label{fig:sec_8_survey_explanations}
\end{figure}

\textbf{Survey Structure} {
Each survey consists of three sections. The \textit{first section} is aimed at gathering participant profiling information. The profiling section aims to collect information related to how proficient the participants are when using Python and AI-assisted tools in code generation tasks. In particular, we asked about their level of expertise in Programming Languages (PL) and how familiar they are with AST structure. Furthermore, we asked about any challenges they encountered when using AI-assisted tools. This information is relevant because we want to control external factors that may influence their perception in validating \astrust. The \textit{second section} aimed to present four code completion scenarios and ask participants to rate their quality. For each scenario, we presented an incomplete Python method as a prompt, the generated code completing the previous method (using \gptII in \tabref{tab:models}), and a specific type of local explanation (\eg \surveySequence, \surveyAstPartial, and \surveyAstComplete in \figref{fig:sec_8_survey_explanations}). {In all four scenarios, the prompt contained either a \textit{semantic error} (\eg using an undefined variable, incorrect condition statement(s), calling an undefined function) or a 
\textit{syntax error} (\eg missing colon, missing parenthesis, incorrect indentation) that participants needed to reason about after considering a given local explanation type. We aimed to capture the participant's perspective regarding the explanation by asking them to describe the cause of a syntax or semantic error from the generated code. To facilitate the analysis, we highlighted the portion of the code generated by the model (see green highlight \figref{fig:local_explaination}). We did not provide details on the \llm used to generate the predictions to avoid introducing bias related to preconceived notions about particular LLMs in the responses. Lastly, the \textit{third section} aim to collect feedback on the effectiveness of the different types of local explanation. Specifically, we asked participants about the complexity of visualizations and potential opportunities for enhancement.}

\textbf{Survey Treatments}

To collect the perception of practitioners about the usability of \astrust, we devised a control survey (\surveyControl) and three treatments with two types of local explanations: sequential explanations (\surveySequence) and AST-based (\surveyAstBased) explanations. \surveyControl represents \textit{the absence of a local explanation} and only collects the participants' perceptions regarding the \textit{correctness} of \llms' output. By contrast, treatment surveys \surveySequence and \surveyAstBased yield syntax-grounded explanations. It is worth noting that {we define \textit{correctness} as the degree to which the generated code reflects the purpose of the algorithm contained in the prompt. In other words, we ask participants to judge whether the model predicted a valid or closely accurate set of tokens given the information context within the prompt.}

{ \figref{fig:sec_8_survey_explanations} depicts all the explanation types considered in the survey. \surveySequence displays the tokens and their corresponding probabilities in a sequential (\ie linear layout). Linear representations are commonly used by feature-importance explainability techniques such as attention-based \citep{mohammadkhani2023explain} and Shapley \citep{Kalai1983OnWS} values. Therefore, \surveySequence serves as a \textit{baseline} to determine how our local explanations \surveyAstBased provide insightful information beyond the linear layout. In contrast, \surveyAstBased uses an AST visualization, comprising two types of local explanations: AST-Complete (\ie \surveyAstComplete) and AST-partial (\ie \surveyAstPartial). \surveyAstComplete represents the entire sample's AST (\ie prompt and generated code) including \astrust confidence performance for all nodes. In contrast, \surveyAstPartial is a filtered \surveyAstComplete representation that only exposes the confidence performance of the generated code and omits the nodes from the prompt.}

\textbf{Survey Metrics} When evaluating the usefulness of our approach to answer \ref{rq:survey_usefulness}, we measure the qualitative features of local explanations depicted in \figref{fig:sec_8_survey_explanations}. More precisely, we propose five qualitative metrics to evaluate the usefulness of our approach: \textit{Information Usefulness}, \textit{Local Explanation Complexity}, \textit{Local Explanation Readability}, \textit{Visualization Usefulness}, and \textit{LLM’s reliability}.} We used a Likert scale with three options to quantitatively measure responses. Specifically for Information Usefulness: \textit{Agree}, \textit{Neutral} and \textit{Disagree}. For Local Explanation Complexity, Local Explanation Readability and Visualization Usefulness: \textit{Useful}, \textit{Slightly useful} and \textit{Not useful}. Finally, for LLM's Reliability: \textit{Not reliable}, \textit{Highly reliable}, and \textit{Impossible to tell}. Each of the survey metrics corresponds to one of the following survey questions.

\textit{{Metric$_1$: Information Usefulness} - `Q: How useful was the information for interpreting the model's decisions?'} In the treatment surveys, we ask the participants to explain the \llm's behavior when completing the code of individual samples, and we gauge their perception regarding the usefulness of the provided information to accomplish this task. We anticipate correlations between the explanation types and perceived usefulness.

\textit{{Metric$_2$: Local Explanation Complexity} - `Q: I found the visualization unnecessarily complex'}. The local explanation complexity refers to the degree of intricacy of its types. The degree of complexity may affect perceptions of usefulness.

\textit{{Metric$_3$: Local Explanation Readability} - `Q: I thought the visualization was easy to read and use'}. We define readability as the degree to which our local explanations are intuitive and easy to understand. We hypothesize that if the explanation fits this criterion, we can consider it useful. Readability accounts for factors such as the amount of information consigned, the arrangement of tokens and categories, and the color scheme.

\textit{{Metric$_4$: Visualization Usefulness} - `Q: I thought the visualization was useful for explaining the model's behavior'}. The visualization is the graphical representation of the local explanation (refer to \figref{fig:local_explaination}). Each treatment survey uses a type of visualization (\ie \surveySequence or \surveyAstBased). We formulate this question to determine which visualization is considered more useful.

\textit{{Metric$_5$: \llm's Reliability} - `Q: What is your perception of the model's reliability in generating code?'}. We define reliability as the degree to which a user trusts the \llm's output based on the outcomes from local explanations. We ask the participants to reflect on the \llm's reliability across our surveys using local explanations. Considering all four code completion scenarios in the surveys include errors, the greater the number of participants in each survey who would not rely on the model, the more valuable the syntax-grounded local explanation.

\textbf{Open Questions} 

{In addition to survey metrics, we formulated several open-ended questions for collecting the participants' perception about the correctness of the predictions (Open$_1$) and the most helpful parts of the visual explanations including potential improvement aspects (Open$_2$). Each of these open metrics corresponds to one or more survey questions. }

{\textit{{Open$_1$: \llm's Prediction Correctness} - `Q: If the generated code is incorrect, can you explain why the model might have made the mistake? Otherwise, If the generated code is correct, can you speculate on why the model may have been able to correctly predict the above snippet?'}. We asked the participants to use the provided information per sample to analyze whether the prompt or the generated code contained any syntax or semantic error. In \surveyControl, we aimed to assess the extent to which participants could reason about the source code correctness without any type of explanation provided. Conversely, in \surveySequence and \surveyAstBased, we inspected if the layout information somehow contributed to detecting and reasoning about the cause of the error.}

{
\textit{{Open$_2$: Importance of visual explanations} - `$Q_1:$ What information from the visualization did you find useful in explaining the model's predictions?', `$Q_2:$ What information from the visualization did you find useful in explaining the model's predictions?', `$Q_3:$ What other information (if any) would you like to see in the visualization?', `$Q_4:$ What elements of the visualization did you like most?', `$Q_5:$ What elements of the visualization did you like least?'}. We asked the participants to provide overall feedback about the type of representation used in the treatment surveys (\surveySequence and \surveyAstBased). We aimed to identify the most and least useful elements, as well as gather potential ideas for improvement.}

{To collect, standardize, and analyze the previous group of open-ended questions, two authors independently gathered and reviewed each survey's responses. Any differences were resolved through discussion to reach a consensus.}

\textbf{Population Profiling}
{The target population consists of software engineering practitioners experienced in using AI tools for code generation (\eg ChatGPT, Copilot). Participants were meant to be knowledgeable in Python and understand how algorithms are structured in programming languages and represented in Abstract Syntax Trees (ASTs). While certain knowledge in Deep Learning architectures used for Text Generation (\eg GPT, BERT, T5) is preferred, it was not required. Individuals of any gender were welcome to participate, with a minimum age requirement of 21 years. Participation was entirely voluntary, and no incentives were offered beyond contributing to our efforts to enhance deep learning interpretability for code generation. Furthermore, participants were informed of the voluntary nature of the study during solicitation \footnote{This study was reviewed by the protection of human subjects committee at the College of William \& Mary under protocol number PHSC-2023-03-03-16218-dposhyvanyk titled \textit{A Survey Research on Code Concepts for Interpreting Neural Language Models}}.}

\textbf{Data Collection}
{We reached out to $50$ potential participants who were unaware of the purpose of this work, from industrial and academic backgrounds with varying levels of expertise in machine learning and Python. Participants were contacted via email invitations. Out of this group, $27$ completed one of the surveys, with the assignment uniformly distributed among the surveys. but we excluded three for low-quality responses, leaving $24$ valid submissions. The study was performed on Qualtrics \citep{noauthor_qualtrics_nodate} and the anonymized survey data can be found in our appendix~\citep{RepoASTrust24}.}

\textbf{Statistical Analysis}
{We use \surveySequence as a baseline for our study. We expose the participants to \astrust with two treatments: \surveyAstPartial and \surveyAstComplete (refer to \figref{fig:sec_8_survey_explanations}). The result of each question is influenced by these two treatments. To compare the influence of \surveyAstPartial and \surveyAstComplete against \surveySequence, we compute the weighted average of the responses from surveys \surveyAstPartial and \surveyAstComplete. We refer to the weighted average as \surveyAstBased. First, we calculate the results of each treatment individually for all the answers. Then, the weight of each answer is estimated by averaging the number of responses per answer across all samples. We then normalize this weight to get the final weighted average for \surveyAstBased. We use this weighted average for all our statistical analyses in the chapter.}

\textbf{Survey Validity}
{To validate the design of the human study, we conducted a pilot experiment with 10 individuals excluded from the pool of participants. Based on this pilot, the quality and appropriateness of the control and treatment surveys were solidified.  Initially, the pilot survey included only the \surveyControl control and the \surveyAstComplete treatment. However, the pilot revealed the need for an intermediate representation serving as a baseline explanation, which is less complex than an AST visualization, to ensure a fair comparison. Consequently, we introduced \surveySequence, inspired by techniques such as SHAP \citep{lundberg2017unified}, as a baseline treatment with a less complex representation. Additionally, we introduced \surveyAstPartial, a partial representation of \surveyAstComplete, as a less complex treatment highlighting only the hierarchical structure of the generated code.}

\subsection{Data Science Study for \astrust  Effectiveness}
\label{sec:global_study_design}
To answer \ref{rq:performance} we implemented a {data science study to globally} interpret 12 \llms' performance described in \tabref{tab:models} on the \galeras dataset. We performed code completion with different input prompts. The input prompt combines the code completion task, a description, and a partial code snippet. Each prompt has a standard maximum size of 1024 tokens for all considered \llms.

We first compute the normalized log-probabilities (Sec.\ref{sec:composable}) or \ntp $\hat{w_i}$ for each \galeras snippet $s \in \mathcal{S}$. These log-probabilities were obtained across the 12 \llms for every token position. The log-probability distributions maintain a consistent vector size $|\mathcal{V}|$ for each token position. Subsequently, these distributions underwent processing to extract the log-probability aligned with the expected token at position $i$. As a result, each token position corresponds to a stored prediction value $\hat{w_i}$ for constructing the \ntp sequence $w_{<=i}$. {As discussed earlier, this experimental setting is based on the premise that token probabilities are well-calibrated to model correctness, which has been confirmed in code completion settings by prior work~\citep{spiess2024quality}. Additionally, we confirm this finding in answering RQ$_3$ by observing a causal link between learning error and the probabilities used within \astrust.}

\begin{marginfigure}
  \begin{center}
    \includegraphics[width=\linewidth]{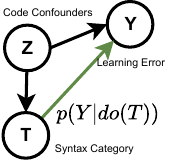}
  \end{center}
  \caption{\scm to estimate Syntax Effect on Learning Error.}
   \label{fig:causal_diagram}
\end{marginfigure}

We used the alignment function $\delta$ to obtain the terminal node $\lambda$ vector (see Def.\ref{def:delta}). Next, we traversed the AST for each terminal node $\lambda$ and clustered them into the corresponding final $\lambda,\alpha$ node and their correspondent \ntp by applying the $\theta$ function (see \secref{ch:decomposition:sec_03}). The clustering was fixed to generate 32 subcategories and their probability values. We estimated a single \textbf{confidence performance} metric (\aka \astrust Interpretability Performance) per model by averaging the subcategories probabilities. The confidence performance per model was bootstrapped with the median (size of 500 samplings) to ensure a fair comparison. Lastly, we mapped the subcategories to the \ascs obtaining a value per Category $\mathcal{C}$ (e.g., Data Structures, Decision, or Scope).

{To provide a baseline comparison, we calculated canonical extrinsic metrics \textit{BLUE-4}~\citep{papineni_bleu_2002} and \textit{CodeBLEU}~\citep{Ren2020codebleu}, and intrinsic performance.
Extrinsic metrics evaluate downstream tasks directly (\ie code completion), while intrinsic metrics assess how well a language model can accurately predict the next word given an incomplete sequence or prompt \citep{Karampatsis2020BigCode}.}

Our analysis also includes a corner case experiment that compares the smaller \gptI to the largest \monoIIII. We contrasted the subcategories for each \llm to obtain a segregated global explanation \figref{fig:largeTreeMap}. Since we mapped the subcategories to categories, we can observe the \astrust probability gaps between \llms at more interpretable levels (see. \figref{fig:global_explanation}). The probability values for subcategories and categories are the bootstrapped median.

\subsection{Causal Inference Study for \astrust  Validity} 

We validate our \astrust approach using causal inference to answer \ref{rq:validity}. To accomplish this, we formulated a \textbf{Structural Causal Model} (\scm) designed to estimate the impact of \asc predictions on the overall learning error of \llms \citep{Pearl2016Causality}. We consider that the learning error (\ie cross-entropy loss) of an \llm is causally impacted by the predicted probabilities of syntax elements. This impact indicates that \ascs influence the quality of an \llm. We conducted a causal inference analysis using the $do_{code}$ technique \citep{palacio_toward_2023} to estimate \ascs influence. Inherently, a developer mentally rationalizes several things such as the \textit{concept} of the \textit{Iteration} category (see Fig.~\ref{fig:largeTreeMap}). If an \llm can make a similar mapping, it suggests that it has \textit{statistically learned} some understanding of the syntax cycle structure.

We calculate the causal influence using the \scm as the Average Treatment Effect (ATE) with the probability $p(Y|do(T))$ for both \gptI and \monoIIII models.  That is, we estimate the \textit{causal effect} of the variable $T$ on $Y$ after controlling for confounders $Z$ (see \figref{fig:causal_diagram}). This probability function is estimated using the \textit{doWhy} tool ~\citep{Sharma2021DoWhyAssumptions,palacio_toward_2023}. The proposed treatments ($T$) embodies Syntax Categories $\mathcal{C}$ such as \textit{Decision}, \textit{Natural Language}, or \textit{Iterative}. 

The first step of this validity evaluation is to obtain the global intrinsic accuracy. We computed the cross-entropy loss for each snippet $s$. After obtaining the cross-entropy loss, we estimate Pearson correlation $\rho$ and ATE for 14 sub-categories ($\lambda$ and $\alpha$ nodes) (\tabref{tab:decomposition:correlations}). Each sub-category and its cross-entropy loss is correlated with four confounding variables (\ie Cyclomatic Complexity, AST Levels, \#AST Nodes, and Sequence Size) calculating the average value from the set of snippets $S$ from \galeras dataset~\citep{daniel23}. The second step is to validate the obtained ATE by testing the \scm robustness (\ie refutation methods ~\citep{palacio_toward_2023}). We limited our exploration to the best and worst models by intrinsic accuracy as we observed similar correlation values across LLMs.

%% file: tables/chap_decomposition/tab1_design_1_models2.tex
\begin{table}[ht]
\centering
\caption{Large Language Models Descriptions.}
\vspace{-0.2cm}
\label{tab:models}

\scalebox{0.90}{%
\setlength{\tabcolsep}{5pt} 

\begin{tabular}{lllll}
\hline
\multicolumn{5}{c}{\textbf{Large Language Models (LLMs)}} \\ \hline
\multicolumn{1}{c}{\textit{Type}} & \textit{ID} & \multicolumn{1}{c}{\textit{Name}} & \multicolumn{1}{c}{\textit{Architecture}} & \multicolumn{1}{c}{\textit{Size}} \\ \hline
\multirow{3}{*}{\textbf{\begin{tabular}[c]{@{}l@{}}Natural L.\\ gpt-3\end{tabular}}} & $M_{1}$ & gpt-neo-125m & \textit{gpt-3} & 125M \\
 & $M_{2}$* & gpt-neo-1.3B & \textit{gpt-3} & 1.3B \\
 & $M_{3}$ & gpt-neo-2.7B & \textit{gpt-3} & 2.7B \\ \hline
\multirow{2}{*}{\textbf{\begin{tabular}[c]{@{}l@{}}Natural L.\\ codegen\end{tabular}}} & $M_{4}$ & codegen-350M-nl & \textit{codegen} & 350M \\
 & $M_{5}$ & codegen-2B-nl & \textit{codegen} & 2B \\ \hline
\multirow{3}{*}{\textbf{\begin{tabular}[c]{@{}l@{}}Multi-\\ Language\end{tabular}}} & $M_{6}$ & codeparrot-small-multi & \textit{gpt-2} & 110M \\
 & $M_{7}$ & codegen-350M-multi & \textit{codegen-350M-nl} & 350M \\
 & $M_{8}$ & codegen-2B-multi & \textit{codegen-2B-nl} & 2B \\ \hline
\multirow{4}{*}{\textbf{\begin{tabular}[c]{@{}l@{}}Mono-\\ Language\end{tabular}}} & $M_{9}$ & codeparrot-small & \textit{gpt-2} & 110M \\
 & $M_{10}$ & codeparrot & \textit{gpt-2} & 1.5B \\
 & $M_{11}$ & codegen-350M-mono & \textit{codegen-350M-multi} & 350M \\
 & $M_{12}$ & codegen-2B-mono & \textit{codegen-2B-multi} & 2B \\ \hline
\end{tabular}

}
\vspace{0.1cm}
{\\ \footnotesize{*The human study was conducted using $M_{2}$ - \gptII}}

\end{table}

%% file: chapters/part_00_chap_03/sec_05_results.tex
\section{Results}
\label{ch:decomposition:sec_05}

{In this section, we present our findings for human, data science, and causal studies. The local analysis is focused on answering \ref{rq:survey_usefulness} by using \astrust to interpret concrete snippets. Similarly, we provide insights into our global analysis to answer \ref{rq:performance} and  \ref{rq:validity}, which incorporates the interpretation of \llms' performance segregated by Syntax Categories, a comparison of edge cases, and a causal assessment of \astrust validity.

Before presenting the results, we point out basic stats about AST data processing: The average tree height of the samples in the empirical study was 30, with an average of 104 tokens and 166 nodes. In the human study, the four samples have distinct complexity levels. The smallest sample has 80 tokens, with 47 AST nodes and a tree of height eight. The biggest sample has a token length of 139, with 117 AST nodes and a tree height of 14.}

\subsection{\texorpdfstring{\ref{rq:survey_usefulness}}{rq1} \astrust Usefulness}

~\label{sec:local1}
Below, we present the results for each survey question as introduced in \secref{sec:local_study_design}. Quantified responses are detailed in \tabref{tab:user_study_quantitative_results}. In addition, we summarize the most relevant feedback received in the open-ended questions. The full human study's results can be accessed in the appendix~\citep{RepoASTrust24}.

\input{tables/chap_decomposition/tab4_user_study_result_vertical3}

\textit{Metric$_1$: Information Usefulness}. The data reveals that $67.48\%$ of participants who evaluated \surveyAstBased explanations, found the presented information useful or slightly useful, with a slight preference for \surveyAstPartial ($67.86\%$) over \surveyAstComplete ($62.5\%$). However, $75\%$ of participants who evaluated \surveySequence felt that it was useful, indicating a stronger preference towards it.

\textit{Metric$_2$: Local Explanation Complexity}. Participants found \surveyAstBased explanations slightly more complex ($44\%$) than \surveySequence ($42\%$). In particular, \surveyAstComplete was found substantially more complex ($67\%$) than \surveyAstPartial. This is not surprising, given that complete ASTs, even for small code snippets can appear complex.

\textit{Metric$_3$: Local Explanation Readability}. Both \surveyAstBased and \surveySequence were found to be similarly readable: $35\%$ participants found \surveyAstBased easy to read and use, compared to $29\%$ for \surveySequence. However, between the two AST types \surveyAstPartial ($57\%$) was far preferred in contrast to \surveyAstComplete ($17\%$), again likely due to the complexity of \surveyAstComplete.

\textit{Metric$_4$: Visualization Usefulness}. \surveySequence visualization was found useful by more than half of the participants who evaluated it ($57\%$). Similarly, $49.8\%$ considered the \surveyAstBased visualizations useful, with an appreciable preference for  \surveyAstComplete ($50\%$) over \surveyAstPartial ($42\%$).

\textit{Metric$_5$: \llm's Reliability}. The high number of participants who judged the \llm as unreliable suggests that all types of explanations helped them assess the quality of the predicted code. However, \surveyAstComplete stood out, with $67\%$ participants favoring it. Meanwhile, $29\%$ participants felt confident about the model based on the information presented by \surveySequence, indicating the potential of formulating incorrect interpretations when using this type of explanation. Interestingly, a high percentage of participants ($43\%$) found that \surveyAstPartial cannot help to conclude whether \llm is reliable.

{\textit{Open$_1$: \llm's Prediction Correctness.} Participants attributed the cause of an incorrect prediction in the model's output to syntax and semantic errors in both treatment and control surveys. The attribution to training bias was prevalent in \surveyControl as evidenced in answers such as \textit{``Model has not seen enough samples to differentiate between [the characters] = and ==''} or \textit{``Maybe the model is trained in problems with similar error''}. However, in \surveySequence and \surveyAstBased those responses included attribution to the low \astrust confidence performance, such as \textit{``The probabilities are very low so the predictions are not correct''} and \textit{``[the character] = has low probability score of 0.0096''}. These results reveal that \astrust explanations provided insightful information for the participants to judge the model's decisions. }

{\textit{Open$_2$: Importance of visual explanations.} Participants favored the color scheme and the \astrust confidence performance associated with each token as the most liked elements in the visual explanations. Conversely, they disfavored the inclusion of certain syntax-related tokens, such as white spaces and punctuation marks, in the interpretability analysis. We also encountered contradictory premises: \surveyAstPartial participants believe the explanation missed important details, while in \surveyAstComplete participants criticized the information overload. Participants also suggested improving the navigation in \surveyAstBased representations by incorporating a mechanism to interactively collapse AST nodes.}

\textit{Profiling.} We found that the participants were well-qualified to take our survey. They all had some background in Machine Learning (Formal or Informal). Similarly, 81.25\% of participants were also familiar with the concept of AST.

\begin{boxK}
\ref{rq:survey_usefulness}: Although Sequence Explanations contain useful information, AST visualizations were viewed most favorably among explanation types. In fact, AST-based Explanations were found most effective to judge \llm reliability.
\end{boxK}

\begin{figure*}[ht]
    \centering
    \includegraphics[width=\linewidth]{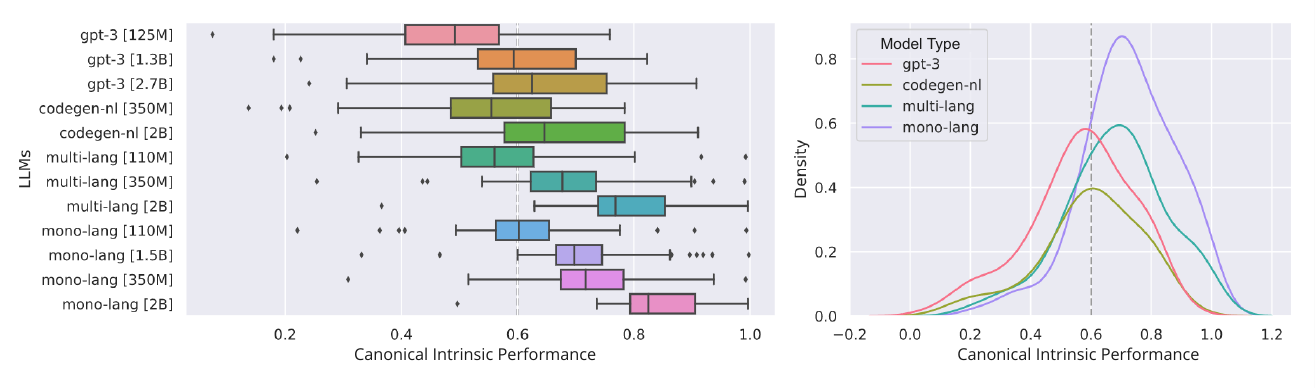}
    \caption{{Canonical intrinsic performance for the models $M1$ to $M12$. Left: box-plots of performance distribution for each model. Right: density plot of performance by model type. }}
    \label{fig:canonical_performance}        
\end{figure*}

\subsection{\texorpdfstring{\ref{rq:performance} }{rq2} \astrust Effectiveness}

\label{sec:global_performance}

{To answer \ref{rq:performance} we computed both the canonical intrinsic performance and the \astrust interpretability performance for 12 \llms (\tabref{tab:models}). \figref{fig:canonical_performance} depicts the canonical intrinsic performance for each \llm (\ie box-plot) and the density canonical intrinsic performance (\ie density plot) by model type (\eg \nlgpt, \nlcodegen, \monolang, and \multilang). The intrinsic performance comprises an aggregated metric that allows us to compare models at a glance. For instance, on average the smallest \monoI ($M_9$) has a similar intrinsic performance as the largest GPT-based \gptIII ($M_3$) model with intrinsic performance of $0.61$ and $0.62$ respectively. After grouping the models by types, we observe that \monolang models excel in the intrinsic performance with the highest density of $0.9$ for performance values between $(0.6 - 0.8)$ and an average intrinsic performance of $\approx 0.7$. Despite the fact canonical intrinsic performance can statistically describe, on average, how the model performs at generating code, these metrics are limited to explaining which categories are being predicted more confidently than others.}

\begin{figure*}[ht]
		\centering
		\includegraphics[width = 0.95\textwidth]{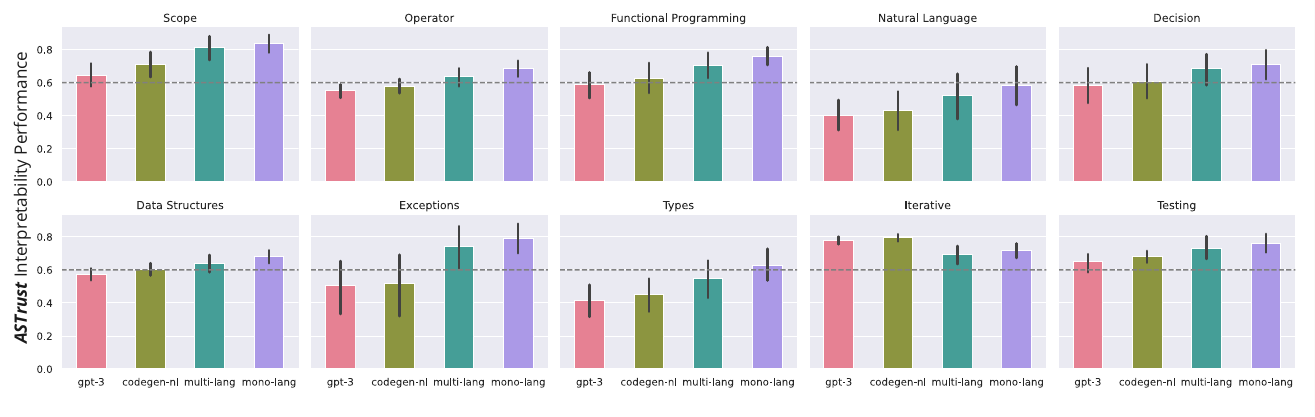}
  \centering
		\caption{Segregated \astrust confidence by Syntax Categories (dotted line is the performance threshold).}
        \label{fig:asc_performance}     
\end{figure*}

{To assess the prediction confidence of each Syntax Category (\asc) for the 12 \llms we present an empirical \astrust interpretability performance value (bootstrapped median columns in \tabref{tab:models_performance}). \figref{fig:asc_performance} illustrates the \astrust interpretability performance segregated by Syntax Categories (\ascs) for each model type. Similarly, \tabref{tab:models_performance} shows bootstrapped median for each model. We set a confidence prediction threshold of $>=0.6$ across all analyses. It is worth noting that this threshold is a tunable parameter that can be modified to obtain tailored interpretations of model performance. We easily identify that \monolang and \multilang surpass our confidence threshold of $0.6$ on all the \ascs but \textit{Data Types}. Conversely, we observe that \textit{GPT-3-type} models face challenges in \textit{Data Types} categories while excelling in \textit{Iteration} categories.}

\input{tables/chap_decomposition/tab2_syntax_concept_evaluation2}

{We found that categories such as \textit{Iteration}, \textit{Except}, and \textit{Scope} surpass our threshold for the majority of our \llms under analysis. For instance, \tabref{tab:models_performance} shows the \textit{Iteration} category consistently surpasses our threshold for all \llms, except for \multiI ($M_6$), which records an average median \astrust of $0.6$ and a highest value of $0.82$ for \codegenII ($M_5$). Notably, our smaller model \gptI ($M_1$) still outperforms the \textit{Iteration} category prediction with an average median of $0.74$. Finally, we note that models trained largely on code, i.e. \codegenII ($M_5$), \monoII ($M_{10}$), \monoIIII ($M_{12}$), could predict the \textit{Data Types} category with and \astrust performance of $0.71$, $0.64$ and $0.83$ respectively.}

By contrast, our \llms struggle to generate good predictions for \textit{Natural language} and \textit{Data Types} categories. We can observe that only \monoIIII ($M_{12}$) surpasses our threshold for \textit{Natural language} with confidence of $0.73$, which is still not an outstanding probability. We attribute poor \astrust performance in certain models to the nature of syntax categories like \textit{Natural Language} and \textit{Data Types}, which demand a larger input context for accurate prediction

\begin{figure}[ht]
\centering
\includegraphics[width=0.9\textwidth]{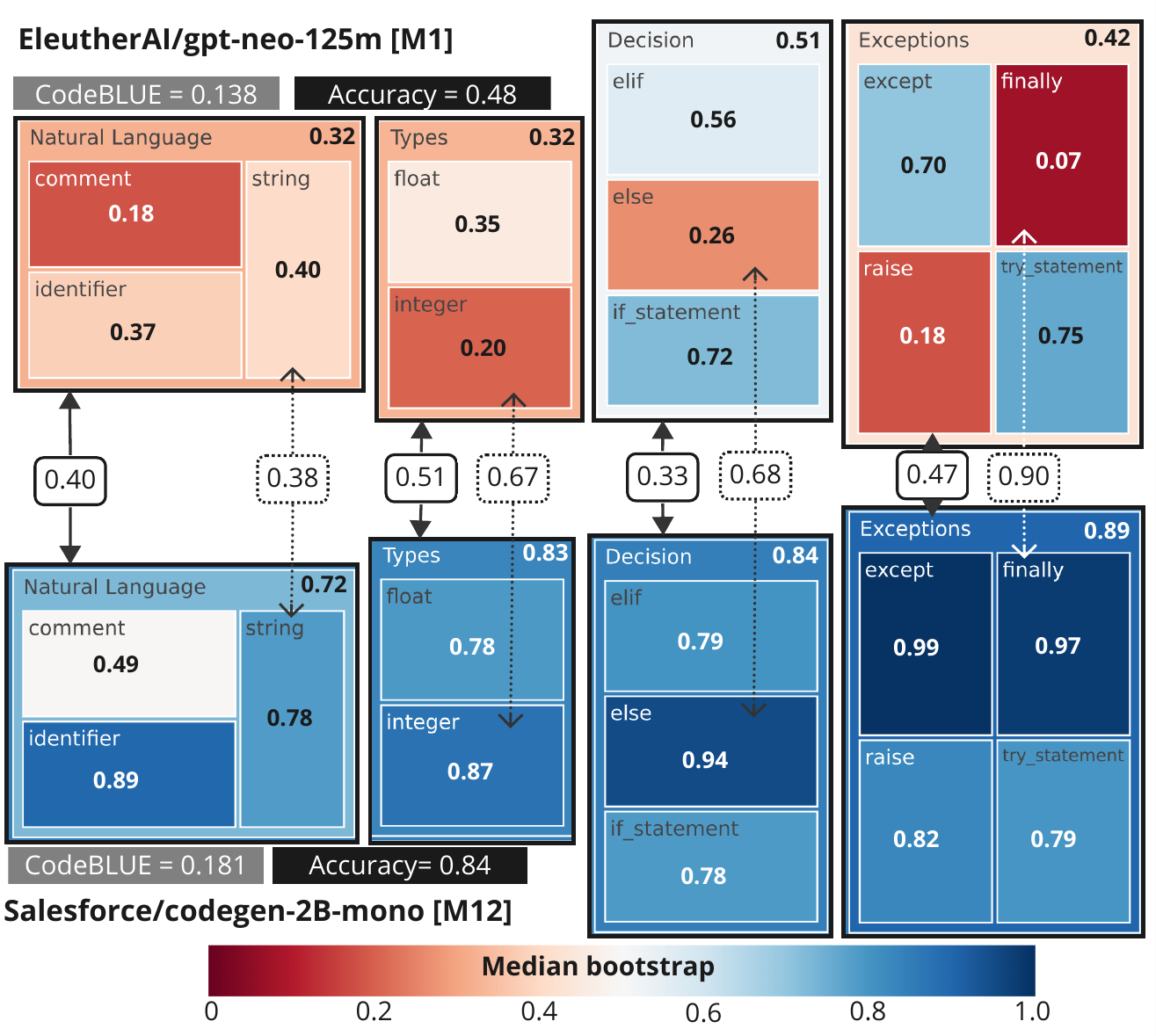}
\caption{\astrust of Syntax Categories and Subcategories (dotted boxes) for corner cases ($M_1$ and $M_{12}$)}

\label{fig:largeTreeMap}
\vspace{1em}
\end{figure}

Our observations indicate that scaling \llms' parameters positively influences the prediction of \ascs. This scaling observation is consistent with canonical scores since our largest models \gptIII ($M_3$), \codegenII ($M_5$), \multiIII ($M_{8}$), and \monoIIII ($M_{12}$) report not only intrinsic accuracy that surpasses our threshold but also \astrust confidence over $0.8$ for categories such as \textit{Exception}, \textit{Iteration}, and \textit{Scope} (see \tabref{tab:models_performance} in blue).  For instance, the largest model, $M_{12}$ exhibits the highest intrinsic accuracy with an avg. median of $0.84$ and exceeds our threshold for each category.

By comparing our \astrust against extrinsic metrics, we observe that $M_{12}$ does not achieve the highest \textit{CodeBLUE} score, recording a $0.181$. Thus, \astrust offers additional insights into the performance of syntax categories. For example, while \monoI ($M_{1}$) outperforms \monoIIII ($M_{12}$) with a \textit{CodeBLUE} of $0.194$, it struggles with inferring \textit{Natural Language} and \textit{Data Types} categories with $0.46$ and $0.47$ respectively (see \tabref{tab:models_performance}). 

{\textit{Corner Case Experiment.} \figref{fig:largeTreeMap}, shows a heatmap with the smallest and largest \llms under analysis, \gptI ($M_1$) and \monoIIII ($M_{12}$) respectively. We selected subcategories and categories with the greatest score jumps for going further into the \asc analysis. For instance, \textit{Data Types} reported the biggest difference with $0.51$ meanwhile the difference between subcategories such as  \texttt{\small `string'} and \texttt{\small `finally'} are $0.38$ and $0.9$, respectively. This difference suggests that model scaling positively impacts the \texttt{\small `finally'} subcategory, while \texttt{\small `string'} or \texttt{\small `if\_statement'} subcategories are slightly affected by the model size. We hypothesize that the poor performance of the \textit{Natural Language} is due to limited context windows in the prompt to predict this category. However, a complementary large-scale exploratory analysis of the proportionality types in the training and testing data is required beforehand to determine other causes of poor performance. We also observe that the \textit{Data Types} is prone to errors, especially as these types may frequently appear at the beginning of code snippets, particularly in Python, where dynamic typing is prevalent. This inter-comparison (\aka across models) would have been infeasible by just using canonical accuracy metrics. }

\begin{boxK}
\textit{\ref{rq:performance}:}
\astrust allows us to segregate the prediction performance of \llms according to Syntax Categories, showing a more interpretable way of comparing models. Syntax-grounded explanations demonstrate, for instance, the struggle of \llms to statistically learn Natural Language nested within code structures. 
\end{boxK}

\subsection{\texorpdfstring{\ref{rq:validity} }{rq3} \astrust Validity}

\label{sec:global_validity}

\vspace{-1em}

We quantitatively demonstrate that cross-entropy loss of \llms tends to be negatively impacted by \astrust probabilities. Therefore, we can explain at syntax category granularity which parts of the code \llms perform poorly (see red boxes in \tabref{tab:models_performance}). We showcase empirical evidence that the previous statement holds for correlations $\rho$ and causal effects $p(y|do(t))$. \tabref{tab:decomposition:correlations} shows, in general, \ascs (\eg Iterative, Scope, or Operator) negatively influence the cross-entropy loss for our best (\ie $M_{12}$) and worst (\ie $M_{1}$) models. Negative effects indicate that the better a syntax category is predicted, the lower the learning error associated.

\input{tables/chap_decomposition/tab3_results_2_correlations2}

The most outstanding finding is that the \textit{Natural Language} category has the largest impact on the cross-entropy loss. For example, the \asc \texttt{\small `identifier'} has a causal effect of $-1.78$ for $M_{1}$ and $-2.89$ for $M_{12}$. In contrast, \textit{Functional Programming} categories present the lowest impact on cross-entropy loss with a subtle \texttt{\small `lambda'} positive causal effect of $0.2$ for $M_{1}$. This subtle positive effect was expected as NL-based \llms have not been fine-tuned on code corpora with \texttt{\small `lambda'} expressions. 

\begin{boxK}
\textit{\ref{rq:validity}: }
The cross-entropy loss of \llms tends to be negatively impacted by \astrust probabilities. This demonstrates that syntax-grounded explanations are indeed representing the syntax learning mechanisms of \llms at segregated granularity. 
\end{boxK}

%% file: tables/chap_decomposition/tab4_user_study_result_vertical3.tex
\begin{table}[t]
\centering
\caption{Survey results for the \astrust Local Study.} 
\label{tab:user_study_quantitative_results}

\scalebox{0.75}{%
\setlength{\tabcolsep}{5pt} 
\begin{tabular}{cllccc}
\hline
\textbf{\textit{Survey Question}} &  & \multicolumn{4}{c}{\textbf{\textit{Results (\% answers)}}} \\ \hline
\textit{} &  &  & \textit{\textbf{Useful}} & \textit{\textbf{Slightly Useful}} & \textit{\textbf{Not useful}} \\
 &  & \textit{\surveySequence} & \cellcolor[HTML]{EFEFEF}\textbf{50.00} & \cellcolor[HTML]{EFEFEF}25.00 & 25.00 \\
 &  & \textit{\surveyAstBased} & 43.57 & \textbf{23.91} & \cellcolor[HTML]{EFEFEF}32.52 \\
 &  & \textit{\surveyAstPartial} & 39.29 & 28.57 & \textbf{32.14} \\
\multirow{-4}{*}{\textit{\begin{tabular}[c]{@{}c@{}}Information\\ Usefulness\end{tabular}}} &  & \textit{\surveyAstComplete} & 41.66 & \textbf{20.84} & \textbf{37.50} \\ \hline \hline
\textit{} &  &  & \textit{\textbf{Agree}} & \textit{\textbf{Neutral}} & \textit{\textbf{Disagree}} \\
 &  & \textit{\surveySequence} & 42.00 & \cellcolor[HTML]{EFEFEF}29.00 & \cellcolor[HTML]{EFEFEF}29.00 \\
 &  & \textit{\surveyAstBased} & \cellcolor[HTML]{EFEFEF}44.00 & \textbf{28.00} & \textbf{28.00} \\
 &  & \textit{\surveyAstPartial} & 14.00 & \textbf{43.00} & \textbf{43.00}\\
\multirow{-4}{*}{\textit{\begin{tabular}[c]{@{}c@{}}Local Explanation\\ Complexity\end{tabular}}} &  & \textit{\surveyAstComplete} &  \textbf{67.00} & 17.00 & 16.00\\ \hline
 &  & \textit{\surveySequence} & 29.00 & \cellcolor[HTML]{EFEFEF}29.00 & 42.00 \\
 &  & \textit{\surveyAstBased} & \cellcolor[HTML]{EFEFEF}\textbf{35.00} & 21.00 & \cellcolor[HTML]{EFEFEF}44.00 \\
 &  & \textit{\surveyAstPartial} & \textbf{57.00} & 29.00 & 14.00 \\
\multirow{-4}{*}{\textit{\begin{tabular}[c]{@{}c@{}}Local Explanation\\ Readability\end{tabular}}} &  & \textit{\surveyAstComplete} & 17.00 & 33.00 & \textbf{50.00} \\ \hline
 &  & \textit{\surveySequence} & \cellcolor[HTML]{EFEFEF}\textbf{57.00} & \cellcolor[HTML]{EFEFEF}29.00 & 14.00 \\
 &  & \textit{\surveyAstBased} & \textbf{49.80} & 27.78 & \cellcolor[HTML]{EFEFEF}22.42 \\
 &  & \textit{\surveyAstPartial} & \textbf{42.00} & 29.00 & 29.00 \\
\multirow{-4}{*}{\textit{\begin{tabular}[c]{@{}c@{}}Visualization\\ Usefulness\end{tabular}}} &  & \textit{\surveyAstComplete} & \textbf{50.00} & 33.00 & 17.00 \\ \hline \hline
 &  &  & \textit{\textbf{Highly Reliable}} & \textit{\textbf{Not Reliable}} & \textit{\textbf{Impossible to Tell}} \\
 &  & \textit{\surveySequence} & \cellcolor[HTML]{EFEFEF}29.00 & \textbf{42.00} & 29.00 \\
 &  & \textit{\surveyAstBased} & 0.00 & \cellcolor[HTML]{EFEFEF}\textbf{62.00} & \cellcolor[HTML]{EFEFEF}38.00 \\
 &  & \textit{\surveyAstPartial} & 0.00 & \textbf{57.00} & 43.00 \\
\multirow{-4}{*}{\textit{\begin{tabular}[c]{@{}c@{}}\llm's \\ Reliability\end{tabular}}} &  & \textit{\surveyAstComplete} & 0.00 & \textbf{67.00} & 33.00 \\ \hline
\end{tabular}
}
{\\ \footnotesize{* \textbf{bold}:Highest \%, background:Highest \% \surveySequence or \surveyAstBased}}
\end{table}

%% file: tables/chap_decomposition/tab2_syntax_concept_evaluation2.tex
\begin{table*}[ht]
\centering
\caption{Syntax Concept Empirical Evaluation Results (bold: best, underlined: worst). }
\label{tab:models_performance}

\begin{adjustbox}{width=\textwidth}

\setlength{\tabcolsep}{5pt}

\begin{tabular}{clcccccccccclccc}
\hline
\multicolumn{1}{l}{} &  & \multicolumn{10}{c}{\textbf{\astrust Interpretable Performance (bootstrapped median)}} &  & \multicolumn{2}{c}{\textbf{Extrinsic}} & \multicolumn{1}{l}{\textbf{Intrinsic}} \\ \cline{3-12} \cline{14-16} 
\multicolumn{1}{l}{\multirow{-2}{*}{\textbf{LLMs}}} &  & \multicolumn{1}{l}{\textit{Data Str.}} & \multicolumn{1}{l}{\textit{Decision}} & \multicolumn{1}{l}{\textit{Except.}} & \multicolumn{1}{l}{\textit{F. Prog.}} & \multicolumn{1}{l}{\textit{Iter.}} & \multicolumn{1}{l}{\textit{NL}} & \multicolumn{1}{l}{\textit{Oper.}} & \multicolumn{1}{l}{\textit{Scope}} & \multicolumn{1}{l}{\textit{Testing}} & \multicolumn{1}{l}{\textit{Data Ts.}} &  & \multicolumn{1}{l}{\textit{BLEU-4}} & \multicolumn{1}{l}{\textit{CodeBLEU}} & 
\textit{Perf.} \\ \hline
$M_1$ &  & 0.50 & 0.52 & \cellcolor[HTML]{FFCCC9}{\color[HTML]{000000} {\ul 0.43}} & \cellcolor[HTML]{FFCCC9}{\color[HTML]{000000} {\ul 0.49}} & 0.74 & \cellcolor[HTML]{FFCCC9}{\color[HTML]{000000} {\ul 0.32}} & \cellcolor[HTML]{FFCCC9}{\color[HTML]{000000} {\ul 0.48}} & 0.51 & 0.59 & \cellcolor[HTML]{FFCCC9}{\color[HTML]{000000} {\ul 0.33}} &  & 0.013 & 0.139 & 0.48 \\
$M_2$ &  & 0.60 & 0.61 & 0.53 & 0.62 & 0.79 & \cellcolor[HTML]{FFCCC9}{\color[HTML]{000000} {\ul 0.43}} & 0.57 & 0.68 & 0.68 & \cellcolor[HTML]{FFCCC9}{\color[HTML]{000000} {\ul 0.44}} &  & 0.016 & 0.151 & 0.59 \\
$M_3$ &  & 0.62 & 0.63 & 0.56 & 0.66 & \cellcolor[HTML]{CACDFA}\textbf{0.81} & \cellcolor[HTML]{FFCCC9}{\color[HTML]{000000} {\ul 0.46}} & 0.60 & 0.74 & 0.70 & \cellcolor[HTML]{FFCCC9}{\color[HTML]{000000} {\ul 0.47}} &  & 0.015 & 0.163 & \textbf{0.62} \\
$M_4$ &  & 0.56 & 0.57 & \cellcolor[HTML]{FFCCC9}{\color[HTML]{000000} {\ul 0.45}} & 0.57 & 0.77 & \cellcolor[HTML]{FFCCC9}{\color[HTML]{000000} {\ul 0.39}} & 0.54 & 0.64 & 0.64 & \cellcolor[HTML]{FFCCC9}{\color[HTML]{000000} {\ul 0.40}} &  & 0.015 & 0.151 & 0.55 \\
$M_5$ &  & 0.65 & 0.65 & 0.58 & 0.68 & \cellcolor[HTML]{CACDFA}\textbf{0.82} & \cellcolor[HTML]{FFCCC9}{\color[HTML]{000000} {\ul 0.48}} & 0.61 & 0.78 & 0.72 & 0.50 &  & 0.016 & 0.155 & \textbf{0.65} \\
$M_6$ &  & 0.54 & 0.55 & 0.64 &  0.60 & {\ul 0.60} & \cellcolor[HTML]{FFCCC9}{\color[HTML]{000000} {\ul 0.40}} & 0.54 & 0.71 & 0.67 & \cellcolor[HTML]{FFCCC9}{\color[HTML]{000000} {\ul 0.42}} &  & 0.010 & 0.189 & 0.57 \\
$M_7$ &  & 0.63 & 0.72 & 0.75 & 0.70 & 0.69 & 0.51 & 0.62 & 0.83 & 0.73 & 0.51 &  & 0.015 & 0.171 & 0.68 \\
$M_8$ &  & 0.74 & 0.79 & \cellcolor[HTML]{CACDFA}\textbf{0.83} & 0.81 & 0.77 & 0.65 & 0.74 & \cellcolor[HTML]{CACDFA}\textbf{0.91} & 0.80 & 0.71 &  & 0.016 & 0.177 & \textbf{0.79} \\
$M_9$ &  & 0.58 & 0.58 & 0.68 & 0.66 & 0.63 & \cellcolor[HTML]{FFCCC9}{\color[HTML]{000000} {\ul 0.46}} & 0.57 & 0.73 & 0.69 & \cellcolor[HTML]{FFCCC9}{\color[HTML]{000000} {\ul 0.47}} &  & 0.011 & 0.194 & 0.61 \\
$M_{10}$ &  & 0.67 & 0.67 & \cellcolor[HTML]{CACDFA}\textbf{0.80} & 0.76 & 0.70 & 0.59 & 0.66 & \cellcolor[HTML]{CACDFA}\textbf{0.82} & 0.74 & 0.64 &  & 0.011 & 0.196 & 0.71 \\
$M_{11}$ &  & 0.68 & 0.76 & 0.78 & 0.76 & 0.73 & 0.57 & 0.68 & \cellcolor[HTML]{CACDFA}\textbf{0.86} & 0.77 & 0.58 &  & 0.014 & 0.179 & 0.73 \\
$M_{12}$ &  & 0.79 & \cellcolor[HTML]{CACDFA}\textbf{0.84} & \cellcolor[HTML]{CACDFA}\textbf{0.90} & \cellcolor[HTML]{CACDFA}\textbf{0.85} & \cellcolor[HTML]{CACDFA}\textbf{0.81} & 0.73 & \cellcolor[HTML]{CACDFA}\textbf{0.82} & \cellcolor[HTML]{CACDFA}\textbf{0.94} & \cellcolor[HTML]{CACDFA}{\color[HTML]{000000} \textbf{0.85}} & \cellcolor[HTML]{CACDFA}\textbf{0.83} &  & 0.016 & 0.181 & \textbf{0.84} \\ \hline
\end{tabular}
\end{adjustbox}

{\footnotesize *Erroneous \astrust values are in red. Confident \astrust scores are in blue. Canonical values serve as a baseline.}

\end{table*}

%% file: tables/chap_decomposition/tab3_results_2_correlations2.tex
\begin{table}[ht]
\centering
\caption{\astrust influence on Learning Error.}
\label{tab:decomposition:correlations}

\scalebox{0.85}{

\setlength{\tabcolsep}{4pt} 

\begin{tabular}{lllrrrr}
\hline
\multicolumn{2}{c}{\textbf{{[}T{]} Syntax Categories}} &  & \multicolumn{4}{c}{\textbf{{[}Y{]} Learning Error}} \\ \cline{1-2} \cline{4-7}
\multicolumn{1}{c}{\textbf{Categories}} & \multicolumn{1}{c}{\textbf{Sub-Categories}} &  & \multicolumn{2}{c}{\textbf{gpt-125}} & \multicolumn{2}{c}{\textbf{mono-2B}} \\  \cline{4-7}
\multicolumn{1}{c}{\textbf{$\mathcal{C}$}} & \multicolumn{1}{c}{\textbf{$\alpha,\lambda$}} &  & \multicolumn{1}{c}{$\rho$} & \multicolumn{1}{c}{\textbf{ATE}} & \multicolumn{1}{c}{$\rho$} & \multicolumn{1}{c}{\textbf{ATE}} \\ \hline
 & \textit{for\_statement} &  & -0.16 & \cellcolor[HTML]{EFEFEF}-0.10 & -0.07 & \cellcolor[HTML]{EFEFEF}-0.01 \\
\multirow{-2}{*}{\textit{\textbf{Iterative}}} & \textit{while\_statement} &  & -0.05 & \cellcolor[HTML]{EFEFEF}-0.11 & -0.03 & \cellcolor[HTML]{EFEFEF}-0.08 \\ \hline
 & \textit{identifier} &  & \textbf{-0.56} & \cellcolor[HTML]{EFEFEF}\textbf{-1.78} & \textbf{-0.80} & \cellcolor[HTML]{EFEFEF}\textbf{-2.89} \\
\multirow{-2}{*}{\textit{\textbf{\begin{tabular}[c]{@{}l@{}}Natural \\ Language\end{tabular}}}} & \textit{string} &  & -0.31 & \cellcolor[HTML]{EFEFEF}-0.36 & \textbf{-0.43} & \cellcolor[HTML]{EFEFEF}\textbf{-0.55} \\ \hline
 & \textit{return\_statement} &  & -0.04 & \cellcolor[HTML]{EFEFEF}-0.09 & -0.22 & \cellcolor[HTML]{EFEFEF}-0.09 \\
 & \textit{{]}} &  & -0.16 & \cellcolor[HTML]{EFEFEF}-0.04 & -0.22 & \cellcolor[HTML]{EFEFEF}-0.10 \\
\multirow{-3}{*}{\textit{\textbf{Scope}}} & \textit{)} &  & -0.37 & \cellcolor[HTML]{EFEFEF}-0.85 & \textbf{-0.54} & \cellcolor[HTML]{EFEFEF}\textbf{-1.49} \\ \hline
\textit{\textbf{Decision}} & \textit{if\_statement} &  & -0.22 & \cellcolor[HTML]{EFEFEF}-0.21 & -0.11 & \cellcolor[HTML]{EFEFEF}-0.11 \\ \hline
 & \textit{comparison\_operator} &  & -0.13 & \cellcolor[HTML]{EFEFEF}0.02 & -0.11 & \cellcolor[HTML]{EFEFEF}0.00 \\
\multirow{-2}{*}{\textit{\textbf{Operator}}} & \textit{boolean\_operator} &  & -0.10 & \cellcolor[HTML]{EFEFEF}0.01 & -0.08 & \cellcolor[HTML]{EFEFEF}-0.09 \\ \hline
 & \textit{for\_in\_clause} &  &  -0.03 & \cellcolor[HTML]{EFEFEF}0.09 & -0.03 & \cellcolor[HTML]{EFEFEF}0.04 \\
 & \textit{if\_clause} &  & -0.01 & \cellcolor[HTML]{EFEFEF}0.19 & 0.01 & \cellcolor[HTML]{EFEFEF}0.13 \\
 & \textit{lambda} &  & -0.04 & \cellcolor[HTML]{EFEFEF}0.20 & -0.05 & \cellcolor[HTML]{EFEFEF}0.06 \\
\multirow{-4}{*}{\textit{\textbf{\begin{tabular}[c]{@{}l@{}}Functional \\ Programming\end{tabular}}}} & \textit{list\_comprehension} &  & -0.04 & \cellcolor[HTML]{EFEFEF}0.06 & -0.03 & \cellcolor[HTML]{EFEFEF}0.04 \\ \hline
\multicolumn{2}{c}{\textbf{Baseline}} &  & \multicolumn{1}{l}{} & \multicolumn{1}{l}{} & \multicolumn{1}{l}{} & \multicolumn{1}{l}{} \\ \cline{1-2}
\textbf{Intrinsic} & Avg. Accuracy &  & -0.38 & \cellcolor[HTML]{EFEFEF}\textbf{-1.60} & -0.38 & \cellcolor[HTML]{EFEFEF}\textbf{-1.78} \\ \hline
\end{tabular}

} %

{\footnotesize{* \textbf{bold}:Highest impact, shadowed: causal effect}}
\vspace{0.5cm}
\end{table}

%% file: chapters/part_00_chap_03/sec_06_discussion.tex
\section{Discussion}
\label{ch:decomposition:sec_06}

{Below, we pose three aspects for discussion: 1) some general insights ($GIs$) from the empirical study, 2) a logical analysis of the connection between trustworthiness and interpretability, and 3) the threats to validity of our approach.}

\subsection{Empirical Study Insights}
\textbf{$GI_1$: Token Predictions Reliability.} { \astrust relies on logit extraction to generate post-hoc explanations as syntax categories. If logits are wrongly predicted (by over/underfitting), our \textit{causal validation process} detects such inconsistency by reducing the Average Treatment Effect (ATE) of syntax categories on the statistical learning error. Our Structural Causal Model (SCM) was designed to test the robustness and fidelity of our approach under models’ misconfigurations or unreliable performance. Also, as stated earlier in the chapter, recent work on calibration for LLMs of code has illustrated that, for code completion (which subsumes the experimental settings in this chapter), LLMs tend to be well calibrated to token probabilities/logits~\citep{spiess2024quality}. This helps to mitigate issues that may arise due to model confidence and correctness being misaligned.}

\textbf{$GI_2$: Syntax Aggregations Improves Explanations.} {Due to its granular nature, token-level predictions are less informative than a hierarchical aggregated level. BPE can leverage the interpretation of individual tokens much more difficult when code-based sequences are split into tokens that may be meaningless. We posit that practitioners can more easily understand syntax categories rather than individual tokens because these categories are \textit{already defined by context-free grammars}, which are semantically rich. Moreover, our human study provides evidence of this claim since AST-based explanations were found to be easy to read and use by participants. AST-based explanations \textit{also capture semantics} by allowing visualization of the full AST structure. This approach helps practitioners evaluate the model's implementation more effectively by providing a clearer, structured view of the code's semantics.}

\textbf{$GI_3$: Natural Language Imbalance.} {Our approach indicates a poor performance on NL sub-categories. We hypothesize this low performance is due to an unbalanced distribution of NL training samples compared to other categories. Before increasing the context window, we believe that a better analysis would be measuring the proportionality of NL sub-categories on the training set and, then, fine-tuning current LLMs to fix possible data bias. Unfortunately, this analysis is currently out of scope since it demands a complementary Exploratory Data Analysis that we envision for future research stages.}

\textbf{$GI_4$: Foundational Interpretability Research.} {\astrust is meant to serve as a more foundational approach required to guide the future development of interpretability tools for users of different backgrounds (\eg researchers, students, and data scientists). We aimed to not only propose a formal methodology to conduct interpretability in our field but also perform a preliminary assessment of \astrust’s usefulness by conducting a control/treatment experiment (\ie with and without the approach) on a visualization technique based on \astrust under clearly defined qualitative metrics.}

\textbf{$GI_5$: Contradictions about the Usefulness of Explanations.} {In our human study, we found that AST-based explanations were preferred over sequential-based ones. Results revealed that the AST-partial representation was considered more useful than AST-Complete, as it presents the AST representation and \astrust confidence performance only for the generated portion of the code. However, the feedback received in the open-ended questions revealed contradictory opinions. Some participants indicated that the AST-partial representation missed important details, while others felt that the AST-Complete representation was excessively detailed. These findings suggest the need for more tailored representations for explanations, aiming to present useful information while maintaining readability. We envision incorporating \astrust into a tool that adds interactivity to navigate the explanations.}

\label{sec:review}

\subsection{Trustworthiness \& Interpretability Connection}
\label{sec:logical}
We outline two premises based on state-of-the-art definitions of trustworthiness and direct observations from our quantitative and qualitative analyses. Then, we use \textit{logical deduction} supported by documented and empirical evidence to link the concept of \textit{trustworthiness} with \textit{\astrust} highlighting the significance of syntax-grounded explanations.

\textbf{Premise$_1$: Interpretability is a cornerstone for trustworthiness in Language Models for Code (LLMs)}. The interpretability field enhances transparency and provides insights into the decision-making process serving as a key factor in fostering practitioner trust and adoption. In the realm of Deep Learning for Software Engineering (DL4SE) ~\citep{watson2020dl4se}, the significance of interpretability in models to engender trust cannot be overstated. Jiaming \etal ~\citep{ji2024ai} underscore interpretability as a pivotal element in aligned models, integral to the RICE principle alongside Robustness, Controllability, and Ethicality. By enhancing transparency in the decision-making process, interpretability plays a crucial role in building trust, a sentiment echoed by Weller \etal, who stress the need to extend transparency beyond algorithms to foster trust ~\citep{weller2019transparency}. The dilemma between accuracy and interpretability, claimed by Lundberg \etal ~\citep{lundberg2017unified}, is magnified by the challenge posed by large, complex models that even experts find difficult to interpret. A user study with UX and design practitioners supports this notion, revealing that explanations are sought to gain deeper insights into AI tool decision-making, providing a remedy for the ``black box'' perception and contributing to user trust and adoption ~\citep{Liao_2020}. Therefore, the indisputable importance of interpretability in \llm decision-making lies in its pivotal role in establishing trustworthiness.

\textbf{Premise$_2$: \astrust improves interpretability.} It is feasible to segregate intrinsic metrics (\ie standard accuracy) into interpretable Syntax Categories revealing the \llms' inner workings concerning code structure and contributing towards interpretability. By conducting extensive qualitative and quantitative studies involving 12 prominent \llms, we have demonstrated the effectiveness of \astrust in enhancing interpretability.  We do not claim that our set of categories is complete; however, we consider that a good \textit{alignment} of the generated categories by the \llm with the ones expected by humans configures a good explanation~\citep{Ghorbani19}. Our \astrust clusters tokens to \textit{meaningful} categories that are easier for human concept association. Furthermore, we uncovered valuable insights, such as the causal influence of AST categories on the cross-entropy loss of \llms after accounting for confounding factors. Our human study participants attested to the usefulness of our \astrust in explaining the predictions of Python code snippets by a \llm \ref{sec:local1}. By breaking down intrinsic metrics into segregated and interpretable terminal and non-terminal nodes, our approach not only enhances the understandability of \llms but also unveils crucial insights into the inner workings of syntax elements.

\textbf{Conclusion.} Given the first premise that interpretability is fundamental for trustworthiness in \llms, supported by several shreds of evidence, and from the second premise asserting that \astrust enhances interpretability by segregating intrinsic metrics into interpretable syntax categories and subcategories, collectively supports the fact that \astrust contributes to the improvement of trustworthiness in \llms for Code using syntax-grounded explanations. 

\subsection{Threats to Validity} \label{sec:threats}

Threats to \textbf{construct validity} concern the intentionality of \astrust in providing useful explanations. Instead of attempting to disentangle information represented between the layers learned by \llms (\ie probing ~\citep{lopez_ast-probe_2022}), \astrust focuses on conceptually mapping \llms' code predictions to present the accuracy in a segregated way. We quantitatively and qualitatively validated the extent to which \astrust is interpretable through causal analyses and a human study.
{While we cannot claim that the results from our study generalize beyond the population of users that participated in our study, our participants represent a diverse range of backgrounds mitigating this threat.  Nonetheless, the purpose of our study was to conduct a preliminary assessment of \astrust representations. As such, all code completion scenarios were designed to include a syntax or semantic error since we assessed how useful our approach is in assisting users in understanding models' incorrect behavior, increasing the reliability of our findings.}

Threats to \textbf{internal validity} refer to the degree of confidence in which the \astrust study results are reliable. Firstly, in our causal study, the potential for unidentified confounders in the code may bias the causal relationship between cross-entropy loss and the Syntax Categories. That is why we ensured the robustness of the Structural Causal Model by performing placebo refutations, which involves simulating unrelated treatments and then re-estimating the causal effects. Secondly, we used rigorous statistical techniques such as bootstrapping to guarantee a consistent comparison between aggregated Token-Level Predictions by syntax elements. 

Threats to \textbf{external validity} represent the extent to which \astrust can be used to contextualize the performance of other \llms or datasets. We excluded GPT-4 based models from our empirical experiments due to the constraints of the current OpenAI API, which restricts access to softmax layers values (a key factor in the intrinsic evaluations). 

While our evaluation relied on decoder-only based models, \astrust can also be used to interpret encoders and other types of auto-regressive architectures. Finally, our \galeras dataset may not contain enough samples to represent all the syntax categories or Python project attributes fairly. Nonetheless, we designed a data mining pipeline to guarantee the diversity of collected samples. 

%% file: chapters/part_00_chap_03/sec_07_conclusions.tex
\section{Lessons Learned \& Conclusions}
\label{ch:decomposition:sec_07}

\textbf{Lesson$_1$: Aggregated metrics may give false impressions about \llms' capabilities.} The research community should incentivize researchers to report AI4SE results in a granular way, as opposed to more traditional aggregated accuracy metrics. After controlling for code confounders, we demonstrated that segregated syntax elements influence the cross-entropy loss of \llms. This influence persists across models at different parameter sizes and fine-tuning strategies. Syntax information is also relevant for any posterior static analysis of code enabling further evaluations of \llms in downstream tasks that entail elements of software design (\eg refactoring).

\textbf{Lesson$_2$: New interpretability methods are required to enable trustworthiness.} In our studies, we have noted an absence of a concrete definition for the term \textit{trust} in the Software Engineering research. However, several researchers have highlighted the importance of establishing trust in work on AI4SE. Research has also shown that interpretability is one of the keys to improving trustworthiness, but at the same time, there is a scarcity of interpretable methods linked to trustworthiness. Despite this limitation, surveyed participants agreed that \astrust was useful to understand why and how a \llm produced certain errors in code-completion tasks. 

\textbf{Lesson$_3$: Grounding model explanations in the relationship between syntactic structures and prediction confidence is useful.} It is feasible to segregate intrinsic metrics (\ie standard accuracy) into interpretable Syntax Categories revealing the \llms' inner workings concerning code structure and contributing towards interpretability. By conducting extensive qualitative and quantitative studies involving 12 prominent \llms, we have demonstrated the effectiveness of \astrust in enhancing interpretability.  We do not claim that our set of categories is complete; however, we consider that a good \textit{alignment} of the generated categories by the \llm with the ones expected by humans configures a good explanation~\citep{Ghorbani19}. Our \astrust clusters tokens to \textit{meaningful} categories that are easier for human concept association. Furthermore, we uncovered valuable insights, such as the causal influence of AST categories on the cross-entropy loss of \llms after accounting for confounding factors. Our human study participants attested to the usefulness of our \astrust in explaining the predictions of Python code snippets by a \llm \ref{sec:local1}. By breaking down intrinsic metrics into segregated and interpretable terminal and non-terminal nodes, our approach not only enhances the understandability of \llms but also unveils crucial insights into the inner workings of syntax elements.

\textbf{Lesson$_4$: The usability of proposed techniques must be further evaluated for industry adoption.} We adapted the non-mathematical definition of \textit{interpretability} by Doshi-Velez \& Kim ~\citep{doshivelez2017rigorousscienceinterpretablemachine}, Molnar ~\citep{Molnar2020InterpretableChallenges} and Miller ~\citep{miller2018explanation} to the field of AI4SE~\citep{watson2020dl4se}. However, as our preliminary human study suggests, \astrust solution is incomplete until being extensively evaluated for industry settings. 

\noindent\textbf{Artifact Availability:} Experimental data, curated datasets, source code, and complementary statistical analysis used in this research are published in an open-source repository~\citep{RepoASTrust24}.

%% file: chapters/part_01_chap_01/bayes.tex
\chapter{On The Causal Nature of Software Information}
\label{ch:bayes}

\lettrine[lines=2]{\textbf{T}}{}he importance of traceability in modern software systems cannot be overstated. Traceability links that connect ``high-level" artifacts such as requirements and use cases to ``low-level'' artifacts written in code help to facilitate crucial components of the software development and maintenance cycle. For instance, linking requirements to code provide visibility into a system by enumerating what has been implemented, whereas linking requirements to test cases helps to indicate that the software is functioning as expected. Additionally, the establishment of trace links aid in facilitating a broad set of developer activities including code comprehension, change impact analysis, and compliance validation~\citep{Cleland-Huang:Springer'12}.  In certain software domains, such as those involving safety-critical systems, traceability is necessarily \textit{mandated} by regulatory bodies to properly demonstrate the safe functioning of a system~\citep{Nejati:IST'12,Rempel:ICSE'14,Cleland-Huang:ICSE'10,Mader:Soft'13}. Furthermore, traceability is increasingly used to help ensure the \textit{security} of a given system~\citep{Nhlabatsi:SST'15}. For example, our industrial partners at Cisco Systems, Inc. require that security-critical requirements are verified by a dedicated group of analysts to avoid software threats and ensure best practices. 

Unfortunately, despite its importance, software traceability is, by its nature, an inherently difficult and error-prone task~\citep{Cleland-Huang:FOSE'14,Mahmoud:ICPC'12,Mader:Soft'13}. This difficulty primarily stems from the need to bridge a logical abstraction gap that exists between different software artifacts, such as requirements written in natural language and code written in ``lower-level'' programming languages.  %
Given the effort required to establish and evolve effective trace links, it is often too costly to manually establish them outside of regulated domains, and in practice the quality of mandated links are often questionable~\citep{Cleland-Huang:FSE'14}.  

The inherent difficulty in establishing trace links has led to research on automated techniques for modeling, establishing, and evolving trace links that primarily rely upon information retrieval (IR)~\citep{Lucia:ICSM'04,Dekhtyar:RE'07,Asuncion:ICSE'10,McMillan:TEFSE'09,Gethers:ICSM'11,DeLucia:ASE'08,DeLucia:EMSE'09,Mahmoud:ICPC'12,Antoniol:ICSE'00,Marcus:ICSE'03,Mills:ICSME18,Jiang:ASE'08,Kuang:SANER'17} and machine learning (ML)~\citep{Mahmoud:RE'16,Guo:MSR'16,Asuncion:ICSE'10,Spanoudakis:SEKE'03,Falessi:EMSE17} techniques which retrieve or predict trace links based upon textual similarity metrics.  However, in large part, current automated approaches for traceability often trade precision for completeness and vice versa, making them difficult to adopt in practice. We observe three major shortcomings of current automated techniques that contribute to their limited effectiveness:

\noindent{\textbf{1) Limited Measures of Artifact Similarity:}} Existing techniques for trace link recovery tend to use a single textual similarity metric to draw relationships between artifacts. This is problematic for several reasons. Perhaps most importantly, it is often difficult or impossible to determine how well a technique that uses a given similarity measure will function on artifacts from a new project without any pre-existing trace links. This so-called ``cold-start'' problem is due to the fact that existing IR/ML techniques for measuring textual similarity often need to be calibrated on a subset of "ground-truth" artifact pairs with pre-existing links. This makes the performance of these techniques difficult to predict when applied to new datasets. Furthermore, in practice, industrial projects often lack pre-existing trace links, as confirmed by our partners at Cisco. Thus, while certain techniques have been shown to perform well on research benchmarks, the efficacy of a similarity measure is often tightly coupled to the underlying semantics of software artifact text~\citep{Lohar:FSE'13,Guo:ICSE'17,Biggerstaff:ACM'94}, and to the configuration of the corresponding IR/ML technique~\citep{Oliveto:ICPC'10}. 

Using only a single textual similarity metric also needlessly restricts the predictive power of a traceability technique. Past work has illustrated the orthogonality of different similarity measures~\citep{Oliveto:ICPC'10}, suggesting that combining \textit{several} different measures could lead to more accurate and robust techniques that function \textit{consistently} well when applied to new projects without pre-existing links.

\noindent{\textbf{2) Inability to Effectively Capture Developer Feedback:}} The rapid pace of modern agile development practices often results in crucial knowledge about a software system being siloed within the expertise of individual developers. Thus, one unstructured development artifact that has gone underutilized by past techniques is \textit{developer feedback}. When an automated traceability model is uncertain about particular trace link pairs, developers can provide critical feedback to help improve trace link inference.

\noindent{\textbf{3) Limited View of Interactions Between Artifacts}}
Existing automated traceability approaches are typically tailored to establish relationships between pairs of specific types of artifacts (\eg user stories and class files). However, information pertaining to the relationship of one type of artifact pair may be contained within other related artifacts. For example, if a piece of source code is linked to a given requirement through textual similarities, and this source code is also intrinsically linked to test code via method calls, then it is likely the requirement is also linked to the test code. However, in this situation, it may be difficult for a textual similarity metric to link the requirements and test code, due to limited test documentation, for example. Thus, in this way, established relationships between certain artifacts may influence the probability of other artifact relationships. In this chapter, we refer to these phenomena as \textit{transitive links}. Existing techniques generally cannot model such interactions between artifacts.

The limitations discussed above stem from both technical and practical limitations of existing traceability techniques, and surfaced during our development of an automated traceability approach in close collaboration with Cisco Systems. In this chapter we introduce a novel technique that overcomes these limitations by constructing a Hierarchical Bayesian Network for inferring a set of candidate trace links. The model that underlies our approach is capable of deriving the probability that a trace link exists between two given artifacts by combining information from multiple measures of textual similarity, while simultaneously modeling transitive relationships and accounting for developer expertise.  We implemented our approach, called \Comet (Hierarchi\textbf{C}al Pr\textbf{O}babilistic \textbf{M}odel for Softwar\textbf{E} \textbf{T}raceability), in both an extensible Python library and as a plugin for the popular Jenkins CI/CD system.  In an extensive set of empirical experiments, we illustrate that \Comet is able to outperform the median precision of \textit{optimally configured} baseline techniques by $\approx$5\% across subjects and $\approx$14\% in the best case. Given that optimal configuration is typically not possible in practice, this illustrates that given a project with no pre-existing trace links, \Comet is likely to perform significantly better than most existing IR/ML techniques. Additionally, we show \Comets potential for integration into the workflows of development teams at Cisco. In summary, this chapter's contributions are as follows: %

\begin{itemize}
	\item{The derivation of a Hierarchical Bayesian Network (HBN) for inferring a candidate set of trace links;}	
	\item{An implementation of this model, called \Comet, as both an extensible Python library and a Jenkins plugin that has been deployed for testing with our industrial partners at Cisco;}
	\item{An extensive evaluation of \Comet on both open source projects and two industrial datasets from one industrial software project, including feedback from professional developers at a major telecommunication software company;}
	\item{An open source, commercial-grade traceability benchmark, developed in coordination with our industrial partner, for the benefit of the research community;}
	\item{An online appendix, including our open source implementation of \Comet and evaluation data for reproducibility~\citep{comet2020}}.
\end{itemize}

\input{chapters/part_01_chap_01/sec_01_approach}

\input{chapters/part_01_chap_01/sec_02_design}

\input{chapters/part_01_chap_01/sec_03_results}

\input{chapters/part_01_chap_01/sec_04_discussion}

%% file: chapters/part_01_chap_01/sec_01_approach.tex
\section{The Hierarchical Bayesian Software Retrieval Model}
\label{sec:approach-hbn}

\begin{marginfigure}
\centering
\includegraphics[width=\linewidth]{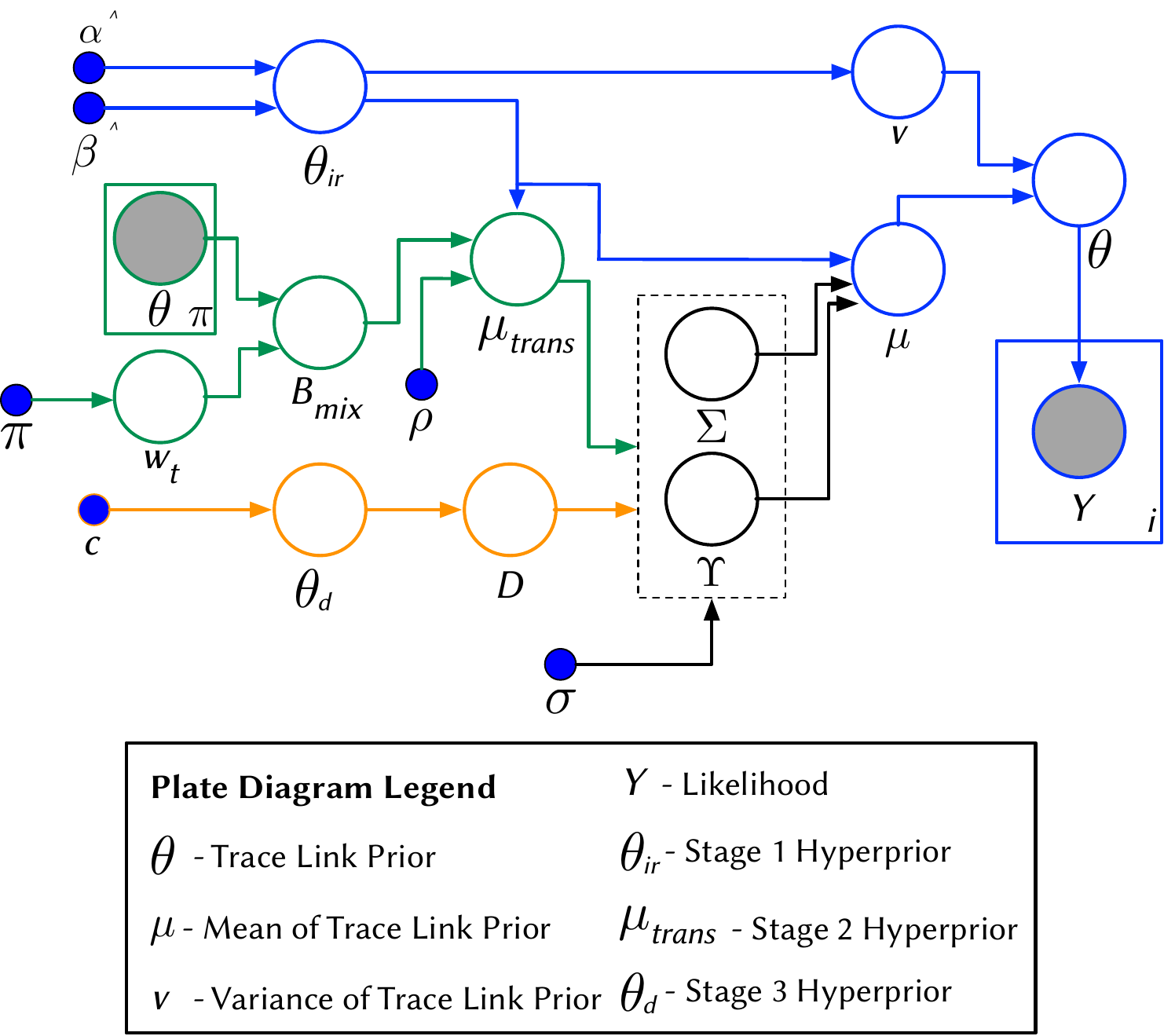}
\caption{Plate Diagram of \Comets HBN}
\label{fig:model-approachI}
\vspace{0.5cm}
\end{marginfigure}

In this section, we provide a formal description of \Comets probabilistic model. To help aid in the comprehension of \Comets underlying model, we provide a graphical representation using plate notation~\citep{Murphy:2012} in \figref{fig:model-approachI}, which we use to guide our introduction and discussion. The model in \figref{fig:model-approachI} is computed on a \textit{per link} basis, that is between all potential links between a set of source ($S$) and target artifacts ($T$). In this section, we will use $S_x$ and $T_y$ to refer to a single source and target artifact of interest respectively. \Comets probabilistic model is formally structured as an HBN, centered upon a \textit{trace link prior} $\theta$ which represents the model's prior belief about the probability that $S_x$ and $T_y$ are linked.  Our model is hierarchical, as the trace link prior is influenced by a number of \textit{hyperpriors}, which are constructed in accordance with a set of \textit{hyper-parameters} that are either derived empirically, or fixed. In \figref{fig:model-approachI}, hyperpriors are represented as empty nodes, and hyper-parameters are represented as shaded blue nodes. In general, empty nodes represent latent, or hidden, variables whereas shaded nodes represent variables that are known or empirically observed quantities. The rectangles, or ``plates'' in the diagram are used to group variables that repeat.

To make our model easier to comprehend, we have broken it down into four major configurations, which we call \textit{stages}, indicated by different colors in \figref{fig:model-approachI}. The first stage of our model (shown in blue at top) unifies the collective knowledge of textual similarity metrics computed by IR/ML techniques. The second stage (shown in orange at the bottom) reconciles expert feedback to improve the accuracy of inferred trace links. The third stage (shown in green in the middle) accounts for transitive relationships among development artifacts, and the fourth stage combines each of the underlying stages. It should be noted that the first stage of our model can be taken as the ``base case'' upon which the other complexities build and is always required to infer the existence of a trace link.  The order of calculation starts with the first stage and proceeds sequentially. The design and parameterization of our model presented in this section is not arbitrary, but instead based on the well-founded theory of \textit{conjugate priors}~\citep{Raiffa:61} which aids in defining appropriate distributions and hyperparameters for a given prior. We center the description of our model first upon the likelihood estimation and then around the estimation of the prior probability distribution as defined by the four stages. After defining the hyperpriors for each of the four stages we briefly discuss the inference techniques we employ to estimate the posterior probability distribution of our model and thus the probability of whether a given link exists. While this section provides an overview of our model, we discuss its instantiation (including utilized IR/ML techniques) in \secref{sub:exp-context}.

\subsection{Estimating the Likelihood}
\label{sub:model-likelihood}

The likelihood function in our HBN models observed data so that these observations can be reconciled with our estimated prior probability distribution to infer a posterior probability.  The likelihood is shown as the observed variable $Y$ (\figref{fig:model-approachI}). The variable $i$ represents the number of observations made. In the context of traceability, we express the likelihood as a discrete Bernoulli distribution, as two artifacts can either be ``linked'' or ``not-linked'':

\begin{equation}\label{eq:likelihood}
Y = p(l_i|\theta_i)= Bern(l_i|\theta_i)
\end{equation}

\noindent where $l_i$ is an observable data point ${0,1}$ for $i$ number of observations. We define an observation as a function of the textual similarity score generated by an IR technique between $S_x$ and $T_y$ and some threshold $k_i$ where any similarity above the threshold is considered an observed link, and any similarity value below this threshold is considered a non-link. The number of IR techniques or configurations utilized corresponds to the number of observations $i$.  Ideally, to capture the most accurate trace link observations from IR techniques, the threshold $k_i$ should be chosen to maximize the chance that each IR technique correctly establishes whether two artifacts are linked.  In other words, $k_i$ should be chosen for each IR technique such that the precision and recall of the technique is maximal across the entire set of considered source and target artifacts $S$ and $T$.  However, this information is not available a-priori without the consultation of a ground truth set of trace links. As we illustrate in \secref{sec:design-hbn} this threshold can often be estimated with surprising accuracy by analyzing the distribution of similarity values an IR technique produces for a given set of artifacts.

\subsection{Stage 1 - Unifying Textual Similarities between Development Artifacts}
\label{sub:model-comp1}

The first ``base'' stage of our model informs the trace link prior, represented as a probability distribution $p(\theta)$, according to the textual similarity measurements of a set of IR techniques.  However, converse to the likelihood estimation, the actual textual similarity values of IR techniques are directly used to estimate a Beta distribution (the conjugate prior of the likelihood's Bernoulli distribution). This Beta distribution is represented as follows: 

\begin{equation}\label{eq:lvl1-dist}
\theta \sim B(\mu, \nu) 
\end{equation}

\noindent where $\mu$ and $\nu$ are parameters of the Beta distribution representing its mean and variance. This prior, and its two parameters are illustrated in the right-most part of the blue segment (\figref{fig:model-approachI}). To inform this Beta distribution, the textual similarity of values of a given number $i$ of IR/ML techniques are normalized according to a sigmoid function centered upon the median of the distribution of similarity values across all $S$ and $T$ in a given dataset. Then a logistic regression is performed upon the normalized similarity values to infer the values $\hat{\alpha}$ and $\hat{\beta}$ which define a hyperprior beta distribution $\theta_{IR}$. This hyperprior with parameters are shown on the left of the blue segment (\figref{fig:model-approachI}). The mean and the variance of this hyperprior distribution then inform $\mu$ and $\nu$ of the base prior $\theta$:

\begin{equation}\label{eq:lvl1-params}
\nu = Var[\theta_{IR}] \quad \mu = Mean[\theta_{IR}]
\end{equation}

Thus, by considering the textual similarity values of a set of IR techniques, our model can effectively reconcile the collective knowledge to ultimately make an informed prediction.

\subsection{Stage 2 - Incorporating Developer Feedback}
\label{sub:model-comp2}

The second stage of our model is capable of leveraging human feedback by influencing the prior distribution introduced in the first stage of our model. To model expert feedback, we estimate hyperpriors $D$ and $\theta_d$, shown in orange in \figref{fig:model-approachI}. To perform this estimation, our model accepts from a developer or analyst, their confidence that a given link exists as a value between $[0,1]$. In \secref{sub:comet-jenkins} we illustrate how such feedback can be collected from developers in a lightweight manner. This confidence value serves as a parameter for estimating the distribution of the first hyperprior:

\begin{equation}
\theta_{d} \sim B(\mu_{d}=c,sd=0.01)	
\end{equation}

\noindent where $\theta_{d}$ is a Beta distribution parameterized by its mean $\mu_d$ set to the confidence value provided by a developer, and standard deviation $sd$ which we set to 0.01 signaling a low variance in the derived Beta distribution. This distribution then parameterizes the second hyperprior $D$, modeled as a Bernoulli distribution.

Now that we have derived a distribution representing developer feedback, we must define how this distribution affects the prior probability of the first stage of our model. To do this, we define reward and penalty functions $\Upsilon$ and $\Sigma$ that are influenced by $\sigma$ which represents a specified \textit{belief factor} between $[0,1]$ that controls the extent to which the feedback influences the trace link prior. The reward function is defined as $\Upsilon = \sigma*D$, whereas the penalty function is defined as $\Sigma = \sigma*(D-1)$. These factors impact the first stage prior Beta distribution by affecting its mean $\mu$:

\begin{equation}\label{eq:mean-affect}
	\mu \sim N(\mu_n= \Sigma + \Upsilon, sd=0.01)
\end{equation}

\noindent where the mean of the first stage prior is represented as a normal distribution parameterized by $\mu_n$ set to the sum of $\Sigma$ and $\Upsilon$, and a standard deviation set to 0.01. Thus, in this manner, expert feedback is utilized to influence the prior distribution that a given trace link exists. The structure of the Stage 2 hyperpriors allows \Comets HBN to effectively consider feedback from multiple developers.

\subsection{Stage 3 - Leveraging Transitive Links}
\label{sub:model-comp3}

\begin{figure}[ht]%
\centering
\includegraphics[width=\columnwidth]{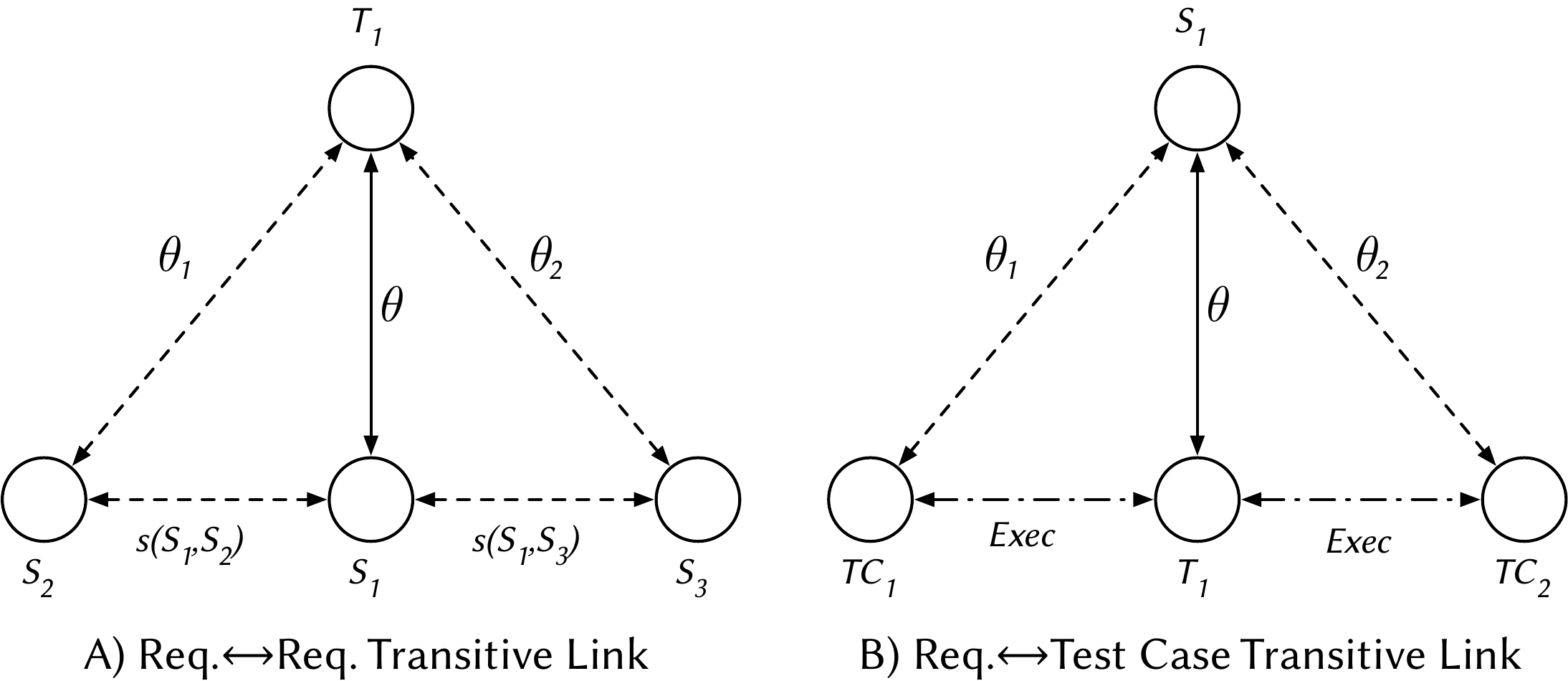}
\caption{Illustration of Transitive Links}
\label{fig:trans-req2req}
\end{figure}

As discussed earlier, the probability that a trace link exists between a source and target artifact can be influenced by \textit{transitive} relationships among varying software development artifacts.  The third stage of \Comets HBN is able to utilize these transitive links to improve the accuracy of its inferred trace link. However, before we describe how our model reconciles this information in a probabilistic manner, it is first important to understand the phenomena of transitive links. At a high level, a transitive link is an inherent relationship between two software artifacts ($A_1$,$A_2$) that may influence the existence of a trace link between either $A_1$ and any other artifact or $A_2$ and any other artifact. \Comet is currently capable of leveraging two types of transitive links, one based on textual-similarities (req. $\leftrightarrow$ req.) and one based on dynamic execution information (req. $\leftrightarrow$ test case). However, \Comet could also be extended to model transitive relationships between other types of artifacts, such as commit messages or issues. \figref{fig:trans-req2req} provides an illustration of both transitive link types, which we detail below. Note that for execution traces, a relationship is considered \textit{strong} between a test method and source method if the test executes the method, and \textit{weak} otherwise. For req$\leftrightarrow$req relationships, it is \textit{strong} if the textual similarity is above the threshold $\tau$, and \textit{weak} if it is below the threshold.

{\textbf{Req. $\leftrightarrow$ Req. Links}} Consider $S_1,S_2,S_3$ as three source artifacts representing three discrete requirements documents and $T_1$ is a potential target document (\ie source code file), where the target relationship being inferred is $S_1\rightarrow T_1$, indicated by the solid line. The connections between nodes denote relationships among the artifacts. Consider the scenario in which the relationship between $S_1$ and $T_1$ is weak, but $S_1$ is highly similar to the other two requirements $S_2$ and $S_3$. Given that we know the three source artifacts are highly related, the relationships between $S_2\rightarrow T_1$ and $S_3\rightarrow T_1$ have a \textit{transitive} influence on the target relationship between $S_1$ and $T_1$. For instance, if $S_2\rightarrow T_1$ and $S_3\rightarrow T_1$ both indicate strong probabilities, then likewise the probability of the target link $S_1\rightarrow T_1$ should be increased to account for these transitive relationships.

{\textbf{Req.$\leftrightarrow$ Test Case Links}} Consider $S_1$ to be a source artifact representing a requirement document, $T_1$ to be a potential target source code file, and $TC_1,TC_2$ to be test cases, where the target relationship being inferred is $S_1\rightarrow T_1$, indicated by the solid line. Consider again the scenario in which the relationship between $S_1$ and $T_1$ is weak, whereas the relationship between $S_1$ and $TC_1,TC_2$ are stronger. If we observe that $TC_1$ and $TC_2$ are related to $T_1$ by execution information (\eg $TC_1$ and $TC_2$ both exercise $T_1$, \textit{Exec} in \figref{fig:trans-req2req}) then, this transitive relationship should influence the probability that a trace link exists between $S_1$ and $T_1$.

{\textbf{Incorporating Transitive Links into C{\footnotesize OMET}'s HBN}} In order for our HBN to incorporate transitive req. $\leftrightarrow$ req. links, it must first derive the set of requirements that are related to a given target requirement $S_x$. To accomplish this, one of several IR techniques can be used to compute textual similarity, or the first stage of our model can be used to derive the relationships, illustrated as $s(S_1,S_2)$ \& $s(S_1,S_3)$ in \figref{fig:trans-req2req}. To incorporate information from req. $\leftrightarrow$ test case links, dynamic information must be collected that provides the $Exec_1$ \& $Exec_2$ relationships illustrated in \figref{fig:trans-req2req}. In either case, a specified threshold $\tau$ signals whether a pair of requirements is related, and the total number of related requirements or test cases is specified by the hyper-parameter $\pi$.  Once the related requirements have been derived, our HBN estimates three hyperpriors, $w_t$, $B_{mix}$ and $\mu_{trans}$. First $w_t$ is formulated as a Dirichlet distribution according to the number of related transitive requirements. Then to estimate $B_{mix}$, the first stage of our HBN is computed between each related requirement and a given target artifact $T_y$. The inferred values for each transitive link, and $w_t$ are used to form a mixture model:

\begin{equation}
	B_{mix} \sim Mix(w_t,\theta_\pi)
\end{equation}

\noindent where $B_{mix}$ is a Beta mixture model parameterized by the 1st stage inference of each transitive link and $\pi$ weights modeled as a Dirichlet distribution parameterized by $\pi$. This Dirichlet distribution is then used to derive a meditated normal distribution $\mu_{trans}$:

\begin{equation}
	\mu_{trans} \sim \rho*B_{mix} + (1-\rho)*Mean[\theta_{IR}]
\end{equation}

\noindent where $Mean[\theta_{IR}]$ represents the mean of the probability distribution of IR similarity values (from stage 1) on the trace link prior and $\rho$ is represents the \textit{belief factor} of the transitive links (\eg the degree to which the transitive relationships should affect overall prior trace link probability). $\mu_{trans}$ can then be utilized to derive the reward and penalty functions introduced earlier where $\Upsilon = \sigma*(1-\mu_{trans})$ whereas $\Sigma = \sigma*\mu_{trans}$. The reward and penalty functions can then in turn be used to influence the mean of trace link prior $\mu$:

\begin{equation}\label{eq:mean-affect-combined}
	\mu \sim N(\mu_n= \mu_{trans} + \Sigma + \Upsilon, sd=0.01)
\end{equation}

\noindent in the same manner as introduced in \equaref{eq:mean-affect}. In this way, our model is capable of incorporating information from transitive links, increasing the overall prior probability if transitive links are strongly connected to the target artifact $T_y$ and decreasing it if they are not strongly connected.

\subsection{Stage 4 - The Holistic Model}
\label{sub:model-comp4}

The holistic model combines all three underlying stages. To accomplish this, the calculations of the reward and penalty functions for affecting the mean $\mu$ of the overall prior are modified to incorporate information from both expert feedback and transitive links:

\begin{align}
\begin{split}
	\Upsilon \sim (1-\mu_{trans})*\sigma*D \\
	\Sigma \sim \mu_{trans}*\sigma*(D-1)
\end{split}
\end{align}

\noindent Then \equaref{eq:mean-affect-combined} can be used to derive the new mean for the overall prior probability distribution of the model.

\subsection{Inferring the Posterior}
\label{sub:model-posterior}

In order to reason about the probability that a trace link exists, we must estimate the posterior probability distribution of our hierarchical model $p(\Theta|L)$ according to the observable data $L$ and prior knowledge of the link $p(\Theta)$. Here $p(\Theta)$ encompasses the trace link prior and all constituent hyperpriors depending upon the stage of the model.  Once the posterior has been estimated, \Comet utilizes the \textit{mean} of the distribution as the general probability that a link exists. We can represent the general calculation of the posterior for our model using using Bayes Theorem as follows:  

\vspace{-0.3cm}
\begin{equation} \label{eq_bayes}
p(\Theta|L) = \dfrac{p(\Theta)p(L|\Theta)}{\int p(\Theta)p(L|\Theta)d\Theta} \propto p(\Theta) \prod\limits_{i=1}^n p(L_i|\Theta_i)
\end{equation}

\noindent where $n$ represents the total number of observations (\ie the number of underlying IR techniques and configurations).  \Comets HBN is non-trivial, and thus the posterior $p(\Theta|L)$ cannot be computed analytically. Therefore, we turn to approximation techniques for estimating the posterior probability distribution. Comet can currently utilize three different techniques including (i) Maximum a Posteriori (MAP) estimation~\citep{Bassett2018MaximumEstimators}, a Markov Chain Monte Carlo (MCMC) technique via the No-U-Turn sampling (NUTS) process~\citep{Hoffman2011TheCarlo}, and a machine learning-based technique called Variational Inference (VI)~\citep{Bishop:2006}. We provide experimental results in \secref{sec:results-hbn} for all techniques for Stage 1 of \Comets model, and NUTS/MAP for Stages 2-4, as VI cannot be applied to more complex stages of the model.

\subsection{ %
The C{\small OMET} Python Library \& Jenkins Plugin
}
\label{sub:comet-jenkins}

\begin{figure}[ht]
	\centering
	\includegraphics[width=\linewidth]{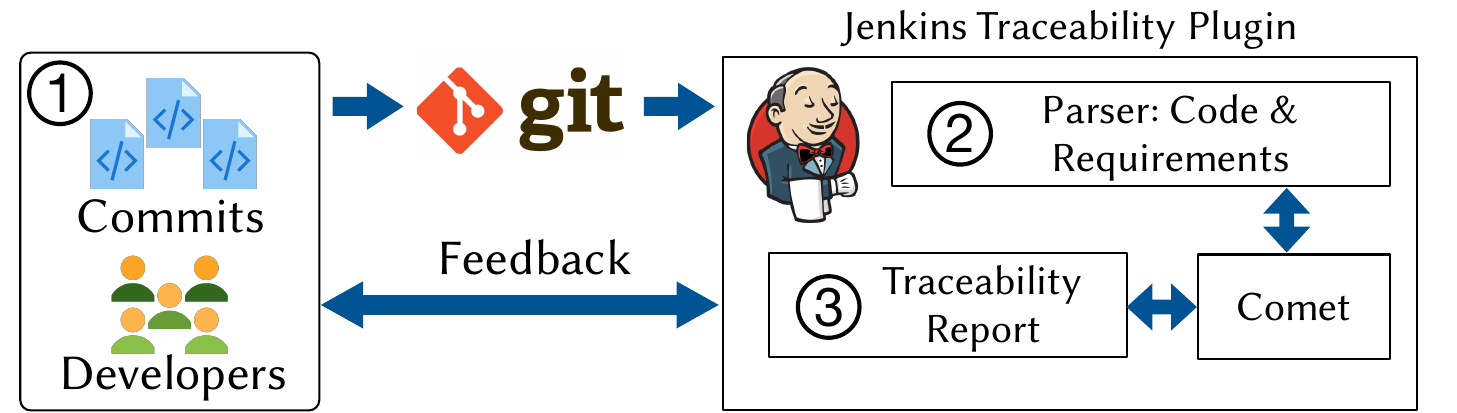}
	\caption{Comet Jenkins Plugin Flow}
	\label{fig:plugin-flow}
\end{figure}

We implemented the four stages of \Comets HBN in an extensible traceability library written in \texttt{\textbf{\small python3}}. In order to explore the practical applicability of \Comet, we implemented the first two stages of \Comets HBN as a Jenkins~\citep{jenkins} plugin.  We did not implement the final two stages due to time limitations, but are actively working on this in partnership with Cisco. This plugin was developed by one of the authors during an internship with Cisco Systems in close collaboration with researchers, engineers and analysts. The code and extensive documentation for the \Comet library and plugin are available in our online appendix~\citep{comet2020}.

\textbf{An Illustrated Use Case of the C{\footnotesize OMET} Plugin} \figref{fig:plugin-flow} provides a general overview of the \Comet plugin architecture. To illustrate the \textit{utility} of the plugin, we describe its workflow from the viewpoint of a developer. As depicted in \figref{fig:plugin-flow}-\textcircled{1} the plugin is triggered when developers commit changes to a given project configured to utilize our traceability plugin in Jenkins. The commit triggers a job that checks out, compiles, and executes the project code according to our industrial partner's existing CI pipeline. As illustrated in \figref{fig:plugin-flow}-\textcircled{2} our plugin parses and preprocesses the source code, test code, and requirement text to be analyzed with \Comet once the normal CI's build flow finishes. The preprocessed corpora of requirements, code, and tests are then passed to the \Comet library where the \textit{first stage} of the HBN (configured according to the tuning datasets described in \secref{sec:design-hbn}) is run to establish an initial set of trace links among artifacts.  Note that our plugin can be configured to run according to varying intervals (\eg every minor or major release). Given that our plugin supports the first two stages of \Comets HBN, it is capable of collecting feedback from developers to improve trace links. To do this, developers select an option from a dropdown likert scale, with each option representing a potential value of $\mu_d$ for stage 2 of \Comets HBN (Strongly agree=1.0, Agree=0.75, Unsure=0.5, Disagree=0.25, \& Strongly Disagree=0.0). To ensure a responsive feedback mechanism to reflect developer feedback, the Stage 2 link probabilities are precomputed. 

\begin{figure}[ht]
\centering
\includegraphics[width=\linewidth]{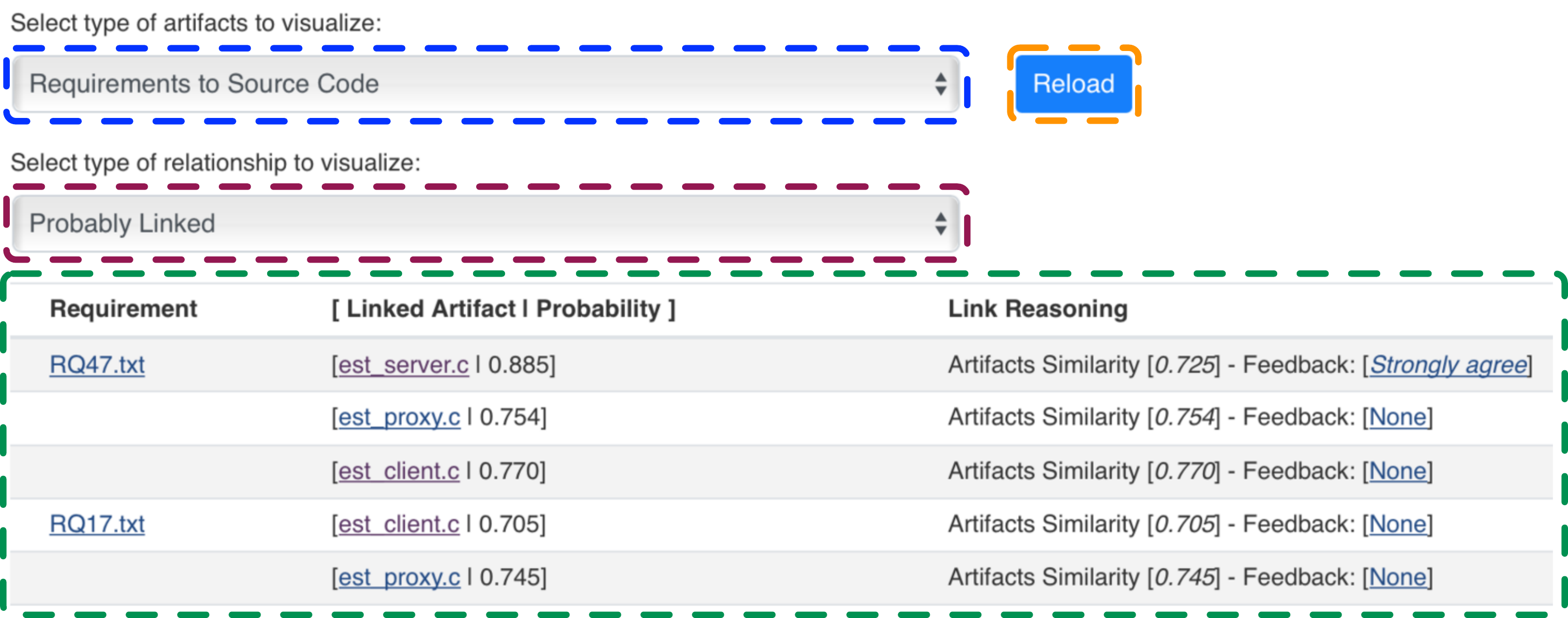}
\caption{Comet Jenkins Plugin Traceability Report}
\label{fig:comet-plugin-ui}
\end{figure}

Once the plugin job finishes, it generates an interactive \textit{traceability page} (\figref{fig:comet-plugin-ui}), that allows a developer or analyst to view inferred trace links and provide feedback. As shown in \figref{fig:comet-plugin-ui} the blue rectangle highlights a drop-down menu that allows a developer to filter inferred trace links by type (\ie req$\leftrightarrow$src, req$\leftrightarrow$test cases, src$\leftrightarrow$test cases, and artifacts not linked). The drop-down menu highlighted in red allows developers to filter trace links according to the inferred probability ranges of $\theta$ in \Comets HBN including: \textit{probably linked} (\ie  $\theta=[0.7,1)$), \textit{probably not linked} (\ie $\theta=[0,0.4)$), and \textit{unsure} representing links where the model is unable to make a confident inference ( \ie $\theta=[0.4,0.7)$). The area highlighted in green lists the types of selected source and target artifacts allowing the user to easily inspect candidate trace links as well as provide feedback on the inferences. When a developer clicks on the \texttt{\small feedback} link for a given pair of artifacts, a popup window allows them to select one of the likert options delineated earlier. Finally, the traceability page also allows developers to view artifacts that are not linked to any other artifacts, and thus may be suspicious. Additional screenshots detailing the workflow of the \Comet plugin are available in our online appendix~\citep{comet2020}.

\textbf{\textbf{C{\footnotesize OMET}'s Complexity \& Scalability.}} We have designed the \Comet plugin and library to facilitate easy integration into modern automated CI/CD systems via a set of higher level APIs that abstracts much of the complexity of the model for most users, while still providing mechanisms for advanced users to tweak parameters. Furthermore, \Comet was able to be successfully deployed into a commercial CI pipeline for a pilot project with our industrial partner. We have also designed the \Comet plugin and library to be highly scalable. Trace link probabilities between each pair of artifacts can be calculated independently, making the process highly parallelizable. Thus, we implemented parallel process management using \texttt{Theano}~\citep{theano} into the \Comet library, allowing computation to scale across modern multi-core machines. To further optimize performance, the \Comet plugin can make use of change analysis in git to only recompute trace link values for artifacts that have been altered since the last computation.

%% file: chapters/part_01_chap_01/sec_02_design.tex
\section{Design}
\label{sec:design-hbn}

To evaluate \Comet, we perform an extensive empirical evaluation with two major \textit{goals}: (i) evaluate the effectiveness of the four stages of \Comets HBN in terms of their ability to effectively infer trace links, and (ii) examine whether \Comet is applicable in industrial workflows. The \textit{quality focus} of our study is \Comets effectiveness, in terms of generating an accurate and complete set of trace links, and practical applicability. We formulate the following set of RQs:

\begin{itemize}

	\item{\textbf{RQ$_1$}: \textit{How effective is \Comet at inferring candidate trace links using combined information from IR/ML techniques?}}

	\item{\textbf{RQ$_2$}: \textit{To what extent does expert feedback impact the accuracy of the candidate trace links of \Comet?}}

	\item{\textbf{RQ$_3$}: \textit{To what extent does information from transitive links improve \Comets trace link inference accuracy?}}

	\item{\textbf{RQ$_4$}: \textit{How effective is the holistic \Comet model in terms of inferring candidate trace links?}}

	\item{\textbf{RQ$_5$}: \textit{Do professional developers and security analysts find our implementation of the \Comet Jenkins plugin useful?}}
	
\end{itemize}

\subsection{Experimental Context}
\label{sub:exp-context}

\textbf{Subject Datasets.} The \textit{context} of this empirical study includes the eight datasets shown in \tabref{tab:datasets-hbn}. Six of these are taken from the open source CoEST community datasets~\citep{coest-datasets}. These datasets represent a set of benchmarks created by the research community and widely used as an effective assessment tool for automated traceability techniques~\citep{Antoniol:ICSE'00,Cleland-Huang:TSE'03,Poshyvanyk:TEFSE'11,Gethers:ICSM'11}. In order to maintain the quality of our experimental subjects, we do not use all available projects in the CoEST repository, as we limited our studied systems to those that: (i) included trace links from requirements or use cases written in natural language to some form of code artifact, (ii) were written in English and/or included English translations, and (iii) had at least 1k LoC.  We utilize two datasets to investigate and tune the hyper-parameters of \Comets HBN, Albergate, and the Rq$\rightarrow$Tests dataset of the EBT project. We utilize the other six datasets for our empirical evaluation. The subject system called ``LibEST'' is an open source networking-related software project, which was created and is actively maintained by engineers at Cisco as an implementation of RFC-7030 ``Enrollment over Secure Transport''. We derived the ground truth set of trace links between Rq$\rightarrow$Src and Rq$\rightarrow$Tests for this dataset in close collaboration with our industrial partner. First, one of the authors carefully created an initial set of trace links. Then, an engineer working on the project reviewed the links and confirmed or denied a subset, based on their availability. The author then revised the links using the engineer's feedback, and this process continued over several months until the ground truth was established. The "LibEST" dataset is available along with all of our experimental data to facilitate reproducibility~\citep{moran_improving_2020}.

\input{tables/chap_07_retrieval-bayes/tab_1_subjects}

\textbf{Studied IR Techniques.} The ``base'' first stage of \Comets HBN is able to utilize and unify information regarding the textual similarity of development artifacts as computed by a set of IR/ML techniques.  While there is technically no limit to the number of IR/ML techniques that can be utilized, we parameterized our experiments using ten IR-techniques enumerated in \tabref{tab:ir-techniques-hbn}. The first five techniques are standalone techniques, whereas the second five are combined techniques utilizing the methodology introduced by Gethers \etal~\citep{Gethers:ICSM'11}. This combined approach normalizes the similarity measures of two IR techniques and combines the similarity measures using a weighted sum. We set the weighting factor $\lambda$ for each technique equal to 0.5, as this was the best performing configuration reported in the prior work~\citep{Gethers:ICSM'11}. We explain the differences between the technique employed by Gethers et. al. and \Comet in \secref{sec:discussion-hbn}. The other parameters for each of the techniques were derived by performing a series of experiments on the two tuning datasets, and using the optimal values from these experiments. For all IR techniques, we preprocessed the text by removing non-alphabetic characters and stop words, stemming, and splitting camelCase. We performed 30 trials for each technique involving LDA, and chose the number of topics that led to optimal performance on our tuning projects. To aid in experimental reproducibility, complete configurations for each technique are listed in our online appendix~\citep{comet2020}.

\subsection{RQ\texorpdfstring{$_1$}{1}: C{\footnotesize OMET} Performance with Combined IR/ML Techniques}

\label{sub:study-rq1}

To answer RQ$_1$, we ran the first stage of \Comets HBN on our six evaluation datasets using the ten IR/ML techniques enumerated in Table \tabref{tab:ir-techniques-hbn}. However, in order to accurately estimate the likelihood function $Y$ we need to choose a threshold $k_i$ for each IR technique that maximizes the precision and recall of the trace links according to the computed textual similarity values. To derive the best method for determining the threshold for each IR technique, we performed a meta evaluation on our two tuning datasets. We examined five different threshold estimation techniques: (i) using the mean of all similarity measures for a given dataset, (ii) using the median of all similarity measures across a given dataset, (iii) using a Min-Max estimation, (iv) a sigmoid estimation, and (v) link estimation (Link-Est), where an estimation of the number of confirmed links for a dataset is made based on the number of artifacts, and a threshold derived to ensure that the estimated number of links is above that threshold.  We performed each of these threshold estimation techniques for all studied IR techniques across our two tuning datasets, and compared each estimation to the known optimal threshold. We used the optimal technique across our two tuning datasets, as reported in \tabref{tab:ir-techniques-hbn}. To aid in reproducibility, we provide a detailed account of these experiments in our online appendix~\citep{comet2020}.

\input{tables/chap_07_retrieval-bayes/tab_2_ir-techniques}

To provide a comparative baseline against which we can measure \Comets performance, we report results for the best-performing and median of the studied IR/ML techniques, optimally configured for each dataset. We chose to optimally configure the baseline techniques, even though such configurations would not be possible in practice due to the absence of a ground truth, in order to illustrate how close Comet can come to the ``best-case baseline scenario''. 

To provide a comprehensive comparison of \Comet to a state of the art technique for candidate trace link generation, we re-implemented the DL-based approach proposed by Guo \etal\citep{Guo:ICSE'17}. However, it should be noted that the intended purpose of this DL approach and Comet differ. The DL technique proposed by Guo \etal was intended to be both trained and evaluated on a single project that contains a set of \textit{pre-existing} trace links the model can be trained upon, and was quite effective in improving the accuracy of trace links in this scenario. However, as pre-existing trace links may not always exist \Comet \textit{does not} require them for analysis. Instead, our experiments aim to illustrate that \Comet can accurately infer trace links when tuned on one small set of projects, and applied to others. Therefore, we design an experimental setup where both techniques are applied on projects without pre-existing trace links. Thus, we train the DL approach on our two tuning projects, using the optimal parameters reported in~\citep{Guo:ICSE'17}. Our main goal in comparing with this DL technique is to illustrate the performance of a recent ML-based technique applied to Comet's intended ``cold-start'' use case.

In order to measure the performance of our studied techniques for inferring trace links, we utilize three main metrics, Precision, Recall, and Average Precision (AP), similar to prior work that evaluates automated traceability techniques \citep{Gethers:ICSM'11,Guo:ICSE'17}. Given that candidate link generation techniques infer a probability or similarity that a trace link exists, a threshold similarity or probability value must be chosen to make the final inference. 

In order to summarize the performance of our studied techniques, we calculate the Average Precision as a weighted mean of precisions per threshold: $AP = \Sigma_n(R_n-R_{n-1})P_n$ where $P_n$ and $R_n$ are the Precision and Recall at the $n$th threshold. Thus, the AP provides a metric by which we can quantitatively compare the performance of different approaches. For the results of \Comet, we report the highest AP achieved by the posterior estimation techniques outlined in Sec. \ref{sub:model-posterior}. In addition to AP, we also provide Precision/Recall (P/R) curves to illustrate the trade-off between precision and recall at different threshold values. Curves further away from the origin of the graph indicate better performance. In lieu of a non-parametric statistical test as suggested by recent work~\citep{Furia:TSE'19}, we perform a confidence interval analysis~\citep{Neyman:37} between our baseline techniques and Stage 1 of \Comet by calculating the standard error across different threshold values, applying bootstrapping where necessary. Thus, if one technique outperforms another within the bounds of our calculated error, it serves as a strong indication of statistical significance.

\subsection{RQ\texorpdfstring{$_2$}{2}: C{\footnotesize OMET} Performance w/Expert Feedback}

\label{sub:study-rq2}

Collecting \textit{actual} developer feedback on trace links for each of our test datasets was not possible  given the time constraints on developers from our industrial partner, and we did not have access to the developers of the other projects. Thus, in order to evaluate Stage 2 of \Comets HBN, we simulated developer feedback by randomly sampling 10\% of the artifact pairs from each studied subject, and used the ground truth to provide a confidence level for each of the sampled links.  To accomplish this, we provided the model with a confidence value $c$ of 0.9 if a link existed in the ground truth, and 0.1, if the link did not exist. However, even trace links derived from experts can be error-prone. Hence, we performed three types of experiments to simulate imperfect links being suggested to our model. That is, for the set of randomly sampled links, we intentionally reversed the confidence values according to the ground truth, for 25\% and 50\% of the sampled links respectively to simulate varying degrees of human error in providing link feedback. In other words, we sampled a small number of trace links from the ground truth, and then used these links to confirm/deny links predicted by \Comet (i.e., if a ground truth link existed, and \Comet predicted it, then it was confirmed). Because developers may not be correct all of the time, we simulated this by randomly flipping the sampled ground truth, which has a similar effect to a developer incorrectly classifying certain predicted links.

We set the value for the \textit{belief factor} of the developer feedback $\sigma=0.5$. For these experiments, we illustrate the impact of developer feedback on AP and P/R curves for \textit{only} the sampled links. In addition to the baseline IR techniques described in the procedure for RQ$_1$, we also compare our results from Stage 2 of the model to Stage 1, to illustrate the relative improvement.

\subsection{RQ\texorpdfstring{$_3$}{3}: C{\footnotesize OMET} Performance w/Transitive Links}
\label{sub:study-rq3}

To measure the impact that transitive links have on the trace link inference performance of Stage 3 of \Comets HBN, we examined the impact of transitive links between requirements as described in \secref{sub:model-comp3}. We utilize transitive requirement links rather than transitive links established by execution traces, as only one of our datasets (LibEST) had executable test cases. To derive the transitive relationships between artifacts, we computed the VSM similarity among all source documents for each dataset (\eg requirements, use cases) and explored two values for the threshold $\tau$, 0.65, and 0.5. We derived these thresholds by examining the total number of transitively linked requirements in our tuning datasets to achieve a balance between too many and too few requirements being linked. We set the \textit{belief factor} $\rho$ for Stage 3 of the HBN equal to 0.5. We report results for these experiments for only those requirements where transitive links impacted \Comets performance.

\subsection{RQ\texorpdfstring{$_4$}{4}: Holistic C{\footnotesize OMET} Performance}
\label{sub:study-rq4}

To evaluate the overall performance of \Comets holistic model, we combined our experimental settings for RQ$_2$ \& RQ$_3$. That is, we randomly sampled 10\% of the links from each dataset and simulated developer feedback with a 25\% error rate. Additionally, we incorporated transitive links between requirements using the same procedure outlined for RQ$_3$. For the transitive links, we set $\tau$ to 0.65, and we set the $\sigma$ and $\rho$ hyper-parameters both equal to 0.5. For this research question, we report results across all links.

\subsection{RQ\texorpdfstring{$_5$}{5}: C{\footnotesize OMET} Industrial Case Study}
\label{sub:study-rq6}

Given that the ultimate goal of designing \Comet is for the approach to automate trace link recovery within industry, we perform a case study with our industrial partner. This case study consisted of two major parts.
First, we conducted a feedback session with six experienced developers who have been contributing to the LibEST subject program. This session consisted of a roughly 15 minute presentation introducing the \Comet Jenkins plugin. Then the developers were asked to use the plugin, which had been configured for LibEST, and evaluate the links and non links for which the model was most confident (\ie the highest and lowest inferred probabilities). Then after using the tool, they were asked a set of likert-based user experience (UX) questions derived from the SUS usability scale by Brooke \citep{Brooke:96}. Additionally, participants were asked free-response user preference questions based on the honeycomb originally introduced by Morville~\citep{Morville:04}. Second, we conducted semi-structured interviews with two groups consisting of roughly 15 engineering managers who specialize in auditing software for security assurance. During these interviews, a video illustrating the \Comet plugin was shown, and a discussion was conducted with the questions illustrated in \figref{fig:LibEST-study}. We report results from both studies.

%% file: tables/chap_07_retrieval-bayes/tab_1_subjects.tex
\begin{table*}[ht]
\caption{Datasets used for \Comets evaluation. Req = Requirement, Src = Source code, UC = Use Case}
\label{tab:datasets-hbn}

\scalebox{0.8}{%

\begin{tabular}{@{}llccccc@{}}
\toprule
\multicolumn{1}{c|}{\textbf{Dataset}} &
  \multicolumn{1}{c|}{\textbf{Language}} &
  \multicolumn{1}{c|}{\textbf{Size (LoC)}} &
  \textbf{\begin{tabular}[c]{@{}c@{}}\#Source \\ Artifacts\end{tabular}} &
  \textbf{\begin{tabular}[c]{@{}c@{}}\#Target\\ Artifacts\end{tabular}} &
  \multicolumn{1}{c|}{\textbf{\begin{tabular}[c]{@{}c@{}}\#Pairs \\ (\#Links)\end{tabular}}} &
  \textbf{Type} \\ \midrule
\multicolumn{7}{c}{\textbf{Training Datasets}}                                                                                                           \\ \midrule
\multicolumn{1}{l|}{Albergate} & \multicolumn{1}{l|}{Java}      & \multicolumn{1}{c|}{10,464} & 55  & 17  & \multicolumn{1}{c|}{935/53}    & Req $\to$ Src  \\
\multicolumn{1}{l|}{EBT}       & \multicolumn{1}{l|}{Java}      & \multicolumn{1}{c|}{1,747}  & 40  & 25  & \multicolumn{1}{c|}{1000/51}   & Req $\to$ Test \\ \midrule
\multicolumn{7}{c}{\textbf{Experimental Datasets}}                                                                                                       \\ \midrule
\multicolumn{1}{l|}{\multirow{2}{*}{LibEST}} &
  \multicolumn{1}{l|}{\multirow{2}{*}{C}} &
  \multicolumn{1}{c|}{\multirow{2}{*}{70,977}} &
  59 &
  11 &
  \multicolumn{1}{c|}{649/204} &
  Req $\to$ Src \\
\multicolumn{1}{l|}{}          & \multicolumn{1}{l|}{}          & \multicolumn{1}{c|}{}       & 59  & 18  & \multicolumn{1}{c|}{1062/352}  & Req $\to$ Test \\
\multicolumn{1}{l|}{eTour}     & \multicolumn{1}{l|}{Java}      & \multicolumn{1}{c|}{23,065} & 58  & 116 & \multicolumn{1}{c|}{6728/308}  & UC $\to$ Src   \\
\multicolumn{1}{l|}{EBT}       & \multicolumn{1}{l|}{Java}      & \multicolumn{1}{c|}{1,747}  & 40  & 50  & \multicolumn{1}{c|}{2000/98}   & Req $\to$ Src  \\
\multicolumn{1}{l|}{SMOS}      & \multicolumn{1}{l|}{Java}      & \multicolumn{1}{c|}{9,019}  & 67  & 100 & \multicolumn{1}{c|}{6700/1044} & UC $\to$ Src   \\
\multicolumn{1}{l|}{iTrust}    & \multicolumn{1}{l|}{Java, JSP} & \multicolumn{1}{c|}{38,087} & 131 & 367 & \multicolumn{1}{c|}{48077/399} & Req $\to$ Src  \\ \bottomrule
\end{tabular}
}

\end{table*}

%% file: tables/chap_07_retrieval-bayes/tab_2_ir-techniques.tex
\begin{margintable}
\centering
\caption{IR/ML Techniques used in the Construction of C{\tiny OMET}'s HBN}
\label{tab:ir-techniques-hbn}
\scalebox{0.7}{%

\begin{tabular}{@{}l|c|c@{}}
\toprule
\multicolumn{1}{c|}{\textbf{\begin{tabular}[c]{@{}c@{}}Information Retrieval \\ Technique\end{tabular}}} &
  \textbf{Tag} &
  \textbf{\begin{tabular}[c]{@{}c@{}}Treshold \\ Technique\end{tabular}} \\ \midrule
Vector Space Model          & VSM     & Link-Est \\
Latent Semantic Indexing    & LSI     & Link-Est \\
Jensen-Shannon Divergence   & JS      & Min-Max  \\
Latent Dirichlet Allocation & LDS     & Min-Max  \\
\begin{tabular}[c]{@{}l@{}}NonNegative \\ Matrix Factorization\end{tabular} &
  NMF &
  Median \\
Combined VSM + LDA          & VSM+LDA & Link-Est \\
Combined JS+LDA             & JS+LDA  & Link-Est \\
Combined VSM+NMF            & VSM+NMF & Link-Est \\
Combined JS+NMF             & JS+NMF  & Link-Est \\
Combined VSM+JS             & VSM+JS  & Min-Max  \\ \bottomrule
\end{tabular}

}
\end{margintable}

%% file: chapters/part_01_chap_01/sec_03_results.tex
\section{Results}
\label{sec:results-hbn}

This section presents the results for our five proposed RQs. We highlight two P/R curves and focus our discussion on the AP results. However, all P/R curves and confidence interval graphs are currently available in our appendix alongside all experimental data~\citep{comet2020}.

\subsection{RQ\texorpdfstring{$_1$}{1} Results: C{\footnotesize OMET} Stage 1 Performance}

\label{sub:results-rq1}

The AP values for Stage 1 of \Comets HBN are provided in \tabref{tab:stage1-4-results} alongside the $p$ values for the Wilcoxon test between Comet and the median IR/ML baseline. The P/R curves for the iTrust dataset are illustrated in \figref{fig:pr1-results}. As \tabref{tab:stage1-4-results} indicates, Stage 1 of \Comet outperforms the median IR/ML baseline across all subjects, to a statistically significant degree according to the confidence intervals. In some cases, such as for iTrust, LibEST, and eTour, Stage 1 of \Comet \textit{significantly} outperforms the median IR/ML baseline, and approaches the performance of the \textit{best} IR/ML baseline. \figref{fig:pr1-results} illustrates the P/R curve for the iTrust project, with performance that outpaces the best IR/ML technique, particularly for lower recall values. \Comet also outperforms the state of the art DL approach across all subjects, likely because the DL approach had difficulty generalizing semantic relationships across datasets.

\input{tables/chap_07_retrieval-bayes/tab_results_stage1}

These results signal remarkably strong performance for \Comets Stage 1 model. Recall that, the Stage 1 model \textit{only} utilizes observations taken from the set of ten IR/ML techniques introduced in Sec. \ref{sub:study-rq1}, thus the fact that the Stage 1 model was able to consistently outperform the median IR/ML baselines and in some cases, nearly match the best IR/ML baseline. This indicates that \Comets HBN is capable of effectively combining the observations from the underlying IR/ML techniques for improved inference power. This is significant, as currently practitioners cannot know a priori which IR/ML technique for traceability will perform best on a given project without pre-existing trace links. Thus, by combining the collective information of several IR techniques \Comets first stage HBN is able to perform \textit{\textbf{consistently well}}, achieving reasonably high performance \textit{\textbf{across projects}}, lending to the credibility of using Comet for projects that do not contain preexisting links.

\subsection{RQ\texorpdfstring{$_2$}{2} Results: C{\footnotesize OMET} Stage 2 Performance}
\label{sub:results-rq2}

The AP for for Stage 2 of \Comet across all subject programs for both 25\% and 50\% error rates is given in Table \ref{tab:stage2-results}. The results indicate that Stage 2 of \Comets HBN is able to effectively incorporate expert feedback to improve the accuracy of its trace link inferences, as the Stage 2 model dramatically outperforms the median (and best) IR/ML techniques as well as the first stage of the model, with a simulated error rate of 25\%.  Even for the larger error rate of 50\%, we see Stage 2 outperform Stage 1 for LibEST (Rq$\rightarrow$Src), LibEST (Rq$\rightarrow$Tests) and EBT, while it slightly underperforms the Stage 1 model for the other subjects. These results illustrate that Stage 2 of \Comets HBN is able to effectively utilize expert feedback to improve its inferences, even in the presence of significant noise.
\input{tables/chap_07_retrieval-bayes/tab_results_stage2}

\input{tables/chap_07_retrieval-bayes/tab_results_stage3}

\subsection{RQ\texorpdfstring{$_3$}{3} Results: C{\footnotesize OMET} Stage 3 Performance}
\label{sub:results-rq3}

The AP results for Stage 3 of \Comet, which incorporates transitive relationships between requirements, for both $\tau=0.55$ and $\tau=0.65$ are given in Table \ref{tab:stage3-results} (There were no transitive links in iTrust for $\tau=0.65$). This table also includes the median of the baseline IR/ML techniques, as well as the Comet Stage 1 model AP results, for the set of links affected by transitive relationships (hence the differing Stage 1 columns). The results show that, in general, for $\tau=0.65$ for \Comets Stage 3 model, the accuracy of \Comets inferred trace links improve, with four of the six datasets showing improvements. For $\tau=0.55$ the results generally exhibit similar or slightly worse performance compared to Stage 1.  The fact that the higher value of $\tau$ led to better performance improvements is not surprising, as this parameter essentially controls the \textit{degree of relatedness} required to consider transitive relationships. Thus, a higher value of $\tau$ means that only highly similar transitive requirement relationships are considered by \Comet's model. Using a lower value for this parameter might introduce noise by incorporating transitive relationships between artifacts that don't have as high a degree of similarity. 

The LibEST (Rq$\rightarrow$Src) dataset exhibited decreased performance for $\tau=0.65$, however this is likely because the requirements for this industrial dataset are based on formal format from the Internet Engineering Task Force (IETF). The somewhat repetitive nature of the language used in these requirements could lead to non-related requirements being transitively linked, leading to a decrease in performance. This suggests leveraging transitive relationships between requirements leads to larger performance gains for more unique language. Overall, our results indicate that \Comets Stage 3 model improves the accuracy of links for a majority of subjects.

\begin{marginfigure}
\centering

\begin{subfigure}{\textwidth}
\includegraphics[clip,width=\textwidth]{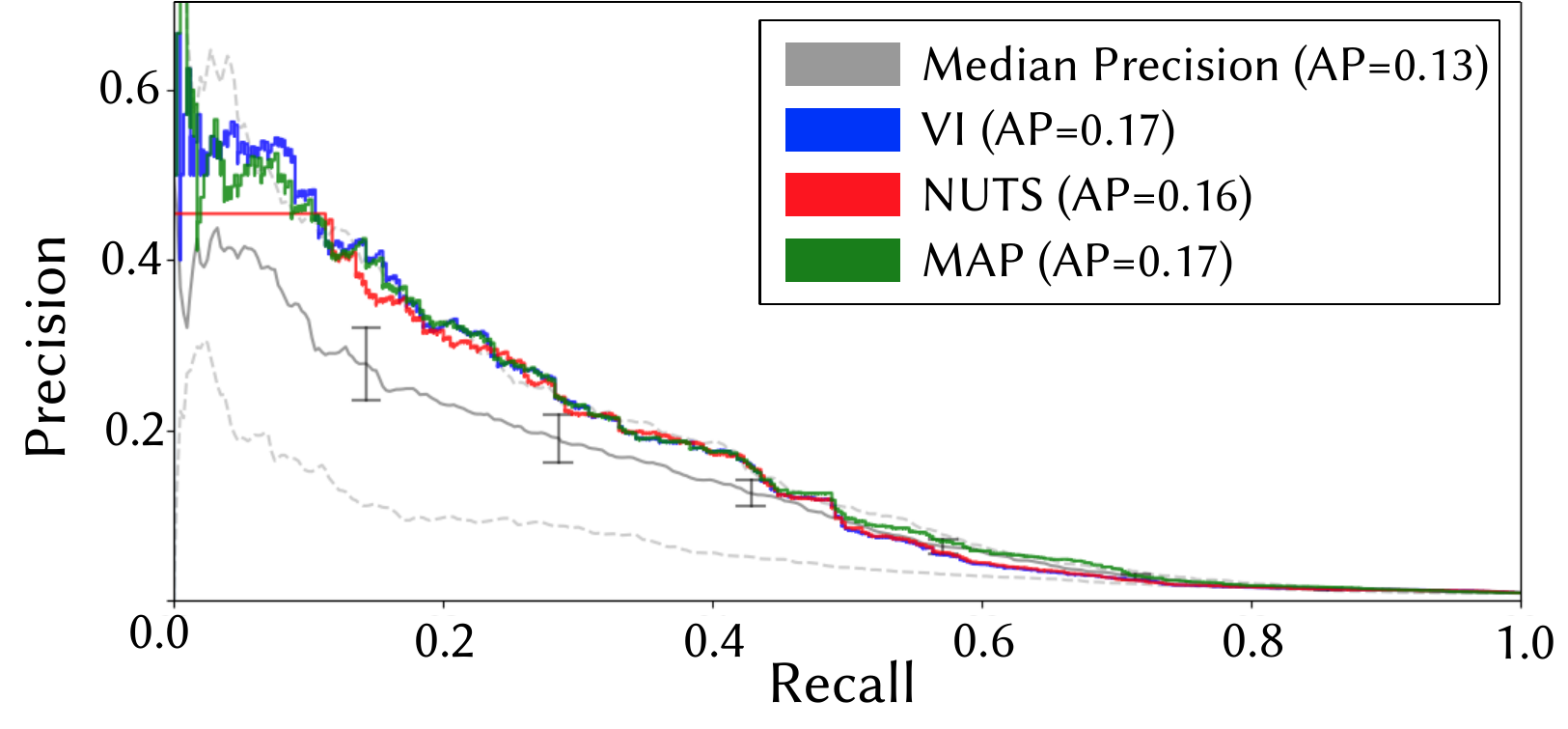}
\caption{\footnotesize P/R Curve for iTrust for Stage 1.}
\label{fig:pr1-results}
\end{subfigure}
\hfill
\begin{subfigure}{\textwidth}
\includegraphics[clip,width=\textwidth]{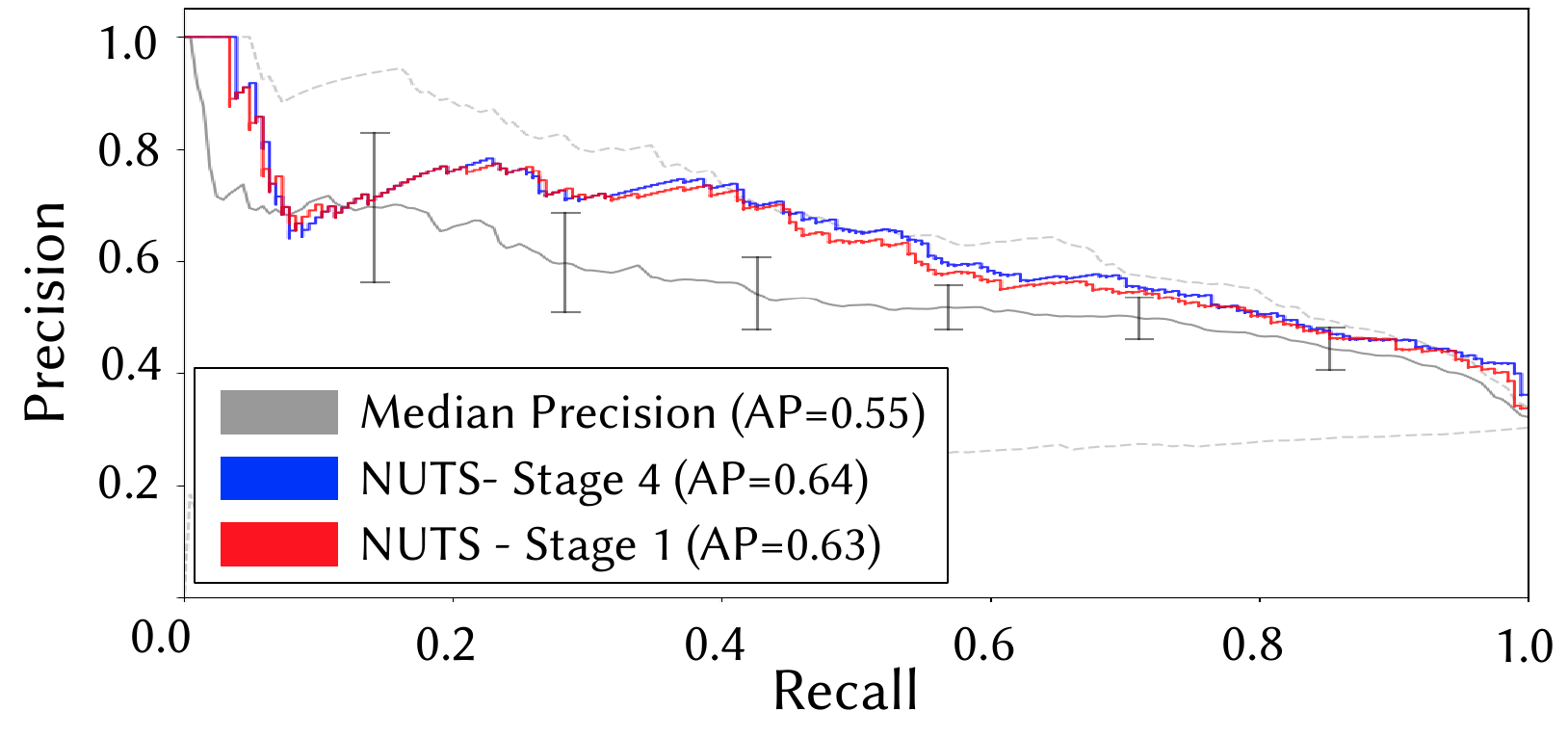}
\caption{\footnotesize P/R Curve for I-Net (Req$\rightarrow$Src) for Stage 4.}
\label{fig:pr4-results}
\end{subfigure}

\caption{Selected P/R Curves for Stage 1 and Stage 4 of \Comet.  Solid grey line is the median of the baseline IR/ML techniques, dotted grey lines are the best and worst performing IR/ML techniques respectively.}
\end{marginfigure}

\subsection{RQ\texorpdfstring{$_4$}{4} Results: C{\footnotesize OMET} Holistic Performance}
\label{sub:results-rq4}

The AP results for the the holistic \Comet (Stage 4) model are given in Table \ref{tab:stage1-4-results}. These results show that \Comets holistic model outperforms the baseline median IR/ML techniques, and Stage 1 for all subject programs. For three subjects (LibEST Req$\rightarrow$Src, EBT, and iTrust), Comet's holistic model matches or outperforms the best baseline IR/ML technique. \figref{fig:pr4-results} illustrates the P/R curve for the LibEST (Req$\rightarrow$Src) dataset, which shows that the performance gains in inference precision extend for a large range of recall values. The results of these experiments demonstrate that \Comets holistic model is able to effectively combine information from multiple sources to improve its trace link inference accuracy.

\subsection{RQ\texorpdfstring{$_5$}{5} Results: Industrial Case Study}
\label{sub:results-rq6}

\begin{marginfigure}
\centering
\includegraphics[width=\columnwidth]{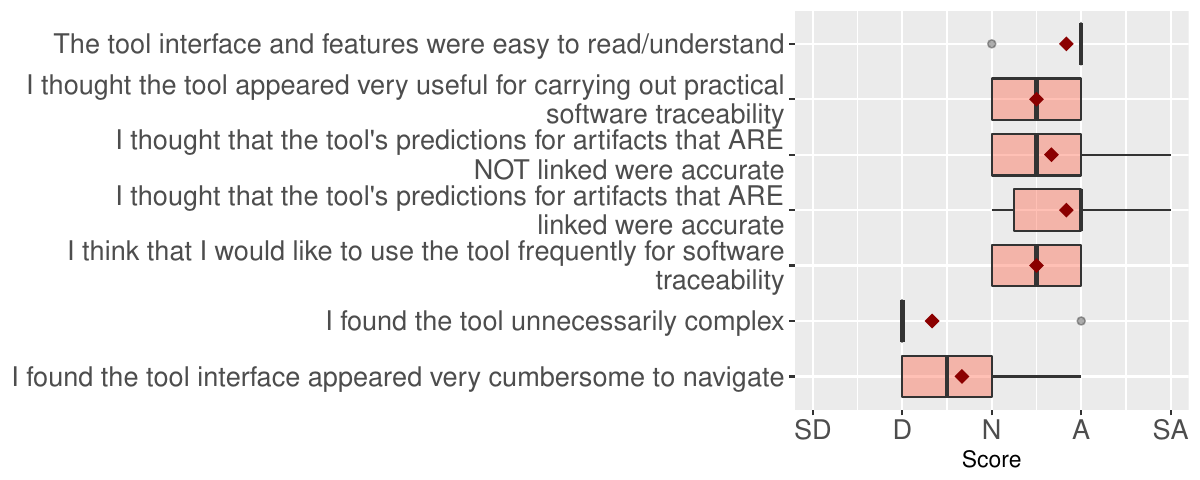}
\caption{Results for LibEST Case Study UX Questions}
\label{fig:LibEST-study}
\end{marginfigure}

\figref{fig:LibEST-study} responds to the likert-based UX questions from the six developers who work on the LibEST project after interacting with the \Comet plugin. Overall, the responses from these developers were quite positive. They generally agreed the \Comet plugin easy to use and understand, but more importantly, generally found the accuracy of the inferred links and non-links to be accurate. Additionally, we highlight representative responses to the user experience questions in this section, and provide the survey questions with response summaries in our online appendix~\citep{comet2020}, in accordance with the NDA established with our industrial partner. Overall the developer responses were encouraging, indicating the practical need for approaches like \Comet. For instance, one developer stated their need for such a tool, \textit{``I really want a tool that could look at test cases and requirements and tell me the coverage. That way the team can know whether we are missing functionality or not.''} Another developer explained the need for a feature that incorporates developer feedback, stating the importance of the \textit{``ability to describe or explain how the code matches up with the code for future reference. Discussion/comments about such explanation as different developers might see links that others don't''}, whereas another developer stated, \textit{``Being able to provide feedback is useful and seeing it update the percentage immediately was nice.''}  This indicates that the support for developer feedback and responsiveness of the \Comet plugin inherently useful. Developers also found the traceability report to be useful, with most criticism suggesting practical UI improvements. For instance, developers appreciated \textit{``The fact that there were the three different options for viewing the traceability between different [artifacts]''}, and \textit{``The ability to bring up the specific requirement quickly in the same window.''}. These responses illustrate the utility that developers saw in the \Comet plugin. Given that these developers had little automated support for traceability tasks, they appreciated any automated assistance. 

We also collected feedback that validated the importance of the practical use cases that the \Comet plugin enabled. In these interviews, the teams generally stated that \Comet would be very useful for code auditing, as one manager stated that it would \textit{``allow compliance analysts to [inspect] links, look at the code and validate [the links]''}. Furthermore, a team responsible for security audits of systems found an interesting use case for \Comet that is often overlooked in traceability analysis. That is, they were interested in code and requirements that are \textit{not linked to any other artifact}, as such artifacts are likely to be suspicious and should be inspected further. In this case, \Comets inferences of non-links would be just as important as the inferences of links. Overall, the interviewed teams saw great promise in \Comet, and expressed interest in adoption.

%% file: tables/chap_07_retrieval-bayes/tab_results_stage1.tex
\begin{table}[ht]
	\caption{AP Results from Stages 1 \& 4 of \Comet. The given $p$ values from the Wilcoxon test measure the significance of performance variations between Stage 1, and Stage 4 of \Comets model compared to the median (Med.) baseline of IR/ML techniques. ``I=Net'' signifies the ``Industry-Net'' dataset.}
	
	\label{tab:stage1-4-results}
    
\begin{adjustbox}{width=\textwidth}
\begin{tabular}{@{}l|c|c|c|c|c|c|c@{}}
\toprule
\multicolumn{1}{c|}{\textbf{Dataset}} &
  \textbf{\begin{tabular}[c]{@{}c@{}}Best\\ Base.\end{tabular}} &
  \textbf{\begin{tabular}[c]{@{}c@{}}Med.\\ Base.\end{tabular}} &
  \textbf{std. Err} &
  \textbf{DL} &
  \textbf{St.1} &
  \textbf{std. Err} &
  \textbf{St.4} \\ \midrule
LibEst (Rq to Src)  & 0.69 & 0.55 & $\pm$ 0.008 & 0.28 & 0.63 & $\pm$ 0.006 & 0.64 \\
LibEst (Rq to Test) & 0.42 & 0.36 & $\pm$ 0.001 & 0.32 & 0.38 & $\pm$ 0.002 & 0.42 \\
eTour               & 0.40 & 0.30 & $\pm$ 0.011 & 0.05 & 0.05 & $\pm$ 0.002 & 0.36 \\
EBT                 & 0.17 & 0.14 & $\pm$ 0.005 & 0.07 & 0.07 & $\pm$ 0.001 & 0.17 \\
SMOS                & 0.29 & 0.25 & $\pm$ 0.003 & 0.16 & 0.16 & $\pm$ 0.001 & 0.27 \\
iTrust              & 0.17 & 0.13 & $\pm$ 0.006 & 0.01 & 0.01 & 0        & 0.17 \\ \bottomrule
\end{tabular}

\end{adjustbox}

\end{table}

%% file: tables/chap_07_retrieval-bayes/tab_results_stage2.tex
\begin{table}[h]
	\centering
	\caption{AP Results from Stage 2 of \Comet with simulated expert feedback with error rates of 25\% and 50\%. The Baseline AP reported in this table is the median of the IR/ML techniques for the sampled links affected by feedback.}
	\label{tab:stage2-results}
	\setlength{\tabcolsep}{0.1em}
    
\begin{adjustbox}{width=\textwidth}	
\begin{tabular}{@{}l|c|c|c|c|c|c@{}}
\toprule
\multicolumn{1}{c|}{\textbf{Dataset}} & \textbf{Baseline} & \textbf{St.1} & \textbf{St.2 (25\%E)} & \textbf{Baseline} & \textbf{St.1} & \textbf{St.2 (50\%E)} \\ \midrule
LibEst (Rq to Src)  & 0.52 & 0.65 & 0.96 & 0.52 & 0.65 & 0.64 \\
LibEst (Rq to Test) & 0.28 & 0.32 & 0.80 & 0.28 & 0.32 & 0.44 \\
eTour               & 0.48 & 0.60 & 0.66 & 0.48 & 0.60 & 0.39 \\
EBT                 & 0.20 & 0.22 & 0.38 & 0.20 & 0.22 & 0.24 \\
SMOS                & 0.18 & 0.17 & 0.39 & 0.18 & 0.17 & 0.17 \\
iTrust              & 0.12 & 0.15 & 0.25 & 0.12 & 0.15 & 0.10 \\ \bottomrule
\end{tabular}
\end{adjustbox}

\end{table}

%% file: tables/chap_07_retrieval-bayes/tab_results_stage3.tex
\begin{table}[h]
	\centering
	\caption{AP Results from Stage 3 of \Comet for transitive links between requirements with $\tau$=0.55 and $\tau$=0.65. The baseline reported in this table is the median of the IR/ML techniques for links affected by transitive relationships.}
	\label{tab:stage3-results}
	\setlength{\tabcolsep}{0.1em}

\begin{adjustbox}{width=\textwidth}
\begin{tabular}{@{}l|c|c|c|c|c|c@{}}
\toprule
\multicolumn{1}{c|}{\textbf{Dataset}} & \textbf{Baseline} & \textbf{St.1} & \textbf{St.3 (tau=.55)} & \textbf{Baseline} & \textbf{St.1} & \textbf{St.2 (tau=.65)} \\ \midrule
LibEst (Rq to Src)  & 0.53 & 0.60 & 0.59 & 0.39 & 0.67 & 0.44 \\
LibEst (Rq to Test) & 0.38 & 0.40 & 0.38 & 0.18 & 0.19 & 0.22 \\
eTour               & 0.33 & 0.40 & 0.42 & 0.37 & 0.48 & 0.48 \\
EBT                 & 0.24 & 0.26 & 0.24 & 0.02 & 0.03 & 0.06 \\
SMOS                & 0.19 & 0.20 & 0.19 & 0.24 & 0.23 & 0.24 \\
iTrust              & 0.11 & 0.14 & 0.15 & -    & -    & -    \\ \bottomrule
\end{tabular}
\end{adjustbox}

\end{table}

%% file: chapters/part_01_chap_01/sec_04_discussion.tex
\section{Discussion}
\label{sec:discussion-hbn}

\subsection{Threats to Validity}
\label{sub:threats}
Similar to past work on automated trace link recovery approaches~\citep{Guo:ICSE'17}, our work exhibits two main threats to validity. The first of these threats is to external validity. We utilize a limited number of systems to carry out our evaluation of \Comet, and thus it is possible that our results may not generalize to other systems. However, the systems utilized in our evaluation have been widely used in past work, and are of varying sizes and domains. We also examine one industrial grade open source project developed by our partner.
	
The second threat to validity that affects experimental evaluation concerns construct validity, and more specifically, the accuracy of the ground truth for our subjects, and our implementation of the DL approach by Guo \etal~\citep{Guo:ICSE'17}. We cannot guarantee that the ground truth links for all of subjects are perfectly accurate. However, the ground truth sets for the CoEST datasets have been accepted by several pieces of prior work~\citep{Antoniol:ICSE'00,Cleland-Huang:TSE'03,Poshyvanyk:TEFSE'11,Gethers:ICSM'11}. The ground truth for LibEST was derived by a team of the authors, and was validated with the help of industrial developers working on the project (see Sec. \ref{sub:exp-context}). We re-implemented Guo \etal's DL approach closely following the details of the paper, although a full replication was not possible due to previous use of a proprietary industrial dataset. We will release our code~\citep{moran_improving_2020} for this approach to aid in reproducibility. Another potential threat to validity is that our simulation of developer feedback in answering RQ$_2$ is not representative of real feedback. Due to constraints on developers time, we could not use real feedback, however, we believe simulating a small number of links using the ground truth, complete with error rates, represents a reasonable approximation.

\subsection{Related Work}
\label{sub:related-work}

We focus our discussion of related work on prior techniques that have, in limited contexts, (i) considered novel or hybrid textual similarity measures, (ii) modeled the effects of multiple types of artifacts, or (iii) incorporated developer expertise. We then conclude with a statement distilling \Comets novelty.

\noindent{\textbf{Novel/Hybird Textual Similarity Measures:}} Guo \etal~\citep{Guo:ICSE'17} proposed an approach for candidate trace link prediction that uses a semantically enhanced similarity measure based on Deep Learning (DL) techniques. However, unlike \Comet, this technique requires pre-existing trace links in order to train the DL classifier.  In contrast, \Comet does not require known links for the projects it is applied to, but rather requires a project to serve as a tuning set. We show that \Comet performs well when tuned and tested on different datasets, outperforming Guo \etals DL-based approach when it is trained in a similar manner. Gethers \etal~\citep{Gethers:ICSM'11}, implemented an approach that is capable of combining information from canonical IR techniques (\ie VSM, Jensen-Shannon) with Topic Modeling techniques. However, their approach can only combine two IR/ML techniques, whereas \Comet can combine and leverage the observations from several IR/ML techniques, and combine this with other information such as expert feedback and transitive links. 

\noindent{\textbf{Modeling of Multiple Artifacts:}} Rath \etal~\citep{Rath:ICSE'18} recently explored linking nontraditional information including issues and commits, and Cleland-Huang \etal~\citep{Cleland-Huang:ICSE'10} have investigated linking regulatory codes to product level requirements.  \Comets model has the potential to improve trace link recovery in these scenarios both through its more robust modeling of textual similarity, and through incorporation of transitive link information. Furtado \etal~\citep{Furtado:RE'16}, explored traceability in the context of agile development, and Nishikawa \etal~\citep{Nishikawa:ICSME'15} first explored the use of transitive links in a deterministic traceability model. Additionally, Kuang \etal used the closeness of code dependencies, to help improve IR-based traceability recovery~\citep{Kuang:SANER'17}. However, none of these approaches is capable of incorporating transitive links while also considering combined textual similarity metrics and developer feedback.

\noindent{\textbf{Incorporation of Developer Expertise:}} De Lucia \etal \citep{DeLucia:ICSM'06} and Hayes \etal~\citep{Hayes:TSE'06} analyzed approaches that use relevance feedback to improve trace link recovery. However, these approaches are either tied to a particular type of model (such as TF-IDF~\citep{DeLucia:ICSM'06}), or require knowledge of the underlying model to function optimally. In contrast, \Comet implements a lightweight, likert-based feedback collection mechanism that we illustrate can improve link accuracy even when only a small amount of feedback is collected.
 
\noindent{\textbf{Summary of Advancement over Prior Work:}} \Comets features facilitate its application to projects without any pre-existing trace links, and as our evaluation illustrates, allow it to perform consistently well across datasets. \Comet is able to combine information from transitive links with both robust textual similarity measures and lightweight developer feedback for improved accuracy. While some aspects of \Comets approach have been considered in limited contexts in prior work -- such as developer feedback~\citep{DeLucia:ICSM'06,Hayes:TSE'06} and restricted combinations of IR/ML techniques~\citep{Gethers:ICSM'11} -- there has never been a framework capable of combining all these aspects in a holistic approach. Our evaluation illustrates that \Comets holistic HBN is able to outperform baseline techniques on average.

%% file: chapters/part_01_chap_02/nature.tex
\chapter{On Interpreting the Effectiveness of Unsupervised Software Traceability with Information Theory}
\label{ch:nature}

\lettrine[lines=2]{\textbf{T}}{}raceability is a cornerstone of modern software development, ensuring system reliability and facilitating software maintenance. While unsupervised techniques leveraging Information Retrieval (IR) and Machine Learning (ML) methods have been widely used for predicting trace links, their effectiveness remains underexplored. In particular, these techniques often assume traceability patterns are present within textual data - a premise that may not hold universally. Moreover, standard evaluation metrics such as precision, recall, accuracy, or F1 measure can misrepresent the model's performance when underlying data distributions are not properly analyzed.
Given that automated traceability techniques struggle to establish links -- even for well-studied datasets -- properly, we need further insight into the information limits related to traceability artifacts. In this chapter, we propose an approach, called \infotheory, for using information theory metrics to evaluate and better understand unsupervised traceability techniques' performance (limits). Specifically, we introduce self-information, cross-entropy, and mutual information (MI) as metrics to measure the informativeness and reliability of traceability links. Through a comprehensive replication and analysis of well-studied datasets and techniques, we investigate the effectiveness of unsupervised techniques that predict traceability links using IR/ML. This application of \infotheory illustrates an imbalance in typical traceability datasets where the source code has on average 1.48 more information bits (\ie entropy) than the linked documentation. Additionally, we demonstrate that an average MI of 4.81 bits, loss of 1.75, and noise of 0.28 bits signify that there are information-theoretic limits on the effectiveness of unsupervised traceability techniques. We hope that these findings spur additional research on understanding the limits and progress of traceability research.

\input{chapters/part_01_chap_02/sec_01_intro}
\input{chapters/part_01_chap_02/sec_02_background}
\input{chapters/part_01_chap_02/sec_03_approach}
\input{chapters/part_01_chap_02/sec_04_design}

\input{chapters/part_01_chap_02/sec_05_results}
\input{chapters/part_01_chap_02/sec_06_case}

\input{chapters/part_01_chap_02/sec_07_conclusion}

%% file: chapters/part_01_chap_02/sec_01_intro.tex
\section{Introduction}
\label{ch:nature:sec_01}

In this chapter, we explore the phenomenon of information transmission in software traceability. Traceability embodies the study of drawing semantic relationships among software artifacts (\eg code, requirements, test cases). These semantic relationships are relevant to facilitate code comprehension \citep{moran_improving_2020}, compliance validation, security tracking \citep{palacio_security,Gadelha2021TraceabilityRB}, and impact analysis \citep{Aung2020ALR}. Usually, information retrieval techniques (IR) (\eg TF-IDF, LSA, or LDA) have been employed to mathematically represent \textit{high-level} (\ie requirements) and \textit{low-level} (\ie source code) software artifacts in compressed tensors \citep{Dit2013,Ge2012,Dasgupta2013,Dit2013a}. These tensors are generated via unsupervised learning and employed to calculate a distance (\eg Euclidean, Cosine, or Word Mover’s Distance) between two artifacts in a derived vector space. The distance defines how semantically close two artifacts are to each other. Unsupervised traceability focuses on finding patterns within high and low-level artifacts to confirm whether a data sequence pair \textit{source-target} is linked.

The effectiveness of unsupervised software traceability is measured in terms of canonical learning metrics such as precision, recall, AUC, accuracy, or F1. However, these metrics can be misleading when data are not properly explored and analyzed. For instance, in software traceability, datasets are generally imbalanced, skewed, and biased as we demonstrated in our empirical study. This observation is supported by studies exploring the importance of more rigorous statistical analysis of software engineering research~\citep{moran_improving_2020,Kitchenham2017RobustEngineering}. As such, we believe that there are \textit{data limitations} related to the underlying software artifacts that traceability techniques operate upon that limit their effectiveness.

This chapter aims to develop techniques that automatically articulate cases where the information captured in these datasets could make unsupervised techniques trace links ineffectively; therefore, resulting in erroneous predictions. For example, consider a situation where a development team needs to assess the impact of incoming new requirements. How should practitioners proceed if we assume they are unfamiliar with the code-based components of the system under analysis? Practitioners most likely will start by reading the associated documentation (\ie code comments) to \textit{trace} the functionality between requirements and code. Nonetheless, \textit{what if requirements are poorly written?  Or what if certain areas of the source code are not completely documented or obfuscated?} Likely, the traceability process --- practitioners would undertake to understand the impact of the new requirements --- would be cumbersome and inefficient. We aim to use information science and automated traceability to identify potential artifacts that might negatively contribute to the understandability of a given system.

By assuming that \textit{data} is the central aspect of learning theory, we can claim insufficient or poorly treated data is a deficiency reflected in the effectiveness of unsupervised algorithms \citep{bishop_deep_2024}. Unfortunately, information retrieval techniques and -more generally- unsupervised models cannot always guarantee that any \textit{predicted} trace link is reliable. Because a data-dependent gap exists among software artifacts, unsupervised traceability is a difficult and error-prone task. Consequently, \textit{bridging the traceability research gap is infeasible under the assumption that data is poorly structured or of low quality}. Traceability research requires efficient statistical approaches that quantify how \textit{reliable} predicted trace links may be. 

Textual artifacts may not contain enough information to allow an unsupervised technique to learn patterns that determine a trace link. We hypothesize that information-theoretic measures (\eg self-information, mutual information, relative entropy, and shared information) can help \textbf{interpret or explain} why unsupervised techniques are \textit{limited for solving the traceability problem}. Unsupervised models (\ie conventional, machine learning, or even Large Language Models) bound traceability effectiveness when data are not informative enough. By performing an information-theoretic analysis, we can estimate the extent unsupervised techniques are robust, reliable, and trustworthy. We named this analysis \infotheory, an interpretability approach that provides a starting point to monitor traceability distances, information-theoretic measures, and the relationship between distances and information measures to understand whether and why unsupervised techniques may be performing poorly in trace link prediction.

This research aims to demonstrate that traceability models are limited when they rely only on unsupervised learning and textual data. By conducting an extensive (information-based) empirical analysis, we show why unsupervised techniques are inadequate for some training configurations and data modalities due to ineffective predictions. Our research is therefore a data-centric analysis oriented to identifying the quality of testbeds employed and reported in the traceability literature. We briefly discuss an overview of our findings from our exploratory analysis below.

For the \cisco system, we found that pull request comments and the associated code often contain contrasting information. The pull request comments, with an entropy of $3.42[0.02]$\bits, and source code comments, with $5.91[0.01]$\bits, describe disparate information. Such information imbalance reflects that code changes could not have accurately captured the information content of the pull request at implementation time. In general, we recommend examining imbalanced links, or information discrepancies, between pull requests and source code to design a refactoring strategy oriented toward reducing the information loss and augmenting the mutual information among the software documentation. Note that the level of mutual information is around $3.21\mathcal{B}$ with an error of $0.02$ for a confidence of 95\%. An ideal range of mutual information can vary according to the system. However, researchers should avoid lower values in the mutual information range. Our full empirical study contains negative results of unsupervised models, a whole analysis of the information transmission on different testbeds, and a correlation study between semantic distance and information metrics.

%% file: chapters/part_01_chap_02/sec_02_background.tex
\section{Background \& Related Work}
\label{ch:nature:sec_02}

Software requirements should be amenable to being \textit{translated} into multiple forms of information such as source code, test cases, or design artifacts. Thus, we refer to these requirements or any initial/raw form of information as the \textit{source} artifacts. Conversely, the information product of a transformation or alteration is considered a \textit{target} artifact. In the software engineering context, a transformation could be any action that software engineers or practitioners apply from those requirements to source code. For instance, implementing a requirement can be seen as a way of \textit{translating} information from the requirements to software components. Our research deals with the intersection of software traceability and interpretability techniques to introduce statistical and information theory methods to explain the ineffectiveness of unsupervised traceability. 

\textbf{Software Traceability Techniques.} Traceability datasets are composed of corpora of documents and code files. These text artifacts are used to establish a link that relates the content of a source document to the content of a target code file. The links have been recovered by employing distinct techniques from \textit{Information Retrieval} (e.g., Vector Space Model, Jensen-Shannon Vectors, or Topic Models)\citep{dit,Dit2013,Dit2013a,Dit2013b} or Machine Learning (\eg \wv, \dv, AutoEncoders, and Transformers) \citep{palacio_security,transformer_trace_21,Lin2022EnhancingAS, Du2020AutomaticTL}. However, these techniques misleadingly assume that the distribution and structure of tokens, \eg words or variable names, present in requirements and source code are similar. Even though source code is essentially \textit{a sequence}, it is more constrained, repetitive, and structured than natural languages \citep{Karampatsis2020BigCode,hindle2012natural}. Ergo, many of the similarities these unsupervised techniques attempt to exploit are flaky and spurious, observed in the wide range of performance (\ie high variability) these techniques have on different datasets. To date, no study has looked specifically at measuring entropy levels in traceability exploratory analysis. Previous research in traceability has largely overlooked the role of entropy or any information metric when evaluating unsupervised models. Nevertheless, we found studies commenting on the importance of modeling software artifacts as probability distributions~\citep{moran_improving_2020}.  

\textbf{Interpretability Techniques.} Interpretability has been used to explain the predictions generated by (deep) learning models, complementing traditional evaluation methods and enhancing our understanding of the decision-making process to reduce uncertainty. Post-hoc interpretability assumes the model is a black box and analyzes its behavior by examining changes in the output derived from the input. In  Machine Learning, most research has focused on improving the plausibility and faithfulness of explanations such as LIME \citep{ribeiro2016why}, DeepLIFT \citep{shrikumar_learning_2019}, and Shapley values \citep{lundberg2017unified}, to the best of our knowledge, our work is the first to evaluate unsupervised techniques for the traceability problem with information theory.

%% file: chapters/part_01_chap_02/sec_03_approach.tex
\section{Information Theory for Traceability}
\label{ch:nature:sec_03}

\begin{marginfigure}
		\centering
    \includegraphics[width=\linewidth]{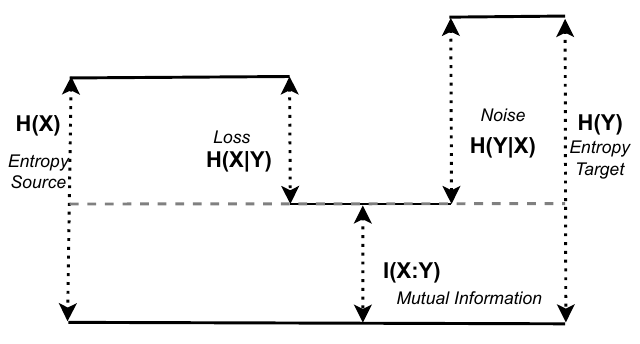}
		\caption{Information Theory Measures in \infotheory.}
        \label{fig:info-theory}
\end{marginfigure}

\begin{figure}[ht]
		\centering
    \includegraphics[width=0.9\textwidth]{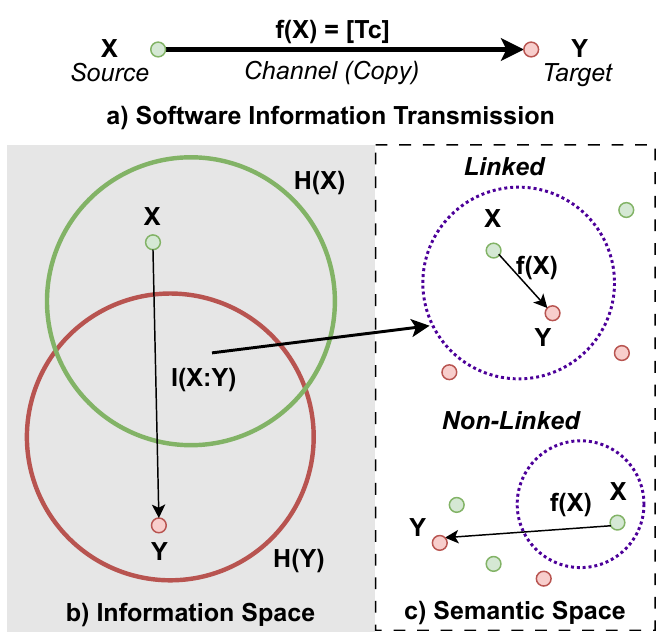}
		\caption{\infotheory: Using Information Theory to Interpret Unsupervised Traceability Models.}
        \label{fig:tracexplainer}
        \vspace{1.2em}
\end{figure}

This section briefly introduces some elements of information theory required to generate advanced data analysis such as unsupervised interpretability. We present our approach \infotheory as a data-centric solution for the analysis of software traceability. \infotheory aims to calculate a set of information measures to complement and explain the limitations of semantic traceability techniques. Understanding such limitations (\ie bounds) will allow us to assess how well traceability algorithms perform for any set of software projects. Our experiments have established that studying the manifold of information measures (\ie information and semantic spaces) might help us detect critical points in the artifacts. For instance, traceability links undetected in overlapping information areas of two artifacts $h(x)$ and $h(y)$ (see \figref{fig:info-theory}). These undetected links can be the product of potentially missed documentation or repetitive tokens that need to be refactored to enhance the effectiveness of traceability algorithms. \infotheory poses a set of information measures and exemplifies their usefulness in software traceability. \figref{fig:tracexplainer} depicts three components of the approach: software information transmission, the information, and the semantic spaces. 

\textbf{Software Information Transmission.} Probability theory and \textit{information theory} are two important frameworks that form the basis of unsupervised interpretability for software traceability. In particular, information theory helps us quantify the amount of information present in software corpora. The software information content can be viewed as the \textit{degree of surprise} on learning the value of a single artifact $x$, a discrete random variable. We receive more information for a less probable event than a more probable one. Consequently, a measure of software information content relies on a probability distribution $p(x)$. This measure is a monotonic function $h(x)$ of the probability $p(x)$ that expresses the information content based on the logarithm $h(x) = -log_2p(x)$. Note that the higher $p(x)$ (\ie more frequent), the lower the information content. Since the information, in this case, has been defined in terms of logarithms to the base of 2, the units of $h(x)$ are \textit{bits} (\ie binary digits $\mathcal{B}$). Otherwise, if the information content is expressed in natural logarithms, the units are now in \textit{nats}~\citep{bishop_deep_2024}.

Artifacts are sequential-based structures (\eg code, requirements, tests) that capture information outcomes from different software data generation processes (\eg programming, eliciting, managing). We can monitor how much information is received when we observe a specific value for the variable $x$. The artifacts can naively be represented as the relative frequency of tokens $t$ contained in a sequence. The frequency of those tokens produces a probabilistic distribution $p(x=t_i) = t_i/|t|$, where $t_i$ is the count for a specific token in a vocabulary $\mathbf{v}$ and $|t|$ is the number of tokens in a sequence. Now consider that a source (\ie sender) wants to transmit the artifact value $x$ to a target (\ie receiver). The average amount of software information, which is transmitted in the process, reflects the expectation of $h(\cdot)$ with respect to the distribution $p(\cdot)$:

 \begin{equation}
H[x] = -\sum_x p(x) \log_2 p(x)
\label{eq:entropy}
\end{equation}

\equaref{eq:entropy} is a quantity named \textit{entropy} of the random variable $x$.
Therefore, we say that a software engineering process can generate an amount $m$ bits of sequence-based artifact. 

Software information can be transmitted by copy $[T_c]$ or transformation $[T_s]$ and measured by \textit{Mutual Information} $I(x:y)$. However, mutual information $I(\cdot)$ cannot distinguish between a scenario where the target is a \textit{copy} or a \textit{transformed} version of the source artifact. Since the basic unit in software artifacts is a set of tokens, the information within the source $x$ can be transmitted by copy if the same number of tokens is found in the target $y$. While the source $x$ can be transmitted by transformation if the target $y$ suffers a systematic modification in its syntax or semantics ~\citep{Kolchinsky2020DecomposingTransformation}. Although software information can be transmitted in both ways, we assume in our empirical experimentation that a function $f(X)$ is by default a \textit{copy} process. On the other hand, addressing the transformation problem requires further investigation in a formal field of \textit{information theory for software engineering}. Moreover, our research contributes to formalizing information measures for software artifacts and bridging the gap between information theory and unsupervised learning for traceability. \figref{fig:info-theory} depicts different interactions of the \textit{entropy} for two artifacts (\ie source and target). These interactions are other related measures such as self-information, loss, noise, and \textit{mutual information}~\citep{MacKay2003InformationAlgorithms}.

\subsection{\texorpdfstring{$[H(X)]$}{Hx}: Source Artifacts Self-Information}

Source self-information is the entropy of source artifacts. Consequently, this measure represents the amount of encoded information (\ie in $\mathcal{B}$ bits) in artifacts representing a sender such as requirements, issues, or pull requests. The following example shows the entropy of a real \cisco, an industrial dataset, pull request:

\begin{itemize}
\item Source artifact content: \colorbox{gray!30}{“B dtimeout”} 
\item $H(x)=0.0\mathcal{B}$ (no info.)
\end{itemize}
\textit{Interpretation.} Generally, having a pull request with few (and repetitive) tokens (\ie words) negatively impacts self-information values. This metric is useful for measuring the content of information in the source. If we track the information content, we will create the conditions to explain \textit{why} traceability occurs. 

\subsection{\texorpdfstring{$[H(Y)]$}{Hy}: Target Artifacts Self-Information}

Target self-information is the entropy of target artifacts. Intuitively, this measure encompasses the information amount in artifacts representing a receiver such as code, test cases, or configuration files. The following example presents the entropy of a real \cisco python file showing a standard exception:

\begin{center}

\begin{minipage}{0.43\textwidth}

\begin{lstlisting}[language=Python]
import sys
import traceback

def fireException(message):   
    try:
        with open("buddy_script_error.txt","w") as file:         
        file.write(str(message))        
        print(message)  
        traceback.print_exc()
        sys.exit(-1)
    except IOError:
        traceback.print_exc()
        sys.exit(-1)
\end{lstlisting} 

\end{minipage}

\end{center}

\textit{Interpretation.} The information amount encompassed in the Python file required $H(y)=4.16\mathcal{B}$. Note that the source content is more diverse than the target, assuming both artifacts are linked. Thus, target self-information is useful when we compare it with the corresponding source. It helps us to detect imbalanced information in a trace link.

\subsection{\texorpdfstring{$[I(X:Y)]$}{Ixy}: Mutual Information }

Mutual Information (\textit{MI})  embodies the information amount that sources $X$ \textit{see} of targets $Y$ (and, complementary, $Y$ \textit{see} of $X$). Therefore, \textit{MI} quantifies the information two artifacts hold in common (\ie the information transmitted from a source to a target). For example, the trace link pull request to code $\{PR_{256} \to binaryfunc.py\}$ has an \textit{MI} of $6.38\mathcal{B}$, which corresponds to the maximum entropy found in \cisco. This value suggests that the content in source artifacts overlaps a large portion of the content in the target artifact. However, the potential link $\{PR_{56} \to binaryfunc.py\} $ has the same  \textit{MI} as the previous link but it is not a \textit{real} link according to the ground truth. When we observe the \textit{Case Study 0} (see \secref{ch:nature:sec_06}), we note both pull requests $PR_{56}$ and $PR_{256}$ exhibit the maximum self-information observed in \cisco as well as the artifact $binaryfunc.py$. Our mutual information analysis suggests the potential link $\{PR_{56} \to binaryfunc.py\} $ could be an information link not considered by the ground truth. 

\subsection{\texorpdfstring{$[H(X|Y)]$}{Hxy}: Information Loss}
Loss is the amount of information that comes into but does not come out of the \textit{channel}. Thus, the information cannot be found in the target since either code files have not been implemented yet or some requirements or code files are poorly documented. For instance, the links $\{PR_{240} \to binaryfunc.py\}$, $\{PR_{240} \to securityfunc.py\}$, and $\{PR_{168} \to auth.py\}$ present a loss of $6.51\mathcal{B}$, $6.48\mathcal{B}$, and $6.3\mathcal{B}$ respectively. These pull requests from \cisco have in common the low value for self-information content. Therefore, we concluded these requests were not well described in natural language (see more cases in~\citep{palacio2023tracexplainer}).

\subsection{\texorpdfstring{$[H(Y|X)]$}{Hyx}: Information Noise}

Noise is the information amount that comes out (or is found in the target) but does not come in (or is never depicted in the source). That is, information implemented by a developer that was never mentioned in the issues or requirements. For instance, the links $\{PR_{194} \to$ \url{__init__.py} $\}$, $\{PR_{177} \to fireException.py \}$, and $\{PR_{56} \to setup.py\}$ exhibit a noise of $1.83\mathcal{B}$, $1.81\mathcal{B}$, and $1.55\mathcal{B}$ respectively. These code files present low information content since they are configuration files. Therefore, an interesting question arises: \textit{why is there a link between a pull request and a configuration file?} (see more cases in~\citep{palacio2023tracexplainer}). 

\subsection{\texorpdfstring{$[Si(X:Y)]$}{Sxy}: Minimum Shared Information}

We introduce the concept of \textit{Minimum Shared of Information (MSI)} for entropy $Si(X:Y)$ and extropy\footnote{Extropy is the inverse function of the entropy. In this case, it measures the missing information in a given software artifact.} $Sx(X:Y)$. \textit{MSI} differs from  \textit{mutual information} because it solely considers the minimum overlapping of tokens between two artifacts to compute the entropy. For instance, an artifact $A$ has the token count of $A=[(for,14), (if,3), (return,10)]$ and artifact $B$ has the token count $B=[(for,10), (if,0), (return,20)]$. The minimum shared vector would be $min(A,B)$ or \\ $AB=[(for,10),(if,0),(return,10)]$. The shared vector represents a probability distribution by which we can compute both minimum entropy and extropy. This minimum entropy is useful for detecting null shared vectors or artifacts whose tokens are completely opposed. Consequently, \textit{MSI} is the minimum number of tokens shared between the source and target set represented as entropy. For instance, \cisco contains 5417 potential links whose shared vector is null. Of those potential links, 68 are real links (\ie ground truth). 

%% file: chapters/part_01_chap_02/sec_04_design.tex
\section{Empirical Design}
\label{ch:nature:sec_04}

Our empirical analysis explores information measures among software artifacts and presents the relationship between these measures and the effectiveness of unsupervised techniques in interpreting the traceability problem. The focus of our study is on using \infotheory to interpret the (negative) effectiveness of traceability techniques and practical applicability. We formulate the following set of RQs: 

\begin{enumerate}[label=\textbf{RQ$_{\arabic*}:$}, ref=\textbf{RQ$_{\arabic*}$}, wide, labelindent=5pt]\setlength{\itemsep}{0.2em}
      \item \label{rq_nature:effectivess} \textit{How effective are unsupervised techniques at predicting candidate trace links using IR/ML representations?}
      \item \label{rq_nature:semantic} \textit{To what extent are semantic metrics imbalanced to the ground truth?}      
      \item \label{rq_nature:exploratory} \textit{How much information is transmitted from source to target artifacts?}
      \item \label{rq_nature:correlation} \textit{To what extent do information metrics correlate with semantic distances?}
\end{enumerate}

\subsection{Experimental Context}

\input{tables/chap_infotheory/tab1_systems_dataset}

The experimental context aims to answer our RQs from the last sub-section and comprises a set of datasets, trained models, and a set of testbeds to apply our approach.

\textit{Datasets and Training Models.}  For evaluating our metrics on traceability recovery we pretrained our models by varying the preprocessing type, the vectorization type, and the pertaining dataset. The preprocessing type could be a conventional unsupervised tokenizer based on NLTK\citep{bird-loper-2004-nltk}, bpe8k, or bpe32k. Both BPE encoders, bpe8k, and bpe32k are subwords units from SentencePiece tokenizer \citep{sentencepiece}. The vectorization technique is skip-gram for \wv and paragraph vector bg of words (pv-bow) for \dv. The pretraining dataset was performed with CodeSearchNet for Java and Python\citep{husain2019codesearchnet}, and the Wikipedia set\citep{tensorflowWikipediaDataset}. The embedding size was 500 and epochs were 20 for each model.

\textit{Testbeds.} We use seven system testbeds, six from public sources and one collected by a Cisco Intern. \libest\citep{libest} is an open-source project maintained by Cisco. \libest includes traceability links between requirements and test cases in C (req2tc). \itrust, \etour, and \smos are widely recognized CoEST datasets for the use case to source code traceability (uc2src). The Event-Based Traceability dataset \ebt links English requirements to Java code, while \albergate focuses on traceability between Java classes in a hotel management system. \tabref{tab:datasets} summarizes the datasets, including artifact counts and traceability links.

\textit{Experimental Setup.} \tabref{tab:experiments} summarizes the experiments depending on the evaluation system testbed, model pretrain parameters, and IR link type. Each configuration is evaluated using both vectorization types word2vec and doc2vec. Our base experiment \exbase compresses the use of a pretrained model with Java and Python sources using conventional preprocessing to recover the requirement. Each experiment aims to introduce a perturbation and measure to what extent the perturbation affects the entropy level. Therefore, the experiments \exfirst and \exsecond variate on the preprocessing tokenizer bpe8k and bpe32k, respectively. The \exthird uses conventional preprocessing but uses a pretrained model with a Wikipedia dataset. Finally, \exfourth variates on the evaluation testbeds using the same model at the \exbase.

\textit{Metrics.} To evaluate the formulated RQs, we used our previously described datasets to include the traceability corpora from industry and research and introduced unsupervised machine learning models for predicting traceability links. Each corpus has multiple types of software artifacts as sources and targets such as requirements, test cases, and source code, as well as their ground truth links. To calculate the different information metrics, we used the Python library dit, which is a popular library for discrete information theory. We need to inform requirements or source artifacts to enhance information content to match target artifacts. The next sub-sections explain the metrics to answer each RQ.

\input{tables/chap_infotheory/tab2_models}

\subsection{\texorpdfstring{\ref{rq_nature:effectivess}}{rq1}: Neural Unsupervised Effectiveness}

To evaluate the effectiveness of neural unsupervised techniques for traceability, we computed the \textit{area under the curve (AUC)} from precision-recall curves and the \textit{receiver operating characteristic curve (ROC)} using the \textit{scikit-learn} API~\citep{sklearn_api}, similar to prior work that evaluates automated traceability techniques~\citep{moran_improving_2020,Dit2013b,Dasgupta2013,Guo2013FoundationsTraceability}.

\subsection{\texorpdfstring{\ref{rq_nature:semantic}}{rq2}: Semantic Traceability Imbalance} 

To answer \ref{rq_nature:semantic}, we estimated the following semantic metrics depending on the neural vectorization technique (\ie \wv or \dv): Euclidean distance (\euc), Soft Cosine Similarity (\scm), Cosine Distance (\cosine), and Word Mover's Distance (\wmd). We also computed the normalized distance inverse for \cosine and \wmd to obtain the \cosine similarity and \wmd similarity. These distances and similarities were computed for pairs of source-target artifacts in each testbed. These pairs are represented in the vector space after unsupervised encoding. We can segregate the distances and similarities by ground truth since we can classify potential links into actual links and non-links. Such segregation depicts imbalances in the semantic space (see \figref{fig:tracexplainer})

\subsection{\texorpdfstring{\ref{rq_nature:exploratory}}{rq3}: Exploratory Information Analysis} 

To compute the set of information measures introduced in \infotheory, we conducted an \textit{Exploratory Data Analysis (EDA)}. Our \textit{EDA} consisted of an exhaustive statistical search of patterns and descriptive statistics within the reported information theory measures. The exploration allows us to interpret how well an unsupervised technique for traceability performs. Our goal is to use information measures to describe and interpret the effectiveness of unsupervised traceability techniques. We proposed two analyses: 

$AN_1$: \textit{Manifold of Information Measures.} This exploration aims to determine the probability distribution of each entropy and similarity metric. We have some assumptions about the shape of the expected distributions. For instance, similarity distributions should be bimodal since we want to observe a link and a non-link. If our assumptions do not match the expected distribution, we can assess the quality of the technique. 

$AN_2$: \textit{Manifold of Information Measures by Ground truth.} This exploration segregates each entropy and similarity metric by the ground truth. The data division by ground truth allows us to interpret the prediction quality for similarity metrics. Additionally, it also allows us to describe how good the ground truth leverage process was for each testbed because we can monitor the information transmission between source and target artifacts with \infotheory.

\subsection{\texorpdfstring{\ref{rq_nature:correlation}}{rq4}: Correlation Analyses} 
We correlate semantic distance and similarity metrics with information theory measures in a scatter matrix using \textit{Pearson coefficient}. This analysis aims to expose highly correlated relationships across unsupervised distances and information measures to explain traceability outputs rationale with the information transmission setup proposed in \infotheory.

%% file: tables/chap_infotheory/tab1_systems_dataset.tex
\begin{margintable}
\centering
\caption{Testbeds}
\label{tab:datasets}
\begin{adjustbox}{width=\linewidth}

\setlength{\tabcolsep}{4pt} 
\begin{tabular}{lrrr}
\hline
\multicolumn{1}{c}{\textbf{}} & \multicolumn{3}{c}{\textbf{Datapoints}} \\
\multicolumn{1}{c}{\textit{\textbf{System}}} & \multicolumn{1}{c}{\textit{\textbf{All}}} & \multicolumn{1}{c}{\textit{\textbf{Links}}} & \multicolumn{1}{c}{\textit{\textbf{Non-Links}}} \\ \hline
\libest\citep{libest} & 1092 & 352 & 740 \\
\cisco & 21312 & 547 & 20765 \\
\albergate\citep{Antoniol:ICSE'00} & 935 & 53 & 882 \\
\ebt\citep{Cleland-Huang:TSE'03} & 2050 & 98 & 1952 \\
\etour\citep{coest-datasets} & 6728 & 308 & 6420 \\
\itrust\citep{coest-datasets} & 47815 & 277 & 47538 \\
\smos\citep{6080780} & 6700 & 1044 & 5656 \\ \hline
\end{tabular}
\end{adjustbox}
\end{margintable}

%% file: tables/chap_infotheory/tab2_models.tex
\begin{table}[ht]
\centering
\caption{Experiment design.}
\label{tab:experiments}
\scalebox{0.85}{
\setlength{\tabcolsep}{4pt} 
\begin{tabular}{cllllll}
\textbf{Experiment} & \multicolumn{3}{c}{\textbf{Evaluation Testbeds}} &  & \multicolumn{2}{c}{\textbf{Information Retrieval}} \\
\textit{\textbf{ID}} & \multicolumn{1}{c}{\textit{\textbf{System}}} & \multicolumn{1}{c}{\textit{\textbf{Language}}} & \multicolumn{1}{c}{\textit{\textbf{Preproc.}}} & \multicolumn{1}{c}{} & \multicolumn{1}{c}{\textit{\textbf{LinkType}}} & \multicolumn{1}{c}{\textit{\textbf{Pretrain ds}}} \\ \hline
\multirow{2}{*}{\exbase} & \libest & english & \multirow{2}{*}{conventional} &  & req2tc & \multirow{2}{*}{Java \& py} \\
 & \cisco & english &  &  & pr2src &  \\ \hline
\multirow{2}{*}{\exfirst} & \libest & english & \multirow{2}{*}{bpe8k} &  & req2tc & \multirow{2}{*}{Java \& py} \\
 & \cisco & english &  &  & pr2src &  \\ \hline
\multirow{2}{*}{\exsecond} &  \libest  & english & \multirow{2}{*}{bpe32k} &  & req2tc & \multirow{2}{*}{Java \& py} \\
 & \cisco & english &  &  & pr2src &  \\ \hline
\multirow{2}{*}{\exthird} &  \libest  & english & \multirow{2}{*}{conventional} &  & req2tc & \multirow{2}{*}{\begin{tabular}[c]{@{}l@{}}wiki \& \\ Java \& py\end{tabular}} \\
 & \cisco & english &  &  & pr2src &  \\ \hline
\multirow{5}{*}{\exfourth} & \albergate & italian & \multirow{5}{*}{conventional} &  & req2src & \multirow{5}{*}{Java \& py} \\
 & \ebt & english &  &  & req2src &  \\
 & \etour & italian &  &  & uc2src &  \\
 & \itrust & english &  &  & uc2src &  \\
 & \smos & italian &  &  & uc2src &  \\ \hline
\end{tabular}
}
\end{table}

%% file: chapters/part_01_chap_02/sec_05_results.tex
\section{Results \& Discussion}
\label{ch:nature:sec_05}

Our empirical evaluation consists of four experiments focused on the following data science tasks: i) describing distance and information measures, iii) predicting trace links, and iii) estimating correlations between semantic distance and information transmission by varying the preprocessing strategy (\ie conventional or bpe), the type of embeddings (\ie \wv or \dv), and the pretraining dataset (\ie CodeSearchNet or Wikipedia). 

\input{tables/chap_infotheory/tab3_efectivenes}
\input{tables/chap_infotheory/tab4_self_information_and_distance}

\input{tables/chap_infotheory/tab5_self_information_by_links}

\subsection{\texorpdfstring{\ref{rq_nature:effectivess}}{rq1}: Traceability Effectiveness Results}

\begin{figure}[ht]
  \vspace{0.2cm}
  \begin{subfigure}{0.5\textwidth}
        \centering
        \includegraphics[width=\linewidth]{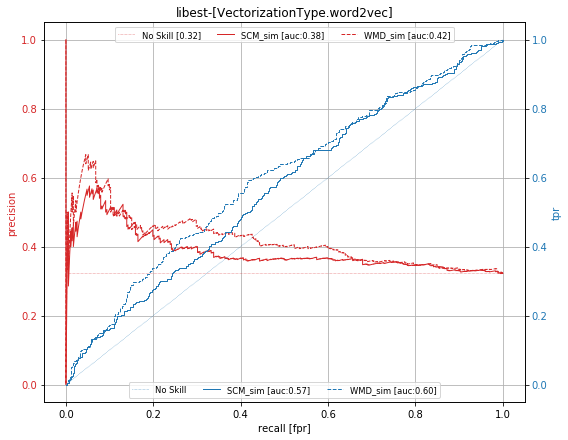}
        \caption{\tiny\libest \wv recall and precision}
        \label{fig:sub11}
    \end{subfigure}
    \begin{subfigure}{0.5\textwidth}
        \centering
        \includegraphics[width=\linewidth]{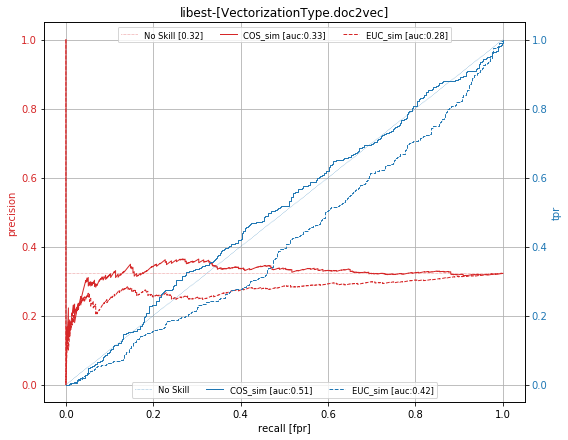}
        \caption{\tiny\libest \dv recall and precision}
        \label{fig:sub12}
    \end{subfigure}
     \caption{Precision and Recall for \wv and \dv using \libest}
    \label{fig:grid1}
    \vspace{1.2em}
\end{figure}

The traceability evaluation consists of measurements of the link recovery accuracy. The link recovery is computed in precision, recall AUC, and ROC. The higher AUC value indicates a better model performance. Nevertheless, \tabref{tab:performance} depicts both unsupervised techniques (\ie \wv and \dv) do not capture semantic similarity efficiently, the skip-gram model for \cisco has better precision than the paragraph distributed model (\ie 0.64 for \cisco at \exthird). A cursory glance at \tabref{tab:performance} reveals that negative results are persistent for both unsupervised techniques due to, most likely, a data limitation. Data is insufficient to represent patterns and features that contribute to the binary classification (\ie traceability generation). The highest ROC values were observed for the \exfourth with \wv at \albergate, \ebt, \etour, and \itrust with 0.77, 0.79, 0.74, and 0.79 respectively. In general \dv fails to retrieve links showcasing its ineffectiveness for the traceability task.

In \figref{fig:sub11} and   \figref{fig:sub12} show the precision and recall behavior for the \scm and \wmd similarity for the \libest testbed using \wv and \dv. The tendency of the curves complements our findings on distance analysis: i) precision falls to 3.8 for \scm similarity and 0.42 for \wmd similarity using \wv; and ii) the \dv on \figref{fig:sub12} depicts low performance.

\textbf{Summary.} The effectiveness observed on the AUC precision-recall for all experiments exhibits low performance when recovering the links at both vectorization types \dv and \wv with a slight improvement for \wv. 

\subsection{\texorpdfstring{\ref{rq_nature:semantic} }{rq2} Semantic Traceability Imbalance Results}

\tabref{tab:information} summarizes the observed results for the similarity soft-cosine (\scm), word mover's distance (\wmd), cosine distance (\cosine), and euclidean distance (\euc). The low values \scm (\ie 0.1 at \cisco and \ebt) indicate low similarity between the source and target. We estimate a maximum of 0.51 \wmd for \libest on the \exfirst, which indicates \wmd is capturing overlap information between the source and target. \tabref{tab:by_links} discriminate semantic metrics by links and non-links. Interestingly, our models' evaluation remains with low \scm values for links indicating non-related information between the source and target. For instance, the \scm for \smos with the ground truth links reports a 0.06 similarity suggesting the neural vectorization technique fails to detect links from the semantics. Similarly, the \wmd for \libest at \exfirst with a value of 0.51 represents an insufficiency of related tokens between source and target. Our evaluated models exhibit similar behavior for \dv with a 0.98 distance for \libest and \cisco for the experiments \exfirst and \exsecond (see \tabref{tab:information}). 

\textbf{Summary.} The ground truth for traceability suffers from extremely imbalanced link classes (\ie actual links and non-links). In addition, cosine (\cosine) distance and Soft-Cosine similarity (\scm) behave better under AUC analysis indicating the unsupervised models identify links with a minimum effectiveness of 0.19 and 0.12 respectively for \cisco testbed.

\subsection{ \texorpdfstring{\ref{rq_nature:exploratory} }{rq3} Exploratory Information Theory Results}

$AN_1$: \textit{Information measures.} \tabref{tab:information} depicts the self-information for source and target artifacts, the noise, loss, mutual information, and the minimum shared information. We observe the self-information $H(X)$ of the source artifacts (or issues) in on average $4.92[1.25]$\bits, while the self-information of the target artifacts (or source code) $H(Y)$ is on average $6.4[0.95]$\bits suggesting that the amount of information of $H(Y)$ is on average 1.48\bits larger than the amount of information in the set of $H(X)$. The last means the source has less information than the target even when the source is more expressive as it uses natural language tokens.

We observe the highest noise at the experiment \exfourth with \albergate with a value of 0.91\bits. Notably, when computing the difference between the self-information $H(X)$ and $H(Y)$ we observe the highest unbalanced information for the \cisco system at the experiments \exbase and \exthird with a value of 2.49\bits. This result indicates non-related information between the source and target and confirms the semantic distance results \scm for the same testbeds and experiments.

The Mutual Information averages $4.61[1.21]$\bits indicating a positive transmission of information from the source to the target. The Minimum Share Information Entropy $Si$ is on average $2.68[1.57]$\bits, and the Minimum Shared Extropy $Sx$ is on average $1.11[0.28]$\bits for all the experiments suggesting the source and target possess information in common. However, the target could not share information with the source as depicted at \exfourth with \ebt testbed where we observe an $Si$ of 0.61\bits and a loss of 2.01\bits. The loss difference of 0.96\bits and similarity of 0.1\bits demonstrates an information imbalance between the source and target. In contrast, the $Si$ value for \libest at the \exfirst reports the highest value with 5.77\bits. In contrast, the loss and noise differences are high with 5.72\bits and 7.21\bits respectively.

The loss and noise are Gaussian distributions with a median of 1.79\bits and 0.31\bits respectively. The loss is larger than the noise by a range of 1.48\bits on average for all the experiments. When the estimated median of the minimum shared entropy is 2.68\bits, we detected a high information amount suggesting the source code can be poorly commented. Furthermore, the noise is barely a bit unit indicating the code is not influenced by an external source of information.

The distance and similarity metrics reflect a non-standard behavior, for instance, when comparing the cosine (for \dv) and the \wmd (for \wv). The average value for conventional preprocessing (\ie \exbase and \exfourth) of the cosine is $0.12[0.05]$, while the value for the \wmd is $0.45[0.01]$. Both distance distributions are unimodal entailing the binary classification does not exist and both distance metrics do not overlap. 

\textbf{Summary.} The maximum information transmission is around 4.52\bits from issues to source code for \cisco testbed at the \exfirst. We recommend that software developers implement inspection procedures to refactor documentation in both requirements and source code to enhance mutual information (and MSI). 

$AN_2$: \textit{Information measures by Ground Truth.} Unfortunately, information measures are not being affected by the nature of the traceability (see \tabref{tab:by_links} information metrics). $H(X)$ and $H(Y)$ are on average similar for links and non-links. That is, information is independent of whether a link between two artifacts exists or not. Nonetheless, all sequence-based artifacts are related somehow -or share some information amount-, this \textit{independent} behavior is not expected in similarity metrics such as \scm, \euc, or \wmd. Neural unsupervised techniques based on skip-gram models are unable to binary classify a link. In other words, data do not have encoded the necessary patterns to determine the classification. We need to employ probabilistic models to intervene in the expectation value of a link\citep{moran_improving_2020} or systematic refactorings on the artifacts.

\textbf{Summary.} Although the information amount in the source code is larger than in the set of issues; the MI, loss, and noise are indistinguishable from confirmed links to non-links. We expect low mutual information values and high loss and noise information amounts for non-related artifacts. 

\subsection{ \texorpdfstring{\ref{rq_nature:correlation} }{rq4} Correlation Results}

\textit{Scatter Matrix for Information Measures.} Correlations help explain variables that we are not easily able to describe just by observing their values. Correlations are useful to interpret the causes or detect similar patterns for a given metric. In this case, we want to study similarity variables by correlating them with other similarity variables and information measures (e.g., MI, Loss, Noise, Entropy, etc). The manifold in \tabref{tab:correlations} depicts correlations and distribution of each information variable. We want to highlight that the \wmd similarity is mostly positively correlated ($\approx$0.74) (\figref{fig:sub23}) with other information metrics, while \cosine similarity has the opposite effect. 
\input{tables/chap_infotheory/tab6_correlations}

\input{graphics/chap_infotheory/libest_composable}

\textit{Mutual Information \& Shared Information Entropy and Extropy.} This analysis consists of computing a correlation between the distance and entropy. Mutual information is negatively correlated with \wmd (positively correlated with \wmd similarity) as observed in \tabref{tab:correlations}. This implies that the larger the amount of shared information, the less distance among artifacts; until we observe that MI is not correlated with the cosine similarity. Is the word vector capturing better semantic relationships than paragraph vectors? Both approaches are underperforming according to the binary classification performance. 

On the other hand, the MSI for entropy is also negatively correlated with the \wmd  (see \tabref{tab:correlations}). The trend is expected after observing the correlation with the mutual information. We show more evidence that \wmd similarity captures better semantic relationships among artifacts. 

\textit{Composable Manifolds.} The composable manifolds are useful for inspecting a third information variable. In this case, we focused on the loss and noise (see \figref{fig:grid2}). We observe the loss is larger when the mutual information and similarity are lower (see \figref{fig:sub23}) for \cisco testbed. However, the noise is more dispersed across the mutual information and similarity  (see \figref{fig:sub24}). We find different clusters of noise in low and larger ranges of MI. These trends indicate that some information amount was injected into the source code but it is independent of the semantic similarity of two artifacts. For \libest testbed, \figref{fig:sub21} and  \figref{fig:sub22}, we observe a dispersed correlation between the \wmd similarity and MI, whereas loss is still high with low MI and noise is high with high MI.

\textbf{Summary.} Loss entropy is associated with low levels of similarity and mutual information. We can intervene in traceability datasets to help improve link classification by reducing the loss. Such interventions (\ie refactorings) can occur because practitioners complete or document artifacts during the software life-cycle. 

%% file: tables/chap_infotheory/tab3_efectivenes.tex
\begin{table}[ht]
\centering
\caption{AUC Performance}
\label{tab:performance}
\vspace{-0.5em}

\begin{adjustbox}{width=\textwidth}

\setlength{\tabcolsep}{3pt} 
\begin{tabular}{cllcccccccc}
\toprule
\textbf{Exp.} & \multicolumn{1}{c}{\textbf{Testbed}} &  & \multicolumn{2}{c}{\textbf{Word2vec (WMD)}} & \multicolumn{2}{c}{\textbf{Word2vec (SCM)}} & \multicolumn{2}{c}{\textbf{Doc2vec (COS)}} & \multicolumn{2}{c}{\textbf{Doc2vec (EUC)}} \\
\textit{\textbf{ID}} & \multicolumn{1}{c}{\textit{\textbf{system}}} &  & \textit{\textbf{AUC}} & \textit{\textbf{ROC}} & \textit{\textbf{AUC}} & \textit{\textbf{ROC}} & \textit{\textbf{AUC}} & \textit{\textbf{ROC}} & \textit{\textbf{AUC}} & \textit{\textbf{ROC}} \\ \cline{1-2} \cline{4-11} 
\multirow{2}{*}{\exbase} & \libest &  & 0.42 & 0.6 & 0.38 & 0.57 & 0.33 & 0.51 & 0.28 & 0.42 \\
 & \cisco &  & 0.04 & 0.62 & 0.04 & 0.62 & 0.05 & 0.64 & 0.02 & 0.4 \\ \cline{1-2} \cline{4-11} 
\multirow{2}{*}{\exfirst} & \libest &  & 0.35 & 0.54 & 0.34 & 0.53 & 0.35 & 0.53 & 0.37 & 0.56 \\
 & \cisco &  & 0.03 & 0.54 & 0.03 & 0.5 & 0.03 & 0.51 & 0.02 & 0.49 \\ \cline{1-2} \cline{4-11} 
\multirow{2}{*}{\exsecond} & \libest &  & 0.34 & 0.53 & 0.3 & 0.49 & 0.31 & 0.49 & 0.32 & 0.5 \\
 & \cisco &  & 0.03 & 0.53 & 0.03 & 0.51 & 0.02 & 0.5 & 0.03 & 0.5 \\ \cline{1-2} \cline{4-11} 
\multirow{2}{*}{\exthird} & \libest &  & 0.59 & 0.41 & 0.56 & 0.38 & 0.4 & 0.59 & 0.31 & 0.46 \\
 & \cisco &  & 0.05 & 0.64 & 0.05 & 0.66 & 0.04 & 0.61 & 0.02 & 0.4 \\ \cline{1-2} \cline{4-11} 
\multirow{5}{*}{\exfourth} & \albergate &  & 0.08 & 0.58 & 0.15 & 0.77 & 0.09 & 0.53 & 0.05 & 0.49 \\
 & \ebt &  & 0.2 & 0.81 & 0.14 & 0.79 & 0.18 & 0.73 & 0.04 & 0.44 \\
 & \etour &  & 0.23 & 0.75 & 0.24 & 0.74 & 0.15 & 0.72 & 0.04 & 0.46 \\
 & \itrust &  & 0.06 & 0.76 & 0.06 & 0.79 & 0.03 & 0.73 & 0 & 0.24 \\
 & \smos &  & 0.2 & 0.57 & 0.21 & 0.6 & 0.17 & 0.54 & 0.12 & 0.41 \\ \bottomrule
\end{tabular}

\end{adjustbox}

{\scriptsize\textit{AUC is a precision-recall area under the curve.}}
\vspace{0.3em}
\end{table}

%% file: tables/chap_infotheory/tab4_self_information_and_distance.tex
\begin{table*}[ht]
\centering

\caption{Self Information, Loss, Mutual Information, and Minimum Shared Information}
\label{tab:information}
\begin{adjustbox}{width=1\textwidth}

\setlength{\tabcolsep}{3pt} 
\begin{tabular}{cllrrrrlcccccccccc}
\toprule
\textbf{Exp.} & \multicolumn{1}{c}{\textbf{Testbed}} &  & \multicolumn{2}{c}{\textbf{Word2vec}} & \multicolumn{2}{c}{\textbf{Doc2Vec}} & \multicolumn{1}{c}{\textbf{}} & \textbf{SI S.} & \textbf{SI T.} &  & \textbf{CI Noise} &  & \textbf{CI Loss} &  & \textbf{MI} & \textbf{MSI} & \textbf{MSI} \\
\textit{\textbf{ID}} & \multicolumn{1}{c}{\textit{\textbf{System}}} &  & \multicolumn{1}{l}{\textbf{SCM}} & \multicolumn{1}{l}{\textbf{WMD}} & \multicolumn{1}{l}{\textbf{COS}} & \multicolumn{1}{l}{\textbf{EUC}} &  & \textit{\textbf{H(X)}} & \textit{\textbf{H(Y)}} & \multirow{-2}{*}{\textbf{D1}} & \textit{\textbf{H(Y|X)}} & \multirow{-2}{*}{\textbf{D2}} & \textit{\textbf{H(X|Y)}} & \multirow{-2}{*}{\textbf{D3}} & \textit{\textbf{I(X:Y)}} & \textit{\textbf{Si[std]}} & \textit{\textbf{Sx[std]}} \\ \cline{1-2} \cline{4-7} \cline{9-18} 
\rule{0pt}{3ex}    
 & \libest &  & 0.28 & 0.49 & 0.17 & 0.01 &  & 5.54 & 7.77 & 2.23 & 0.18 & 7.59 & 2.4 & 3.14 & 5.37 & 3.82[0.66] & 1.37[0.03] \\
\multirow{-2}{*}{\exbase} & \cisco &  & \cellcolor[HTML]{FFCCC9}0.1 & 0.45 & 0.09 & 0.02 &  & 3.42 & 5.91 & \cellcolor[HTML]{DAE8FC}2.49 & 0.21 & 5.7 & 2.7 & \cellcolor[HTML]{DAE8FC}0.72 & 3.21 & 1.45[1.14] & 0.87[0.54] \\
\cline{1-2} \cline{4-7} \cline{9-18} 
\rule{0pt}{3ex}
 & \libest &  & 0.42 & 0.51 & 0 & \cellcolor[HTML]{FFCCC9}0.98 &  & 6.58 & 7.33 & 0.75 & 0.12 & 7.21 & 0.86 & 5.72 & 6.46 & \cellcolor[HTML]{DAE8FC}5.77[0.6] & 1.42[0.02] \\
\multirow{-2}{*}{\exfirst} & \cisco &  & 0.28 & 0.47 & 0 &\cellcolor[HTML]{FFCCC9} 0.98 &  & 4.68 & 6.6 & 1.92 & 0.17 & 6.43 & 2.08 & 2.6 & 4.52 & 3.17[1.44] & 1.24[0.35] \\
\cline{1-2} \cline{4-7} \cline{9-18} 
\rule{0pt}{3ex}
 & \libest &  & 0.37 & 0.49 & 0 &\cellcolor[HTML]{FFCCC9} 0.98 &  & 6.54 & 7.47 & 0.93 & 0.14 & 7.33 & 1.07 & 5.47 & 6.4 & 5.42[0.64] & 1.42[0.01] \\
\multirow{-2}{*}{\exsecond} & \cisco &  & 0.18 & 0.46 & 0 & \cellcolor[HTML]{FFCCC9}0.98 &  & 4.42 & 6.56 & 2.14 & 0.26 & 6.3 & 2.4 & 2.02 & 4.16 & 2.63[1.41] & 1.15[0.41] \\
\cline{1-2} \cline{4-7} \cline{9-18} 
\rule{0pt}{3ex}

 & \libest &  & 0.23 & 0.49 & 0.26 & 0.01 &  & 5.54 & 7.77 & 2.23 & 0.18 & 7.59 & 2.4 & 3.14 & 5.37 & 2.82[0.66] & 1.37[0.03] \\
\multirow{-2}{*}{\exthird} & \cisco &  & \cellcolor[HTML]{FFCCC9} 0.08 & 0.45 & 0.17 & 0.02 &  & 3.42 & 5.91 & \cellcolor[HTML]{DAE8FC}2.49 & 0.21 & 5.7 & 2.7 & \cellcolor[HTML]{DAE8FC}0.72 & 3.21 & 1.45[1.14] & 0.87[0.54] \\
\cline{1-2} \cline{4-7} \cline{9-18} 
\rule{0pt}{3ex}
 & \albergate &  & \cellcolor[HTML]{FFCCC9} 0.07 & 0.45 & 0.22 & 0.02 &  & 6.62 & 6.18 & 0.44 & 0.91 & 5.27 & 0.47 & 6.15 & 5.7 & 2.81[0.96] & 1.27[0.19] \\
 & \ebt &  & \cellcolor[HTML]{FFCCC9}0.1 & 0.45 & \cellcolor[HTML]{FFCCC9}0.1 & 0.03 &  & 2.97 & 4.73 & 1.76 & 0.25 & 4.48 & 2.01 & \cellcolor[HTML]{DAE8FC}0.96 & 2.72 & \cellcolor[HTML]{FFCCC9}0.61[0.72] & 0.5[0.54] \\
 & \etour &  & 0.16 & 0.46 & 0.07 & 0.02 &  & 5.23 & 5.77 & 0.54 & 0.53 & 5.24 & 1.07 & 4.16 & 4.7 & 2.1[1.13] & 1.02[0.34] \\
 & \itrust &  & 0.14 & 0.46 & 0.09 & 0.02 &  & 3.93 & 5.56 & 1.63 & 0.33 & 5.23 & 1.96 & 1.97 & 3.6 & 1.33[1.02] & 0.85[0.52] \\
\multirow{-5}{*}{\exfourth} & \smos &  & 0.05 & 0.45 & 0.13 & 0.02 &  & 5.09 & 5.7 & 0.61 & 0.55 & 5.15 & 1.16 & 3.93 & 4.54 & 1.44[0.71] & 1.02[0.34] \\
\bottomrule
\end{tabular}
\end{adjustbox}
\vspace{0.5em}

{\scriptsize\textit{D1: $H(X)-H(Y)$, D2: $H(Y)-H(Y|X)$, D3: $H(X)-H(X|Y)$, S.: Source, T.: Target}
}
\end{table*}

%% file: tables/chap_infotheory/tab5_self_information_by_links.tex
\begin{table*}[hb]
\centering

\caption{Self Information, Loss, Mutual Information and Minimum Shared Information by Links}
\label{tab:by_links}
\begin{adjustbox}{width=1\textwidth}

\setlength{\tabcolsep}{3pt} 
\begin{tabular}{cllllllllllllllllllllllllll}
\toprule
\multicolumn{1}{l}{} &  &  & \multicolumn{2}{c}{\textbf{Similarity}} &  & \multicolumn{6}{c}{\textbf{Distance}} &  & \multicolumn{14}{c}{\textbf{Information}} \\
\textbf{Exp.} & \multicolumn{1}{c}{\textbf{Testbed}} &  & \multicolumn{2}{c}{\textbf{SCM}} &  & \multicolumn{2}{l}{\textbf{WMD}} & \multicolumn{2}{c}{\textbf{COS}} & \multicolumn{2}{c}{\textbf{EUC}} &  & \multicolumn{2}{c}{\textbf{H(X)}} & \multicolumn{2}{c}{\textbf{H(Y)}} & \multicolumn{2}{c}{\textbf{H(Y|X)}} & \multicolumn{2}{c}{\textbf{H(X|Y)}} & \multicolumn{2}{c}{\textbf{I(X:Y)}} & \multicolumn{2}{c}{\textbf{Si(X:Y)}} & \multicolumn{2}{c}{\textbf{Sx(X:Y)}} \\
\textit{\textbf{ID}} & \multicolumn{1}{c}{\textit{\textbf{System}}} &  & \multicolumn{1}{c}{\textit{\textbf{Link}}} & \multicolumn{1}{c}{\textit{\textbf{NoL}}} &  & \multicolumn{1}{c}{\textit{\textbf{Link}}} & \multicolumn{1}{c}{\textit{\textbf{NoL}}} & \multicolumn{1}{c}{\textit{\textbf{Link}}} & \multicolumn{1}{c}{\textit{\textbf{NoL}}} & \multicolumn{1}{c}{\textit{\textbf{Link}}} & \multicolumn{1}{c}{\textit{\textbf{NoL}}} &  & \multicolumn{1}{c}{\textit{\textbf{Link}}} & \multicolumn{1}{c}{\textit{\textbf{NoL}}} & \multicolumn{1}{c}{\textit{\textbf{Link}}} & \multicolumn{1}{c}{\textit{\textbf{NoL}}} & \multicolumn{1}{c}{\textit{\textbf{Link}}} & \multicolumn{1}{c}{\textit{\textbf{NoL}}} & \multicolumn{1}{c}{\textit{\textbf{Link}}} & \multicolumn{1}{c}{\textit{\textbf{NoL}}} & \multicolumn{1}{c}{\textit{\textbf{Link}}} & \multicolumn{1}{c}{\textit{\textbf{NoL}}} & \multicolumn{1}{c}{\textit{\textbf{Link[std]}}} & \multicolumn{1}{c}{\textit{\textbf{NoL[std]}}} & \multicolumn{1}{c}{\textit{\textbf{Link[std]}}} & \multicolumn{1}{c}{\textit{\textbf{NoL[std]}}} \\ \cline{1-2} \cline{4-5} \cline{7-12} \cline{14-27} 
\rule{0pt}{3ex}   
 & \libest &  & 0.30 & 0.27 &  & 0.49 & 0.48 & 0.17 & 0.17 & 0.01 & 0.01 &  & 5.70 & 0.00 & 7.73 & 7.78 & 0.19 & 0.17 & 2.22 & 2.48 & 5.51 & 5.30 & 3.95[0.7] & 3.76[0.63] & 1.37[0.04] & 1.37[0.03] \\
\multirow{-2}{*}{\exbase} & \cisco &  & \cellcolor[HTML]{FFCCC9}0.14 & 0.10 &  & 0.46 & 0.45 & 0.13 & 0.09 & 0.02 & 0.02 &  & 3.80 & 3.41 & 6.23 & 5.90 & 0.20 & 0.21 & 2.63 & 2.71 & 3.60 & 3.20 & 1.98[1.11] & 1.44[1.14] & 10.7[0.43] & 0.86[0.54] \\
\cline{1-2} \cline{4-5} \cline{7-12} \cline{14-27} 
\rule{0pt}{3ex}   
 & \libest &  & 0.43 & 0.42 &  & \cellcolor[HTML]{FFCCC9}0.51 & 0.51 & 0.00 & 0.00 & 0.98 & 0.98 &  & 6.80 & 6.48 & 7.31 & 7.33 & 0.13 & 0.11 & 0.64 & 0.97 & 6.67 & 6.37 & 5.94[0.53] & 6.68[0.062] & 1.42[0.01] & 1.42[0.02] \\
\multirow{-2}{*}{\exfirst} & \cisco &  & 0.28 & 0.28 &  & 0.47 & 0.47 & 0.00 & 0.00 & 0.98 & 0.98 &  & 4.94 & 4.68 & 6.72 & 6.59 & 0.16 & 0.17 & 1.93 & 2.08 & 4.79 & 4.51 & 3.59[1.26] & 3.16[1.44] & 1.31[0.26] & 1.24[0.35] \\
\cline{1-2} \cline{4-5} \cline{7-12} \cline{14-27} 
\rule{0pt}{3ex}   
 & \libest &  & 0.37 & 0.37 &  & 0.49 & 0.49 & 0.00 & 0.00 & 0.98 & 0.98 &  & 6.71 & 6.46 & 7.44 & 7.49 & 0.15 & 0.13 & 0.88 & 1.16 & 6.56 & 6.33 & 5.58[0.57] & 5.35[0.65] & 1.42[0.01] & 1.41[0.01] \\
\multirow{-2}{*}{\exsecond} & \cisco &  & 0.18 & 0.18 &  & 0.46 & 0.46 & 0.00 & 0.00 & 0.98 & 0.98 &  & 4.66 & 4.41 & 6.71 & 6.56 & 0.26 & 0.26 & 2.31 & 2.41 & 4.40 & 4.15 & 3.04[1.26] & 2.62[1.41] & 1.25[0.31] & 1.15[0.41] \\
\cline{1-2} \cline{4-5} \cline{7-12} \cline{14-27} 
\rule{0pt}{3ex}   
 & \libest &  & 0.25 & 0.23 &  & 0.49 & 0.48 & \cellcolor[HTML]{DAE8FC}0.28 & \cellcolor[HTML]{FFCCC9}0.26 & 0.01 & \cellcolor[HTML]{FFCCC9}0.01 &  & 5.70 & 5.47 & 7.73 & 7.78 & 0.19 & 0.17 & 2.22 & 2.48 & 5.51 & 5.30 & 3.95[0.7] & 3.76[0.63] & 1.37[0.04] & 1.37[0.03] \\
\multirow{-2}{*}{\exthird} & \cisco &  & \cellcolor[HTML]{FFCCC9}0.12 & 0.08 &  & 0.46 & 0.45 & 0.19 & 0.16 & 0.01 & 0.02 &  & 3.80 & 3.41 & 6.23 & 5.90 & 0.20 & 0.21 & 2.63 & 2.71 & 3.60 & 3.20 & 1.98[1.11] & 1.44[1.14] & 1.07[0.43] & 0.86[0.54] \\
\cline{1-2} \cline{4-5} \cline{7-12} \cline{14-27} 
\rule{0pt}{3ex}   
 & \albergate &  & \cellcolor[HTML]{FFCCC9}0.11 & 0.07 &  & 0.46 & 0.45 & 0.23 & 0.22 & 0.02 & 0.02 &  & 6.70 & 6.61 & 6.19 & 6.18 & 0.93 & 0.91 & 0.42 & 0.48 & 5.77 & 5.70 & 3.07[0.91] & 2.8[0.96] & 1.3[0.1] & 1.27[0.2] \\
 & \ebt &  & 0.20 & 0.09 &  & 0.47 & 0.45 & 0.16 & 0.10 & 0.03 & 0.03 &  & 2.94 & 2.97 & 5.11 & 4.71 & 0.15 & 0.26 & 2.32 & 2.00 & 2.78 & 2.71 & 1.2[0.74] & 0.56[0.7] & 0.9[0.46] & 0.46[0.53] \\
 & \etour &  & 0.24 & 0.15 &  & 0.47 & 0.46 & 0.11 & 0.07 & 0.02 & 0.02 &  & 5.28 & 5.22 & 5.79 & 5.76 & 0.49 & 0.53 & 1.00 & 1.07 & 4.79 & 4.69 & 2.48[0.91] & 2.08[1.03] & 1.22[0.2] & 1.13[0.35] \\
 & \itrust &  & 0.28 & 0.14 &  & 0.48 & 0.46 & 0.15 & 0.09 & 0.01 & 0.02 &  & 4.07 & 3.92 & 6.28 & 5.55 & 0.12 & 0.33 & 2.33 & 1.95 & 3.95 & 3.60 & 2.51[1.07] & 1.32[1.01] & 1.21[0.27] & 0.85[0.52] \\
\multirow{-5}{*}{\exfourth} & \smos &  & \cellcolor[HTML]{FFCCC9}0.06 & 0.05 &  & 0.45 & 0.45 & 0.13 & 0.13 & 0.02 & 0.02 &  & 5.08 & 5.09 & 5.82 & 5.67 & 0.45 & 0.57 & 1.19 & 1.15 & 4.63 & 4.52 & 1.62[0.66] & 1.41[0.71] & 1.1[0.2] & 1[0.36] \\
\bottomrule
\end{tabular}

\end{adjustbox}

\end{table*}

%% file: tables/chap_infotheory/tab6_correlations.tex
\begin{table}[ht]
\centering
\caption{Correlation between the distance and entropy}
\label{tab:correlations}

\begin{adjustbox}{width=\textwidth}

\setlength{\tabcolsep}{3pt} 
\begin{tabular}{cccccccccccc}
\textbf{Exp.} & \textbf{Testbed} & \multicolumn{2}{c}{\textbf{Mutual Information}} & \multicolumn{2}{c}{\textbf{Loss}} & \multicolumn{2}{c}{\textbf{Noise}} & \multicolumn{4}{c}{\textbf{Minimum Shared Entropy}} \\
\textit{\textbf{ID}} & \textit{\textbf{system}} & \textit{\textbf{WMD}} & \textit{\textbf{COS}} & \textit{\textbf{WMD}} & \textit{\textbf{COS}} & \textit{\textbf{WMD}} & \textit{\textbf{COS}} & \textit{\textbf{WMD}} & \textit{\textbf{SCM}} & \textit{\textbf{COS}} & \textit{\textbf{EUC}} \\ \hline
 & \textit{libest} & 0.44 & 0.28 & -0.16 & -0.13 & 0.12 & 0.14 & 0.69 & 0.33 & 0.41 & -0.19 \\
\multirow{-2}{*}{\exbase} & \textit{csc} & \cellcolor[HTML]{DAE8FC}0.74 & 0.01 & -0.63 & -0.12 & 0.49 & 0.05 & 0.63 & 0.5 & 0.22 & -0.37 \\ \hline
 & \textit{libest} & 0.62 & 0.02 & -0.52 & -0.03 & 0.59 & 0.01 & \cellcolor[HTML]{DAE8FC}0.71 & 0.67 & 0.01 & 0.05 \\
\multirow{-2}{*}{\exfirst} & \textit{csc} & \cellcolor[HTML]{DAE8FC}0.87 & 0 & \cellcolor[HTML]{DAE8FC}-0.81 & 0 & 0.58 & 0 & \cellcolor[HTML]{DAE8FC}0.87 & \cellcolor[HTML]{DAE8FC}0.81 & -0.01 & -0.03 \\ \hline
 & \textit{libest} & 0.66 & 0.02 & -0.53 & 0.01 & \cellcolor[HTML]{FFCCC9}0.62 & 0 & \cellcolor[HTML]{DAE8FC}0.75 & 0.52 & 0.03 & -0.05 \\
\multirow{-2}{*}{\exsecond} & \textit{csc} & \cellcolor[HTML]{DAE8FC}0.86 & 0 & \cellcolor[HTML]{DAE8FC}-0.79 & 0 & \cellcolor[HTML]{FFCCC9}0.62 & 0 & \cellcolor[HTML]{DAE8FC}0.83 & 0.76 & 0 & -0.03 \\ \hline
 & \textit{libest} & 0.47 & 0.15 & -0.2 & -0.1 & 0.17 & 0.06 & \cellcolor[HTML]{DAE8FC}0.71 & 0.4 & 0.46 & -0.12 \\
\multirow{-2}{*}{\exthird} & \textit{csc} & \cellcolor[HTML]{DAE8FC}0.78 & 0.34 & -0.67 & -0.27 & 0.53 & 0.18 & 0.65 & 0.5 & 0.42 & -0.31 \\ \hline
 & \textit{albergate} & 0.25 & \cellcolor[HTML]{DAE8FC}0.35 & \cellcolor[HTML]{FFCCC9}0.43 & \cellcolor[HTML]{FFCCC9}0.41 & -0.2 & -0.37 & 0.56 & 0.45 & 0.59 & \cellcolor[HTML]{FFCCC9}-0.72 \\
 & \textit{ebt} & 0.36 & -0.04 & -0.04 & -0.19 & -0.14 & 0.05 & 0.26 & 0.22 & 0.21 & -0.29 \\
 & \textit{etour} & 0.39 & 0.25 & 0.07 & 0.05 & \cellcolor[HTML]{DAE8FC}-0.23 & -0.13 & 0.59 & 0.45 & 0.44 & -0.56 \\
 & \textit{itrust} & 0.54 & 0.13 & -0.32 & -0.12 & 0.13 & -0.01 & 0.61 & 0.44 & 0.3 & -0.44 \\
\multirow{-5}{*}{\exfourth} & \textit{smos} & \cellcolor[HTML]{FFCCC9}0.19 & \cellcolor[HTML]{FFCCC9}-0.09 & -0.39 & 0.14 & 0.31 & 0.2 & 0.4 & 0.48 & 0.2 & -0.28 \\ \hline
\end{tabular}

\end{adjustbox}

\end{table}

%% file: graphics/chap_infotheory/libest_composable.tex
\begin{figure*}[ht]
    \centering
    \begin{subfigure}{0.4\textwidth}
        \centering
        \includegraphics[width=\linewidth]{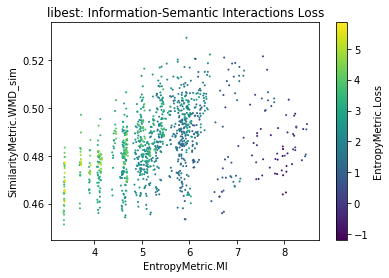}
        \caption{\tiny\libest \wmd similarity, MI and Loss }
        \label{fig:sub21}
    \end{subfigure}
    \begin{subfigure}{0.4\textwidth}
        \centering
        \includegraphics[width=\linewidth]{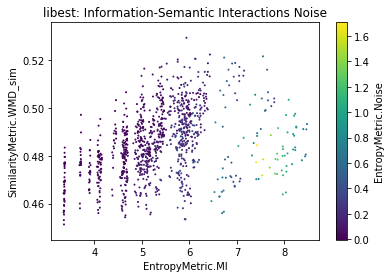}
        \caption{\tiny\libest \wmd similarity, MI and Noise}
        \label{fig:sub22}
    \end{subfigure}
    \vspace{0.2cm}
    \begin{subfigure}{0.4\textwidth}
        \centering
        \includegraphics[width=\linewidth]{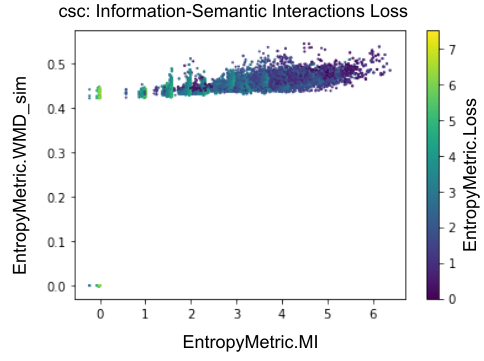}
        \caption{\tiny\cisco \wmd similarity, MI and Loss}
        \label{fig:sub23}
    \end{subfigure}
    \begin{subfigure}{0.4\textwidth}
        \centering
        \includegraphics[width=\linewidth]{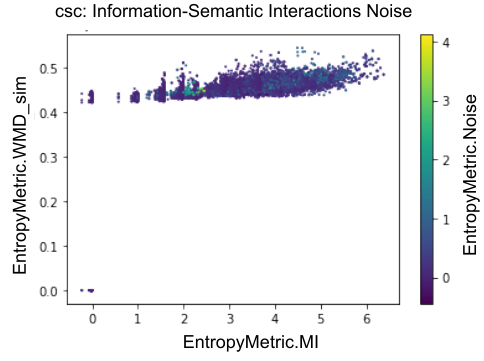}
        \caption{\tiny\cisco \wmd similarity, MI and Noise}
        \label{fig:sub24}
    \end{subfigure}
    \caption{Similarity Vs Mutual Information Vs Noise and Loss for \libest and \cisco.}
    \label{fig:grid2}
    \vspace{1.2em}
\end{figure*}

%% file: chapters/part_01_chap_02/sec_06_case.tex
\section{A Case Study in Industry}
\label{ch:nature:sec_06}

This section shows four information science cases from processing the \cisco testbed. Experiments and samples can be found in our online appendix~\citep{palacio2023tracexplainer,RepoTraceXplainer24}. 

\textit{$Case_0$: Self-Information.} This study highlights the information imbalance between the source and target artifacts. Artifacts with low entropy struggle to generate traceability links since unsupervised techniques rely on concise descriptions in natural language. By contrast, high levels of entropy struggle with other conditions like loss or noise. Refactoring operations need to be applied to source and target artifacts to combat the information imbalance. At least, a semantic idea expressed as a clause is required in both artifacts to guarantee a link. When we preprocess the source code, we transform the code structure into a sequence structure to extract those clauses. In other words, source code is not treated as a structured language but as a regular text file. The amount of information lost by preprocessing source code as text has not yet been computed.

\textit{$Case_1$: Minimum and maximum loss.} This study presents edge cases for entropy loss to identify poorly documented target artifacts. Edge cases of loss are useful to detect starting points for general refactoring in the target documentation. The max case shows us the pull request content composed of just one word. The tokens that represent this word were not found in the target artifact, which is one of the highest entropy files. Whereas, the min case shows us the pull request has a complete description not found in the target. Note the 99\% quartile for positive links. The PR content cannot be easily found in the source code with low entropy, even if a practitioner tries to extract links manually. %

\textit{$Case_2$: Minimum and maximum noise.} This study presents edge cases for the entropy noise to identify poorly documented source artifacts. Edge cases of noise are useful to detect refactorings in documentation found in the source documentation. In the max case, the PR content is high, but the target artifact is empty. This suggests that the target is not well documented. Something similar occurs in the min case, where the PR content is repetitive and less expressive.

\textit{$Case_3$: Orphan informative links.}  This study points out a set of informative links not found in the ground truth. Orphan links not only exhibit inconsistencies in the ground truth file but also suggest potential positive links independent of the employed unsupervised technique.

%% file: chapters/part_01_chap_02/sec_07_conclusion.tex
\section{A Case Study in Industry}
\label{ch:nature:sec_07}

The core idea of using Information Measures with Software Traceability is to interpret, determine, and quantify the boundaries of the effectiveness of Unsupervised techniques. We demonstrated traceability data is insufficient to derive a trace pattern when information measures are not in optimal ranges. Hence, unsupervised techniques are limited approaches to producing reliable links. However, we can also determine to what extent information is lost during the transmission process source-target (\ie ${PullRequest \to Code}$). Loss and noise could be relevant information measures for security purposes \ie why a given piece of source code is not covered by the corresponding requirements? 

We sought to understand the information amount in state-of-the-art traceability datasets to determine why unsupervised techniques (\ie \wv and \dv) recover links ineffectively. To accomplish this, we evaluated four main information metrics using \infotheory: self-information, loss, noise, and mutual information. Complementary, we propose an additional metric, \textit{shared information}, where we determine the overlap entropy across source and target artifacts. Finally, we conducted an empirical analysis to show how our approach contributes to interpreting unsupervised model predictions.  

Our findings highlight an important area for future research: bridging the gap between unsupervised interpretability and the problem of automated software traceability. A viable direction to move beyond our research is to use \infotheory measures to help guide practitioners toward writing better artifacts because \infotheory quantifies the bits required to represent sets of source and target artifacts and their transmission. For example, a requirement contains ``\texttt{bucle}'', a natural language token, as the source, while the target contains a code token ``\texttt{for}''. In this scenario, \textit{mutual information} (MI) between these tokens is zero despite they have similar semantic spaces. The difference lies in the fact that the source operates within the natural language semantic space, whereas the target belongs to the code semantic space. Although unsupervised models can identify semantically related tokens, as in our previous example, our empirical analysis demonstrates that the semantic matching ability is insufficient for handling imbalanced data. Estimated MI and MSI for ground truth links show, therefore, no significant differences.

\textit{Artifacts.} We published an online appendix~\citep{palacio2023tracexplainer,RepoTraceXplainer24}, which contains documented notebooks for practitioners, experimental data of our exploration, unsupervised models, etc.

%% file: chapters/part_01_chap_03/explanations.tex
\chapter{Code-Based Explanations}
\label{ch:rationales}

\lettrine[lines=2]{\textbf{I}}{}n recent years, Language Models for Code (\lmsc) have significantly changed the landscape of software engineering (SE) on downstream tasks, such as code generation, by making software development more efficient. Therefore, a growing interest has emerged in further evaluating these Language Models to homogenize the quality assessment of generated code. As the current evaluation process can significantly overreact on accuracy-based metrics, practitioners often seek methods to \textit{interpret} \lmsc outputs beyond canonical benchmarks. While the majority of research reports on code generation effectiveness in terms of expected ground truth, scant attention has been paid to LLMs' explanations. In essence, the decision-making process to generate code is hard to interpret. To bridge this evaluation gap, we introduce \textit{code rationales} (\codeRational), a technique with rigorous mathematical underpinning, to identify subsets of tokens that can explain individual code predictions. We conducted a thorough \textit{Exploratory Analysis} to demonstrate the method's \textit{applicability} and a \textit{User Study} to understand the \textit{usability} of code-based explanations. Our evaluation demonstrates that \codeRational is a powerful interpretability method to explain how (less) meaningful input concepts (\ie natural language particle `at') highly impact output generation (\ie code conditionals). Moreover, participants of this study highlighted \codeRational's ability to show a causal relationship between the input and output of the model with readable and informative explanations on \textit{code completion} and \textit{test generation} tasks. Additionally, \codeRational also helps to uncover model rationale, facilitating comparison with a human rationale to promote a fair level of trust and distrust in the model.

\input{chapters/part_01_chap_03/sec_01_intro}
\input{chapters/part_01_chap_03/sec_02_background}
\input{chapters/part_01_chap_03/sec_03_approach}
\input{chapters/part_01_chap_03/sec_04_experimental}
\input{chapters/part_01_chap_03/sec_05_exploration}
\input{chapters/part_01_chap_03/sec_06_user_study}
\input{chapters/part_01_chap_03/sec_07_results}
\input{chapters/part_01_chap_03/sec_08_related}
\input{chapters/part_01_chap_03/sec_09_lessons}

%% file: chapters/part_01_chap_03/sec_01_intro.tex
\section{Introduction}
\label{ch:rationales:sec_01}

Research on Artificial Intelligence for Software Engineering (AI4SE) has evolved the complexity of algorithms in four development stages~\citep{liang_holistic_2022}. The first stage is \textit{Statistical Language Models (SLM)} in which $n$-grams were employed to enhance code representation by predicting the next token based on the most recent context (\ie Markov assumption)~\citep{nguyen_statistical_2013,hindle2012natural}. The second stage is \textit{Neural Language Models (NLM)} where neural models were introduced to capture the distribution representation of tokens with recurrent neural networks (RNNs)~\citep{White:MSR15,White2016,Chen2019}. The third stage is \textit{Pre-trained Language Models (PLM)}, which poses two phases: pre-training and then fine-tuning the network. CodeBERT~\citep{feng2020codebertpretrainedmodelprogramming} and CodeT5~\citep{wang-etal-2021-codet5} are among the most relevant PLMs employed for code completion. The last stage is Large Language Models (LLM) where pre-trained models are scaled to a range of billions of parameters showcasing emergent abilities. One of these abilities is impressive code prediction capability with GTP-3~\citep{brown2020languagemodelsfewshotlearners}, Claude3~\citep{TheC3}, CodeGen~\citep{Nijkamp2022CodeGenAO}, StarCoder~\citep{li2023starcodersourceyou}, and Llama3~\citep{dubey2024llama3herdmodels} among the others.

Despite years of research on \lms for code (\lmsc), researchers primarily report evaluations in terms of accuracy-based metrics, which encompass \textit{Exact Match} and \textit{F1} for general evaluation; \textit{ROUGE} for code summarization; CodeBLEU for code synthesis~\citep{Ren2020codebleu}; HumanEval with \textit{pass@k} and correctness rate for code reasoning~\citep{Chen2021}; APPS for testing the correctness of generated code~\citep{hendrycksapps2021}. Measuring only accuracy has become a \textit{de facto} standard for evaluating Language Models for code. Although Language Models (\lms) have been one of the most powerful statistical techniques for representing (code) syntax and semantics, model evaluation has overly relied on computing the notion of \textit{correctness}. Simply put, this measure of correctness often comprises a difference between a predicted and an expected (code) token~\citep{liang_holistic_2022}. Practitioners cannot rely purely on accuracy to interpret code predictions since it often overestimates model's performance\citep{ribeiro2020checklist,burnell_rethink_2023}. 

This chapter challenges the conventional wisdom in AI4SE research of only reporting standard accuracy evaluations \textit{without considering any interpretability aspect of complex models (e.g., PLMs and LLMs)}. Furthermore, our work aligns well with the vast majority of research in the area of model evaluation, pointing to the need for a more holistic way of measuring the quality of LMs~\citep{liang_holistic_2022}, introducing behavioral testing for LMs' capabilities~\citep{ribeiro2020checklist}, and enhancing the rigor of statistical evaluation analyses \citep{watson2020dl4se}.

We believe that such evaluation rigor can be achieved by making Language Models \textit{interpretable} to reduce the barrier to their adoption~\citep{molnar2025}. We endorse the definition of \textit{interpretability} as the \textit{ability to explain (code) predictions by \lms in understandable terms to a human}~\citep{doshivelez2017rigorousscienceinterpretablemachine}. Our main hypothesis is that we can complement the evaluation process beyond accuracy by \textit{rationalizing} \lms. Rationalizing involves uncovering the statistical relationships between the input and output of \lms to build model explanations based on the idea that a practitioner can predict model outputs by observing patterns in the input context~\citep{molnar2025,vafa_rationales_2021}.

\begin{boxK}
 \textit{Rationalizing} Language Models contribute to understanding code predictions by searching a set of concepts that best interpret the relationships between input and output tokens, thus complementing the standard accuracy evaluation. 
\end{boxK}

Software researchers cannot easily establish to what extent features (\eg tokens or code concepts) in the input context (\ie prompt) influence code predictions (\ie generated code). \textit{How do practitioners determine if, for instance, a given \lms is detecting syntax-based structures in the prompt to generate code?} A computational solution is required to assess and quantify the extent to which neural nets (\eg PLMs and LLMs) can capture meaning from the input context. This chapter lays out a mathematical framework based on the approach of rationales~\citep{vafa_rationales_2021} and syntax decomposition~\citep{palacio2024trustworthyinterpretablellmscode} to introduce an interpretability method, namely \textit{code rationales} (\codeRational), that assists researchers and practitioners in understanding neural code generation. This method was designed and developed as a practical approach for identifying underlying (code) concepts that explain code predictions. In summary, the main contributions of our work are as the following:
\begin{itemize}
    \item We developed an interpretability method \codeRational based on three mathematical components: rationalization, mapping, and reduction. The method's output is an \textit{interpretability tensor} that embeds information about relevant concepts influencing code prediction of a given \lmsc. This tensor has novel mathematical properties, which are of considerable software research interest (\eg heatmap of influential concepts, frequency of rationales, and distribution of rationale probability).  
    \item We performed an \textit{exploratory analysis} that embodies three statistical techniques demonstrating the \textit{applicability} of \codeRational. The first technique estimates the statistical dependencies between input context and generated code. The second technique computes the frequency of concept rationales. The third technique reconstructs the probability distribution of each rationale. 
    \item We conducted a \textit{user study} to measure the \textit{usability} of \codeRational. Participants found \codeRational to be informative, readable, and useful for showing statistical correlations and interpreting code predictions. \codeRational also helped show a gap between human and model reasoning, indicating a need for future research.
    \item We published an online appendix~\citep{CodeQ2025}, which contains documented notebooks for practitioners, experimental data of our exploration, source code, models, and statistical analysis of the user study results.  
    
\end{itemize}

%% file: chapters/part_01_chap_03/sec_02_background.tex
\section{Motivation \& Background}
\label{ch:rationales:sec_02}

\begin{figure}[ht]

\centering
\includegraphics[width=0.99\linewidth]{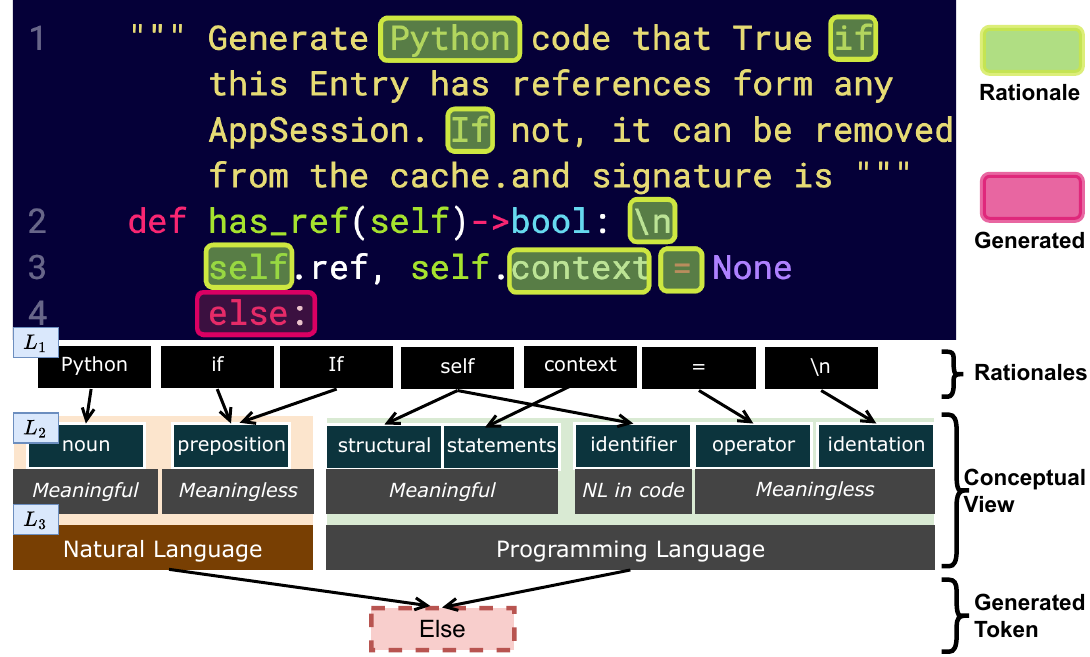}
\caption{Dependency Map for Local Explanations}
\label{fig:dependency_map}
\end{figure}

Consider a practitioner who is interested in identifying groups of tokens in the input context that can \textit{explain} the generation of the keyword \fbox{\texttt{\small else}} as depicted in \figref{fig:dependency_map}. Note that the input context can contain tokens from both the prompt and generated sequence up to that step. A valid \textit{local explanation} embodies the tokens' set that most influences the generated keyword. This set of tokens is known as \textit{rationales}, a subset of input tokens that leads to the model's prediction. In this example, the rationales that most influence the code prediction \fbox{\texttt{\small else}} are highlighted in green: \fbox{\texttt{\small Python}}, \fbox{\texttt{\small if}}, \fbox{\texttt{\small If}}, \fbox{\texttt{\small self}}, \fbox{\texttt{\small \textbackslash n}}, \fbox{\texttt{\small context}}, and \fbox{\texttt{\small =}}.

To obtain the set of rationales, the neural model is exposed to a pipeline of rationalization. This pipeline is based on a greedy technique named \textit{sequence rationales}\citep{vafa_rationales_2021} that finds the minimum set of tokens that most contribute to specific token predictions. Now that a set of rationales has been identified, a practitioner found that some rationales contain less \textit{meaningful} semantic information than others. For instance, the rationale \fbox{\texttt{\small if}} is more intuitive to generate the keyword \fbox{\texttt{\small else}} than a newline rationale \fbox{\texttt{\small \textbackslash n}}. The more rationales tend to have less semantic information, the more the neural model is using spurious or irrelevant information in the context window to generate a prediction (\ie  \textit{overinterpretation}), an undesired phenomenon studied in \lms for code~\citep{li2023}.

Note that inspecting the set of rationales at the token level can be a less interpretable explanation for the practitioner than grouping rationales by natural or programming language concepts. A concept embodies a group of tokens that enables a more human-understandable unit of explanation than fine-grained individual tokens. Therefore, these units are expected to be meaningful (\ie a concept should capture similar meaning), coherent (\ie a concept should be comparable), and important (\ie a concept should be relevant and related to the context)~\citep{Ghorbani19}. \figref{fig:dependency_map} shows a dependency map that traces the semantic relationships from a lower token level granularity to a conceptual view. Instead of a long isolated list of rationales, the practitioner has access to a set of human-understandable code concepts that explain the same keyword prediction. The practitioner can turn this local post-hoc explanation for a single snippet into a more sophisticated \textit{global explanation} that identifies influential concepts in an LM for code. \codeRational leverages the idea of sequential rationales to build local and global explanations based on a semantic mapping to hierarchical code and natural language structures (\ie code concepts). Our method enables researchers and developers to produce explanations in practical settings.   

A model explanation is often expected to be understandable for a human and faithful to a complex model \citep{ribeiro2016why}. Explanations can contribute to the assessment of \lmsc in facilitating model debugging, detecting model bias, providing recourse to practitioners, and enhancing trustworthiness and confidence \citep{palacio2024trustworthyinterpretablellmscode,kim_interpretability_2018,lipton_mythos_2017}. As we mentioned before, in software research, local and global explanations can help us reduce and detect overinterpreted models \citep{li2023}. That is why, although \textit{interpretability} is a young field in AI4SE, we must explicitly establish a desiderata for assessing code-based explanations. The two subsections below extend the notion of AI4SE desiderata and introduce the theory of (code) rationales as the core technique for post-hoc interpretability. 

\subsection{Interpretability Desiderata}

We introduce interpretability desiderata that software researchers should consider when explaining a \lmsc. The desiderata comprise trust, informativeness, and practicability properties. These properties are often reported in the machine learning literature \citep{lipton_mythos_2017,kim_interpretability_2018}. Although this list is not exhaustive, we believe our filtered desiderata is a first approximation to help guide and motivate our work to assess and validate interpretability methods for \lmsc rigorously.

\begin{enumerate}[label= {[D$_{\arabic*}$]:}, ref=D$_{\arabic*}$, wide,labelindent=5pt]\setlength{\itemsep}{0.2em}
\item \label{D1} \textbf{Trust.} Trust is defined as the attitude that an agent will help achieve an individual's goal in a situation characterized by uncertainty and vulnerability\citep{trustincollab}. Interpretability helps to provide a holistic view of the model's behavior beyond canonical accuracy metrics\citep{molnar2025}. Increasing the model's interpretability has been shown to increase trust and confidence in models and thus help in their adoption\citep{doshivelez2017rigorousscienceinterpretablemachine}. Lipton argues \citep{lipton_mythos_2017} that if we can show that the model is accurate when humans are accurate and make mistakes where humans tend to make mistakes then it may be considered trustworthy. In SE context, we seek a similar alignment. 

\item \label{D2}\textbf{Informativeness.} In the context of \lmc, explanations must be informative\citep{lipton_mythos_2017}. Interpretability methods should provide insights and information beyond simply explaining the model's predictions\citep{DesiderataSokol}. For example, \figref{fig:dependency_map} depicts two \textit{IFs} rationales from the input context in natural language. Thus, these two \textit{IFs} do not represent conditional structures. When \codeRational generates the explanation for the token \fbox{\texttt{\small else}}, we expect the prediction was motivated by a code conditional \textit{if}, which is not the case. Exposing the context of the rationale (\eg NL or Code) can be more informative than just identifying the rationale itself.   

\item \label{D3}\textbf{Usability.} An interpretability method for \lmc must be \textit{effective, efficient, and satisfy the context of use} inspired in the definition of usability by \textit{ISO9241}~\citep{speicher2022usability}. Furthermore, evaluating these methods should be carried out with real human developers \citep{doshivelez2017rigorousscienceinterpretablemachine}. An interpretability method must derive actionable insights to improve models' usability~\citep{linardatosExplainable,  management_solutions_xai}.

\end{enumerate}

\subsection{Theory of (Code) Rationales}

The \textit{sequential rationales} is an explainability method that comprises extracting the most relevant input tokens from the context window that can \textit{explain} individual model prediction\citep{vafa_rationales_2021}. The sequential rationales method provides an automated and effective way to extract practical interpretability insights from \lmsc generative tasks. 

The rationale extraction consists of finding the smallest subset of input tokens that would predict the same token as the full sequence\citep{vafa_rationales_2021}. To find the global optimum subset of rationales, Vafa \etal \citep{vafa_rationales_2021} introduce a \textit{greedy rationalization} strategy to approximate the \textit{best rationale} since enumerating all possible subsets is intractable. 

Consider the code sequence  $w_{1:T}$, where the token $w_t$ is predicted given the context window $w_{<t}$ at any token position $t$ up to sequence size $T=|w|$. The purpose of code rationales is to find a subset $r_w \in \mathcal{R}$ that achieves the same prediction as the context $w_{<t}$. Such rationales should be as small as possible $\hat{r_w}$ so that the interpretation of the token $w_t$ becomes clearer. The rationales of a sequence $s$ is defined as: \begin{equation}
r(w_{1:T}) = \arg \min_{r_w \in \mathcal{R}} |r| : \arg \max_{w'_t} P_{\theta}(w'_t|w_{r_w}) = w_t
\label{eq:combinatorial}
\end{equation}

Take into consideration that predicting a token $w_t$ depends upon the context $w_{<t}$, while predicting a token $w'_t$ depends on the optimal rational $w_{r_w}$. Thus, code rationales are formulated as a combinatorial problem. The \textit{gready rationalization} algorithm starts with the empty set, adding iteratively the rationale that most contributes to the probability of $w_t$ at each step $t$ in Eq~\ref{eq:combinatorial}. Note that a rational is the set of tokens that \textbf{covers} a given sequence $s$ if such rational predicts a target token $w_t$.

To use the greedy rationalization the \lmc must be \textit{compatible} with a given combination of subsets of the input context. To make the model \textit{compatible} $P_{\theta}(w_t|w_r)$ is approximated using $F_{\theta}(w_t|w_r)$. The function $F_{\theta}$ in Eq~\ref{eq:llm2} is a concrete model (\eg Transformer, Recurrent Nets, or n-grams) trained on complete subsets $w_{<t}$. Compatibilization is required since models are not trained on incomplete context $w_r$.

\begin{equation}
P_{\theta}(\mathcal{S})  = P_{\theta}(w_{1:T}) 
                         = F_{\theta}(w_1)\prod_{t = 2}^{T} F_{\theta}(w_t | w_{<t} )
\label{eq:llm2}
\end{equation}

The increasing adoption of LLM4SE has led to significant advancements in automated code generation, completion, and summarization. However, despite their widespread success, these models remain fundamentally opaque. Practitioners cannot systematically understand how these models arrive at their predictions, making it difficult to assess whether a model is learning meaningful patterns or relying on spurious correlations. Current evaluation techniques overwhelmingly focus on accuracy-based metrics, such as pass@k, BLEU, and CodeBLEU, which provide an overall measure of correctness, but do not reveal the reasoning or biases underlying the model. Consequently, the software engineering community faces critical challenges in debugging, improving, and trusting these models, particularly in high-stakes applications where reliability and interpretability are essential. 

Various interpretability techniques have been proposed to address these challenges, yet they remain limited in scope, primarily providing explanations at the token level rather than uncovering model-wide trends. Shapley Values (SHAP), a game-theoretic approach, assigns importance scores to individual tokens but is computationally expensive and infeasible for large-scale model evaluations. Saliency maps use gradient-based methods to highlight influential tokens in a prediction, but their sensitivity to minor input perturbations makes them inconsistent and unreliable. Attention scores, commonly used in Transformer models, indicate which tokens a model “focuses on” but do not provide causal explanations—high attention does not necessarily mean a token is important for the prediction. These existing methods operate on single predictions and fail to offer global, post-hoc explanations that reveal how a model behaves across thousands of predictions. They do not answer fundamental questions such as whether a model systematically overuses certain syntactic patterns, exhibits biases toward specific token types, or demonstrates overinterpretation of irrelevant input tokens.

To bridge this gap, we introduce \codeRational, a global post-hoc interpretability framework that extends beyond traditional token-level methods. \codeRational extracts the minimal subset of input tokens necessary for a model’s prediction and aggregates these rationales across multiple predictions to generate a global view of model behavior. Unlike existing approaches, \codeRational provides insights at both the token level and the concept level, mapping individual tokens to higher-order programming structures. This enables practitioners to not only interpret single-instance predictions but also diagnose systematic trends across an entire dataset. By shifting the focus from isolated explanations to a concept-aware, model-wide analysis, \codeRational offers a more robust and scalable approach to understanding LLM4Code.

%% file: chapters/part_01_chap_03/sec_03_approach.tex
\section{Interpretability Approach}
\label{ch:rationales:sec_03}

The \textit{interpretability approach} enables the practitioners to examine which parts of the prompt mostly contribute to generating individual code prediction. The relationship between the prompt and output can be evaluated for a snippet scope (\ie using dependency maps \figref{fig:dependency_map}) as a \textit{local explanation} or can be evaluated for a complete dataset in a \textit{global explanation}. \codeRational is comprised of four steps described below.

\begin{figure*}[ht]

\centering
\includegraphics[width=\linewidth]{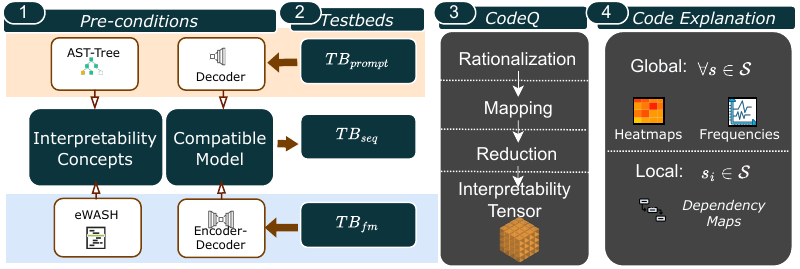}
 \caption{\codeRational: An Interpretability Approach}
    \label{fig:interpretability-pipeline}
\end{figure*}

\begin{enumerate}[label= {\textbf{Step$_{\arabic*}$:}}, ref=Step$_{\arabic*}$, wide,labelindent=5pt] \setlength{\itemsep}{0.2em}

\item \label{step1}\textbf{Stating Preconditions.} The first step consists of preparing the preconditions to use \codeRational to interpret the \lmc. The first precondition is making the model compatible using the algorithm described in Eq~\ref{eq:llm2}. The second precondition is to structure the \textit{interpretable-concepts} $\mathcal{C}$.

The interpretability concepts aim to introduce understandable concepts associated with the \lmc input-output. Most \lmsc operate upon features, such as token prediction, that do not \textit{inherently} match high-level concepts a human can easily understand.  A concept depends on its context (\secref{ch:rationales:sec_02}), therefore, the Intepretability-concepts depend on the SE tasks. We propose two types of interpretability concepts. The first one relies on the Abstract Syntax Tree (AST) for code generation (\figref{fig:rationales_taxonomy}). The second one is based on focal methods proposed at \ewash~\citep{clement_long-range_2021} for test case generation (\figref{fig:ewash_taxonomy}).

\item \label{step2}\textbf{Constructing Tesbets.} The second step involves testbed construction by curating and configuring the input to the model. The input depends on the SE task and the interpretability concepts. For instance, a set of prompts form a testbed $TB_{prompt}$ for code generation (\figref{fig:interpretability-pipeline}). Since the compatible model will generate code, we concatenate the generated code to the corresponding prompt and obtain a new testbed $TB_{seq}$ to apply \codeRational. We refer to $TB_{seq}$ as a generated set of snippets $\mathcal{S}$.

\item \label{step3} \textbf{Building Interpretability Tensors.} The third step consists of applying our interpretability method \codeRational. \codeRational is a novel method based on sequential rationales to interpret \lmc prediction. \codeRational works with encoder-decoder models such as BART~\citep{lewis2019bartdenoisingsequencetosequencepretraining} for test generation and decoder-only models such as Code-Parrot\citep{codeparrot} for code generation. \codeRational introduces three mathematical components to transform a set of tokens $w_t$ from a snippet $\mathcal{S}$ to a \textit{interpretability tensor} $\Phi$. 

\begin{figure}[hb]
    \centering
    \includegraphics[width=\linewidth]{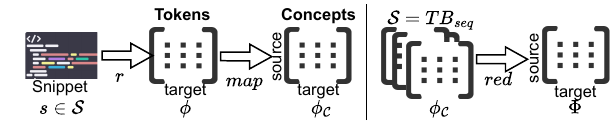}
     \caption{\codeRational Intepretability Tensors}
        \label{fig:interpretability-tensors}
\end{figure}

\begin{enumerate}[label= {\textit{Component$_{\arabic*}$:}}, ref=component$_{\arabic*}$, wide,labelindent=5pt]\setlength{\itemsep}{0.2em}
    \item \label{stet:ratialization} {\textit{Rationalization.}} The first step is the rationalization extraction $r(\mathcal{S})$ using  Eq~\ref{eq:combinatorial} for the greedy rationalization. Therefore, the step starts with the empty set $r^{(0)}=\emptyset$. The first rationales set is defined as  $r^{(1)} = \arg \max_{j \in \{ [t-1] \} } P_{\theta}(w_t|w_j)$. The rationalization process continues by choosing the sequence that maximizes the probability of the token $w_t$ at each step. %
\begin{equation}
r^{(k+1)} = r^{(k)} \cup  \arg \max_{j \in \{[t-1] \setminus r^{(k)} \} } P_{\theta}(w_t|w_{ \{ r^{(k)}\cup j \} })
\label{eq:greedy}
\end{equation}

The rationalization process has $O(T^2)$ complexity when the number of rationales is the same as the sequence $w_{1:T}$ with a token size $T$. We organize the rationales into a \textit{interpretability matrix} $\phi_{[src,tgt]}$, where $src$ are the source tokens and $tgt$ are the target tokens; $[src,tgt]$ has a size of $[T,T]$. Each cell of the matrix $\phi$ represents an estimated probability $r(s)$ for a given source and target tokens.

\item \textit{{Mapping.}} This step consists of assigning the corresponding label $c\in\mathcal{C}$ to each token $w_t$ in the matrix  $\phi_{[src,tgt]}$ by a change of dimension from $T$ to $|C|$. Formally, $map: \phi_{[T,T]} \to \phi_{[|C|,|C|]}$.

\begin{equation}
\phi_C = \phi_{[|C|,|C|]} = map(\phi_{[T,T]}, \mathcal{C})
\label{eq:mapping}
\end{equation}

The matrix $\phi_\mathcal{C}$ was named the \textit{interpretability matrix}. This tensor could adopt many dimension forms depending on the definition of the mapping function and, therefore, it may not capture all the aspects of its input domain. Also, notice that $\phi_\mathcal{C}$ loses the cardinality (\ie the sequence position order) from the initial rationale matrix $\phi_{[T,T]}$. 

\item \textit{{Reduction.}} The previous step generates a tensor $\phi_\mathcal{C}$  of interpretability concepts for a single snippet $s \in \mathcal{S}$. The reduction step introduces a reduction function $red$ when the analysis involves a set of snippets such as the testbed $TB_{seq}$ for a global analysis. The reduction function is defined as $ red: \Vec{\phi_{C}}, C \to \Phi_{[tgt,src]} $ with dimension $[C,C]$, therefore:

\begin{equation}
\Phi = red( \Vec{\phi_{C}} )=  \forall_{c} \in C, \forall_{s} \in S : g(\phi_{c}^{s})
 \label{eq:reduce}
\end{equation}

The tensor $\Phi$ was named the \textit{interpretability tensor}. The reduction function receives a \textit{vector} of \textit{interpretability matrices} $ \Vec{\phi_{C}}$ with a size $|S|$ that represents each sequence in a given testbed $S = TB_{seq}$. Note that $g$ is a function that computes the summarized rationale probability iterating first over all the concepts $c \in C$ and, later, over all the sequences $s \in S$. This $g$ summarization function could be defined in terms of average, median, maximum, or any particular aggregating function the user decides to implement for a tailored analysis.  

\end{enumerate}

\item \label{step4} \textbf{Code-based Explanation.} Finally, the interpretability approach considers the use of $\Phi$ to generate \textit{local post-hoc explanation} like the \textit{dependency maps}. \figref{fig:dependency_map}
shows three levels of human interpretable concepts: $L_1$) fine-grain level rationales, $L_2$) concept rationales, and $L_3$) modality. In addition, the interpretability tensor can be further explored to give rise to post hoc global explanations (refer to \secref{ch:rationales:sec_05} for specific statistical analyses). 
\end{enumerate}

%% file: chapters/part_01_chap_03/sec_04_experimental.tex
\section{Interpretability Setup}
\label{ch:rationales:sec_04}

This section outlines the methodology employed to calibrate our interpretability approach. We conducted an exploratory analysis  and a user study to explore the following RQs:

\begin{enumerate}[label=\textbf{RQ$_{\arabic*}$}, ref=\textbf{RQ$_{\arabic*}$}, wide, labelindent=5pt]\setlength{\itemsep}{0.2em}
    
    \item \label{rq:aplicability}\textbf{[Applicability]:} \textit{How applicable is \codeRational to globally interpret code generation?} {We define applicability as the ability to use \codeRational in crafting understandable explanations.  This RQ is focused on exploring the experience of using \codeRational to explain the behavior of \lmsc in code generation tasks.  We hypothesize that through greedy rationalization, we can explain the most important rationales of an input that lead to a certain prediction. The explanation with rationales would provide useful information about prediction dependencies captured by the model.}

     \item \label{rq:utility} \textbf{[Usability]:} \textit{How useful is  \codeRational in practical settings?} We validate the extent to which \codeRational is useful in practical settings in a user study. We measure the usability of our approach with qualitative metrics such as usefulness, reliability, readability of \codeRational, and the extent to which \codeRational helps in assessing the alignment of \lmsc.
        
\end{enumerate}

We followed our approach from \secref{ch:rationales:sec_03} for the code generation task to answer \ref{rq:aplicability} and both, code and test generation tasks to answer \ref{rq:utility}. Thus, our approach setup incorporates a \lmc, curated testbeds, and the application of \codeRational.

\begin{figure}[ht]
\centering
\includegraphics[width=\linewidth]{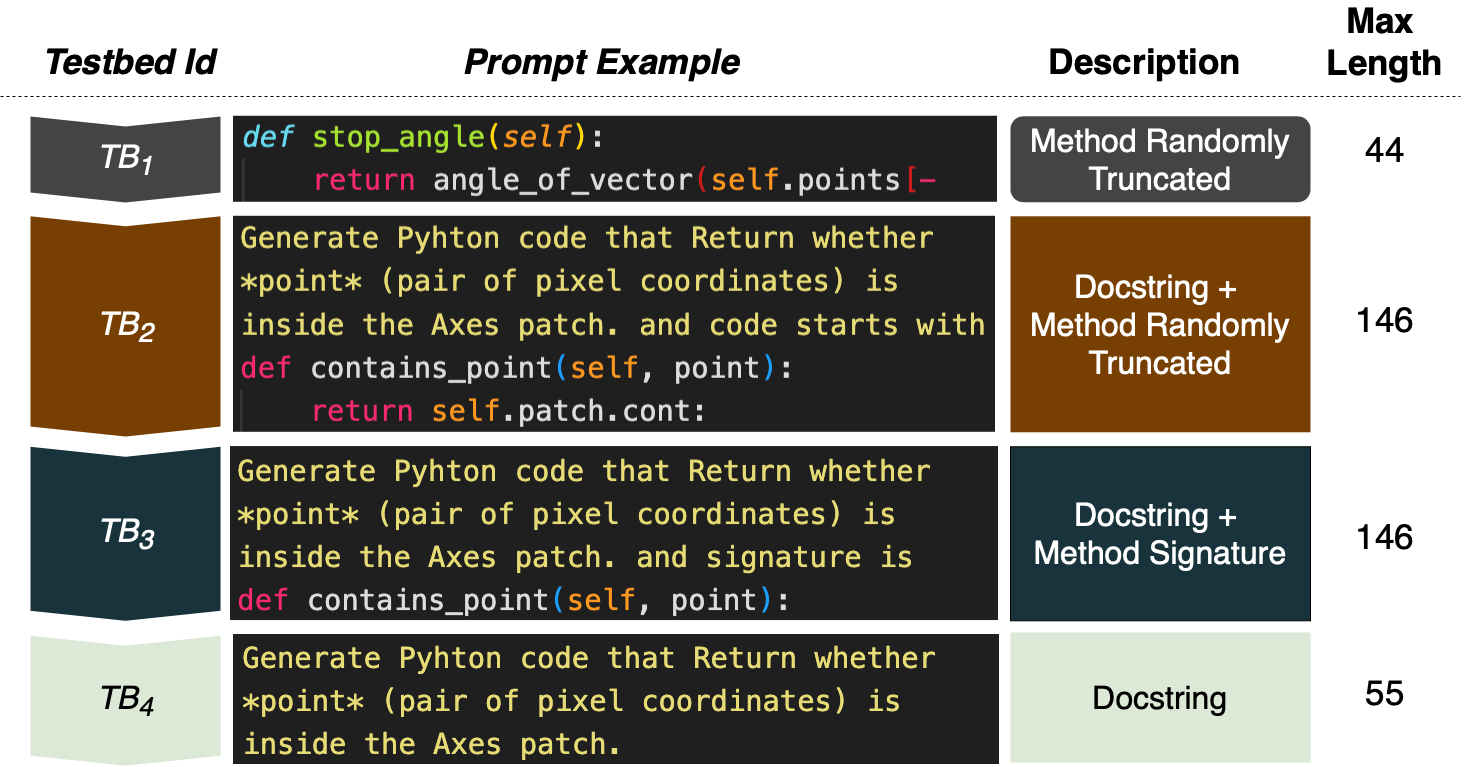}
\caption{{Code Completion Testbeds}}
\label{fig:code_completion_experiments}
\end{figure}

\subsection{ \texorpdfstring{\ref{step1} }{s1} Preconditions} \label{sec:design_preconditions}

\textbf{Model.} To conduct our experiments, we used \codeparrot \citep{codeparrot}, a decoder-only transformer model with $110M$ parameters and a maximum input size of $1,024$ tokens. The selected model was trained on the \textit{codeparrot-clean} dataset \citep{codeparrotDataset}, which includes approximately $5M$ Python samples from GitHub. We used the same codeparrot-clean dataset for the model's compatibilization process (\secref{ch:rationales:sec_03}). Throughout this process, we employed the Adam optimizer with a learning rate of $1e-5$ and a word dropout rate of $0.5$. The training involved $60K$ optimization steps, using a single batch per training step. In addition, we maintained the original settings and hyperparameters of the model without any modification.

\begin{figure}[ht]
\centering
\includegraphics[width=\linewidth]{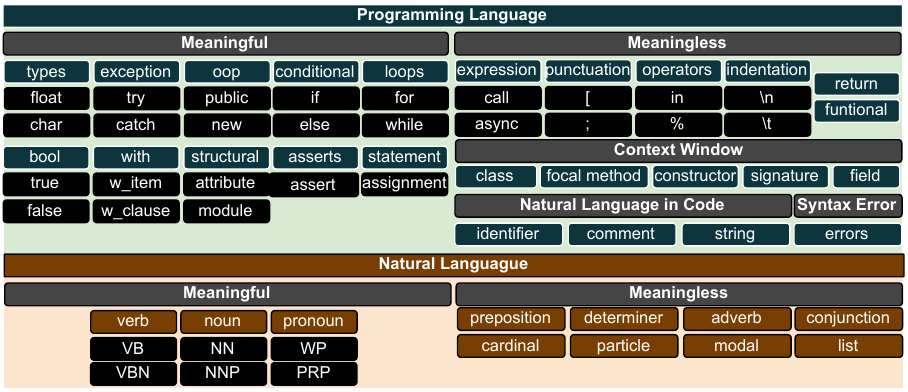}
\caption{Interpretability Concepts $\mathcal{C}$ for Java and Python}
\label{fig:rationales_taxonomy}
\end{figure}

\begin{figure}[ht]
\centering
\includegraphics[width=\linewidth]{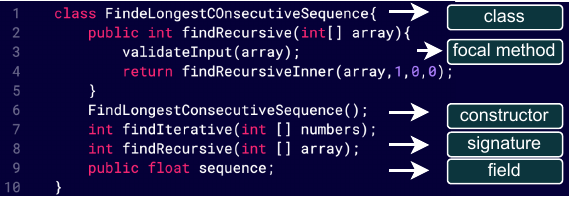}
\caption{Focal Method Concepts $\mathcal{C}$ }
\label{fig:ewash_taxonomy}
\vspace{2em}
\end{figure}

\textbf{Intepretability concepts.} 
We propose two types of taxonomies $\mathcal{C}$, the first one for code generation and the second one for test generation. The first taxonomy evokes the structure of the  ASTs. The AST structure depicts several concept levels from the leaf to the root. Traversing the AST we can associate the token to the corresponding Object Oriented Programming (OOP) concept (\figref{fig:rationales_taxonomy}). Notice that we aggregated natural language (NL) concepts to our $\mathcal{C}$  (using NLTK \citep{nltk}). The NL concepts enable the mapping and explanation of AST nodes such as comments, strings, descriptions, or identifiers. The second type of taxonomy is based on context windows from \ewash~\citep{clement_long-range_2021}. A context window embodies the scopes of a Java file, starting from the class. To generate a test the input should include the focal method and any other additional context such as the constructor of the class or class fields~\figref{fig:ewash_taxonomy}.

\subsection{\texorpdfstring{\ref{step2} }{s2} Testbed construction}

We mined approximately $50K$ unique Python snippets using Galeras\citep{daniel23}, all sourced from the most popular Python GitHub repositories between 01/01/22 and 01/01/23. From the mined snippets, we built four distinct evaluation testbeds. Each testbed was generated by different prompts types \figref{fig:code_completion_experiments}. Each prompt indicates a specific context to generate the code (\eg part of the body method code, signature, method description). 

The first testbed \sgbd, incorporates prompts with the method \textit{signature} and the method \textit{body}. The body was randomly truncated one code line after the signature to ensure sufficient code tokens. In the second testbed \dcsgbd, we added the method description \textit{docstring} to the prompt, along with the signature and partially truncated body. The third testbed \dcsg contains the docstring and the signature only. Finally, in the fourth testbed \dc, the prompt asked the model to complete code using only the docstring, without any partially written code.
After asking the \lmc to complete the code we append the generated code to the prompt. The prompt with the source tokens and the generated code with the target tokens configure the snippet $s$. Each testbed compiles 100 unique sequences.

Our interpretability design aims to evaluate the applicability and usability of our interpretability approach. Therefore, we want to provide recourse to statistical evaluations. Thus, we sample the rationales for the same sequence 30 times. This process resulted in $3K$ sequences per testbed. We generated a total of $12K$ sequences with their respective generated code.

\subsection{ \texorpdfstring{\ref{step3} }{s3} Interpretability Tensors}

{\textbf{Rationalization}. We applied greedy rationalization from \ref{stet:ratialization} on each testbed. In other words, we computed the rational $r(\mathcal{S})$ for each snippet $s \in \mathcal{S}$ to obtain the matrices~$\phi$.

\textbf{Mapping}. In this step, we used the $map$ function and the concepts $\mathcal{C}$ for code generation defined in the preconditions. Each token is mapped to the corresponding AST leaf node similar to ASTrust\citep{palacio2024trustworthyinterpretablellmscode}. The AST node type from tree-sitter \citep{tree-sitter} the associated OOP concept.

\textbf{Reduction.} {The reduction step in our interpretability design consist of reducing $100$ $\phi$ or $\phi_\mathcal{C}$ tensors depending of the type of explanation level. The reduction of $\phi$ explains the rationales at the token level meanwhile the reduction of $\phi_\mathcal{C}$ explains at the concept level.  

The reduction step generates the interpretability code tensor $\Phi$ that is used as an input to our code base explanation for the \ref{step4} from our approach. The explanation could be of a single snippet $s$ as a local explanation (used in our user study \secref{ch:rationales:sec_05}) or the complete testbed $\mathcal{S}$ as a global explanation (used in our exploratory analysis \secref{ch:rationales:sec_06}). 
}

%% file: chapters/part_01_chap_03/sec_05_exploration.tex
\section{On the Reproducibility of Code Rationales}
\label{ch:rationales:sec_05}

\subsection{Exploratory Methodology}
Our exploratory methodology compresses three statistical analyses to evaluate the applicability of \codeRational. Each statistical analysis aims to provide recourse to understand \codeparrot's behavior for code generation. This subsection describes the statistical analyses presented in \figref{fig:code_generation_results}.

{\textbf{Analysis$_1$: Concept dependencies.} We use heatmaps to explore the dependency between the source rationales and generated target concepts. To configure this analysis, we used the $100$ unique sequences from testbeds and the $30$ samples. Therefore, our analysis compresses $30$ $\Phi$ interpretability tensors of $100$ $\phi$ matrices. We computed the median of the rationale probability across all the trials. Any new target or rationale not present in the previous tensor during the iteration of all trials was appended to the result. 

We aim to provide a confidence value on the rationale probability to explain the relationship between the rationale and the target concepts. Thus, we applied bootstrapping to ensure at least $100$ values per concept. Area \circled{1} in \figref{fig:code_generation_results} presents the generated heatmap for the \sgbd testbed.} 

\textbf{Analysis$_2$: Frequency of Rationales} We created frequency maps to analyze how \codeRational can identify the most relevant rationales for code generation. To build this analysis, the reduction uses a $g$ function to collect the rational probabilities per concept and count the frequency. The process is computed for each $\Phi$ trial and combined into a single tensor for all the testbeds. Then, we created a tree-map representation of the resulting tensor to visually inspect the frequency of rationale values per concept, as detailed in area \circled{2} of \figref{fig:code_generation_results}. We also included the mean and standard deviation values of the rationale probabilities and the proportionality for each concept. Note that the color and size of the boxes in \circled{2} correspond to the frequency of the concept presented on the logarithmic scale.

{\textbf{Analysis$_3$: Distribution of Rationale Probability.} We created density plots to inspect the distribution shape of rationale probabilities using \codeRational. In a similar way to the previous analysis, the reduction function collected the probabilities of each rationale per concept for each testbed. Then, we created density plots to describe the tendency and shape of the distribution for each concept, as well as the differences across the four testbeds. Area \circled{3} of \figref{fig:code_generation_results} shows the density plots for some of the most, moderately, and least frequent concepts across all testbeds.}

\subsection{Exploratory Results}

\begin{figure*}[ht]
\caption{\codeRational Explanations for Code Generation}
\centering
\includegraphics[width=\textwidth]{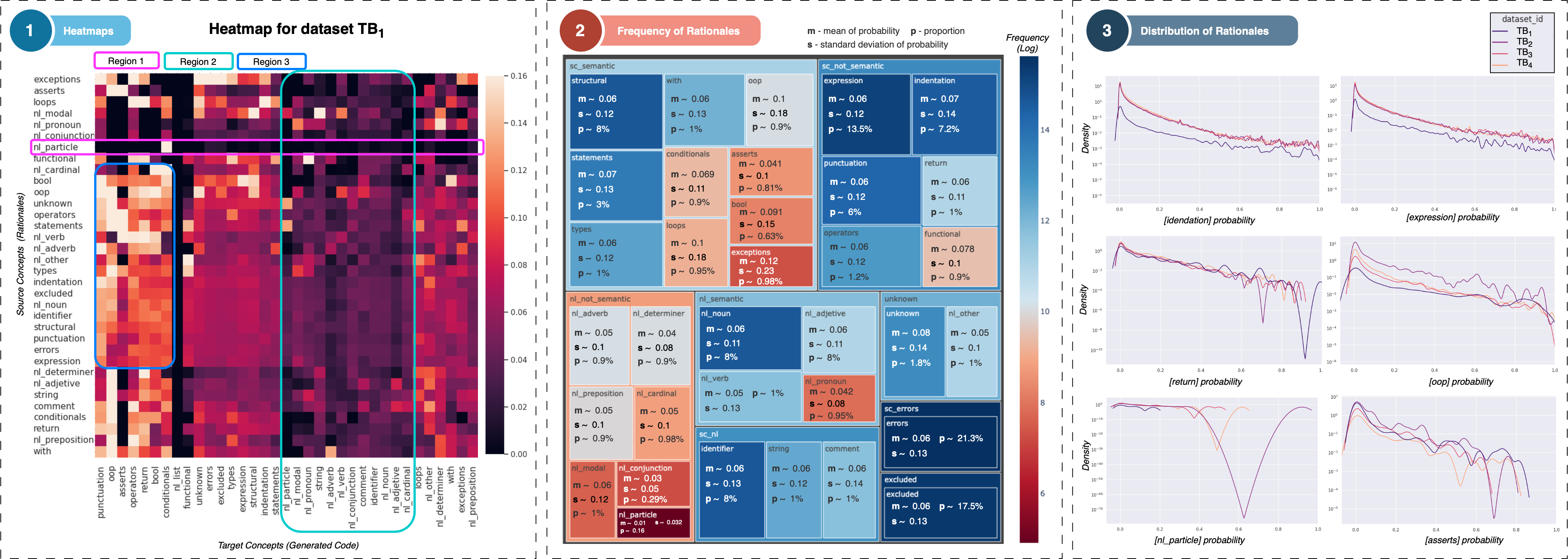}
\label{fig:code_generation_results} 
\end{figure*}

\textbf{Analysis$_1$.} {Heatmaps (\circled{1} in \figref{fig:code_generation_results}) highlighted areas of influence of rationales on the output for \sgbd. Specifically, \textit{Region 1} showed that \texttt{\small{[nl\_particle]}} had minimal impact. Similarly, \textit{Region 2} indicated that natural language elements in the output were hardly affected by any rationale. Lastly, \textit{Region 3} revealed that  \texttt{\small{[oop]}}, \texttt{\small{[operators]}}, \texttt{\small{[punctuation]}}, and \texttt{\small{[return]}} were significantly influenced by meaningful and meaningless concepts.}

\begin{boxK}
{Heatmaps allow the global identification of regions with the minimum and the most influenced (concept) rationales.}
\end{boxK}

\textbf{Analysis$_2$.} {Frequency maps (\circled{2} in \figref{fig:code_generation_results}) revealed that the most frequent rationales used by \codeparrot in code generation were \texttt{\small{[errors]}}, \texttt{\small{[excluded]}}, \texttt{\small{[expression]}}, \texttt{\small{[identation]}}, \texttt{\small{[punctuation]}}, \texttt{\small{[structural]}}, \texttt{\small{[statements]}}, \texttt{\small{[nl\_noun]}} and \texttt{\small{[identifier]}}. Conversely, the least frequent types of rationales include \texttt{\small{[nl\_particle]}}, \texttt{\small{[nl\_modal]}},\\ \texttt{\small{[nl\_conjuction]}}, \texttt{\small{[nl\_pronoun]}} and \texttt{\small{[exceptions]}}. We also note that most of the means of the rationale probabilities were around $0.06$, with a standard deviation lower than $0.13$.}

\begin{boxK}
{Frequency maps allow identifying the most and least frequent types of (concept) rationales that guide code predictions.}
\end{boxK}

\textbf{Analysis$_3$.} {Density plots of rationale probabilities (\circled{3} in \figref{fig:code_generation_results}) showed similar trends for the most frequent concepts (\eg \texttt{\small{[indentation]}}, \texttt{\small{[expression]}}) and medium frequent concepts (\eg \texttt{\small{[return]}}, \texttt{\small{[oop]}}), with most probabilities concentrated around $0.01$. The least frequent concepts showed high variability, particularly with \dcsgbd, which had the highest variability for \texttt{\small{[nl\_particle]}} but with most values around $0.2$. Overall, the probabilities for the most frequent rationales were very low, with the 75th percentile at $0.05$ and a maximum of $0.079$. In contrast, the less frequent rationales had slightly higher probabilities, with the 75th percentile at $0.064$ and a maximum of $0.144$.}

\begin{boxK}
Distribution plots reveal how similar is the distribution of rationales probabilities across all testbeds in terms of \textit{interpretable-concepts}.
\end{boxK}

\textbf{Discussion.}
Heatmaps indicated that overall, source code elements, whether meaningful or meaningless, do not show dependencies on NL rationales. We attribute this behavior to the specific nature of \codeparrot, which was exclusively trained to predict source code.  Heatmaps show occasional instances of self-dependencies for some types of rationales. For example, in the \sgbd dataset, rationales such as \texttt{\small{[oop]}}, \texttt{\small{[loops]}}, and \texttt{\small{[exception]}} significantly influenced the generation of the same types of target tokens. This phenomenon can be attributed to the mapping function $map$, which assigns each token to a concept based on the code's AST. As a result, the rationale of a particular token can fall into the same category as the generated token.

We observe a tendency for certain code elements such as \texttt{\small{[oop]}}, \texttt{\small{[operators]}}, \texttt{\small{[punctuation]}}, and \texttt{\small{[return]}} to be more dependent on a broad subset of rationales than others, as depicted in region 2 of area \circled{1} in \figref{fig:code_generation_results}. We attribute this behavior to the location of these elements within a function's corpus, as they appear after the signature and sometimes at the end of the body. Consequently, they are not frequently included as part of the prompts in the four datasets we studied. Conversely, as depicted in region 3 of area \circled{1}, natural language target elements such as \texttt{\small{[nl\_adjective]}} and \texttt{\small{[nl\_list]}} exhibit low dependency.

Evidence from the frequency maps suggests that \codeparrot may suffer from \textit{overinterpretation}, as it often relies on meaningless rationales (\ie \texttt{\small{[identation]}}, \texttt{\small{[expression]}}, \texttt{\small{[punctuation]}}, \texttt{\small{[errors]}} and \texttt{\small{[unknown]}}) rather than meaningful rationales (\eg \texttt{\small{[bool]}}, \texttt{\small{[conditional]}}, and \texttt{\small{[assert]}}) for code generation across all testbeds.

Finally, our findings indicate no clear correlation between the frequency of a rationale and its probability distribution. Specifically, the probability values of the most frequent rationales were smaller than those of the least frequent rationales.

\subsection{Threats to validity}

We considered all the sequences as part of the code including comments and method descriptions. Therefore, the complete sequences are parsed at the AST. Usually, AST nodes such as code comments, strings, and identifiers are mapped to an NL concept. Descriptions such as the docstring are AST errors. We made an extra evaluation by splitting code from docstring and adding docstring as a natural language. We found the number of concepts was similar and did not impact the interpretability-concept rationale distribution for global analysis. This can be explained due to the \codeparrot training dataset with only comments.

%% file: chapters/part_01_chap_03/sec_06_user_study.tex
\section{User Study Design}
\label{ch:rationales:sec_06}

We aim to gather information about the usability of our technique in practical settings from the practitioners. To this end, we need insights from domain experts with ML and SE knowledge. Therefore, our RQs are aimed at evaluating the \codeRational usability. To enhance the quality of our findings we followed \textit{purposive sampling approach}~\citep{baltes_sampling_2021} and carefully reached out to experts in our network.

We contacted practitioners of varying backgrounds, including students, academic and industrial researchers, data scientists, and developers who use \lmc for multiple downstream SE tasks. 
\subsection{Structure}

The survey consists of four different sections\footnote{The full list of questions refer to \citep{CodeQ2025} in the appendix to see questions asked in each section.} The \textit{First Section} is designed to gauge the participant’s proficiency in both Python and Java and their familiarity with utilizing language models for code-related tasks. The participants were asked their use cases for each programming language. They were also asked how long they have been using ML for code completion and test case generation tasks.  This information helps us to qualify participant's fit to our user study. The \textit{Second Section} presents ten samples, five for code completion in Python (similar as shown at \figref{fig:dependency_map}) and five for Test case generation in Java, where users were asked to pinpoint tokens responsible for generating the predicted tokens. All the participants were presented with the same samples. The users were also asked to identify whether the generated code was correct. 

The \textit{Third Section} presents our \codeRational visualizations to explain the generation of the tokens posed in the second section. In this section, users were asked to rate the visualizations generated using \codeRational. We asked the participants to assess the quality of rationales by indicating how strongly they agree or disagree with rationals generated by \codeRational (\ie Agree, Mostly Agree, Neutral, Mostly Disagree, and Disagree). In the \textit{Final Section,} we wrap up our survey with some general questions to measure participants' views on the Usefulness of our approach based on the overall survey experience. 

\subsection{Qualitative Metrics} \label{survey_metric}

\begin{enumerate}[label= \textbf{[m$_{\arabic*}]$:}, ref=m$_{\arabic*}$, wide,labelindent=5pt]\setlength{\itemsep}{0.2em}

\item \textbf{Usefulness.}  \label{user_study_usefulness}  \textit{How useful is code-based explanations in practical settings?} Usefulness measures the quality of \codeRational to be useful in various downstream SE tasks. As mentioned in \ref{D3}, usable and practical interpretable tool for evaluating \lmsc is highly important \citep{linardatosExplainable, management_solutions_xai} due to the impact on efficiency and adaptability to the real scenarios where the model is evaluated \citep{trustincollab}.

We expose \codeRational technique to show interpretable code-based explanations for two downstream tasks.  Based on these explanations we prompt the participants to gauge the usefulness of our method in various model development activities. In addition, we also asked an open-ended question to understand participants' views on the usefulness of our technique. 

\item \textbf{Readability.} \label{user_study_readability} \textit{How informative is our code-based explanation, and how easy are our visualizations to understand?} The interpretability method must be informative and readable (\ie easy to associate with concepts). We assess the readability of our technique with a user study among practitioners and aligned with the informativeness desideratum \ref{D2}. We explicitly ask the users the degree to which they find our technique readable and informative. We do this for each downstream task we present in our study.  We ask participants to reflect upon the readability of our explanations presented in AST representations and Context Level dependency maps. A good quality, readable explanations are informative explanations.

\item \textbf{Reliability.} \label{user_study_reliability}\textit{How reliable is \codeRational in identifying rationales?} We want to validate to what extent the user can validate the correctness, and consistency of the model \citep{liu2024trustworthyllmssurveyguideline} using \codeRational. We ask the users to reflect upon the reliability of \codeRational on two downstream tasks and to generalize its reliability on any code generation task.

\item \textbf{Alignment.}\label{user_study_alignment} \textit{How useful is \codeRational in assessing \lmc alignment?} We define alignment as the degree to which the model’s rationales sync with human rationales. We ask users how they agree with code rationales generated by our technique. In addition to this, we also asked participants to report rationales manually for given predictions. We then calculate the Jaccard Similarity between \codeRational rationales and human rationales. Jaccard Similarity ranges from 0 to 1. A higher Jaccard Similarity (closer to 1) would mean the model rationale is highly aligned with the human rationale and vice-versa. Note, we assume that the rationalization process is accurate and \codeRational is indeed reflective of model decisions.
\end{enumerate}

\subsection{Sample Selection}

The survey asks participants to evaluate five different samples each for both code completion and test case generation tasks consisting of distinct approaches for each use case namely with AST-concepts and Context-levels \secref{sec:design_preconditions}. Each case contains at least one correct sample and a mixture of samples containing various semantic and syntactic errors produced as documented in \citep{synerror}\citep{synerror1}\citep{synerror2}. Two researchers that contribute to this research independently made a comprehensive list of these documented errors. One of those authors then manually went through the generated samples from both Test Case generation and Code Completion. These samples were then mapped to documented error categories. The collected samples were then verified by an author not involved in the sample categorization process. Choosing samples with semantic and syntactic errors of different kinds helps us cover multiple scenarios. We also added samples with no error in our study to understand model behavior on correct predictions. 

\subsection{Data Collection}

 We reached out to 86 potential participants from both industrial and academic sectors with mixed experience in Python, Java, and Machine learning, who were not involved in or aware of the purpose of this work. Participation in the user study was completely voluntary. Out of these groups, 43 participants completed the survey. However, we had to discard the 6 participants due to the incomplete or poor quality of their responses leaving 37 valid responses. Of the valid responses, 39\% of the participants were researchers and 61\% were developers. 91\% of the practitioners were Intermediate or Experts in Python, and 79\% were intermediate or experts in Java. The study was performed using Qualtrics \citep{noauthor_qualtrics_nodate}.

\subsection{Validity}
To solidify our survey, we initially conducted a pilot study with five participants not included in the main survey. Based on the results of this pilot, we eliminated any leading questions, added varying error cases to both downstream cases, and corrected all minor errors. We also contacted an external expert who is knowledgeable in conducting qualitative studies to assess the validity of our survey. In addition, two authors carried out the survey data analysis to avoid bias and human error. For the open-ended questions, two authors gathered and coded responses from the survey independently. Any differences were resolved through discussion to reach a consensus

%% file: chapters/part_01_chap_03/sec_07_results.tex
\section{User Study Results \& Discussion}
\label{ch:rationales:sec_07}

The summarized responses are detailed in \tabref{tab:survery_metrics} with the metrics discussed in \ref{survey_metric}. We also present results from open-ended questions.
\subsection{Results}

\textit{Usefulness \ref{user_study_usefulness}:}  The data reveals that practitioners find \codeRational useful in various model development activities. 89\% of the practitioners believe \codeRational can be useful in fine-tuning the model with no participants disagreeing with this potential use case. This is indeed a great observation as \codeRational can reveal code concepts that the model is struggling with. Thus, practitioners can fine-tune models with such missing code concepts. Similarly, 84\% agreed that it can be used to interpret model output. 81\% agreed that our tool can be useful for inferring causal relationships between the input and output of the model. Overall, practitioners believe \codeRational is useful across the board. On the open-ended question, participants favored the ability of \codeRational to show a causal relationship between the input and output of the model, they were appreciative of \codeRational in helping them interpret model behavior. Similarly, participants also saw the potential use case of the technique to help debug and design better models. One user stated, \textit{``Useful for identifying what input data leads to generated code ''}. Similarly, another user quoted, \textit{``I think it is useful to get a more detailed look at what the model is doing. It would be useful for debugging and understanding the model better.''}
\begin{boxK}
 \codeRational can be useful in model development activities such as fine-tuning, debugging, or causal relationships between input and output.
\end{boxK}

\input{tables/chap_rationales/tab1_survey}

\textit{Readability \ref{user_study_readability}:} 78\% of the participants found AST-based code explanations to be informative for explaining code completion. Moreover, none of the participants disagreed with the informativeness of AST-based explanations. Similarly, 79\% of the participants found context-level-based explanations to be informative for explaining test-case generations. However, only 52\% found AST-based graphical representations easy to read and 59\% found context-level graphical representations easy to read. This explains that our proposed taxonomy for code-based explanations of code-completion and test-case generation is solid, however, there is an opportunity for HCI researchers to improve the readability of graphical representations.

\begin{boxK}
Participants highlighted its ability to show a causal relationship between the input and output of the model with readable and informative explanations for code completion and test case generation tasks.
\end{boxK}

\textit{Reliability \ref{user_study_reliability}:} 68\% of the participants found that code-based explanations for test case generation were mostly reliable. Similarly, 54\% found it to be mostly reliable for code completion tasks. However, only 38\% agreed that our technique would be reliable for any code generation task. This is not surprising, as we only provided two use cases in our survey, and further work in other downstream tasks is needed to demonstrate the reliability of the technique for all code generation tasks. This is echoed by 40\% of the participants showing neutral responses to our technique’s reliance on any code generation task.

\begin{boxK}
    \codeRational is fairly reliable for the tasks presented in our case study. However, with further exploration and development, the technique has the potential to enhance reliability in a wider array of code generation tasks.
\end{boxK}

\textit{Alignment \ref{user_study_alignment}:}
Our analysis revealed significant disparities between model-generated rationales and human reasoning for code-related tasks. For code completion samples, the Jaccard Similarity score was a mere 0.074, while for test-case generation samples, it was 0.306. These low scores indicate a substantial difference between the rationales provided by the model and those given by humans.

This gap in reasoning is further evidenced by participant agreement rates with the generated rationales. Only 53.6\% of participants concurred with the model's explanations for code completion tasks, while 62\% agreed with the rationales for test-case generation tasks. These figures underscore a misalignment between the model and human reasoning processes.

These findings highlight an important area for future research: bridging the gap between the model reasoning and human thought processes in code generation tasks.

\begin{boxK}
 \codeRational helps uncover model rationale, facilitating comparison with a human rationale to promote a fair level of trust and distrust in the model.
\end{boxK}

\subsection{Discussion}

We find that \codeRational is highly useful in practical settings. Perhaps the most important finding is the result of our alignment study~\ref{user_study_alignment}. Trust \ref{D1} is one of the major components while using \lmc. However, misplaced trust can lead to catastrophic consequences such as data leaks, vulnerabilities, security risks, and more \citep{thorneInterplay}, while lack of it can hinder innovation \citep{balayn2024empiricalexplorationtrustdynamics}. Alignment~\ref{user_study_alignment} is one of the key components in fostering appropriate trust in \lmc~\citep{sun2024trustllmtrustworthinesslargelanguage}. We have shown that \codeRational can produce rationale behind the prediction of tokens by \lmc in two downstream tasks. This can be applied on a larger scale to study the alignment of \lmc with human counterparts. These findings can be used to develop appropriate trust and foster appropriate distrust, often defined as lack of trust \citep{distrust2023} in models. A well-aligned model would promote trust while an ill-aligned model would promote distrust among practitioners. Recent work has also shown that trust and distrust are distinct concepts, and fostering appropriate levels of both can lead to careful and critical evaluation of \lmc, thereby, potentially improving decision-making and reducing errors \citep{distrust2023}. The practitioners were a bit skeptical about the reliability of \codeRational, however, this is expected as we did not provide the internal details of the model behind our technique in the survey. However, our code is publicly available and our method is based on the greedy rationalization technique that has already been applied in \lms~\citep{vafa_rationales_2021}.

\subsection{Threats to Validity}

\textit{Internal.} Some major threats might arise from our syntax taxonomy and sample selection in code completion and test-case generation. To mitigate the bias in sample selection, we utilized the open coding process in constructive grounded theory \citep{kathy2006}. Four authors reviewed each selected sample. We followed a similar approach to derive our taxonomy as well. We have also released all our data relating to our taxonomy and sample selection in our online appendix ~\citep{CodeQ2025}.

\textit{External.} One potential threat to generalizability of our results could be the selection of participants for the study. We followed a purposive sampling strategy for our study. We carefully reached out to participants with a wide range of experiences in the industry and academia. We also added general filtering questions to ensure participants had both DL and SE experiences. We released all the anonymized responses we gathered in our online repository for transparency~\citep{CodeQ2025}. 

%% file: tables/chap_rationales/tab1_survey.tex
\begin{table}[ht]
\centering
\caption{User Study Results}

\label{tab:survery_metrics}

\scalebox{0.79}{%
\setlength{\tabcolsep}{3pt} 
\renewcommand{\arraystretch}{1}

\begin{tabular}{clllccc}
\textbf{\begin{tabular}[c]{@{}c@{}}Metric\\ ID\end{tabular}} &
  \textbf{\begin{tabular}[c]{@{}c@{}}Downstream \\ Task\end{tabular}} &
  \multicolumn{1}{c}{\textbf{Use Case}} &
   &
  \multicolumn{3}{c}{\textbf{Results(\%Answers)}} \\ 
 &
   &
   &
   &
  \textit{\textbf{\begin{tabular}[c]{@{}c@{}}Mostly \\ Agree\end{tabular}}} &
  \textit{\textbf{Neutral}} &
  \textit{\textbf{\begin{tabular}[c]{@{}c@{}}Mostly \\ Disagree\end{tabular}}} \\\cline{1-3} \cline{5-7} \\
\multirow{5}{*}{\ref{user_study_usefulness}} &
  \multirow{5}{*}{\textit{Model related}} &
  Debugging a \lmc &
   &
  56.0 &
  22.0 &
  22.0 \\
 &
   &
  Fine-tunning a \lmc &
   &
  {\ul 89.0} &
  11.0 &
  {\ul 0.0} \\
 &
   &
  Curating datasets (tr/tst/val) &
   &
  68.0 &
  21.0 &
  11.0 \\
 &
   &
  Inferring causal relationship &
   &
  {\ul 81.0} &
  11.0 &
  8.0 \\
 &
   &
  Interpreting \lmc output &
   &
  {\ul 84.0} &
  13.0 &
  3.0 \\
\multicolumn{1}{l}{} &
   &
   &
   &
  \multicolumn{1}{l}{} &
  \multicolumn{1}{l}{} &
  \multicolumn{1}{l}{} \\ %
\multirow{5}{*}{\ref{user_study_readability}} &
  \textit{Code Comp.} &
  AST-based informativeness &
   &
  {\ul 78.0} &
  22.0 &
  {\ul 0.0} \\
 &
  \textit{Test Case Gen.} &
  Context-level informativeness &
   &
  {\ul 79.0} &
  11.0 &
  10.0 \\
 &
  \textit{Code Comp.} &
  AST-based readability &
   &
  52.0 &
  27.0 &
  21.0 \\
 &
  \textit{Test Case Gen.} &
  Context-level readability &
   &
  59.0 &
  14.0 &
  27.0 \\
\multicolumn{1}{l}{} &
   &
   &
   &
  \multicolumn{1}{l}{} &
  \multicolumn{1}{l}{} &
  \multicolumn{1}{l}{} \\  
 &
   &
   &
   &
  \textit{\textbf{\begin{tabular}[c]{@{}c@{}}Mostly \\ Reliable\end{tabular}}} &
  \textit{\textbf{Neutral}} &
  \textit{\textbf{\begin{tabular}[c]{@{}c@{}}Mostly \\ Unreliable\end{tabular}}} \\\cline{1-3} \cline{5-7} \\
\multirow{3}{*}{\ref{user_study_reliability}} &
  \textit{Code Comp.} &
  Reliability &
   &
  54.0 &
  25.0 &
  21.0 \\
 &
  \textit{Test Case Gen.} &
  Reliability &
   &
  {\ul 68.0} &
  16.0 &
  6.0 \\
 &
  \textit{Any} &
  Prediction of any Token &
   &
  38.0 &
  {\ul 40.0} &
  16.0 \\ \cline{1-3} \cline{5-7} 
\end{tabular}

}
\vspace{0.5cm}
\end{table}

%% file: chapters/part_01_chap_03/sec_08_related.tex
\section{Related Work}
\label{ch:rationales:sec_08}

Interpretability methods have been used to explain the predictions generated by deep learning models, complementing traditional evaluation methods and enhancing our understanding of the decision-making process to reduce uncertainty. In the context of Machine Learning and closer to the goals of our study, Lage \etal \citep{lage_human_2019} conducted controlled human subject experiments to study the degree to which the complexity of decision trees as surrogate models affects users' perceptions of response time, accuracy, and difficulty of use. Similarly, Poursabzi-Sangdeh \etal \citep{poursabzi-sangdeh_manipulating_2021} acknowledge the lack of consensus around defining, quantifying, or measuring the interpretability of machine learning models.

Post-hoc interpretability assumes the model is a black box and analyzes its behavior by examining changes in the output derived from the input. In  Machine Learning, most research has focused on improving the plausibility and faithfulness of explanations such as LIME \citep{ribeiro2016why}, DeepLIFT \citep{shrikumar_learning_2019}, and Shapley values \citep{lundberg2017unified}, to the best of our knowledge, our work is the first to evaluate a feature importance-based interpretability method qualitatively for code predictions. Incorporating human feedback into evaluating the desiderata and faithfulness of interpretability techniques is crucial \citep{chen_what_2022}. \codeRational has been evaluated through a user study to measure the utility of explanations for code generation from the user's perspective.

In the context of AI4SE, post-hoc methods have been adopted for different explanation types such as counterfactual~\citep{cito_counterfactual_2022} and causal~ \citep{palacio_toward_2023} for code generation. In addition, methods based on \textit{self-attention} have been broadly adopted~\citep{mohankumar_towards_2020}. For instance, Mohammadkhani \etal~\citep{mohammadkhani2023explain} propose an XAI method based on attention for three downstream tasks (\eg generation, refinement, and translation). However, it has been shown that attention-based explanations are generally not faithful (\ie do not accurately reflect how a model makes a prediction) \citep{jacovi2020faithfully, jain_attention_2019, serrano_is_2019}. Attention-based interpretability typically averages attention weights across all layers and heads. The average of attention weights is not a faithful score to reflect the importance of individual tokens~\citep{brunner2020identifiability}. Basting \etal~\citep{bastings_elephant_2020} suggest that general input saliency methods are better suited to provide explanations than attention. As demonstrated by Vafa \etal~\citep{vafa_rationales_2021}, \textit{rationales} contributes to more faithful and plausible explanations than gradient and attention-based methods.

%% file: chapters/part_01_chap_03/sec_09_lessons.tex
\section{Lessons Learned}
\label{ch:rationales:sec_09}

\textbf{Lesson$_1$. Accuracy is neither a sufficient nor a complete group of metrics to assess Language Models (LMs).} The reported accuracy of \codeparrot on the code completion task using HumanEval was $3.8\%$ for pass@1 and $12.78\%$ for pass@100. Although low accuracy indicates the model's overall performance for code completion, it was insufficient to determine dependencies between the input context and predicted tokens or detect issues such as overinterpretation. We have shown that \codeRational provides valuable information to complement the reported accuracy scores.

\textbf{Lesson$_2$. \codeRational facilitates the prompt engineering optimization.} Our approach is concentrated on explaining the input-output relationship. Rationalization aims to elicit the minimum input prompt that generates a good answer. Ergo, \codeRational constitutes a good candidate to optimize and evaluate prompt engineering strategies. 

\textbf{Lesson$_3$. \codeRational enables new research opportunities in Human-Computer Interaction (HCI) for \lmsc.} Although \codeRational was considered useful for detecting dependencies and causal relationships between the context input and the model's predictions, only $59\%$ of participants in our user study agreed that the visualization is easy to read.

\textbf{Lesson$_4$. \codeRational allows a fair model comparison.} Our exploratory analysis provides insights into applying \codeRational for code generation and crafting explanations through rigorous statistical analysis of interpretability tensors. A systematic and more extensive empirical analysis of other models might promote a clearer understanding and fair comparison of how these models perform on different tasks.

\textbf{Lesson$_5$. \codeRational is computationally expensive.} Computing the rationales is highly resource-consuming in both steps: model compatibilization and greedy rationalization. The compatibilization requires fine-tuning with the same training set using word dropout. The greedy rationalization is $O(n^2)$ when all the rationales impact the same number of target tokens.

%% file: chapters/part_02_chap_01/do_code.tex
\chapter{\texorpdfstring{\docode}{docode}, A Causal Interpretability Approach}
\label{ch:docode}

\epigraph{\scriptsize (...) causation is not merely an aspect of statistics; it is an addition to statistics, an enrichment that allows statistics to uncover workings of the world that traditional methods alone cannot (...)}{\scriptsize\textit{Judea Pearl \citep{Pearl2016Causality}}}

\begin{figure}[ht]
		\centering
		\includegraphics[width=1\textwidth]{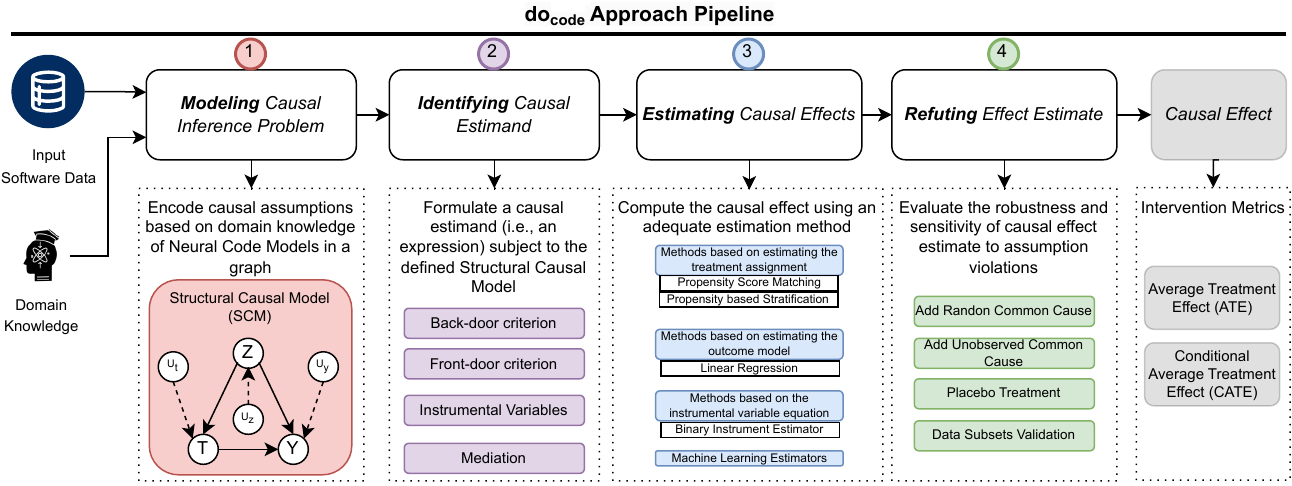}
		\caption{Overview of the \docode Approach: Each numeral represents a step in the process of generating causal interpretations. First, a Structural Causal Model (\scm) that frames the explanation hypothesis is formulated. Second, \docode executes graph surgery on the \scm to isolate a targeted estimand of the causal effect. Third, the causal effect is assessed based on the targeted estimand. Finally, the estimated effect is scrutinized through refutation techniques to confirm its validity.}
        \label{fig:pipeline}
\end{figure}

\docode comprises a set of statistical and causal inference methods to \textit{generate (and evaluate) post hoc interpretations} of \nlms. In the simplest of terms, our proposed interpretability approach asks \textit{Why does a \nlm make a given code prediction?} -- and provides a framework for answering this broad question by \circled{1} \textbf{modeling} the inference problem to encode causal assumptions, which rely on domain knowledge of \nlms, in a graph representation; \circled{2} \textbf{identifying} the causal estimand based on the previous graph representation; \circled{3} \textbf{estimating} the causal effect based on probabilistic and machine learning methods that operate on observable software data; and \circled{4} \textbf{refuting} the obtained estimate or causal effect using different sensitivity and robustness techniques (\eg placebo, random common cause, data subsets validation) \includegraphics[height=2ex]{graphics/chap_decomposition/fig6_ASC_Causal3.pdf}. 

\figref{fig:pipeline} depicts the four steps that consolidate our post hoc interpretability approach (\ie after models have been trained). We provide a high-level explanation of how each step of \docode pipeline functions (modeling: \secref{ch:docode:sec_01}, identifying: \secref{ch:docode:sec_02}, estimating: \secref{ch:docode:sec_03}, and refuting: \secref{ch:docode:sec_04}) before describing the case study design in detail in \chapref{ch:case}. In addition to the previous pipeline and to help bridge the gap between a given \nlms low-level code prediction and human-understandable categories, we propose a \textbf{syntax clustering} criterion that aims to group well-known categories of programming languages from individual code tokens. This criterion is not explicitly defined in the pipeline but it is necessary for posterior analysis. Therefore, the syntax clustering is introduced in \secref{ch:case:sec_04}.

While our syntax clustering provides the building blocks for explaining global model behavior when applied to predict different types of software data, it may be difficult to determine whether any correlations in performance are \textit{causal}, \ie resulting from the intervention, or \textit{spurious}, \ie potentially caused by confounding factors. To enable an analysis of whether a change in model performance is truly causal or not, allowing for the generation of \textit{accurate explanations}, we propose an analysis based on causal inference which we describe below.

\textbf{Step One: Modeling Causal Inference Problem.} A variable $T$ is a \textit{cause} of $Y$ if the variable $Y$ depends on $T$ to determine its value. However, there are also variables $U$ that represent \textit{unmodeled} factors \citep{Scholkopf2022}. This causal relationship can be precisely formulated via a causal model that uses directed acyclic graphs (DAGs) to describe direct parent-child relationships, instead of probabilistic dependencies, among random variables such as $T$ and $Y$. It also allows us to introduce a graphical definition of causation  \citep{Scholkopf2022,Pearl2009Causality}. Therefore, Structural Causal Models (SCMs) are graphical models responsible for enabling us to quantitatively estimate the results of an action or intervention simulated in the graph. In other words, SCMs provide a framework for \textit{counterfactual reasoning}. 

\textbf{Step Two: Identifying Causal Estimand.} In this step, \docode formulates a causal estimand, which is a mathematical expression, subject to the defined Structural Causal Model (SCM). The method employs graph-based algorithms and do-calculus to draw potential paths of identifying an expected causal effect. We use the doWhy library to perform identification including criteria such as back-door criterion, front-door criterion, instrumental variables, and mediation analysis \citep{dowhy}. 

\textbf{Step Three: Estimating Causal Effects.} After obtaining an expression using any identification criteria, our method uses a suitable or proper estimation method to compute the causal effect. The estimation method depends on the nature of the SCM's variables (\eg binary, discreet, continuous). The library doWhy supports methods based on estimating the treatment assignment, the outcome model, the instrumental variable equation, and machine learning estimators \citep{dowhy}. 

\textbf{Step Four: Refuting Effect Estimate.} An important step in causal analysis is the validation of our causal estimates. To perform this validation, we employ \textit{refutation methods} that calculate the robustness of the causal estimate. These methods test the robustness of our assumptions through various types of perturbations to see the impact on our Average Treatment Effect ($ATE$) estimation. There are multiple refutation methods, but in this chapter, we focus on four: adding a random common cause or covariate $\mathcal{R}_1$, adding an \textit{unobserved} common cause or covariate $\mathcal{R}_2$, replacing the treatment with a random variable or placebo $\mathcal{R}_3$, and removing a random subset of the data $\mathcal{R}_4$. Depending on the method, the resulting value should either be the same as the original $ATE$ or close to zero.

\textbf{Enabling Causal Interpretability for NCMs.} Consider the scenario in which we want to understand how a given model operates between buggy and fixed code. To do this \docode constructs a SCM composed of an intervention variable $T=Buggy$ and potential outcome $Y=Perf$ (representing a measure of model effectiveness or performance in terms of our defined code categories). In addition, we identify confounding factors $Z$ for interventions and potential outcomes. Next, we want to perform the intervention wherein the model is applied to fixed code instead of buggy code. \docode then constructs a parallel graph for this intervention. Using notions from causal inference theory (\ie modeling, identifying, estimating, refuting), \docode is then able to determine how the intervention affected model performance in terms of code categories, and whether the change in performance is a true causal effect, or spurious, resulting from effects of covariates.

While the above scenario illustrates the intuition of our causal analysis, it is important to define how causal effects, and ultimately interpretations, will be generated. First, we need methods by which we can measure model performance. To do this we use Cross-Entropy loss, or the difference in distribution of the model's prediction and the ground truth for given token sequences, and average Next Token Predictions, or probabilities assigned by models to individual tokens, across token sequences as these measures are relevant values produced at inference time that reflect the prediction effectiveness. By relating these values to SE-based interventions $T$ (\ie buggy/non-buggy intervention), we can gain an understanding of how well a studied model is \textit{generating code} under these treatments. To calculate causal effects for binary treatments similar to the ones we used in $T_{[data]}$ (\eg Buggy/Fixed, Commented/Uncommented, and Clone1/Clone2), we first calculate association correlations (Def.~\ref{def:js}) and then causal effects (Def.~\ref{def:ate}).

If we want to estimate how our buggy/non-buggy SE-based intervention affects the performance of a \nlm, we just need before to \textit{identify} the effect of the treatment using the \textit{adjustment formula} of causal inference (see Def.~\ref{def:effect}). After the causal effect is calculated, we aim to generate natural language explanations from our analysis via structured templates that relate code token category to the intervention: \eg \textit{\texttt{\small [code token category]} performed worse by \texttt{\small [change in performance]}, due to a change in model application from \texttt{\small [intervention]} to \texttt{\small [intervention]}, with a causal analysis Average Treatment Effect of \texttt{\small [ATE value]}}.

\input{chapters/part_02_chap_01/sec_01_step1}

\input{chapters/part_02_chap_01/sec_02_step2} 
\input{chapters/part_02_chap_01/sec_03_step3} 
\input{chapters/part_02_chap_01/sec_04_step4} 

%% file: chapters/part_02_chap_01/sec_01_step1.tex
\section{Step One: Modeling Causal Problem}
\label{ch:docode:sec_01}

\begin{marginfigure}
		\centering
		\includegraphics[width=\textwidth]{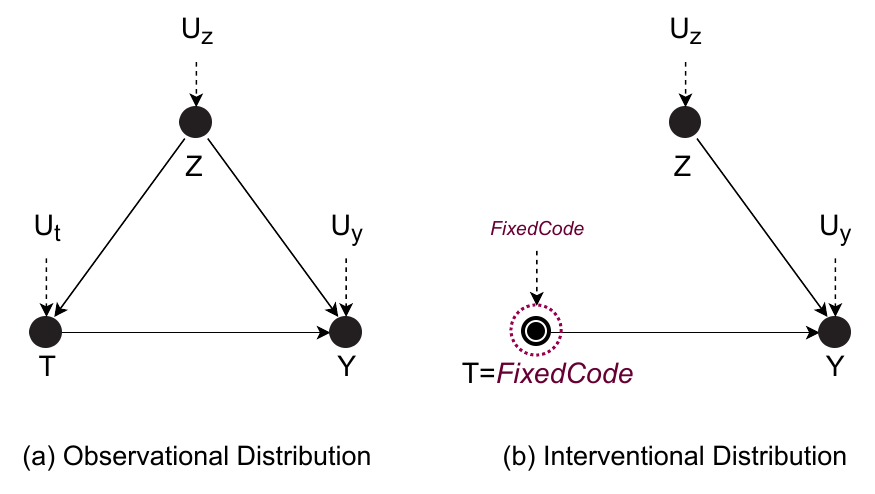}
		\caption{(a) \textit{Structural Causal Model} representing cause-effect relationships of program repair in \nlms. (b) The \textit{SCM} after intervening the treatment with \textit{Fixed Code}.}
        \label{fig:scm}
\end{marginfigure}

In the first pipeline step, assumptions about relationships among data are defined in a Structural Causal Model similar to Fig.~\ref{fig:scm}, which is later processed in the next step to compute $p(Y|do(T))$. Causal assumptions must be made explicit, which means defining the interventions $T$ (\ie binary: buggy/non-buggy, discrete: layer modifications), potential outcomes $Y$ (\ie Cross-Entropy or NTPs performance), and the common causes or confounders $Z$ that can affect the interventions and potential outcomes. For all our SCMs in this study, we assumed that confounders are SE quality metrics since they have the potential to influence models' code predictions, \ie a \nlm may be influenced by more or less for loops, as well as influence interventions, \ie more code is correlated with more bugs \citep{4027145}. The following example showcases a practical scenario in which performance can be explained by the influence of having buggy code SE-based intervention. 

\begin{exmp}
\label{exmp:association}
Consider a Bayesian network created to explain the performance of a given \nlm, represented as an arrow that links buggy input code to model performance $Buggy\to Perf$. The variables $Buggy$ and $Perf$ are dependent, both variables embody a \textit{program repair intervention}. Therefore, if we want to define the joint distribution $p(Buggy,Perf)$ to represent the network, we must specify by Bayes' rule using the prior $p(Buggy)$ and the conditional probability $p(Perf|Buggy)$. Nonetheless, such a joint distribution can be computed as well in the opposite direction  $Perf\to Buggy$ using the same Bayes' rule for a prior $p(Perf)$ and conditional $p(Buggy|Perf)$. The fact that the joint distribution can be represented with both networks is non-intuitive since we know from experience and software understanding that the model performance cannot give rise to a bug in the original input. In other words, the relationship between these variables is asymmetric or \textit{causal}. Hence, we expect that buggy snippets affect the performance of \nlms, not the other way around.
\end{exmp}

A first attempt to address the relationship in Ex.~\ref{exmp:association} would be computing a correlation coefficient $\rho_{TY}\approx p(Y|T)=p(Perf|Buggy)$, where $T$ is a binary \textit{intervention} that represents the \textit{debugging} process and $Y$ is a \textit{potential outcome} that corresponds to the model performance. This coefficient, however, is still symmetric: if $T$ is correlated with $Y$, then $Y$ is equally correlated with $T$. \textit{Causal networks} allow us to model causal asymmetries where directionality goes beyond probabilistic dependence. These causal models represent the mechanism by which data \textit{were generated} \citep{Pearl2016Causality}. Instead of testing whether $Buggy$ and $Perf$ are conditionally dependent, \textit{causation} asks which variable \textit{responds} to the other one: $Buggy$ to $Perf$ or $Perf$ to $Buggy$? \citep{Pearl2016Causality}. Therefore, we can formally introduce a definition of causation:

\begin{definition}
\label{def:causation}
  \textbf{Causation.} A variable $T$ is a \textit{cause} of $Y$ if the variable $Y$ depends on $T$ to determine its value. Formally, the value of $Y$ was \textit{assigned} based on what is known about $T$. In other words, the value of $Y$ is determined by a \textit{\textbf{structural equation}} $Y=f_y(T,U_y)$ and the arrow $T\to Y$. The $U$ variables in these equations represent \textit{unmodeled} variables that are exogenous to the causal network but disturb the functional relationship between the outcome and its treatment \citep{Scholkopf2022}.    
\end{definition}

Similarly, we can define a structural function for the treatment $T=f_t(U_t)$ that depends only on $U$ disturbances assuming that no \textit{common causes} exists between potential outcomes and interventions. A common cause (or confounder) is a random variable $Z$ that causally influences two variables that are initially perceived as statically dependent ($T \not\!\perp\!\!\!\perp Y$). However, this dependency can be explained by the underlying influence of $Z$ on the effects, making the effects conditionally independent ($T \perp\!\!\!\perp Y | Z$). Therefore, there exist more complex causal relationships between treatments and outcomes that we can model with structural equations that correspond to a \textit{\textbf{Structural Causal Model}} (SCM). We can formally introduce a definition for Structural Causal Models as graphical models that capture causal assumptions:

\begin{definition}
\label{def:scm}
  \textbf{Structural Causal Models.} These directed acyclic graphs (DAGs) describe direct parent-child relationships, instead of probabilistic dependencies, among random variables $X_i$. The value $x_i$ of each variable $X_i$ is defined by the structural equations $x_i = f_i(PA_i,U_i)$ where $PA_i = {X_j: X_j \to X_i}$ denotes the set of parents or direct causes of $X_i$. This model allows us to introduce a graphical definition of causation  \citep{Scholkopf2022, Pearl2009Causality}.
\end{definition}

In summary, this step consists of \textit{setting down assumptions} about the causal relationships of software data employed to interpret \nlms. SCMs help us to describe the relevant features of the software data and how they interact with each other. In the following subsection, we formally define each component of the Structural Causal Models (\ie interventions, potential outcomes, and common causes or confounders).

\subsection{SE-Based Counterfactual Interventions} 
\label{sec:seintervention}

\begin{marginfigure}
		\centering
		\includegraphics[width=\textwidth]{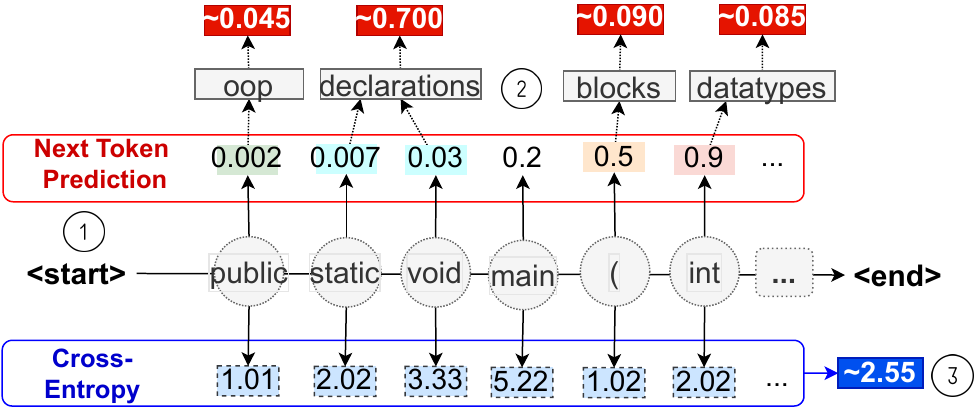}
		\caption{Potential Outcomes are Code Prediction of  \nlms: Cross-Entropy $(Y_g)$, Next Token Predictions $(Y_l)$, or Distance metrics $(Y_d)$.}
        \label{fig:performance}
\end{marginfigure}

\nlms are notorious for behaving differently in different datasets \citep{abu-mastafa}. For example, if a model trained on a well-commented dataset is applied to predict segments of poorly commented code, this mismatch could potentially impact its performance. As such, we assert that observing model performance across datasets with different characteristics can aid in understandability and interpretability. Hence, we defined \textit{SE-Based Counterfactual Interventions} $T$ to better understand model performance across different settings. We formulate these interventions based on domain knowledge from observable testbeds (\ie datasets) organized in sample pairs \textbf{treatment} $T=0$ (\ie BuggyCode) and \textbf{control} $T=1$ (\ie FixedCode). Note that we define testbeds according to different applications often described in SE research. The general process comprises of identification of some specific \textit{intervention} (\ie program repair) and 2) construction or collection of the necessary data via mining repositories or other means that contain these observed interventions. More complex interventions will likely be more challenging to prepare. 

In \docode, \textit{counterfactual interventions} produce explanations motivated by both semantic perturbations $T_{[data]}$ to our SE Application Settings (\eg Buggy/Fixed, Commented/Uncommented, Clone1/Clone2) and model hyper-parameter variations $T_{[hyp]}$ on \nlms (\eg layers, units, or heads). Although hyperparameter variations are NOT data perturbations based on SE settings, we include them to extend the analysis of possible causes of models' predictions beyond data interventions. 

\subsection{Potential Outcomes / Code Predictions} 
Cross-Entropy ({Fig.~\ref{fig:performance}-\circled{3}}), Next Token Prediction ({Fig.~\ref{fig:performance}-\circled{2}}), and Distant Metrics (\ie Jaccard, Levenshtein, and Sorence-Dicen) are relevant values produced at inference time that reflect the effectiveness of a model at \textit{predicting code}. By relating these values to counterfactual interventions $T$ (\ie program repair), we can gain an understanding of how well a studied \nlm is \textit{generating code} under these SE-based interventions. 

\textbf{Cross-Entropy.} \label{out:global}
We refer to Cross-Entropy loss as a measure of a model's \textit{coarse-grained performance} $Y_g: w \to - \sum_{t \in |w|} P(w_t | d_t) \log Q(w_t | w_{<t})$ as these losses capture the overall performance of a \nlm over an entire sequence of tokens $w$. Due to the discrete nature of the data, the expression $P(w_t | w_{t-1:1} )$ can be estimated using a classifier. The classifier, in our particular case, is a \nlm \citep{Bengio2003AModel}. Hence, rather than using \textit{n}-grams or Markov Models to approximate $P(w_t | w_{t-1:1})$ \citep{Karampatsis2020Open-VocabularyAbstract}, it is convenient to use a latent model $P(w_t | w_{t-1:1} ) \approx P(w_t | d_t )$, where $d_t$ is known as a \textit{hidden state} that embeds the sequence information from past observations up to the time step $t$. Depending on \textit{how} the sequence is processed, the hidden state $d_t$ can be computed using either an autoregressive network (\ie such as a Transformer ($GPT$)~\citep{vaswani2017transformers}) or a Recurrent Neural Network ($RNN$). 

\textbf{Next Token Prediction (\ntp).} \label{out:local} 
Conversely, \ntp values signal \textit{fine-grained performance} $Y_l:w_{<t} \to P(w_t | w_{t-1:1})$ within token-level contexts. NTPs capture local predictions for individual tokens that are affected by complex interactions in \nlms and are equivalent to the estimated predicted value (or softmax probability) $\sigma(k)_t$ for each token. Bear in mind that the size of the vector $\sigma(k)_t$ is the vocabulary $|\mathcal{V}|$, in which $k$ represents the non-normalized log probabilities for each output token $t$. NTPs capture the value of the expected token $w_t$ instead of the maximum value estimated in the vector $\sigma(k)_t$.

\textbf{Distance Predictions.} \label{out:distance}

{Similarity distance scores play a crucial role in assessing the model's \textit{distance performance}, defined by the expression $Y_d: s' \to \nabla(s, s')$. Function $\nabla$ represents the similarity coefficient used for the pairwise comparison of two finite sample sets, $s$ and $s'$. For instance, if set $A$ denotes Node Types in the AST (Abstract Syntax Tree) of a ground-truth code sample, and set $B$ represents Node Types in the AST of a predicted code sample, we can quantify their similarity using metrics such as the Jaccard Index, Levenshtein distance, and Sorensen-Dice coefficient. The resulting similarity scores, which range from $0$ to $1$, effectively measure the degree of similarity between the sets, with $0$ indicating no similarity and $1$ meaning that the sets are identical.}

\subsection{Common Causes or SE-based Confounders}\label{sec:confounders}
In past work, several factors such as code duplication~\citep{Allamanis19} have been illustrated to affect model predictions. As such, we derive a list of potential SE-based confounders that could influence a model's prediction of our clustered syntax categories beyond our interventions defined in the previous section. Our initial set of confounders (also called common causes) include McCabe's complexity, LoC (Lines of code), number of:

\begin{multicols}{2}
    \begin{itemize}
        \item \texttt{\small returns}
        \item \texttt{\small loops}
        \item \texttt{\small comparisons}
        \item \texttt{\small try/catches}
        \item \texttt{\small parenthesized expressions}
        \item \texttt{\small string literals}
        \item \texttt{\small variables}
        \item \texttt{\small max nested blocks}
        \item \texttt{\small anonymous classes}
        \item \texttt{\small inner classes}
        \item \texttt{\small lambda expressions}
        \item \texttt{\small unique words}
        \item \texttt{\small log statements}
        \item \texttt{\small modifiers}
    \end{itemize}
\end{multicols}

Practitioners and researchers can extend the search of potential SE-based confounders based on their domain knowledge, empirical analysis on \nlms, or observations of \nlms' behavior in production.

%% file: chapters/part_02_chap_01/sec_02_step2.tex
\section{Step Two: Identifying Causal Estimand}
\label{ch:docode:sec_02}

In the second pipeline step, once the SCM (similar to \figref{fig:scm}) is constructed, \docode applies various estimation methods (\ie backdoor-criterion or instrumental variables) to formulate a correct estimand by adjusting for confounders using Eq.~\ref{eq:effect}. For our case study, the causal effect $p(Y|do(T))$ is computed, which will be done in the next step, based on the following causal graph or SCM: data-based $T_{[data]}$ and parameter-based $T_{[hyp]}$ interventions; SE confounders $Z \in SE_{metrics}$; and potential outcomes $Y_l, Y_g, Y_d$.

Structural Causal Models (SCMs) are stable mechanisms that remain invariant to local changes unlike probabilities computed by Bayesian networks \citep{Pearl2016Causality}. This characteristic is responsible for enabling us to estimate quantitatively the results of an intervention in the graph without actually performing it in controlled settings (\ie randomized experiments). The mathematical tool employed to perform these interventions is the $do(\cdot)-operator$ \citep{Pearl2009Causality}. For instance, if we want to estimate how fixed code affects the performance of a \nlm, we just need to compute the action $p(Perf|do(Buggy=False))$ (Eq.~\ref{eqn:do-1}). These types of actions are \textit{interventional} distributions since we \textbf{set} the value of $Buggy$ to $False$. Note that this interventional distribution is \textbf{not} the same as the distribution $p(Perf|Buggy=False)$ (Eq.~\ref{eqn:do-2}). This latter distribution is \textit{observational} as we are \textit{conditioning} the performance on the value of the $Buggy$ variable. Intervening on a variable in a SCM means fixing its value and, therefore, changing the value of other variables of the network as a result. Conversely, conditioning on a variable means narrowing the cases that the outcome takes once we assign a value to the intervention. 

\begin{exmp} 
\label{exmp:scm}
Consider Fig.~\ref{fig:scm} a generalization of the buggy influence on a deep model's performance. The first graph is a SCM composed of an intervention variable $T=Buggy$ and potential outcome $Y=Perf$. In addition, we identified some common causes $Z=SE_{Metrics}$ for interventions and potential outcomes. These common causes (or confounders) can be mapped to Software Engineering quality metrics (\eg Lines of Code, McCabe Complexity, Size of Methods). Now we want to perform the intervention $do(Buggy=False)$, which is the same as $do(T=FixedCode)$. The second graph depicts this program's repair intervention. 
\end{exmp}

Note that fixing the value of $T$ makes the SCM change by eliminating the effect or influence arrow of the confounder $Z$ to the intervention. The disturbance $U_t$ is also eliminated. This elimination process of input arrows to fixed variables is formally known as \textit{graph surgery}. Both $do(\cdot)-operator$ and graph surgery allow us to untangle causal relationships from mere correlations \citep{Pearl2009Causality}. The law of total probabilities and invariance principle are required to compute the observational and interventional distributions \citep{Scholkopf2022}:

\begin{subequations}
\begin{align}
p(Y|do(t=FixedCode)) &=\sum_{z \in _{metrics}}p(Y|t,z)p(z)\label{eqn:do-1} \\
p(Y|t=FixedCode) &= \sum_{z \in _{metrics}}p(Y|t,z)p(z|t) \label{eqn:do-2} 
\end{align}
\label{eqn:do-all-lines}
\end{subequations}

Note that Eq.~\ref{eqn:do-1} differs from  Eq.~\ref{eqn:do-2}: the prior $p(z)$ in contrast to $p(z|t)$, which is precisely the link that is eliminated in the SCM. Eq.~\ref{eqn:do-1} is formally known as the \textbf{\textit{adjustment formula}}. This formula is one of the building blocks in causal inference since it helps us to adjust common causes or control for confounders to allow the estimation of \textit{causal/treatment effects} \citep{Pearl2009Causality,Pearl2016Causality}.

\begin{definition}\label{def:effect}
\textbf{Treatment Effects.} Given a Structural Causal Model where a set of variables $PA$ denotes the parents of $T$, the treatment effect of T on the potential outcome Y is given by
\begin{subequations}
    \begin{align}
    p(Y=y|do(T=t)) &=  \label{eq:effect-1}\\
    \Sigma_z p(Y=y|T=t,PA=z)p(PA=z)  &= \label{eq:effect-2}\\
    \Sigma_z p(T=t,Y=y,PA=z)/p(T=t|PA=z) \label{eq:effect-3}
    \end{align}
\label{eq:effect}
\end{subequations} 
  
\end{definition}

In our initial causal statement, we generally accept that buggy code \textit{causes} a \nlm to predict poorly. Although the causal statement is true, it is not guaranteed that ``every buggy snippet'' is certain to make a model predict poorly. Therefore, causal relationships are \textbf{uncertain}. This uncertainty is captured by employing conditional probabilities described in Eq.~\ref{eq:effect-2}. Note that Eq.~\ref{eq:effect-2} is a generalization of the adjustment formula Eq.~\ref{eqn:do-1}. To compute treatment effects, we need to connect observational data with our interventional distribution. A standard way of connecting data with the interventional distribution is by employing a summation described in Eq.~\ref{eq:effect-3}. Note Eq.~\ref{eq:effect-3} is obtained with the application of Bayes' rule and algebraic manipulation once we multiply and divide Eq.~\ref{eq:effect-2} by the term $p(T=t|PA=z)$. This term is a conditional probability known as the \textit{propensity score}. This propensity score and the joint probability of all the nodes are distributions that can be \textit{obtained directly from data} \citep{Pearl2016Causality}. %

%% file: chapters/part_02_chap_01/sec_03_step3.tex
\section{Step Three: Estimating Causal Effects}
\label{ch:docode:sec_03}

In the third pipeline step, \docode estimates the causal effect using statistical and ML methods based on the adjustment formula from the previous step. \docode computes \textit{Propensity Score Matching} for binary SE interventions (\ie Buggy/Fixed) and \textit{Linear Regressions} for SE discrete interventions (\eg layers, units, or heads). We refer interested readers to the \textit{doWhy} documentation for full estimation methods details \citep{dowhy}. For completeness, we will solely show how to estimate a causal effect assuming a binary intervention as an example. We can start by explaining the notion of \textit{Average Treatment Effect (ATE)}. ATE is simply the average score of all treatment effects (see Def.~\ref{def:effect}) computed for a population. In our case, an individual of the population is just a code snippet $x$. 

\begin{definition}
\label{def:ate}
\textbf{Average Treatment Effect (ATE)} Defining the first intervention as $do(T=1)$ and the second by $do(T=0)$, the Average Treatment Effect is the population average of the difference of causal effects of each code snippet $x$.  
\begin{subequations}
    \begin{align}
     ATE = \mathbb{E}_{x\sim p(x)}[Y=y|x,do(T=t)]  &= \label{eq:ate-1}\\
     \mathbb{E}_{x\sim p(x)}[ \mathbb{E}[Y|x,do(T=1)] - \mathbb{E}[Y|x,do(T=0)] ] &= \label{eq:ate-2}\\
     \mathbb{E}_{x\sim p(x)}[ \mathbb{E}[Y^{1}|x,T=1] - \mathbb{E}[Y^{0}|x,T=0] ] &= \label{eq:ate-3}
    \end{align}
\label{eq:ate}
\end{subequations} 
\end{definition}

The previous Eq.~\ref{eq:ate} shows the formal definition of an ATE. We can derive the final expression by applying the law of total expectations and the ignorability assumption $Y \perp\!\!\!\perp  Z|T$, where the potential outcomes $Y$ are independent of intervention assignments conditioned on covariates $Z$ \citep{Pearl2009Causality}. That is, the effects of the hidden confounders $Z$ and missing data are ignored. In Eq.~\ref{eq:ate}, the term $\mathbb{E}[Y^1|x,T=1]$ represents the expected value of a potential outcome under an observable intervention (\ie FixedCode). Similarly, the term $\mathbb{E}[Y^0|x,T=0]$ represents an expected value of a potential outcome under an observable intervention (\ie BuggyCode). Both terms are quantities that can be \textit{estimated from data}. Covariate adjustment (in Eq.~\ref{eqn:do-1}), propensity score (in Eq.~\ref{eq:effect-3}), and linear regression are some of the estimation methods that we employ to approximate $ATEs$. Their usage depends upon the type of the intervention variable (\ie binary, discrete, or continuous) and causal graph assumptions.

%% file: chapters/part_02_chap_01/sec_04_step4.tex
\section{Step Four: Refuting Effect Estimate}
\label{ch:docode:sec_04}

In the fourth pipeline step, the obtained causal effect can be validated using \textit{refutation methods} that calculate the robustness and sensitivity of the estimate. In essence, the refutation methods apply random perturbations to the original Structural Causal Model to test for the robustness of the estimated causal effect or $ATE$. {Although researchers can design their tailored methods to address this sensitivity analysis, we chose four baseline methods for our case study:

\textbf{Adding a random common cause or covariate $\mathcal{R}_1$}. We evaluate if the estimation method changes its estimate after adding an independent random variable as a confounder to the dataset. 

\textbf{Adding an \textit{unobserved} common cause or covariate $\mathcal{R}_2$}. We evaluate how sensitive the estimated causal effect is if we add a confounder to the dataset. This artificial confounder is correlated to the treatment $T$ and the potential outcome $Y$.

\textbf{Replacing the treatment with a random variable or placebo $\mathcal{R}_3$}. We evaluate to what extent the estimated causal effect changes if we replace the true treatment variable $T$ with an independent random variable of the same nature.  

\textbf{Removing a random subset of the data $\mathcal{R}_4$}. We evaluate if the estimation method changes its estimate significantly after replacing the given dataset with a randomly selected subset.}

For robustness, we expected that $\mathcal{R}_1$, $\mathcal{R}_2$, and $\mathcal{R}_4$ values were close to the estimated $ATE$ (Eq.~\ref{eq:ate}). Conversely, the placebo $\mathcal{R}_3$ should tend to zero.

%% file: chapters/part_02_chap_02/case.tex
\chapter{A Case Study on Deep Code Generation }
\label{ch:case}

\epigraph{\scriptsize(...) Correlation is thus an epiphenomenon, the byproduct of a causal process (...)}{\scriptsize\textit{Sch{\"o}lkopf \& K{\"u}gelgen \citep{Scholkopf2022}}}

\begin{figure}[ht]
		\centering
		\includegraphics[width=1\textwidth]{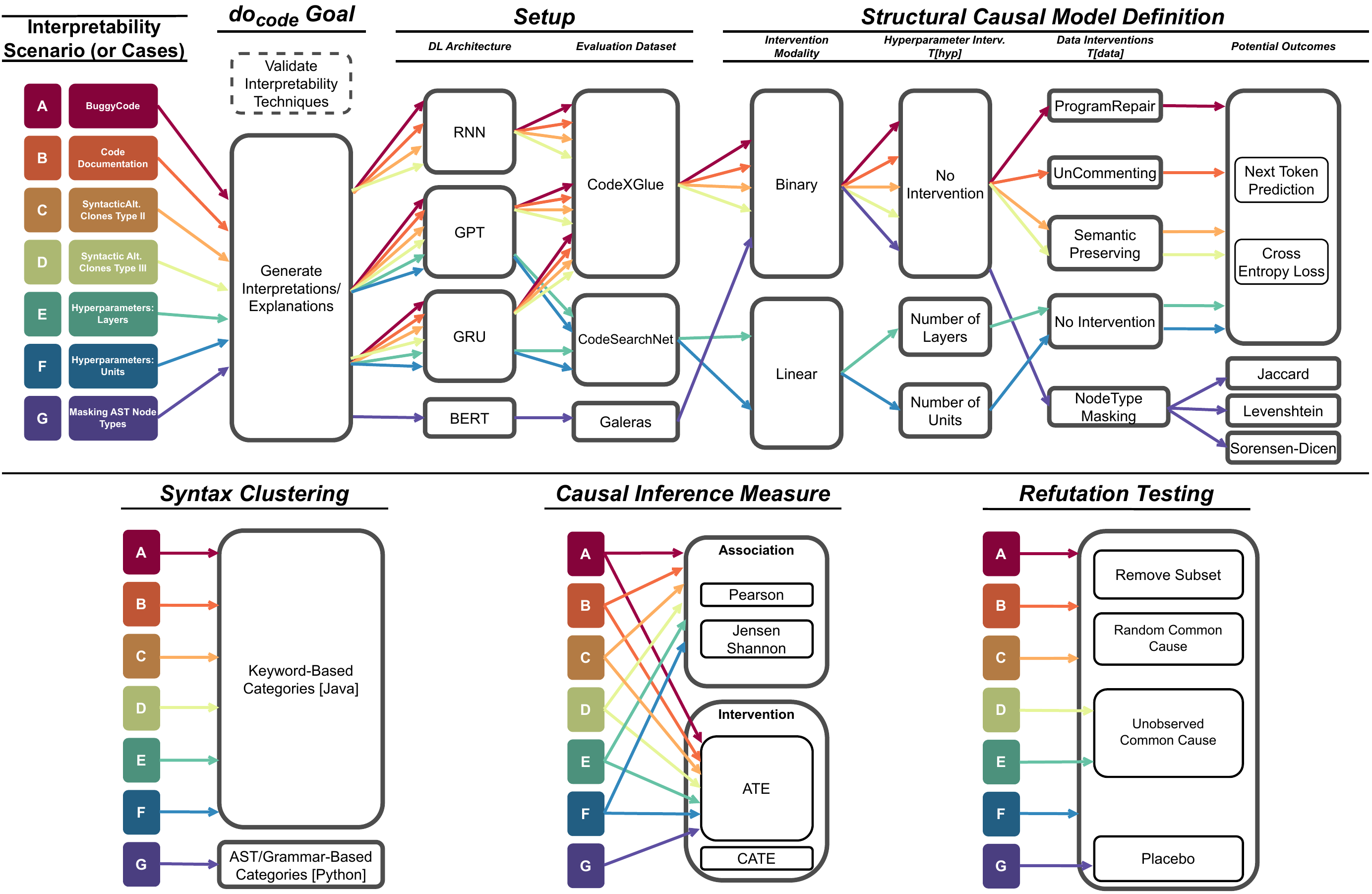}
		\caption{Configuration criteria for enabling causal interpretability. Seven scenarios are proposed for this case study. However, new interpretability cases can be formulated using a different permutation or by extending the criteria.}
        \label{fig:case_studies}
\end{figure}

\lettrine[lines=2]{\textbf{I}}{}n this section, we aim to illustrate how \docode is applied to \textit{enable causal interpretations} based on different interpretability scenarios. Fig. \ref{fig:case_studies} depicts an overview of seven scenarios (or cases from A to G) and six essential criteria that comprise the case study. Each scenario can be unfolded into the following criteria: (i) the goal of \docode (\ie generating interpretations or validating an interpretability technique), (ii) the setup (\ie the deep learning architecture under analysis and the evaluation dataset or testbed) (iii) the definition of the Structural Causal Model (\scm) from domain knowledge (\ie intervention modality, hyperparameter interventions, data interventions, and potential outcomes), (iv) a syntax clustering strategy for grouping potential outcomes based on token predictions, (v) the usage of causal inference measure whether the value is an association (\eg Pearson, Jensen Shannon) or an intervention (\eg ATE, CATE), and (vi) the refutation testing method employed to validate the robustness of the interpretations (\eg placebo, unobserved common cause). 

Note that practitioners and researchers should not necessarily have to stick to our seven cases, the configuration criteria can be extended to assess unexplored interpretability scenarios. Various permutations of the criteria outlined in \figref{fig:case_studies} can be formulated depending on the research goal of the causal analysis. The subsections below detail the criteria that we propose for this study.

\input{chapters/part_02_chap_02/sec_01_goal} 
\input{chapters/part_02_chap_02/sec_02_setup}

\input{chapters/part_02_chap_02/sec_03_scm_design}
\input{chapters/part_02_chap_02/sec_04_clustering}

\input{chapters/part_02_chap_02/sec_05_measures}
\input{chapters/part_02_chap_02/sec_06_refutation}
\input{chapters/part_02_chap_02/sec_07_questions}
\input{chapters/part_02_chap_02/sec_08_results} 

%% file: chapters/part_02_chap_02/sec_01_goal.tex
\section{\texorpdfstring{\docode}{docode} Goal}
\label{ch:case:sec_01}

The goal of \docode is to enable \nlms' interpretations of code predictions by estimating the causal effects of specific SE-based interventions. On top of that, \docode also facilitates causal inference for \textit{validating} interpretability methods. Users of \docode should define their goals in alignment with this premise as these goals determine how the \scm is formulated. For example, \docode can be employed to interpret the effects of code smells on code predictions produced by given \nlm, the SE-based intervention would be a binary treatment of samples with and without smells. On the other hand, we decided not to introduce a scenario in which \docode validates an interpretability method as it is out of the scope of this study. However, we offer sufficient guidance on assembling the criteria for this purpose in this section.

%% file: chapters/part_02_chap_02/sec_02_setup.tex
\section{Setup}
\label{ch:case:sec_02}

Applying \docode requires selecting both the deep learning architecture (\ie \nlm) and the evaluation dataset or testbed for our interpretability scenario. 

\textbf{About Deep Learning Architectures.} Because causal interpretability is agnostic to any peculiar \nlm, \docode supports (but is not limited to) architectures such as RNN, GPT, GRU, and BERT. Formally, \docode conceptualizes an \nlm as a probability distribution $P(x_t | x_1, x_2, \ldots, x_{t-1})$, where the output $x_t$ at time step $t$, given the input sequence $(x_1,x_2,..,x_{t-1})$, is inferred through the conditional probability $ P(x_t | h_t)$. The hidden state $h_t$ encapsulates the properties of the preceding context. Recurrent Neural Networks (RRNs) update the hidden state $h_t$ using the current input and the previous hidden state, thus $h_t = f(h_{t-1}, x_t)$. RNNs can take the form of a Gated Recurrent Unit (GRU) \citep{cho_properties_2014}, which uses reset $r_t = \sigma(W_r \cdot [h_{t-1}, x_t])$, and update $u_t = \sigma(W_u \cdot [h_{t-1}, x_t])$ gates to control how it combines the previous hidden state $h_{t-1}$ with a new candidate state $h_t' = \tanh(W_h \cdot [r_t \odot h_{t-1}, x_t])$ to obtain the updated hidden state $h_t = (1 - u_t) \odot h_{t-1} + u_t \odot h_t'$. 

Furthermore, GPT and BERT are deep architectures known as Transformers \citep{vaswani2017transformers}. In transformer models, the hidden state $h_t$ in each time step is updated through a multi-head self-attention mechanism and a feed-forward network. In simpler terms, the expression: $$h_t = \text{Attention}(Q_t, K_t, V_t) = \text{softmax}({{Q_iK_i^T}}/{\sqrt{d_k}})V_i$$ where $Q_i, K_i$, and $V_i$ represent the queries, keys, and values for the time step $t$, and $d_k$ is the dimension of the key vectors. Transformers adopt the form of Encoder-based (\eg BERT) or Decoder-based (\eg GPT).

We employed RNNs and Transformers for all our scenarios as RNNs are widely used in SE \citep{watson2020dl4se}, while Transformers have gained popularity in the SE/NLP domain for their high performance ~\citep{Mastropaolo2021StudyingTasks}. In \figref{fig:case_studies}), {we opted to train the \nlms for cases $[A-F]$ to gain more control over the training data. This was particularly important given our need to manipulate the tokenizer, allowing us to assign special tokens to specific code categories (\ie syntax clustering). However, \docode is not confined to specialized models trained by the user as demonstrated in the $G$ case, in which we used an in-the-wild BERT model. Therefore, \docode operates effectively on already pre-trained \nlms as long as there exists a way to group fine-grained predictions (\ie BPE tokens) to human-understandable categories (\eg AST/Grammar-based or keyword-based).}

Table~\ref{tab:method} provides a detailed overview of the training and testbed configurations for each scenario. In $[A-F]$ cases, the \nlms were trained using the Java portion of the \training dataset~\citep{husain2019codesearchnet}, which consists of a diverse array of methods (mts) from GitHub \citep{github}. We partitioned \training into training, validation, and test sets. For model development, we employed Tensorflow and Pytorch \citep{tensorflow2015-whitepaper, pytorch}, along with Huggingface's Transformers library \citep{wolf2020transformers}. To mitigate overfitting, training was terminated if cross-entropy did not improve by at least $1\text{e-2}$ over $5$ epochs. Inputs were standardized with a \textit{start of sentence} token and adjusted to a length of $300$. This process was executed on an Ubuntu 20.04 system with an AMD EPYC 7532 32-Core CPU, an A100 NVIDIA GPU with 40GB VRAM, and 1TB RAM. For the $G$ case, we used a pre-trained BERT model: codebert-base-mlm \citep{feng2020codebertpretrainedmodelprogramming}, trained specifically for a Masking Language Model (MLM) objective and also using the \training dataset.

\input{tables/chap_4_tab1_cases}

\textbf{About Evaluation Datasets.} To build the data intervention testbeds for our interpretability scenarios, we obtained \textit{Java} samples from \textit{CodeXGLUE} \citep{lu2021codexglue} and \textit{Python} samples from \textit{Galeras} \citep{daniel23}. In  $[A-F]$ cases, parallel corpora were used to examine the impacts of buggy code (\BuggyTB: 64,722 methods or mts), code documentation (\CommentsTB: 6,664 mts), and syntactic alterations in semantically similar snippets, specifically focusing on type II (\BigCloneIITB: 666 mts) and type III (\BigCloneIIITB: 8,097 mts) clones from \BigCloneTB. For the $G$ case, we conducted binary interventions by masking tokens corresponding to each AST element in the \Galeras samples (8,299 mts), covering all node types in the Python grammar as defined by Tree-sitter \citep{tree-sitter}. In the control group, a comparable number of tokens were randomly masked in each sample.

\textbf{About Code Tokenization.} For all scenarios, Byte Pair Encoding (BPE) tokenization \citep{sennrich2015neural} was applied to the testbeds before processing them with the \nlms. Known for its efficacy in training \nlms on code, BPE significantly mitigates the \textit{out-of-vocabulary} problem \citep{Karampatsis2020BigCode}. For $[A-F]$ cases, we developed a BPE tokenizer trained on $10\%$ of our training data and with a vocabulary size of 10K. Conversely, for the $G$ case, we employed the pretrained BPE tokenizer from the BERT-selected model. Nonetheless, the BPE tokenization process sometimes resulted in sub-tokens that either combined multiple reserved keywords or split keywords across different tokens. This splitting issue presented a significant challenge for our interpretability analysis as our method relies on accurately grouping token predictions into semantic categories, a criterion that we named \textit{syntax clustering}. To address this, we developed clustering functions to ensure the tokens are correctly aligned with our defined categories, which will be discussed in \secref{ch:case:sec_04}.

%% file: tables/chap_4_tab1_cases.tex
\begin{table*}[ht]
\centering
\caption{SE-based Interventions Experiment Overview (left). \nlms training specifications for Recurrent Neural Networks (\ie Vanilla RNN, GRU), and Transformers (\ie GPT-2, BERT) (right). 
}
\label{tab:method}

\begin{adjustbox}{width=1\textwidth}

\begin{tabular}{llllll|llll}
\hline
\multicolumn{6}{c|}{\multirow{2}{*}{\textbf{\begin{tabular}[c]{@{}c@{}}Counterfactual\\ Interventions\end{tabular}}}} &
  \multicolumn{4}{c}{\textbf{\nlms Training}} \\ \cline{7-10} 
\multicolumn{6}{c|}{} &
  \multicolumn{1}{c}{\textit{\textbf{RNNs}}} &
  \multicolumn{1}{c|}{\textit{\textbf{Transformers}}} &
  \multicolumn{1}{c}{\textit{\textbf{Hyper.}}} &
  \multicolumn{1}{c}{\textit{\textbf{Val.}}} \\ \hline
\multicolumn{1}{c}{\textit{Type}} &
  \multicolumn{1}{c}{\textit{Interv.}} &
  \textit{Case Id} &
  \multicolumn{1}{c}{\textit{Intervention}} &
  \multicolumn{1}{c}{\textit{Associated Dataset}} &
  PL &
  \textit{\nlm $_{lyr,unt}$} &
  \multicolumn{1}{l|}{\textit{\nlm$_{lyr,hds}$}} &
  \multirow{8}{*}{\begin{tabular}[c]{@{}l@{}}dropout RNN\citep{Karampatsis2020BigCode}\\ dropout TF\\ optimizer\citep{Kingma2015AdamAM}\\ learning rate\\ beta1,beta2\\ epsilon\\ epochs\\ batch RNN|TG\end{tabular}} &
  \multirow{8}{*}{\begin{tabular}[c]{@{}l@{}}0.5\\ 0.l\\ adam\\ 1e-3\\ 0.9\\ 1e-7\\ 64\\ 512,128\end{tabular}} \\ \cline{1-8}
\multirow{5}{*}{$T_{[data]}$} &
  $T_{A}$ &
  A &
  \datainterI &
  \BuggyTB\citep{Tufano2019LearningBugFixes} &
  Java &
  \multirow{7}{*}{\begin{tabular}[c]{@{}l@{}}\rnn\\ \gru\\ \grui\\ \gruii\\ \gruiii\\ \gruiv\end{tabular}} &
  \multicolumn{1}{l|}{\multirow{7}{*}{\begin{tabular}[c]{@{}l@{}}\tf\\ \tfi\\ \tfii\\ \bert\end{tabular}}} &
   &
   \\
 &
 $T_{B}$ &
  B &
  \datainterIII &
  \CommentsTB\citep{husain2019codesearchnet} &
  Java &
   &
  \multicolumn{1}{l|}{} &
   &
   \\
 &
  $T_{C}$ &
  C &
  \datainterII &
  \BigCloneIITB\citep{Svajlenko2015EvaluatingBigCloneBench}&
  Java &
   &
  \multicolumn{1}{l|}{} &
   &
   \\
 &
  $T_{D}$ &
  D &
  \datainterII &
  \BigCloneIIITB\citep{Svajlenko2015EvaluatingBigCloneBench} &
  Java &
   &
  \multicolumn{1}{l|}{} &
   &
   \\
 &
  $T_{G}$ &
  G &
  \datainterIV &
  Galeras \citep{daniel23} &
  Python &
   &
  \multicolumn{1}{l|}{} &
   &
   \\ \cline{1-6}
  $T_{[hyp]}$ &
  $T_{E}$ &
  E &
  \modelinterI &
  \training\citep{husain2019codesearchnet}&
  Java &
   &
  \multicolumn{1}{l|}{} &
   &
   \\
 &
  $T_{F}$ &
  F &
  \modelinterII &
  \training &
  Java &
   &
  \multicolumn{1}{l|}{} &
   &
   \\ \hline
\end{tabular}

\end{adjustbox}

\vspace{0.45cm}

\end{table*}

%% file: chapters/part_02_chap_02/sec_03_scm_design.tex
\section{Structural Causal Model Design}
\label{ch:case:sec_03}

\begin{figure}[ht]
		\centering
		\includegraphics[width=1\textwidth]{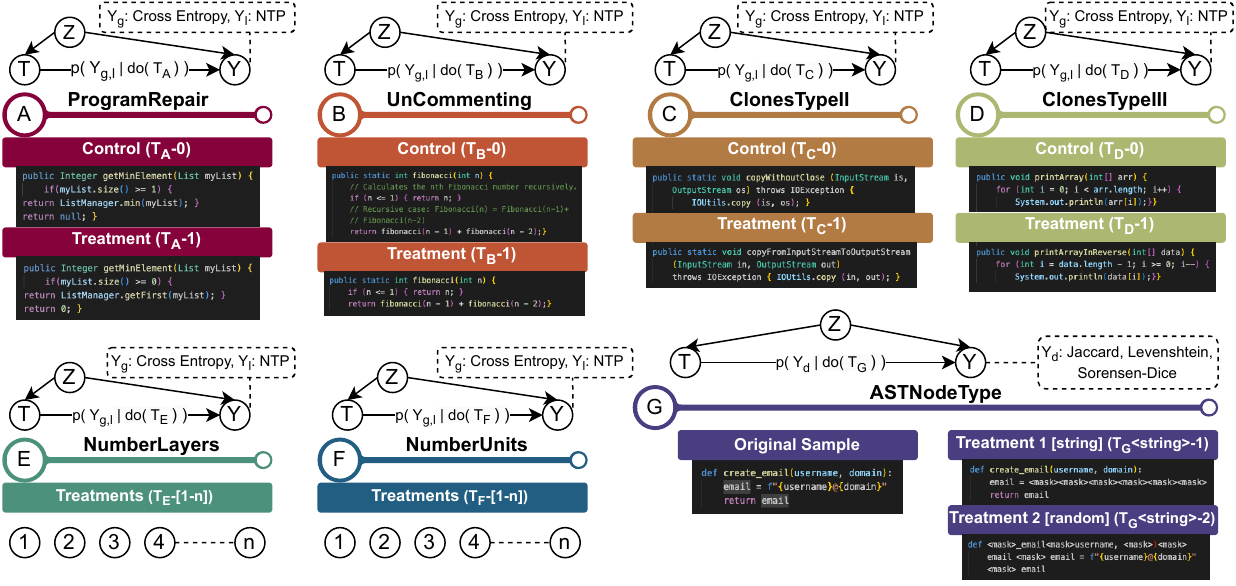}
		\caption{Structural Causal Models' Definition. Each \scm presents the causal estimand, the potential outcomes, the intervention type, and examples of interventions for each interpretability scenario.}
        \label{fig:causal_modes}
\end{figure}

\docode users must design the Structural Causal Model (\scm) based on their domain knowledge and available data. This criterion involves choosing the intervention modality, specifying the type of the intervention, and computing the potential outcome. The SE-based interventions, contingent upon the objectives of \docode, primarily fall into two categories: Data Interventions $T_{[data]}$ and Hyper-parameter Interventions $T_{[hyp]}$. Data interventions occur within the testbed, whereas Hyper-parameter interventions involve model training parameters (\eg Learning rate, Batch Size, Number of Epochs, Number of Hidden Layers/Units, dropout rate), which are not embedded in the data per se. Moreover, potential outcomes cover a variety of measures such as Next Token Prediction (\ntp), Cross Entropy Loss, BLEU, CODEBLEU, and distance similarity scores including Jaccard, Levenshtein, and Sorensen-Dice. We aimed to estimate the causal effect of $T_{[data]}$ and $T_{[hyp]}$ interventions on different potential outcomes. Fig. \ref{fig:causal_modes} describes seven SCMs, the expected potential outcomes with their corresponding treatments $T_{A-G}$, and examples of interventions per scenario.

\textit{For $A$ and $B$ cases}, we examined the impact of buggy code and inline comments on \nlm's predictions. The $A$ case assessed how the presence of bugs influences the cross-entropy and \ntp. Similarly, the $B$ case explored the effect of including inline comments on the model's performance.

\textit{For $C$ and $D$ cases}, we assessed how \nlms respond to minor and major syntactic variations in semantically equivalent code snippets, focusing on the effects of semantic-preserving changes on code generation. Due to the absence of a natural split in clone types in the \BigCloneTB dataset, a typical control and treatment approach for causal analysis was impractical. The selection between \textit{function$_1$} and \textit{function$_2$} methods, as categorized by \BigCloneTB, is arbitrary. Instead, we concentrated on how syntactic differences in methods performing identical functions impact our models. Hence, we used the differences between these function sets as our primary confounders and intervention. Specifically, we employed the Levenshtein Distance – a metric quantifying the necessary edit operations (insert, modify, remove) to transform one sequence into another – to simulate a 'refactoring' treatment. This approach approximated the number of edits needed to convert one method into another and allowed us to bypass the need for a natural split between \textit{function$_1$} and \textit{function$_2$}.

\textit{For $E$ and $F$ cases}, as indicated in \tabref{tab:method}, our case study involved exploring two distinct deep learning architectures. We particularly examined how variations in hyper-parameters such as layers and units within each architecture influence the prediction performance.

{\textit{For the $G$ case}, our goal was to generate interpretations for the prediction of AST node types using BERT. To achieve this, we essentially constructed the dataset by implementing two treatments: masking tokens corresponding to AST Node types (Treatment 1) and randomly masking an equivalent number of tokens (Treatment 2). We then calculated the normalized similarity distances (\ie Jaccard, Levenshtein, and Sorensen-Dice) between the AST of the predicted code and the ground truth samples, which served as the potential outcomes in the \scm.}

%% file: chapters/part_02_chap_02/sec_04_clustering.tex
\section{Syntax Clustering of Code Predictions}
\label{ch:case:sec_04}

Consider the situation where a developer inserts a \fbox{\texttt{\small `('}} character after the \fbox{\texttt{\small `main'}} keyword in a function declaration in Java ({\circled{1}}-Fig.~\ref{fig:performance}). Inherently, a developer mentally rationalizes several things such as the concept of a function declaration and expected Java syntax. If a \nlm can make a similar prediction, it suggests to us that it has \textit{statistically learned} some understanding of the concept of a function declaration and corresponding syntax. Therefore, we assert that by associating human-interpretable categories (\eg Programming Languages Keywords or AST nodes) to model predictions and then analyzing the statistical properties of those predictions, we can begin to learn how well a given \nlm reflects human knowledge.

A major conjecture in interpretability research is that \nlms are more understandable when they \textit{reflect human knowledge} \citep{kim_interpretability_2018}. One way of determining whether a model reflects human knowledge is testing it to see whether or not it operates (or predicts) \textit{similar to how a human would operate}. \docode accomplishes this by grouping code token predictions of \nlms to human interpretable categories.

To help bridge the gap between a given \nlms token-level representation of code and human-understandable categories, \docode aggregates individual tokens to well-known syntactic categories from programming languages. Source code tokens can be clustered to any number of syntactic elements, and particularly for \docode, we focus on aggregating tokens to different syntactic categories, which \textit{\textbf{do not}} require manual labeling. This syntax clustering mitigates the cost involved with large-scale data labeling and still provides explanations rooted in categories with which most programmers and researchers are likely familiar.

{The syntax clustering comprises high-level properties of code using a \textbf{clustering function} defined by $\phi_{\mathcal{H}}: \vec{w} \to \vec{h}$, in which the vector $\vec{w}$ corresponds to tokens from a vocabulary $\mathcal{V}$. Thus, each token in a sequence $w$ is assigned to a specific syntax-understandable category $h$. Note that \docode allows the definitions of any potential clustering function, users are not forced to use our clustering category system $\mathcal{H}$. The proposed categories in our system  $\mathcal{H}$ are classified into keyword-based or grammar-based. Below, we pose a separate clustering function for Java and another one for Python.}

\begin{figure}[ht]
		\centering
		\includegraphics[width=1\textwidth]{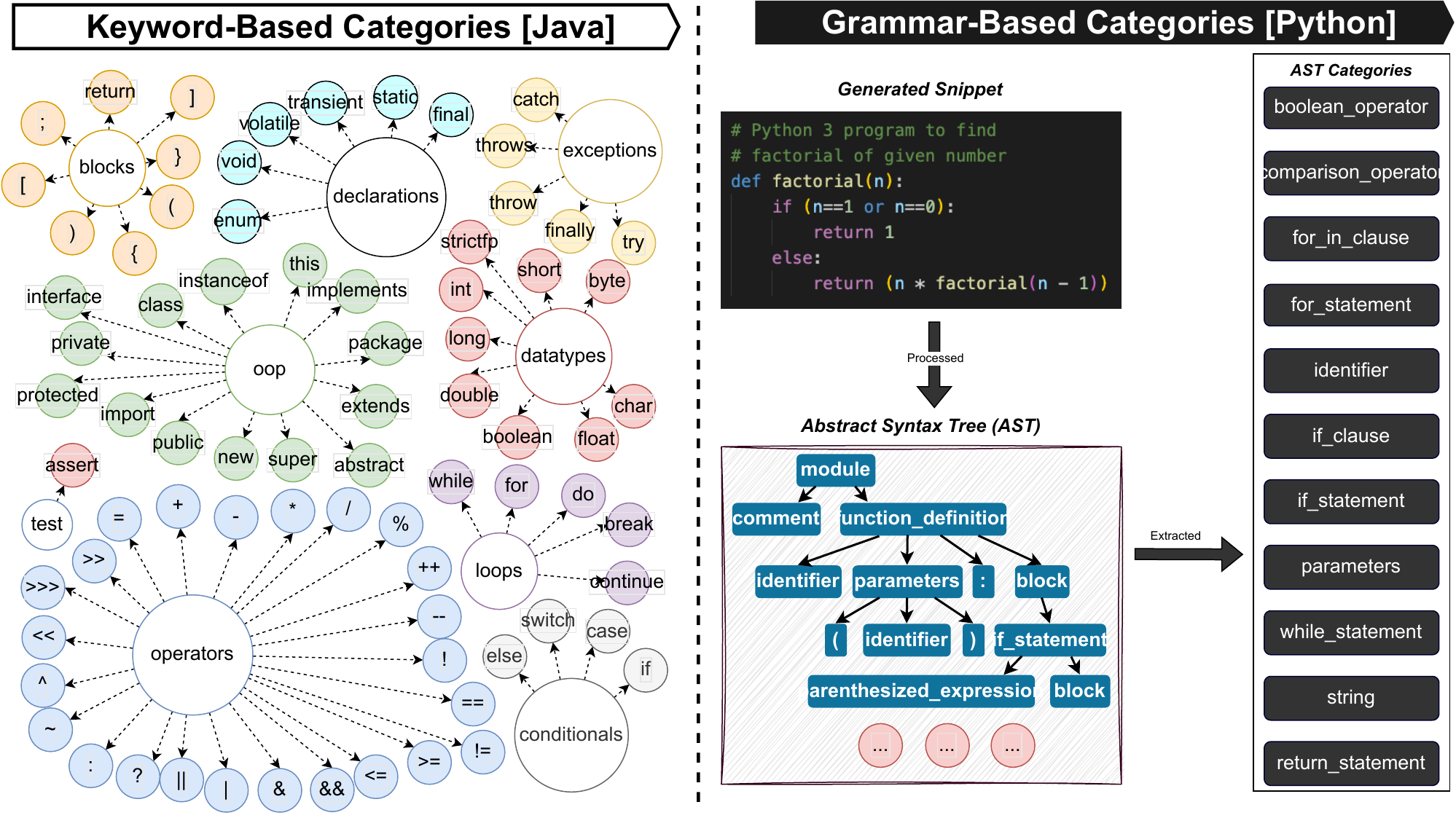}
		\caption{Code Syntax Clustering. Code predictions are grouped into Keyword-based or AST/Grammar-based categories.}
        \label{fig:syntactic_aggregations}
\end{figure}

\subsection{Keyword-Based Aggregation}\label{sec:keyword_aggregation}
In programming languages, different types of tokens retain different semantic meanings. For instance \fbox{\texttt{\small `='}} and \fbox{\texttt{\small `<'}} are common \operators. Therefore, tokens can be grouped into semantically meaningful keyword-based \textit{categories} $\mathcal{H}$. We can establish a clustering function $\phi_{\mathcal{H}}: \vec{w} \to \vec{h}$, in which a token $w$ in a snippet $s$ is clustered to a corresponding keyword category $h$. We propose an initial set of nine categories for Java. Figure~\ref{fig:syntactic_aggregations} illustrates the categories and associated keyword tokens. For each category, we group keywords by functionality collected from the official Oracle Documentation and Manuals \citep{OracleJavaKeywords}. These keywords remain consistent across Java versions from Java 7 onwards.

\subsection{AST/Grammar-Based Aggregation}\label{sec:grammar_aggregation}
{Every expression in a programming language is determined by production rules defined in its Context-Free Grammar defined by the expression $\mathcal{H} = (\alpha, \lambda, \omega, \beta)$, in which $\alpha$ denotes the finite set of non-terminal nodes, $\lambda$ the finite set of terminal nodes, $\omega$ the finite set of production rules, and $\beta$ the start node. The set of production rules $\omega$ for any type of statement (\eg conditional, assignation, operator) is expressed in terms of the terminal and non-terminal nodes $(\alpha, \lambda)$. In programming languages, terminal and non-terminal nodes retain different semantic meanings. Following this, we can establish a clustering function $\phi_{\mathcal{H}}: \vec{w} \to \vec{h}$, where each token $w$ in a snippet $s$ is aggregated and assigned to a corresponding node $h$. We extracted a total of $196$ node types (\ie terminal and non-terminal), from tree-sitter Python's grammar \citep{tree-sitter}. This set of categories was then used to group tokens from code snippets. Figure~\ref{fig:syntactic_aggregations} offers a visual example of this clustering.}

%% file: chapters/part_02_chap_02/sec_05_measures.tex
\section{Causal Inference Measures}
\label{ch:case:sec_05}

Below, we explain association and intervention metrics that we estimate to validate the reliability of an interpretation. Association metrics serve as baselines that we need to confirm for confounding after computing intervention metrics, which represent actual causal effects.  

\begin{figure}[ht]
  \centering
  \begin{subfigure}[b]{0.46\linewidth}
    \includegraphics[width=\linewidth]{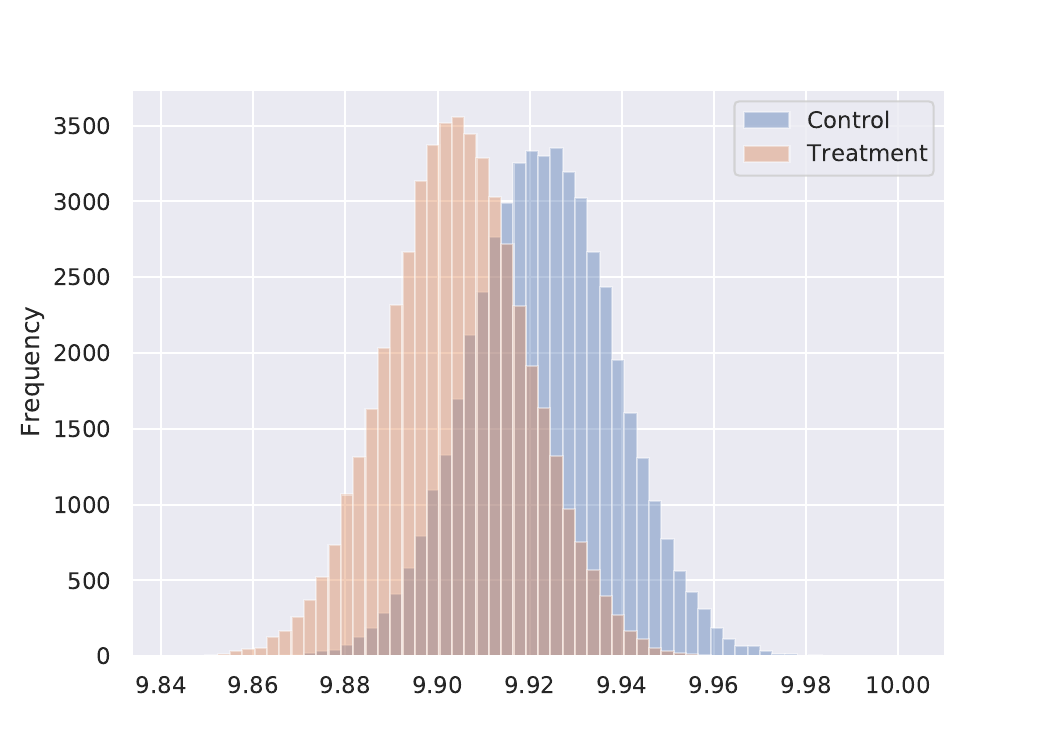}
  \end{subfigure}
  \begin{subfigure}[b]{0.46\linewidth}
    \includegraphics[width=\linewidth]{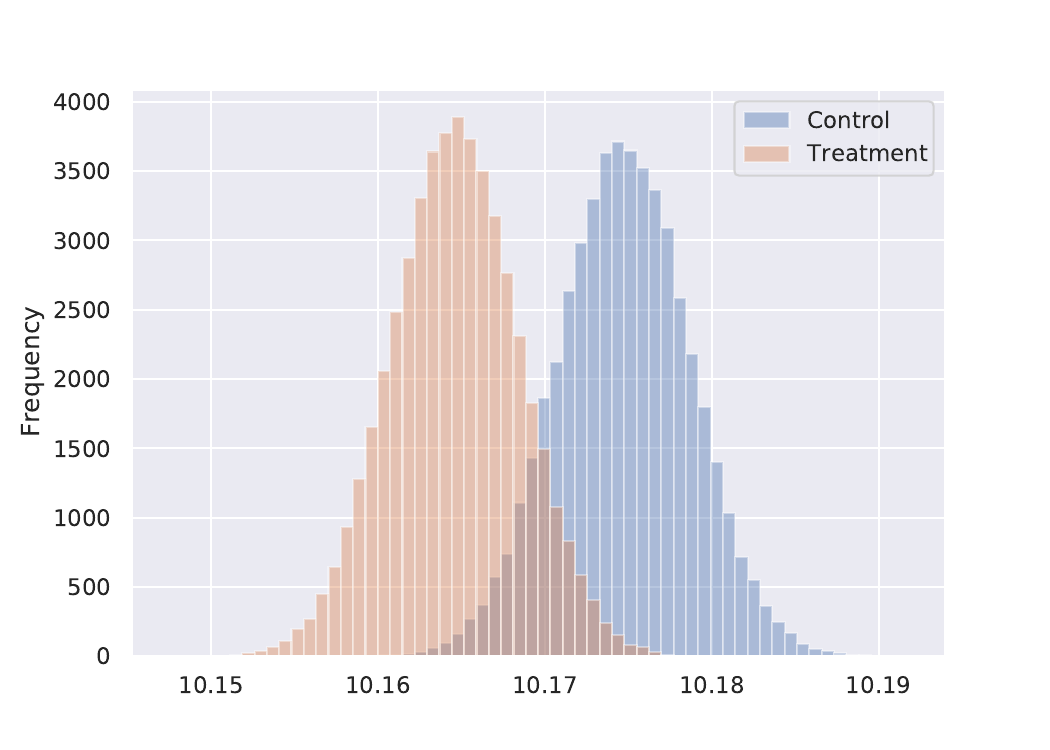}
  \end{subfigure}
  \begin{subfigure}[b]{0.46\linewidth}
    \includegraphics[width=\linewidth]{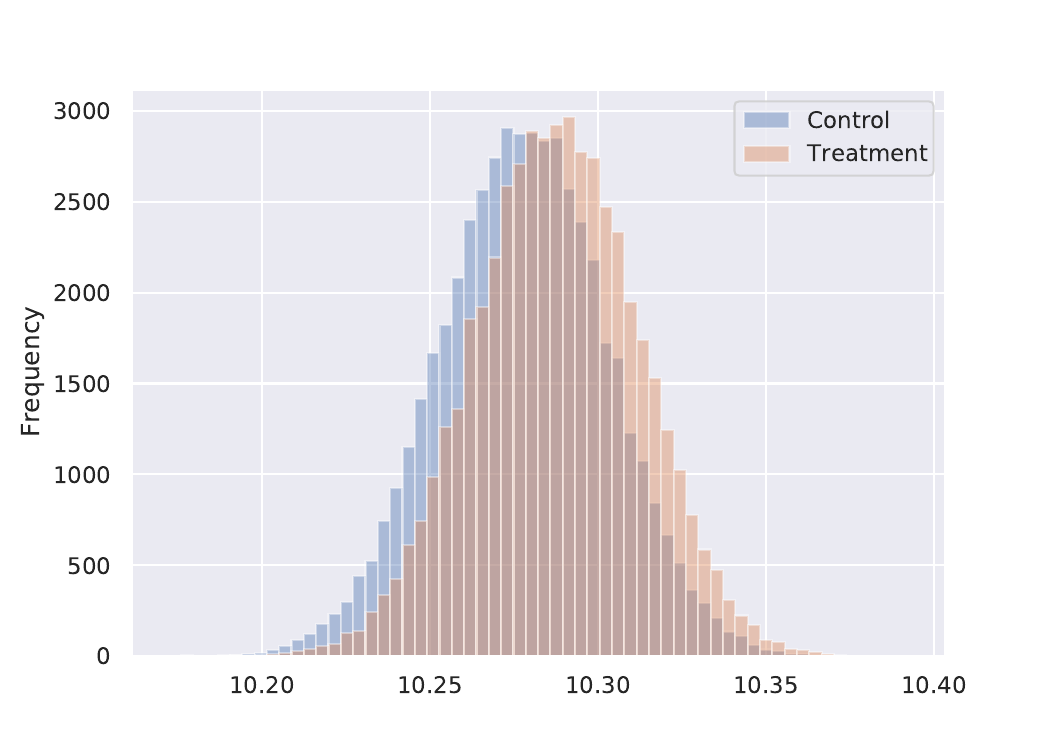}
  \end{subfigure}
  \begin{subfigure}[b]{0.46\linewidth}
    \includegraphics[width=\linewidth]{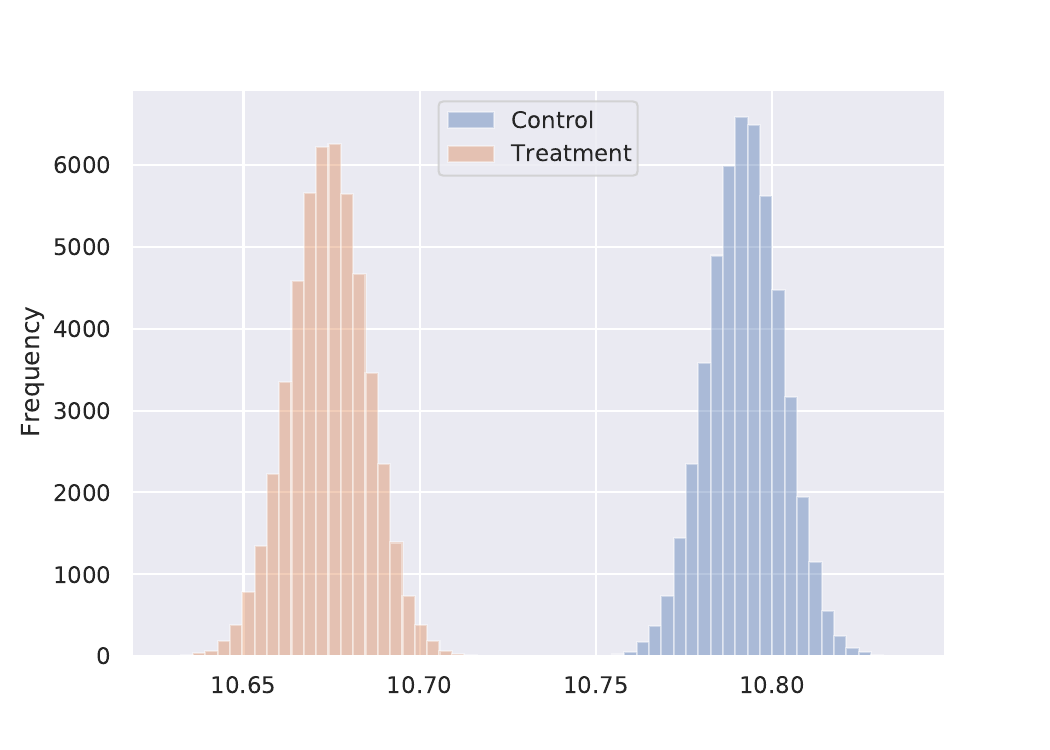}
  \end{subfigure}
  \vspace{0.2cm}
  \caption{\datainterII Intervention (\BigCloneIIITB) for Global Performance: Bootstrapped Cross-Entropy. Top Left: \rnn ($JS=0.3$), Top Right: \gru ($JS=0.8$), Bot Left: \tf ($JS=0.6$), Bot Right: \tfi ($JS=1$).}
  \label{fig:jssimilarity}
  \vspace{0.4cm}
\end{figure}

\textbf{Association Metrics.} For $[A-F]$ cases, we employed two methods to empirically estimate the association distributions $p(Y_g|T)$ and $p(Y_l|T)$. These two methods are the classic Pearson correlation and the Jensen-Shannon distance. For the latter, imagine we wish to understand the correlation between syntactic changes, \ie variable renaming, alterations in white space, \etc and a \nlms performance. One way we can study this is through computing the association of Cross-Entropy values $Y$ under two treatments, the first $T=0$, would be an unaltered code snippet and the second $T=1$ would be its Type III clone. Computing this association can be done using the Jensen Shannon distance $p(Y|T)\approx JS(Y^0,Y^1)$ as defined in Def.~\ref{def:js} for four models. Fig.~\ref{fig:jssimilarity} shows the distributions of $Y^0$ and $Y^1$ with their distances after applying bootstrapping as an example. 

\begin{definition}
\label{def:js}
\textbf{Jensen-Shannon Distance (JS).} The Jensen-Shannon divergence (JSD) overcomes the asymmetric computation of the KL divergence and provides a measure of the difference between distributions Eq.~\ref{eq:js_divergence}. The JS distance is the square of the JS divergence $p(Y|T)\approx JS(Y^{T=0},Y^{T=1}) = JSD(Y^{0},Y^{1})^2$. JS is proportional to the influence of $T$ on $Y$, which measure the separation of the distributions $Y^{0},Y^{1}$. The notation $Y^{T=0}$ refers to the potential outcomes observed under the treatment $T=0$. 

\begin{subequations}
    \begin{align}
     JSD(Y^{0}=y^0,Y^{1}=y^1) &= \label{eq:js_divergence-1}\\
     \frac{1}{2}\left[ D_{KL}\left(y^0||\frac{y^0+y^1}{2}\right) + D_{KL}\left(y^1||\frac{y^0+y^1}{2}\right)\right] &= \label{eq:js_divergence-2}
    \end{align}
\label{eq:js_divergence}
\end{subequations} 

\end{definition}

\textbf{Intervention Metrics.} Conversely, the interventional distributions $p(Y_g|do(T))$, $p(Y_l|do(T))$, and $p(Y_d|do(T))$ are estimated in terms of the Average Treatment Effect (ATE), as previously introduced in Eq~.\ref{eq:ate}. \docode can potentially estimate other types of causal effect metrics such as \textit{Conditional Average Treatment Effect (CATE)}. 

%% file: chapters/part_02_chap_02/sec_06_refutation.tex
\section{Refutation Testing}
\label{ch:case:sec_06}

After \docode estimates causal effects, these effects are validated for robustness. Ergo, \docode incorporates various refutation methods to evaluate the sensitivity of the causal estimations as we defined in \secref{ch:docode:sec_04}. In our case study, we employ the following refutations: introducing a random common cause (\ie independent common causes added randomly should not impact the causal estimates), applying placebo treatments (\ie causal effect should go zero if the true treatment is replaced by an independent random variable), considering unobserved common causes (\ie causal estimates should not be too sensitive when we add an independent common cause correlated with the treatment and outcome), and performing data subset validations (\ie the causal effect should not change when the dataset is replaced by a random subset).

%% file: chapters/part_02_chap_02/sec_07_questions.tex
\section{Research Questions}
\label{ch:case:sec_07}

Inspired by the notion of Pearl's Ladder of Causation \citep{Pearl2016Causality,Sharma2021DoWhyAssumptions}, we framed two research questions to explore \docode's potential in enabling causal interpretability for \nlms. \texttt{\ref{rq:inference}} computes the feasibility of some SE-based Interventions and \texttt{\ref{rq:robustness}} performs a sensitivity analysis to validate the robustness of previously computed causal effects.

\begin{enumerate}[label=\textbf{RQ$_{\arabic*}$:}, ref=\textbf{RQ$_{\arabic*}$}, wide, labelindent=5pt]\setlength{\itemsep}{0.2em}
      \item \label{rq:inference} {\textbf{Causal Inference}: \textit{To what extent do SE interventions affect code prediction?}} 
      \item \label{rq:robustness} {\textbf{Robustness}: \textit{How robust are the treatment effects based on SE interventions?}}
\end{enumerate}

%% file: chapters/part_02_chap_02/sec_08_results.tex
\section{Results}
\label{ch:case:sec_08}

In this section, we present the results for both research questions per each interpretability scenario introduced in previous \secref{ch:case:sec_07}. \tabref{tab:results_A_global} and \ref{tab:results_B_global} offer an overview of the associative and interventional metrics observed across our diverse \nlms and datasets, specifically cross-entropy (\ie $Y_g$ \ref{out:global}). Note that the association results in these tables contain both Pearson correlation and Jensen-Shannon distance. Similarly, \tabref{tab:resultsLocalJS} and \ref{tab:resultsLocalPearson} focus on presenting the causal inference measures in terms of \ntp (\ie $Y_l$ \ref{out:local}). Lastly, Table \ref{tab:resultsLocalPearsonMLM} showcases potential outcomes based on distance predictions (\ie $Y_d$ \ref{out:distance}). In addition, to test the robustness of the estimated causal effects, we used the four baseline refutation methods introduced in \secref{ch:docode:sec_04}.

\input{tables/chap_4_tab_2_A_results_global}
\input{tables/chap_4_tab_2_B_results_global}
\input{tables/chap_4_tab_3_results_local_jensen}

\subsection{Interpretability Scenario \texorpdfstring{$A$}{a}: BuggyCode}

\textbf{To what extent does \datainterI affect the predictions of \nlms?}
For the $A$ case, focusing on the \datainterI intervention, both \rnn and \tf models exhibit considerably high JS distances of $0.73$ and $0.67$, indicating a strong correlation between the \datainterI intervention and the cross-entropy: $P(Y_g|T_{A})$. Additionally, we observed a similar pattern in \ntp outcomes for $Y_{l\blocks}$ with the \gru model, which had the highest correlation at $0.626$ for \datainterI.

However, when we controlled for SE covariates, the correlations initially observed turned out to be misleading for causation. Across the three models (\rnn, \gru, and \tf), the ATEs were very small, both in cross-entropy $p(Y_g|do(T_{A}))$ with values of $-3E-4$, $-2.33E-05$, and $-2E-4$, and in \ntp outcomes $p(Y_{l\blocks}|do(T_{A}))$ with a value of $-5.25E-4$. {Having null effects after observing high correlations confirms the presence of confounding bias for this case. Therefore, the prediction performance is being affected by other confounders different from the actual buggy code intervention.}

\textbf{How robust are the causal effect estimations?}
To verify the sensitivity of the ATEs for $Y_g$, we computed four robustness tests: $\mathcal{R}_1$, $\mathcal{R}_2$, $\mathcal{R}_3$, and $\mathcal{R}_4$. The results for $\mathcal{R}_1$, $\mathcal{R}_2$, and $\mathcal{R}_4$ aligned closely with the ATEs we found for the three models. Furthermore, the results for $\mathcal{R}_3$ were nearly zero. This indicates that the estimated causal effects were robust. Similarly, $\mathcal{R}_1$ values for $Y_l$ results were close to the obtained ATEs.

\begin{boxK}
\textbf{BuggyCode Finding $F_A$:} The presence or absence of buggy code does not appear to causally influence (or explain) the prediction performance of our \nlms even under measured high correlation.
\end{boxK}

\subsection{Interpretability Scenario \texorpdfstring{$B$}{b}: Code Documentation}
\textbf{To what extent does \datainterIII (or inline comments) intervention affect the predictions of \nlms?}
In the \datainterIII interventions, there were no strong correlations in cross-entropy $P(Y_{g}|T_{B})$ for the \rnn, \gru, and \tf models, with values of $0.18$, $0.22$, and $0.25$ respectively. However, the \ntp outcomes showed significant associations in specific categories: for \gru, a strong correlation in \datatype ($P(Y_{l\datatype}|T_{B})=0.749$), and for \tf, notable correlations in \operators ($P(Y_{l\operators}|T_{B})=0.998$), \conditionals ($P(Y_{l\conditionals}|T_{B})=0.821$), and \tests ($P(Y_{l\tests}|T_{B})=0.724$).

Once we adjusted for covariates, the ATEs in both $p(Y_g|do(T_{B}))$ and $p(Y_l|do(T_{B}))$ were close to zero, showing a trend towards no causal effects. This suggests that actions like removing comments from code have little to no causal impact on cross-entropy or \ntp values.

\textbf{How robust are the causal effect estimations?}The obtained results across refutation tests indicated unstable causal effects. Specifically, the outcomes for $\mathcal{R}_1$, $\mathcal{R}_2$, and $\mathcal{R}_4$ did not align closely with the ATEs. This discrepancy could stem from several factors: 1) a lack of sufficient data samples, 2) inaccuracies in our causal diagram assumptions (\eg confounders, instrumental variables, and effect modifiers), and/or 3) the treatment is inadequate.

\begin{boxK}
\textbf{Code Documentation Finding $F_B$:} Despite observing strong correlations between the removal of comments and \ntp, we cannot causally interpret code predictions from inline comments since measured ATEs are not robust after refutations. This suggests that other hidden confounders are influencing the estimation of this causal effect.
\end{boxK}

\input{tables/chap_4_tab_3_results_local_pearson} 
\input{tables/chap_4_tab_3_results_local_pearson_mlm}

\subsection{Interpretability Scenario \texorpdfstring{$C$}{c}: Clones Type II}
\textbf{To what extent does \datainterII intervention (Clones Type II) affect the predictions of \nlms?}
In the \datainterII intervention, we observed a positive correlation between Levenshtein ``edit'' distance in Type II clone pairs and cross-entropy values (\ie $p(Y_g|T_{C})$) across \rnn, \gru, and \tf models. The correlation values were $0.45$, $0.6$, and $0.452$ for each model respectively, as detailed in Table~\ref{tab:results_B_global}.

Furthermore, the interventions significantly impacted \ntp across different categories, as shown by our keyword-based categories $p(Y_l|T_{C})$. Notably, within the \gru model, we found a positive correlation between Type II clones' Levenshtein distance and \textit{operators}: $p(Y_{l\operators}|T_{C})\approx0.56$.

In our study, we found weak causal effects between clone type variations and \ntp across all \nlms in the nine categories analyzed. For instance, the performance of \oop tokens and \operators was causally affected with ATEs of $p(Y_{l\oop}|do(T_{C}))=0.388\pm0.241$ and $p(Y_{l\operators}|do(T_{C}))=0.316\pm0.202$. However, the high standard deviation observed across the \nlms in these categories indicates a wide variance in what the models statistically learned. Unfortunately, most of the ATEs could not be computed for each keyword category because it was not possible to create a linear model to estimate the effects due to the shape of the data (\ie input variables have the same values).

\textbf{How robust are the causal effect estimations?} For testing the robustness of the ATEs, we could not compute the refutation methods  \textit{Placebo} and \textit{Remove Subset} due to limited data. However, the other two methods, \textit{Random Comm. Cause} and \textit{Unobserved Comm. Cause}, showed stability, which reinforces our confidence in the ATEs for global results.

\begin{boxK}
\textbf{Clones Type II Finding $F_C$:} The presence of Clones Type II impacts (or causally explains) the cross-entropy on \rnn, \gru, and \tf, as obtained correlations and causal effects suggest an influence of some interventions in identifiers, literals, types, white spaces, layout, and comments.
\end{boxK}

\subsection{Interpretability Scenario \texorpdfstring{$D$}{d}: Clones Type III}
\textbf{To what extent does \datainterII intervention (Clones Type III) affect the predictions of \nlms?}
By contrast, our analysis showed a different pattern for Type III clones. We detected no strong correlations between the Levenshtein ``edit'' distance and the cross-entropy in any of our \nlms. This is evident from the marginal correlation values for the three models—\rnn, \gru, and \tf: $-0.056$, $0.14$, and $-0.14$, respectively.

In our models, the causal effect on cross-entropy was generally small but showed interesting variations. For the \tf model, this effect was negative, with $P(Y_g|do(T_{D}))\approx-0.274$, whereas, for the \gru model, it was positive at $P(Y_g|do(T_{D}))\approx0.11$. Additionally, the causal effect on \ntp outcomes for \blocks was insignificant: $$p(Y_{l\blocks}|do(T_{D}))=2.80E-05$$.

\textbf{How robust are the causal effect estimations?}
Similar to the findings with clone type II, testing the robustness of the ATEs presented challenges. Due to limited data, we were unable to apply $\mathcal{R}_3$ and $\mathcal{R}_4$. However, the use of the other two methods, $\mathcal{R}_1$ and $\mathcal{R}_2$, demonstrated stability in our results.

\begin{boxK}
\textbf{Clones Type III Finding $F_D$:} The presence of Clones Type III does not consistently impact (or causally explain) the cross-entropy and \ntp, as revealed by negative and positive ATEs obtained across the \nlms under analysis.
\end{boxK}

\subsection{Interpretability Scenario \texorpdfstring{$E$}{e}: Layers}
\textbf{To what extent does \modelinterI intervention affect the predictions of \nlms?}
We observed that \modelinterI interventions are negatively correlated with cross-entropy. This trend is evident in the values of $p(Y_g|T_{E})$ for \gru and \tf models: $-0.093$ and $-0.485$ respectively. This suggests that as the number of layers increases, the cross-entropy tends to decrease. Conversely, we found strong correlations for \ntp outcomes across keyword-based categories ($p(Y_l|T_{E})$): the categories \blocks, \conditionals, and \extra showed notably high correlations, with values of $0.725$, $0.682$, and $0.606$ respectively.

Nonetheless, we detected confounding bias in almost all categories of our intervention analysis for \ntp outcomes. This was indicated by the extremely low ATEs, such as $p(Y_{\blocks}|do(T_{E}))=0.018$ and $p(Y_{l\extra}|do(T_{E}))=0.0145$. A plausible explanation for this phenomenon can be traced to SE metric confounders. For instance, the size of the methods appears to have a significant influence on both cross entropy and \ntp results across the \nlms.

\textbf{How robust are the causal effect estimations?}
To verify the sensitivity of the ATEs for $Y_g$, we estimated $\mathcal{R}_1$, $\mathcal{R}_2$, $\mathcal{R}_3$, and $\mathcal{R}_4$. The results for $\mathcal{R}_1$, $\mathcal{R}_2$, and $\mathcal{R}_4$ aligned closely with the ATEs we found for \gru and \tf. Furthermore, $\mathcal{R}_3$ results were nearly zero. Thus, the estimated causal effects were robust. Similarly, $\mathcal{R}_1$ values for $Y_l$ results were close to the obtained ATEs, reinforcing the robustness of our findings.

\begin{boxK}
\textbf{Layers Finding $F_E$:} Although it is observed strong correlations between the number of layers and \ntp, which might suggest a causal interpretation of code predictions, the reported ATEs close to zero demonstrate the presence of confounding bias.
\end{boxK}

\subsection{Interpretability Scenario \texorpdfstring{$F$}{f}: Units}
\textbf{To what extent does \modelinterII intervention affect the predictions of \nlms?}

In a similar vein, \modelinterII interventions tend to be negatively correlated with the cross-entropy. The number of units showed a negative effect with the \gru model, with $p(Y_g|T_{F})\approx-0.084$.

Likewise, we found negative values for $P(Y_l|T_{F})$, suggesting that increasing the number of units negatively impacts the \ntp outcomes across all keyword-based categories. However, these reported values were relatively low, making it challenging to draw definitive conclusions. For example, the highest negative correlation observed was $P(Y_{l\exceptions}|T_{F})\approx -0.253$ for \ntp.

Negative correlations observed for the cross-entropy were consistent with the corresponding negative ATEs estimated. However, the negative correlation observed for \blocks was not in fact causal: $P(Y_{l\blocks}|do(T_{F}))\approx \text{-5E-06}$. Similarly, for the other categories, the ATEs were so minimal that they can be considered as causal null effects.

\textbf{How robust are the causal effect estimations?} 
Just as with \modelinterI, we conducted robustness tests $\mathcal{R}_1$, $\mathcal{R}_2$, $\mathcal{R}_3$, and $\mathcal{R}_4$ for $Y_g$. The outcomes of $\mathcal{R}_1$, $\mathcal{R}_2$, and $\mathcal{R}_4$ closely matched the ATEs we observed in the \gru model. Furthermore, the $\mathcal{R}_3$ values were nearly zero, reinforcing the robustness of the estimated causal effects. Similarly, the $\mathcal{R}_1$ values for $Y_l$ are also closely aligned with the corresponding ATEs, further confirming the robustness of the results.

\begin{boxK}
\textbf{Units Finding $F_F$:} Intervening the number of units tends to be negatively correlated with the cross-entropy; in fact, measured causal effects on \ntp are null suggesting that this intervention does not explain code predictions.
\end{boxK}

\subsection{Interpretability Scenario \texorpdfstring{$G$}{g}: Masking AST Nodes}

\textbf{To what extent does \datainterIV intervention affect the predictions of \nlms?}
We computed the distance outcomes $Y_d$ for some AST node types, as detailed in \secref{sec:grammar_aggregation}. However, in \tabref{tab:resultsLocalPearsonMLM}, we specifically focus on a subset of both terminal and non-terminal nodes. This subset was chosen for its familiarity with developers and includes elements such as conditional statements, identifiers, and repetition statements. The intervention demonstrated strong negative correlations between AST node types and the distance outcomes $Y_d$, consistent across all distance metrics. For example, $P(Y_{d\jaccard}|T_{G\forinclause})=-0.4232$, $P(Y_{d\levenshtein}|T_{G\forinclause})=-0.3724$, and \\ $P(Y_{d\soren}|T_{G\forinclause})=-0.3568$.

The correlation results are further supported by the estimated ATEs $P(Y_d|do(T_{G\node}))$, demonstrating no confounding bias. Although the causal effects are lower than the estimated correlations, they reveal the presence of negative causation.

\textbf{How robust are the causal effect estimations?}
We calculated $\mathcal{R}_1$, and we obtained the same values for ATEs indicating that the causal effects are robust. Furthermore, we also calculated  $\mathcal{R}_4$, and the values were close to zero.

\begin{boxK}
\textbf{Masking AST Nodes Finding $F_G$:} The intervention of masking random tokens has more impact on code predictions than masking grammar-based categories. This suggests that \bert does not entirely capture the nodes' information of Abstract Syntax Trees (ASTs).
\end{boxK}

%% file: tables/chap_4_tab_2_A_results_global.tex
\begin{table}[ht]
\centering
\caption{Causal Interventions $p(Y_g|do(T))$ and Associations $p(Y_g|T)$ of Cross-Entropy across models and datasets.}
\label{tab:results_A_global}

\begin{adjustbox}{width=1\textwidth}
\begin{tabular}{l|rrr|rrr|rr}
\hline
\multirow{2}{*}{\textbf{\begin{tabular}[c]{@{}l@{}}Counterfactual  \\ Interventions $T$\end{tabular}}} &
  \multicolumn{3}{c|}{\multirow{2}{*}{\textbf{\begin{tabular}[c]{@{}c@{}}\datainterI (BuggyTB)\\ $T_{dat0}$\end{tabular}}}} &
  \multicolumn{3}{c|}{\multirow{2}{*}{\textbf{\begin{tabular}[c]{@{}c@{}}\datainterIII (CommentsTB)\\ $T_{dat1}$\end{tabular}}}} &
  \multicolumn{2}{c}{\multirow{2}{*}{\textbf{\begin{tabular}[c]{@{}c@{}}\modelinterI (CodeSearchNet)\\ $T_{hyp0}$\end{tabular}}}} \\
                                    & \multicolumn{3}{c|}{}                      & \multicolumn{3}{c|}{}                      & \multicolumn{2}{c}{}          \\ \hline
\nlm &
  \multicolumn{1}{c}{\textit{\rnn}} &
  \multicolumn{1}{c}{\textit{\gru}} &
  \multicolumn{1}{c|}{\textit{\tf}} &
  \multicolumn{1}{c}{\textit{\rnn}} &
  \multicolumn{1}{c}{\textit{\gru}} &
  \multicolumn{1}{c|}{\textit{\tf}} &
  \multicolumn{1}{c}{\textit{\gru}} &
  \multicolumn{1}{c}{\textit{\tf}} \\ \hline
\textit{Association \assoJS/ \assoPR} & \cellcolor[HTML]{EFEFEF}$0.730_{JS}$ & $0.230_{JS}$ & \cellcolor[HTML]{EFEFEF}$0.670_{JS}$ & $0.180_{JS}$ & $0.220_{JS}$ & $0.250_{JS}$ & $-0.093_{PR}$ & $-0.485_{PR}$ \\
\textit{Causal Eff. ATE}            & -0.0003      & -2.33E-05    & -0.0002      & 0.0023       & 2.90E-05     & 0.0026       & -0.0058       & -0.0124       \\
\textit{Random Comm. Cause}         & -0.0003      & -2.45E-05    & -0.0002      & 0.0011       & -0.0004      & 0.0015       & -0.0058       & -0.0124       \\
\textit{Unobserved Comm. Cause}     & -0.0003      & 1.54E-05     & -0.0001      & 0.0002       & -0.0001      & 0.0007       & -0.0050       & -0.0108       \\
\textit{Placebo}                    & 0.0001       & 1.44E-05     & 0.0001       & 0.0006       & -1.33E-05    & 0.0006       & -0.0001       & -2.77E-05     \\
\textit{Remove Subset}              & -0.0003      & -3.68E-05    & -0.0002      & 0.0012       & -0.0003      & 0.0016       & -0.0058       & -0.0124       \\ \hline
\end{tabular}

\end{adjustbox}

\end{table}

%% file: tables/chap_4_tab_2_B_results_global.tex
\begin{table}[ht]
\centering
\caption{Causal Interventions $p(Y_g|do(T))$ and Associations $p(Y_g|T)$ of Cross-Entropy across Semantic datasets.}
\label{tab:results_B_global}

\begin{adjustbox}{width=\textwidth}
\begin{tabular}{l|rrrrrr}
\hline
\multirow{2}{*}{\textbf{\begin{tabular}[c]{@{}l@{}}Counterfactual  \\ Interventions $T$\end{tabular}}} &
  \multicolumn{6}{c}{\textbf{\begin{tabular}[c]{@{}c@{}}\datainterII (BigClone2/3TB)\\ $T_{dat2}$\end{tabular}}} \\ \cline{2-7} 
                                & \multicolumn{3}{c|}{\textbf{BigClone2TB}}              & \multicolumn{3}{c}{\textbf{BigClone3TB}} \\ \hline
\nlm &
  \multicolumn{1}{c}{\textit{\rnn}} &
  \multicolumn{1}{c}{\textit{\gru}} &
  \multicolumn{1}{c|}{\textit{\tf}} &
  \multicolumn{1}{c}{\textit{\rnn}} &
  \multicolumn{1}{c}{\textit{\gru}} &
  \multicolumn{1}{c}{\textit{\tf}} \\ \hline
\textit{Association \assoJS/ \assoPR} &
  $0.45_{PR}$ &
  $0.598_{PR}$ &
  \multicolumn{1}{r|}{$0.452_{PR}$} &
  $-0.056_{PR}$ &
  $0.14_{PR}$ &
  $-0.14_{PR}$ \\
\textit{Causal Eff. ATE}        & 0.6288 & \cellcolor[HTML]{EFEFEF}\textbf{0.8713} & \multicolumn{1}{r|}{0.5635} & -0.1042      & 0.1085      & -0.2739     \\
\textit{Random Comm. Cause}     & 0.6297 & 0.8720          & \multicolumn{1}{r|}{0.5651} & -0.1043      & 0.1084      & -0.2741     \\
\textit{Unobserved Comm. Cause} & 0.2950 & 0.4257          & \multicolumn{1}{r|}{0.2737} & -0.0800      & 0.0830      & -0.2168     \\
\textit{Placebo}                & -      & -               & \multicolumn{1}{r|}{-}      & -            & -           & -           \\
\textit{Remove Subset}          & -      & -               & \multicolumn{1}{r|}{-}      & -            & -           & -           \\ \hline
\end{tabular}

\end{adjustbox}
\end{table}

%% file: tables/chap_4_tab_3_results_local_jensen.tex
\begin{table}[ht]
\centering
\caption{\ntp Association Results $p(Y_l|T)$ are \textit{Jensen-Shannon Dist}. Causal Effects are ATEs $p(Y_l|do(T))$ (\textbf{bold}:strong corr., background:best effect)}
\label{tab:resultsLocalJS}

\begin{adjustbox}{width=1\textwidth}

\begin{tabular}{@{}l|rrrrrr|rrrrrr@{}}
\toprule
\textbf{\begin{tabular}[c]{@{}l@{}}Counterfactual  \\ Interventions $T_{data}$\end{tabular}} &
  \multicolumn{6}{c|}{\textbf{\begin{tabular}[c]{@{}c@{}}\datainterI \\ $T_{dat0}$\end{tabular}}} &
  \multicolumn{6}{c}{\textbf{\begin{tabular}[c]{@{}c@{}}\datainterIII\\ $T_{dat1}$\end{tabular}}} \\ \midrule
\nlm &
  \multicolumn{3}{c}{\textit{\gru}} &
  \multicolumn{3}{c|}{\textit{\tf}} &
  \multicolumn{3}{c}{\textit{\gru}} &
  \multicolumn{3}{c}{\textit{\tf}} \\ \midrule
\textbf{Categories} &
  \multicolumn{1}{c}{\textit{\begin{tabular}[c]{@{}c@{}}Association\\ \assoJS\end{tabular}}} &
  \multicolumn{1}{c}{\textit{\begin{tabular}[c]{@{}c@{}}Causal Eff.\\ ATE\end{tabular}}} &
  \multicolumn{1}{c}{\textit{\rfi}} &
  \multicolumn{1}{c}{\textit{\begin{tabular}[c]{@{}c@{}}Association\\ \assoJS\end{tabular}}} &
  \multicolumn{1}{c}{\textit{\begin{tabular}[c]{@{}c@{}}Causal Eff.\\ ATE\end{tabular}}} &
  \multicolumn{1}{c|}{\textit{\rfi}} &
  \multicolumn{1}{c}{\textit{\begin{tabular}[c]{@{}c@{}}Association\\ \assoJS\end{tabular}}} &
  \multicolumn{1}{c}{\textit{\begin{tabular}[c]{@{}c@{}}Causal Eff.\\ ATE\end{tabular}}} &
  \multicolumn{1}{c}{\textit{\rfi}} &
  \multicolumn{1}{c}{\textit{\begin{tabular}[c]{@{}c@{}}Association\\ \assoJS\end{tabular}}} &
  \multicolumn{1}{c}{\textit{\begin{tabular}[c]{@{}c@{}}Causal Eff.\\ ATE\end{tabular}}} &
  \multicolumn{1}{c}{\textit{\rfi}} \\ \midrule
\blocks &
  \cellcolor[HTML]{EFEFEF}\textbf{0.626} &
  -0.0005 &
  -0.00052 &
  0.206 &
  -0.0001 &
  -0.000115 &
  {0.133} &
  0.0003 &
  0.000488 &
  {0.052} &
  -0.0006 &
  -0.000352 \\
\exceptions &
  {0.107} &
  -1.00E-06 &
  1.00E-06 &
  \cellcolor[HTML]{EFEFEF}\textbf{0.651} &
  \cellcolor[HTML]{EFEFEF}-1.20E-05 &
  -1.10E-05 &
  {0.165} &
  -4.80E-05 &
  -4.80E-05 &
  {0.091} &
  \cellcolor[HTML]{EFEFEF}-1.20E-05 &
  -4.20E-05 \\
\oop &
  {0.048} &
  2.00E-05 &
  1.20E-05 &
  {0.058} &
  -1.30E-05 &
  -9.00E-06 &
  {0.070} &
  -7.00E-06 &
  -4.70E-05 &
  \cellcolor[HTML]{EFEFEF}{0.090} &
  \cellcolor[HTML]{EFEFEF}-4.90E-05 &
  -2.60E-05 \\
\tests &
  0.581 &
  9.00E-06 &
  8.00E-06 &
  \cellcolor[HTML]{EFEFEF}\textbf{0.997} &
  1.00E-05 &
  9.00E-06 &
  0.595 &
  1.90E-05 &
  4.00E-05 &
  \textbf{0.724} &
  \cellcolor[HTML]{EFEFEF}-1.70E-05 &
  2.00E-06 \\
\declarations &
  {0.071} &
  4.00E-06 &
  4.00E-06 &
  {0.054} &
  1.00E-06 &
  1.00E-06 &
  {0.061} &
  -2.10E-05 &
  -0.000145 &
  \cellcolor[HTML]{EFEFEF}{0.148} &
  \cellcolor[HTML]{EFEFEF}4.60E-05 &
  6.00E-05 \\
\conditionals &
  0.345 &
  -3.90E-05 &
  -3.90E-05 &
  {0.176} &
  -8.00E-06 &
  -1.00E-05 &
  {0.111} &
  8.00E-06 &
  7.40E-05 &
  \cellcolor[HTML]{EFEFEF}\textbf{0.821} &
  \cellcolor[HTML]{EFEFEF}-0.000596 &
  -0.000603 \\
\loops &
  {0.130} &
  -2.00E-06 &
  -3.00E-06 &
  \cellcolor[HTML]{EFEFEF}0.323 &
  \cellcolor[HTML]{EFEFEF}2.10E-05 &
  1.90E-05 &
  {0.027} &
  -1.70E-05 &
  1.50E-05 &
  {0.085} &
  1.80E-05 &
  3.00E-06 \\
\operators &
  {0.123} &
  -6.00E-06 &
  -5.00E-06 &
  {0.039} &
  1.50E-05 &
  1.10E-05 &
  0.278 &
  0.0002 &
  0.000345 &
  \cellcolor[HTML]{EFEFEF}\textbf{0.998} &
  \cellcolor[HTML]{EFEFEF}0.0079 &
  0.008991 \\
\datatype &
  0.253 &
  -1.00E-05 &
  -1.00E-05 &
  {0.168} &
  9.00E-06 &
  1.00E-05 &
  \cellcolor[HTML]{EFEFEF}\textbf{0.749} &
  \cellcolor[HTML]{EFEFEF}0.0003 &
  0.000328 &
  0.432 &
  0.0002 &
  0.000111 \\
\extra &
  {0.169} &
  2.60E-05 &
  2.00E-05 &
  {0.121} &
  5.70E-05 &
  6.00E-05 &
  0.368 &
  0.0014 &
  0.001016 &
  \cellcolor[HTML]{EFEFEF}0.526 &
  0.0024 &
  0.002004 \\ \bottomrule
\end{tabular}

\end{adjustbox}

\end{table}

%% file: tables/chap_4_tab_3_results_local_pearson.tex
\begin{table}[ht]
\centering
\caption{\ntp Association Results $p(Y_l|T)$ are \textit{Pearson Corr}. Causal Effects are ATEs $p(Y_l|do(T))$.
}
\label{tab:resultsLocalPearson}

\begin{adjustbox}{width=1\textwidth}

\begin{tabular}{l|rrrrrr|rrrrrr}
\hline
\textbf{\begin{tabular}[c]{@{}l@{}}Counterfactual  \\ Interventions $T$\end{tabular}} &
  \multicolumn{6}{c|}{\textbf{\begin{tabular}[c]{@{}c@{}}\datainterII (Type III)\\ $T_{dat2}$\end{tabular}}} &
  \multicolumn{6}{c}{\textbf{\begin{tabular}[c]{@{}c@{}}\modelinterI\\ $T_{hyp0}$\end{tabular}}} \\ \hline
\nlm &
  \multicolumn{3}{c}{\textit{\gru}} &
  \multicolumn{3}{c|}{\textit{\tf}} &
  \multicolumn{3}{c}{\textit{\gru}} &
  \multicolumn{3}{c}{\textit{\tf}} \\ \hline
\textbf{Categories} &
  \multicolumn{1}{c}{\textit{\begin{tabular}[c]{@{}c@{}}Association\\ \assoPR\end{tabular}}} &
  \multicolumn{1}{c}{\textit{\begin{tabular}[c]{@{}c@{}}Causal Eff.\\ ATE\end{tabular}}} &
  \multicolumn{1}{c}{\textit{\rfi}} &
  \multicolumn{1}{c}{\textit{\begin{tabular}[c]{@{}c@{}}Association\\ \assoPR\end{tabular}}} &
  \multicolumn{1}{c}{\textit{\begin{tabular}[c]{@{}c@{}}Causal Eff.\\ ATE\end{tabular}}} &
  \multicolumn{1}{c|}{\textit{\rfi}} &
  \multicolumn{1}{c}{\textit{\begin{tabular}[c]{@{}c@{}}Association\\ \assoPR\end{tabular}}} &
  \multicolumn{1}{c}{\textit{\begin{tabular}[c]{@{}c@{}}Causal Eff.\\ ATE\end{tabular}}} &
  \multicolumn{1}{c}{\textit{\rfi}} &
  \multicolumn{1}{c}{\textit{\begin{tabular}[c]{@{}c@{}}Association\\ \assoPR\end{tabular}}} &
  \multicolumn{1}{c}{\textit{\begin{tabular}[c]{@{}c@{}}Causal Eff.\\ ATE\end{tabular}}} &
  \multicolumn{1}{c}{\textit{\rfi}} \\ \hline
\blocks &
  0.026 &
  -1.50E-05 &
  -1.50E-05 &
  0.186 &
  -2.80E-05 &
  -2.80E-05 &
  {\color[HTML]{333333} -0.102} &
  {\color[HTML]{333333} -0.010559} &
  {\color[HTML]{333333} -0.010559} &
  \cellcolor[HTML]{EFEFEF}\textbf{0.725} &
  \cellcolor[HTML]{EFEFEF}{\color[HTML]{333333} 0.018004} &
  {\color[HTML]{333333} 0.018004} \\
\exceptions &
  0.017 &
  - &
  - &
  0.002 &
  - &
  - &
  {\color[HTML]{333333} -0.070} &
  {\color[HTML]{333333} -} &
  {\color[HTML]{333333} -} &
  {\color[HTML]{333333} 0.349} &
  {\color[HTML]{333333} -} &
  {\color[HTML]{333333} -} \\
\oop &
  0.049 &
  - &
  - &
  0.012 &
  - &
  - &
  {\color[HTML]{333333} 0.019} &
  {\color[HTML]{333333} -} &
  {\color[HTML]{333333} -} &
  {\color[HTML]{333333} 0.255} &
  {\color[HTML]{333333} -} &
  {\color[HTML]{333333} -} \\
\tests &
  - &
  - &
  - &
  - &
  - &
  - &
  {\color[HTML]{333333} -0.130} &
  {\color[HTML]{333333} -} &
  {\color[HTML]{333333} -} &
  {\color[HTML]{333333} 0.174} &
  {\color[HTML]{333333} -} &
  {\color[HTML]{333333} -} \\
\declarations &
  0.375 &
  - &
  - &
  0.034 &
  - &
  - &
  {\color[HTML]{333333} -0.257} &
  {\color[HTML]{333333} -} &
  {\color[HTML]{333333} -} &
  {\color[HTML]{333333} 0.405} &
  {\color[HTML]{333333} -} &
  {\color[HTML]{333333} -} \\
\conditionals &
  0.274 &
  - &
  - &
  -0.087 &
  - &
  - &
  {\color[HTML]{333333} -0.009} &
  {\color[HTML]{333333} -} &
  {\color[HTML]{333333} -} &
  \cellcolor[HTML]{EFEFEF}\textbf{0.682} &
  {\color[HTML]{333333} -} &
  {\color[HTML]{333333} -} \\
\loops &
  0.024 &
  - &
  - &
  0.111 &
  - &
  - &
  {\color[HTML]{333333} 0.042} &
  {\color[HTML]{333333} -} &
  {\color[HTML]{333333} -} &
  {\color[HTML]{333333} 0.275} &
  {\color[HTML]{333333} -} &
  {\color[HTML]{333333} -} \\
\operators &
  0.099 &
  - &
  - &
  0.062 &
  - &
  - &
  {\color[HTML]{333333} -0.032} &
  {\color[HTML]{333333} -} &
  {\color[HTML]{333333} -} &
  {\color[HTML]{333333} 0.389} &
  {\color[HTML]{333333} -} &
  {\color[HTML]{333333} -} \\
\datatype &
  0.037 &
  - &
  - &
  -0.069 &
  - &
  - &
  {\color[HTML]{333333} 0.002} &
  {\color[HTML]{333333} -} &
  {\color[HTML]{333333} -} &
  {\color[HTML]{333333} 0.275} &
  {\color[HTML]{333333} -} &
  {\color[HTML]{333333} -} \\
\extra &
  0.192 &
  -3.00E-05 &
  -3.00E-05 &
  -0.017 &
  5.40E-05 &
  5.40E-05 &
  {\color[HTML]{333333} 0.092} &
  \cellcolor[HTML]{EFEFEF}{\color[HTML]{333333} 0.014588} &
  {\color[HTML]{333333} 0.014588} &
  \cellcolor[HTML]{EFEFEF}\textbf{0.606} &
  {\color[HTML]{333333} 0.014525} &
  {\color[HTML]{333333} 0.014525} \\ \hline
\end{tabular}

\end{adjustbox}

\end{table}

%% file: tables/chap_4_tab_3_results_local_pearson_mlm.tex
\begin{table}[ht]
\centering
\caption{Masking AST Node Types. Distance Association Results $p(Y_d|T_{G})$ are \textit{Pearson Corr} and Causal Effects are ATEs $p(Y_d|do(T_{G}))$. (\textbf{bold}:strong corr., background:best effect)
}
\label{tab:resultsLocalPearsonMLM}

\begin{adjustbox}{width=1\textwidth}

\begin{tabular}{l|lllllllll}
\hline
 &
  \multicolumn{9}{c}{\textbf{\bert}} \\ \cline{2-10} 
\multirow{-2}{*}{\textbf{\begin{tabular}[c]{@{}l@{}}Counterfactual  \\ Interventions\end{tabular}}} &
  \multicolumn{3}{c}{\textit{\jaccard}} &
  \multicolumn{3}{c}{\levenshtein} &
  \multicolumn{3}{c}{\soren} \\ \hline
\textbf{\datainterIV} &
  \multicolumn{1}{c}{\textit{\begin{tabular}[c]{@{}c@{}}Association\\ \assoPR\end{tabular}}} &
  \multicolumn{1}{c}{\textit{\begin{tabular}[c]{@{}c@{}}Causal Eff.\\ ATE\end{tabular}}} &
  \multicolumn{1}{c}{\textit{R4}} &
  \multicolumn{1}{c}{\textit{\begin{tabular}[c]{@{}c@{}}Association\\ \assoPR\end{tabular}}} &
  \multicolumn{1}{c}{\textit{\begin{tabular}[c]{@{}c@{}}Causal Eff.\\ ATE\end{tabular}}} &
  \multicolumn{1}{c}{\textit{R4}} &
  \multicolumn{1}{c}{\textit{\begin{tabular}[c]{@{}c@{}}Association\\ \assoPR\end{tabular}}} &
  \multicolumn{1}{c}{\textit{\begin{tabular}[c]{@{}c@{}}Causal Eff.\\ ATE\end{tabular}}} &
  \multicolumn{1}{c}{\textit{R4}} \\ \hline
\textit{boolean\_operator} &
  \textbf{-0.3294} &
  \cellcolor[HTML]{EFEFEF}-0.1508 &
  0.0052 &
  -0.2845 &
  \cellcolor[HTML]{EFEFEF}-0.1363 &
  0.0147 &
  -0.2809 &
  -0.0947 &
  0.0009 \\
\textit{comparison\_operator} &
  -0.2648 &
  -0.0272 &
  -0.0066 &
  -0.2071 &
  -0.0183 &
  -0.0031 &
  -0.2228 &
  -0.0154 &
  -0.0019 \\
\textit{for\_in\_clause} &
  \textbf{-0.4232} &
  -0.0532 &
  -0.0004 &
  \textbf{-0.3724} &
  -0.0449 &
  0.0013 &
  \textbf{-0.3568} &
  -0.0291 &
  0.0002 \\
\textit{for\_statement} &
  \textbf{-0.4006} &
  -0.1011 &
  0.0341 &
  -0.2525 &
  -0.0411 &
  -0.0124 &
  \textbf{-0.3949} &
  -0.0832 &
  -0.0037 \\
\textit{identifier} &
  0.0024 &
  -0.0752 &
  0.0230 &
  -0.0375 &
  -0.0734 &
  -0.0017 &
  0.0243 &
  -0.0387 &
  0.0068 \\
\textit{if\_clause} &
  \textbf{-0.3904} &
  -0.0407 &
  0.0002 &
  \textbf{-0.3666} &
  -0.0365 &
  5.9143 &
  \textbf{-0.3395} &
  -0.0220 &
  0.0011 \\
\textit{if\_statement} &
  \textbf{-0.3667} &
  \cellcolor[HTML]{EFEFEF}-0.1211 &
  -0.0328 &
  -0.2409 &
  -0.0933 &
  -0.0130 &
  \textbf{-0.3624} &
  -0.0954 &
  -0.0131 \\
\textit{parameters} &
  \textbf{-0.3287} &
  -0.0481 &
  0.0071 &
  \textbf{-0.3230} &
  -0.0457 &
  -0.0104 &
  -0.2951 &
  -0.0294 &
  0.0032 \\
\textit{return\_statement} &
  -0.2450 &
  \cellcolor[HTML]{EFEFEF}-0.1211 &
  0.0062 &
  -0.2250 &
  \cellcolor[HTML]{EFEFEF}-0.1127 &
  -0.0158 &
  -0.2261 &
  -0.0756 &
  -0.0083 \\
\textit{string} &
  \textbf{-0.3651} &
  \cellcolor[HTML]{EFEFEF}-0.1683 &
  -0.0041 &
  -0.2961 &
  \cellcolor[HTML]{EFEFEF}-0.1450 &
  0.0274 &
  \textbf{-0.3199} &
  \cellcolor[HTML]{EFEFEF}-0.1161 &
  0.0036 \\
\textit{while\_statement} &
  -0.2677 &
  -0.0964 &
  -0.0197 &
  -0.1383 &
  -0.0091 &
  -0.0027 &
  -0.2979 &
  0.0113 &
  0.9199 \\ \hline
\end{tabular}

\end{adjustbox}

\end{table}

%% file: chapters/part_02_chap_03/benchmarking.tex
\chapter{Looking For Code Confounders, A Primer in Benchmarking}
\label{ch:benchmarking}

\epigraph{\scriptsize (...) causal questions can never be answered from data alone (...)}{\scriptsize\textit{Judea Pearl \citep{Pearl2018Causality}}}

\lettrine[lines=2]{\textbf{O}}{}ne of the most common solutions adopted by software researchers to address code generation is by training Large Language Models (\llms) on massive amounts of source code. \llms are rooted in the concept of emergent capabilities in which machines statistically learn complex patterns from code data. Although several studies have shown that \llms have been effectively evaluated on popular accuracy metrics (\eg BLEU, CodeBleu), previous research has largely overlooked the role of Causal Inference as a fundamental component of the interpretability of \llms' performance. Existing benchmarks and datasets are meant to highlight the difference between the expected and the generated outcome, but do not take into account confounding variables (\eg lines of code, number of tokens, prompt size) that equally influence the accuracy metrics. The fact remains that, when dealing with generative software tasks by \llms, no benchmark is available to tell researchers how to quantify neither the causal effect of SE-based treatments nor the correlation of confounders to the model's performance. In an effort to bring statistical rigor to the evaluation of \llms, this chapter introduces a benchmarking strategy named \galeras comprised of curated testbeds for three SE tasks (\ie code completion, code summarization, and commit generation) to help aid the interpretation of \llms' performance. 

We illustrate the insights of our benchmarking strategy by conducting a case study on the performance of \chatgpt under distinct prompt engineering methods. The results of the case study demonstrate the positive causal influence of prompt semantics on \chatgpt's generative performance by an \textit{average treatment effect} of $\approx 3\%$. Moreover, it was found that confounders such as prompt size are highly correlated with accuracy metrics ($\approx 0.412$). The result of our case study is to showcase causal inference evaluations, \textit{in practice}, to reduce \textit{confounding bias}. By reducing the bias, we offer an interpretable solution for the accuracy metric under analysis.      

\input{chapters/part_02_chap_03/sec_01_intro}
\input{chapters/part_02_chap_03/sec_02_pipeline}

\input{chapters/part_02_chap_03/sec_03_study}
\input{chapters/part_02_chap_03/sec_04_discussion}

%% file: chapters/part_02_chap_03/sec_01_intro.tex
\section{Introduction}
\label{ch-06:sec-01-wm}

Deep Learning for Software Engineering (\dlse) is an emerging research area in the field of software maintainability that entails a paradigm shift in the form by which machines statistically learn complex patterns from code data. To support actionable downstream SE tasks (\eg code completion, code summarization, or commit generation), ample evidence supports that \dlse approaches in the form of Language Models are able to generate code conditioned on a well-defined prompt \citep{watson2020dl4se, zhao_survey_2023, zan_large_2023}. While essential, \dlse approaches have been reduced to a group of large and self-supervised neural architectures (\ie Large Language Models or simply LLMs) comprised of multiple self-attention layers that perform linear transformations to extract salient features from programming and natural language data. In particular, Large Language Models for Code (\llmc) have led to a renewed interest in the automation of software engineering tasks. Most of this automation is a generative process in which underlying code and natural language features interact with each other to auto-complete\citep{austin2021program, hendrycksapps2021, chen_generation_2021,White:MSR15,Ciniselli.TSE,Ciniselli.MSR}, summarize\citep{Hussain2020DeepTL, leclair_ensemble_2021,Moran.SANER.2022}, review \citep{Mastropaolo2021StudyingTasks, Tufano:icse2021, Tufano.ICSE.2022}, trace \citep{moran_improving_2020} and translate code \citep{Nguyen:ICSE15}; generate test cases \citep{white_reassert_2020, Raychev2014CodeCW,Watson:ICSE20}, detect cone clones \citep{White2016,Tufano.MSR.2018} or fix bugs \citep{Tufano2019LearningBugFixes,zhou_devign_nodate,SANER.2019,Tufano:icse2019,Tufano2018,Chen2019,CanWeFix}. In fact, \llmc have been deployed in large-scale solutions to provide code generative services. Tools such as \chatgpt and GitHub Copilot, which are based on the \textit{gpt} architecture, exhibit good performance at the aforementioned tasks \citep{zhao_survey_2023}. 

Therefore, an increased interest has emerged in further evaluating these \llmc \citep{liu2023code,liu2023improving,xu_systematic_2022,Chen2021} to standardize the quality assessment of the generated code. Unfortunately, the current evaluation process overly relies on accuracy metrics leaving no consensus as to what other features or properties are impacting the code generation process. In other words, we are required to control for factors that influence the performance of \llmc if our goal is to \textit{interpret} models' output. Few studies have sought to examine accuracy metrics from a causal perspective to interpret \llmc \citep{palacio_toward_2023}. Ergo, the problem remains that, when attempting to understand the prediction performance of \llmc, no benchmarks are available to articulate causal queries. 

Previous research has largely overlooked the role of causal inference in evaluating \llmc. In fact, existing benchmarks are not without flaws to detect \textit{confounding bias}, which refers to the statistical ability to control for variables that can influence models' performance beyond the SE treatments under study (\ie evaluating the best prompting method). That is, we study causation because we need to understand not only \textit{what} but also \textit{why} \llmc arrive at performance decisions. To overcome these challenges, we pose a code-based benchmarking strategy, named \galeras, to interpret \llmc concentrated on answering causal queries of interest. \galeras enables SE researchers to explain \llmc performance decisions from a curated set of code-based confounders, which are associated with a given SE treatment under study. \galeras is comprised of three parts: 1) seven testbeds for evaluating distinct SE downstream tasks free of sampling bias and data snooping, 2) a set of confounders to compute causal effects, and 3) a pipeline to curate data from open repositories. 

To illustrate how to exploit \galeras to interpret \llmc, we conducted a causal study to quantify the impact of confounding variables on \chatgpt's prediction performance to assess whether certain types of \textit{prompt engineering} methods are excelling at automating code completion tasks. Prompt engineering is associated with the emergent ability of \llms to learn from prompts (\ie in-context learning). This ability comprises a set of techniques that manipulates the structure of a \llm's input sequence to attain better and less computationally expensive outputs than applying other downstream methods such as fine-tuning \citep{liu2023improving}. We organize our study around two RQs that are fundamentally centered on the problem of \textit{prompt engineering} for code:

\begin{enumerate}[label=\textbf{RQ$_{\arabic*}$}, ref=\textbf{RQ$_{\arabic*}$}, wide, labelindent=5pt]\setlength{\itemsep}{0.2em}
      \item \label{rq:exploratory} { \textbf{Exploratory Analysis:} \textit{How different is the distribution of tokens between the generated and ground-truth code?}} 
      \item \label{rq:causal} {\textbf{Causal Analysis:} \textit{To what extent the type of Prompt Engineering is influencing the code completion performance?}}
\end{enumerate}

The achieved results show that prompt engineering methods indeed causally impact the accuracy of the model by an \textit{Average Treatment of Effect} (ATE) of 3\% between the semantics of the prompt and the accuracy metric. Hence, choosing an adequate prompting strategy can positively influence the code completion performance of \chatgpt. To summarize, our key contributions are 1) A filtered testbed with non-contaminated code snippets for \llmc benchmarking; 2) a set of (confounding) features (\eg Cyclo Complexity, \# of AST levels)  included in the testbed; 3) a pipeline to generate new testbeds for a given SE task; and 4) a causal inference benchmarking to interpret \llmc.

%% file: chapters/part_02_chap_03/sec_02_pipeline.tex
\section{Testbed Curation}
\label{ch-06:sec-02-wm}

This section considers our proposed pipeline to structure and collect required testbeds for the comparative causal evaluation of \llmc. \galeras is a benchmarking strategy that entails a software architecture solution for the curation process.

\subsection{Structuring Testbed's Features}
\galeras testbeds are sets of Python methods that serve as evaluative data points. Each data point comprises four dimensions. The first dimension corresponds to snippets' identification,  which includes the \commitID (\ie commit hash), \textit{repository} name, \textit{path}, \textit{file\_name}, and \funName. The second dimension corresponds to snippets' documentation, which includes the \commitMessage and \docstring. The \docstring belongs to a JSON object that is extended to complementary natural language features such as \textit{n\_words, vocab\_size, language,} and \whitespaces. The third dimension corresponds to the snippet's syntactic information, which includes the actual code base, \ASTerrors, \ASTlevels, \ASTnodes,\textit{n\_words}, \textit{vocab\_size}, \tokenCount, and \whitespaces. Finally, the fourth dimension corresponds to canonical software metrics, which include \nloc, \complexity, and \identifiers. 

\begin{figure}[ht]
		\centering
	\includegraphics[width=\textwidth]{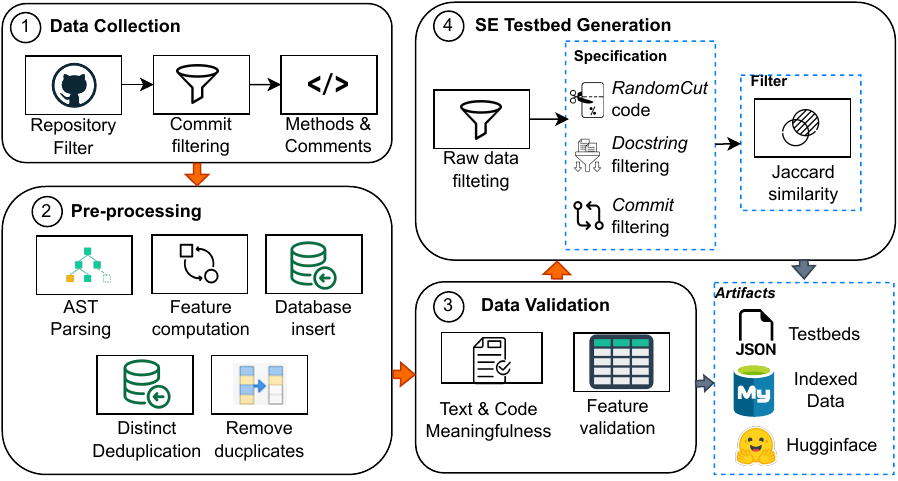}
		\caption{Testbed Curation Pipeline of \galeras}
        \vspace{-0.5cm}
        \label{fig:collection}
\end{figure}

\subsection{Collecting Code Samples}

\figref{fig:collection} describes a 4-step pipeline that \galeras uses to collect code samples. In the first step (Fig.~\ref{fig:collection}-\circled{1}), we filtered the most popular Python Github repositories using the following query: $language:Python$, $fork:false$, $size:>=30,000$,  $pushed:>2021-12-31$, $stars:>1,000$. From the last paper report of \chatgpt \citep{openai2023gpt4}, we assumed \chatgpt and other \llmc under analysis were not trained on commits from Jan 2, 2022 to Jan 1, 2023. Therefore, we claim that our testbeds help to avoid \textit{data snooping}, which is the misuse of data points to evaluate statistical hypotheses using training samples. Then, we collected a set of brand-new methods for each commit. This step resulted in $\approx338k$ data points. For each data point, we also collected its corresponding documentation without considering inline comments.

In the second step (\figref{fig:collection}-\circled{2}), we engineered and preprocessed both code and documentation related features from collected data points. Then we parsed the AST variables for our data points by employing the Tree-Sitter library. To guarantee efficient data management and once the previous features were engineered and extracted, we stored raw and preprocessed data points in a relational database. Next, we removed duplicated samples using a distinct query, reducing the testbed size to $\approx 227K$ data points for code (\textit{RawData} in tab.~\ref{tab:statistical-description}). Of these reduced data points, $\approx 77K$ contains a valid \docstring (\textit{RawDataDocstring} in tab.~\ref{tab:statistical-description}). A \docstring is valid when its text is larger than 3 words.

\input{tables/chap_6_tab1_statistical_analyisis}

In the third step \ref{fig:collection}-\circled{3}), we manually validated $960$ out of $\approx 227K$ data points. These validated data points were randomly selected from \textit{RawData} and \textit{RawDataDocstring}.  The remaining data points were automatically validated. Our validation process ensures the date of each pushed commit is within the range of dates stated in the original query. We also validated that the methods attached to each commit were indeed updated within the same range of dates. In addition, we validated the meaningfulness of the \docstring and \commitMessage by inspecting the consistency of the natural language descriptions with the actual code implementation, removing $\approx1.9\%$ \textit{RawDataDocstring} obtaining $\approx57K$ datapoints (tab. \ref{tab:statistical-description}). Lastly, \complexity was validated using the Codalyze plugin in Visual Studio Code. For the sake of simplicity, we omit explaining all considered fine-grained validation steps in this chapter. However, the reader can consult our online appendix for more information\citep{daniel23}.

In the final step \figref{fig:collection}-\circled{4}), we sampled $3k$ data points from \textit{RawData} testbed to build five additional testbeds, each one for a specific SE task. \galeras comprises  \randomCut,  \withDocstring and \fromDocstring for \textit{code completion}; \commitGen for \textit{code generation}; and \summarizationGen for \textit{code summarization}. These additional testbeds are described in Tab.~\ref{tab:statistical-description}. To build \randomCut, we chose data points with more than $10$ tokens or $100$ characters. Next, the data point is randomly cut after the method signature. To build \summarizationGen and \commitGen, we filtered the \textit{RawDataDocstring} data points with more than 10 words or 50 characters. After building the five testbeds, we removed duplicated snippets using the Jaccard similarity on preprocessed data points with BPE HuggingFace tokenizer.  Because the de-duplication between training and test sets was discarded (\ie no multiset threshold), we set $0.7$ as the similarity threshold for our testbeds \citep{Allamanis19,wang_neural_2019}. Table.~\ref{tab:dedupe} shows the SE Task associated with each curated testbed, the percentage rate of detected duplicates, and the final size.

\input{tables/chap_6_tab2_dedupe}

%% file: tables/chap_6_tab1_statistical_analyisis.tex
\begin{table*}[ht]
\centering
\caption{Descriptive Analysis $[avg \pm std]$ of \galeras' Testbeds and Code Features.}
\label{tab:statistical-description}

\setlength{\tabcolsep}{7pt} 
\begin{adjustbox}{width=\textwidth}
\begin{tabular}{lclccclcccccc}
\hline
 &
  \multicolumn{1}{l}{} &
   &
  \multicolumn{3}{c}{\textbf{Confounders*}} &
   &
  \multicolumn{6}{c}{\textbf{Effect modifiers}} \\ \hline
\multicolumn{1}{c}{\textbf{Testbed}} &
  \textbf{Dedup} &
   &
  \textbf{n\_whitespaces} &
  \textbf{nloc} &
  \textbf{token\_counts} &
   &
  \textbf{n\_ast\_errors} &
  \textbf{ast\_levels} &
  \textbf{n\_ast\_nodes} &
  \textbf{complexity} &
  \textbf{token\_counts} &
  \textbf{n\_identifiers} \\ \hline
\textit{RawData} &
 277226 &
   &
  $259.23\pm902.22$ &
  $21.16\pm47.46$ &
  $137.38\pm262.59$ &
   &
  $0.09\pm0.42$ &
  $11.85\pm3.5$ &
  $221.91\pm438.23$ &
  $3.25\pm6.98$ &
  $137.38\pm262.59$ &
  $17.94\pm16.45$ \\
\textit{RawDataDocstring} &
  57045 &
   &
  $206.98\pm453.06$ &
  $18.89\pm30.98$ &
  $112.50\pm183.78$ &
   &
  $0.10\pm0.73$ &
  $11.57\pm3.45$ &
  $184.53\pm436.76$ &
  $3.42\pm6.61$ &
  $112.50\pm183.78$ &
  $15.96\pm14.48$ \\ \hline
\textit{RandomCut} &
  2931 &
   &
  $229.24\pm479.38$ &
  $18.27\pm26.98$ &
  $126.54\pm177.19$ &
   &
  $0.10\pm0.30$ &
  $12.25\pm3.06$ &
  $207.55\pm259.46$ &
  $3.16\pm6.09$ &
  $126.54\pm177.19$ &
  $17.70\pm13.42$ \\
\textit{WithDocstring} &
  2926 &
   &
  $208.48\pm414.67$ &
  $18.08\pm20.65$ &
  $111.98\pm122.27$ &
   &
  $0.08\pm0.58$ &
  $12.22\pm3.10$ &
  $188.99\pm400.74$ &
  $3.78\pm4.33$ &
  $111.98\pm122.27$ &
  $16.61\pm11.50$ \\ \hline
\textit{FromDocsting} &
  2937 &
   &
  $167.96\pm244.56$ &
  $16.68\pm20.91$ &
  $100.13\pm118.36$ &
   &
  $0.10\pm0.59$ &
  $11.38\pm3.44$ &
  $156.39\pm180.71$ &
  $3.48\pm4.48$ &
  $100.13\pm118.36$ &
  $14.62\pm11.94$ \\
\textit{CommitGen} &
  2919 &
   &
  $179.62\pm363.64$ &
  $16.75\pm21.37$ &
  $101.66\pm128.65$ &
   &
  $0.09\pm0.36$ &
  $11.51\pm3.36$ &
  $160.30\pm201.11$ &
  $3.28\pm4.93$ &
  $101.66\pm128.65$ &
  $15.07\pm11.90$ \\
\textit{SummarizationGen} &
  2924 &
   &
  $212.08\pm415.90$ &
  $18.66\pm21.63$ &
  $114.38\pm128.94$ &
   &
  $0.07\pm0.32$ &
  $12.22\pm3.16$ &
  $197.05\pm532.69$ &
  $3.85\pm5.36$ &
  $114.38\pm128.94$ &
  $16.71\pm12.16$ \\ \hline
\end{tabular}

\vspace{2cm}

\end{adjustbox}

{ \small *The confounder \promptSize was omitted due to its treatment dependency.}
\vspace{0.5cm}
\end{table*}

%% file: tables/chap_6_tab2_dedupe.tex
\begin{table}[ht]
\centering
\caption{Jaccard Similarity De-Duplication}
\label{tab:dedupe}

\begin{adjustbox}{width=0.9\textwidth}

\setlength{\tabcolsep}{7pt} 

\begin{tabular}{llcccc}
\hline
\multicolumn{1}{c}{\textbf{SE Task}} & \multicolumn{1}{c}{\textbf{Testbed}} & \multicolumn{1}{c}{\textbf{I/O}} & \textbf{Dupes} & \textbf{Dupe \%} & \textbf{size} \\ \hline
 & \randomCut & code $\Rightarrow$ code & 69 & 2.30\% & 2931 \\
 & \withDocstring & code-text $\Rightarrow$ code & 74 & 2.47\% & 2926 \\
\multirow{-3}{*}{\textbf{Code completion}} & \fromDocstring & text $\Rightarrow$ code & 63 & 2.10\% & 2937 \\
\textbf{Code generation} & \commitGen & code $\Rightarrow$ text & 81 & 2.70\% & 2919 \\
\textbf{Summarization} & \summarizationGen & code $\Rightarrow$ text & 76 & 2.53\% & 2924 \\ \hline
\end{tabular}

\end{adjustbox}

\end{table}

%% file: chapters/part_02_chap_03/sec_03_study.tex
\section{Causal Study: Interpretable Code Completion}
\label{ch-06:sec-03-wm}

To demonstrate how to employ \galeras for causal analysis, in practice, we design a study in which we evaluate \chatgpt's performance for two prompt engineering methods $T_1$ and $T_2$ based on Liu \etal \citep{liu2023improving}. Prompt engineering is the activity of optimizing the input space of a given \llm in order to generate better outcomes without giving rise to expensive fine-tuning. The goal of our case study is to compare these two prompting methods after controlling for confounding features.  

\subsection{Evaluation Methodology}
The evaluation methodology of the case study is divided into three parts. The first part addresses the exploratory analysis of \galeras testbeds. We employed the BPE tokenizer to normalize the vocabulary of each treatment $T$ and outcome $Y$ sentence. The token count categorized by taxonomy is presented in Fig.\ref{fig:descriptive-analysis}. Tokens within each sentence were classified based on their taxonomy, \ie  \textit{`try'} and \textit{`catch'} are classified as \textit{exceptions} and  \textit{`if'} and \textit{`else'} as conditionals. Since the analysis focused solely on Python, keywords related to data types were classified as \textit{casting} tokens.

\begin{marginfigure}
		\centering
		\includegraphics[width=\linewidth]{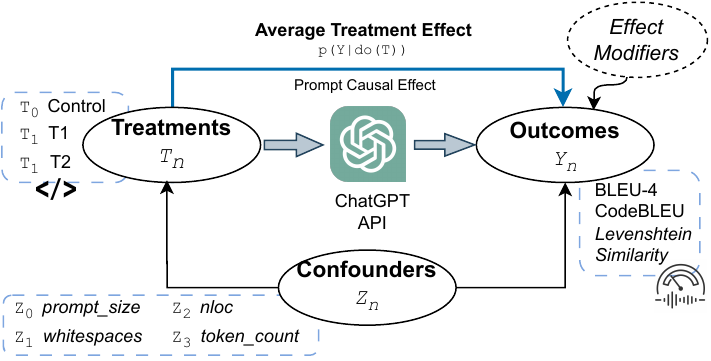}
		\caption{\galeras Structural Causal Model Benchmarking}
        \label{fig:case-study}
\end{marginfigure}

The second part canonically evaluates \chatgpt using our testbed \withDocstring. CodeBLEU was computed with a default parameter value of $0.25$. In addition, BLUE was computed with a 4-gram parameter. On the other hand, we computed the Levenshtein distance and similarity for a local evaluation (see Tab~.\ref{tab:correlation}-Performance Metrics).

The third part estimates the causal effect of prompt engineering methods and \chatgpt performance. Figure \ref{fig:case-study} illustrates our Structural Causal Models for the prompt engineering case of \chatgpt. We use \galeras to compare the performance of two different treatments. The first treatment \Ta is one prompt, which contains a command (\eg \textit{Complete the following a Python code, return only code and complete method: `\{partial code\}'} ) followed by the actual input code to be completed. The second treatment \Tb comprises two prompts. The first one is a context prompt that entails both the \docstring and the incomplete cut code. The second one is a \textit{processing prompt} that contains sentences asking for removing comments and optimizing code (\eg Remember you have a Python function named `\{ \funName\}', the function starts with the following code `\{code\}'. The description for the function is: `\{ \docstring\}' ). We used the previous treatments against a \control group. The \control is a \textit{task prompt} that encompasses an action word or verb followed by the incomplete code input (\eg Complete the following python method: `\{partial code\}'). To evaluate whether treatments $T$ are impacting \chatgpt performance $Y$, we controlled for confounding features $Z$. Our confounders \promptSize, \whitespaces, \tokenCount, and \nloc were selected due to their high correlation ($[0.4-0.8]$) with the Levenstein distance in control and treatment groups. Although \ASTnodes has a high correlation with the Levenstein distance, we assumed that structural features are ignoring the treatments. Hence, AST-based features are effect modifiers. The potential outcomes $Y_2$,$Y_1$,$Y_0$ are observed under the treatments \Ta,\Tb,\control. Next, we approximate the \textit{Average Treatment Effect} $p(Y|do(T)$ using the SCM defined in \figref{fig:case-study}. 

\subsection{Results}

\textit{Exploratory Analysis.}  The purpose of the exploratory analysis is to expose and understand the testbeds' feature distribution grouped by prompt engineering methods $T$. Table \ref{tab:statistical-description} depicts the average and standard deviation for each code feature. We observed high variability in \whitespaces ($902.22$) and \tokenCount ($262.59$), which implies the method sizes are not homogeneous across the testbeds. While the descriptive analysis showcases high variability for all code features, our testbeds are a representative sub-sample of open repositories. For instance, the \complexity feature has an average value of $3.25$ suggesting that the code has a reasonable number of loops, conditionals, and operators. Therefore, our collected methods exhibit that our pipeline process guarantees data point diversity.   

\begin{figure}[ht]
		\centering		\includegraphics[width=0.95\textwidth]{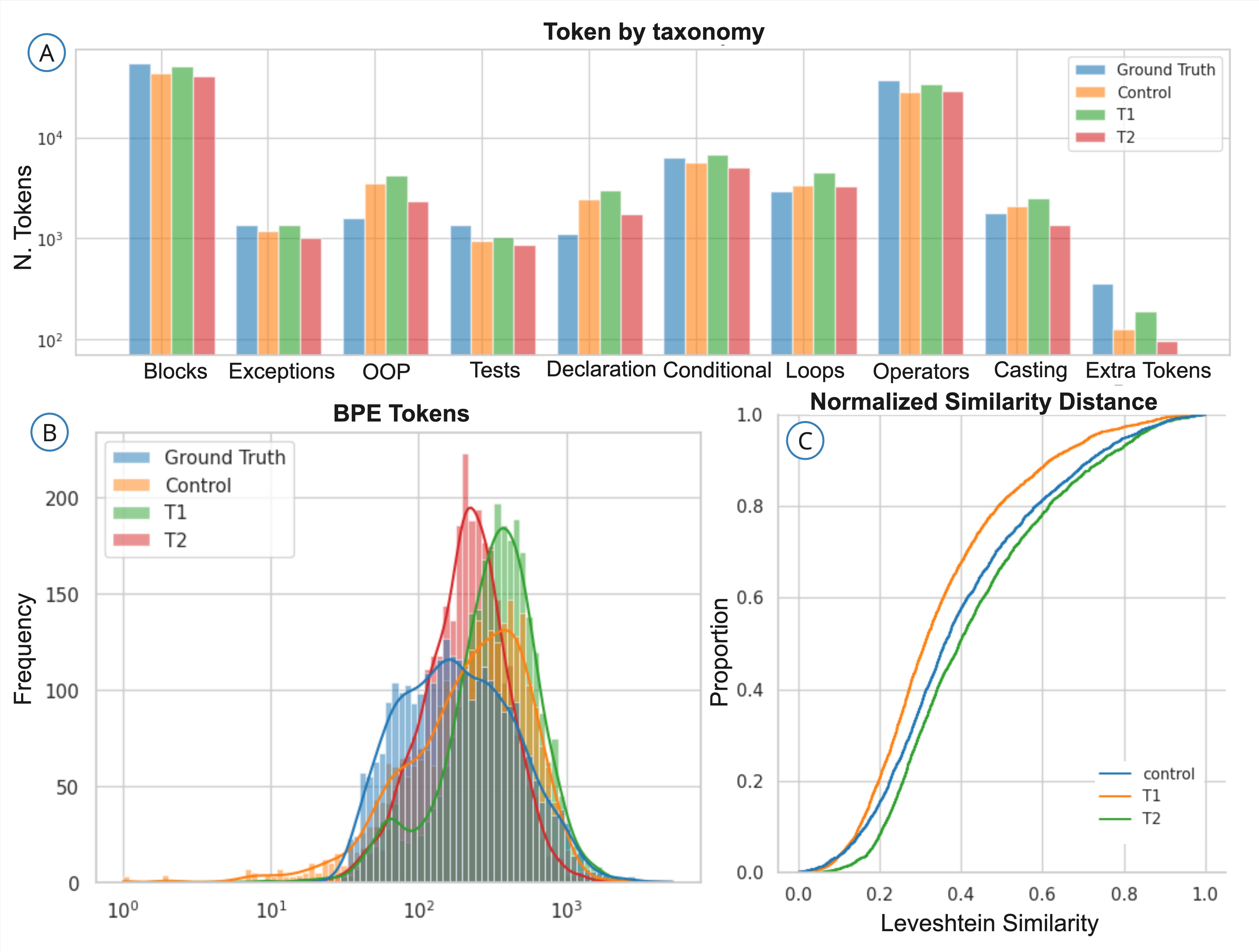}
		\caption{Descriptive Analysis: Top graph the token count for each testbed, bottom left the token frequency distribution, bottom right the similarity proportion score.}
        \label{fig:descriptive-analysis}
\end{figure}

We observed no significant differences in the counting of tokens among potential outcomes (including the \control) and the ground truth (see Fig.~\ref{fig:descriptive-analysis}-A). For instance, \control and \Tb on declarations (with a diff. around $550$ tokens) and loops (with a diff. around $600$ tokens) are relatively small. However, \Ta outcome exhibited high difference and excessive use of OOP, declarations, and loops with a diff. around $2.6k$, $2k$, and $1.5k$  tokens respectively. Figure \ref{fig:descriptive-analysis}-B showcases the token distribution for each testbed. We detected that the two prompt engineering methods were generating a similar amount of tokens (\ie green and red distributions) compared to the \control and ground truth. This suggests that sophisticated prompts tend to generate repetitive tokens. Figure \ref{fig:descriptive-analysis}-C depicts the Levenshtein similarity distance between the \chatgpt outputs, generated with both prompt engineering methods and the \control, and the ground truth. We can observe from the proportion curve that \Ta similarity performs the worst compared to the \control and \Tb.

\begin{boxK}
\ref{rq:exploratory} Exploratory Analysis: Grouped by taxonomy the ground truth does not repeat the same tokens as much as the treatments. The \Ta outcome seems to have notable intense use of keywords for OOP, declarations, and loops; \Tb obtains better performance with the highest similarity average of $0.43$
\end{boxK}

\ref{rq:causal} \textit{Causal Analysis.} 
For two basic prompt engineering methods, code completion performance of \chatgpt is mainly affected by the following confounders: number of white spaces, lines of code, tokens in the outcome, and tokens in the prompt with a maximum correlation of $0.80$ with the Levenstein distance (see Tab.~\ref{tab:correlation}-Correlations). This suggests that after controlling for confounders, the \textit{Average Treatment Effect} (ATE) the prompt engineering method$_1$, represented by $T_1$, has a negative causal effect $p_1(Y|do(T)) = E[Y_1 - Y_0] \approx -5.1\%$ compared to a positive causal effect $p_2(Y|do(T)) = E[Y_2 - Y_0] \approx 3.3\%$ of method$_2$, represented by $T_2$ (see Tab.~\ref{tab:correlation}-Causal Effects). This indicates that method$_1$ is negatively affecting the Levenshtein similarity (\ie poor performance) across \withDocstring testbed, while method$_2$ is actually enhancing \chatgpt prediction performance. These results are consistent with the previous section in which we demonstrated that \Tb performs better than \Ta. After controlling for the confounding effect of the code features such as the prompt size and token counts, we can claim that the reason why \Tb is performing better than \Ta is \textit{purely} due to the information contained in the prompt. 

To validate the robustness of computed ATEs and proposed SCM, we refuted our effects using the following methods: \textit{Placebo, Random Common Cause (RCC)} and \textit{Subet} (see DoWhy refuters in \citep{Sharma2021DoWhyAssumptions}). We found that, for the ATEs computed with score matching, their corresponding refutation values are not stable. That is, the placebo value for $Y_1$ similarity is far from zero with $2.98$, while the RCC value differs by around $212$ in $Y_2$ distance. 

\input{tables/chap_6_tab3_correlation_and_causal_effect}

\begin{boxK}
\ref{rq:causal} Causal Analysis: The prompt engineering method$_1$ (treatment \Ta) has a negative causal impact on the \chatgpt performance with an ATE estimation of $-5\%$. Conversely, the prompt engineering method$_2$ (treatment \Tb) has a subtle positive influence on the same performance with an ATE of $3\%$. This suggests that after controlling for prompt size, white spaces, \# of tokens, and nlocs; prompt engineering strategies are indeed affecting the quality of code completion.
\end{boxK}

In summary, this study used a qualitative technique to analyze the causal effect of SE-oriented treatments on the performance of \llmc. Such a technique is embedded into a benchmarking strategy named \galeras. Our benchmarking enables researchers to interpret \textit{why} a given \llmc is reporting a particular accuracy metric. We curated two raw Python testbeds: \textit{RawData} with only mined code and \textit{RawDataDocstring} with the corresponding documentation from GitHub. We also provide five SE Python testbeds for three SE tasks (\ie code completion, code summarization, and commit generation), we proposed a pipeline for collecting testbeds from git repositories. Finally, we conducted a rigorous evaluation of code completion with \chatgpt. Our causal study suggests that \chatgpt's performance is not only affected by the prompt size but also by the prompt semantics.  Future research will focus on determining whether other unmeasured confounders are affecting \llmc's prediction by augmenting the number of testbeds.

%% file: tables/chap_6_tab3_correlation_and_causal_effect.tex
\begin{table}[ht]
\centering
\caption{Code Completion Testbed Results: 
Performance Metrics, Correlations, and Causal Effects.}
\label{tab:correlation}

\begin{adjustbox}{width=\textwidth}

\setlength{\tabcolsep}{5.5pt} 

\begin{tabular}{llcccccc}
\hline
\multicolumn{2}{l}{\textit{\textbf{Treatments}}} & \multicolumn{2}{c}{\textbf{Control}} & \multicolumn{2}{c}{\textbf{T1}} & \multicolumn{2}{c}{\textbf{T2}} \\ \hline
\multicolumn{8}{l}{\textbf{Performance Metrics}} \\ \hline
 \multirow{1}{*}{\textit{\textbf{Distance}}}& Bleu & \multicolumn{2}{c}{0.444} & \multicolumn{2}{c}{0.45} & \multicolumn{2}{c}{0.42} \\
 & CodeBleu & \multicolumn{2}{c}{0.441} & \multicolumn{2}{c}{0.438} & \multicolumn{2}{c}{0.469} \\
 \textit{\textbf{Similarity}}& Avg. Lev. & \multicolumn{2}{c}{0.40$\pm$0.20} & \multicolumn{2}{c}{0.35$\pm$0.18} & \multicolumn{2}{c}{0.43$\pm$0.20} \\ \hline
\multicolumn{2}{l}{\textbf{Correlations (vs Levenshtein)}} & \textbf{Dist.} & \textbf{Sim.\%} & \textbf{Dist.} & \textbf{Sim.\%} & \textbf{Dist.} & \textbf{Sim.\%} \\ \hline
\multirow{1}{*}{\textit{\textbf{Confounders}}} & \promptSize & 0.45 & 25.6\% & 0.40 & 41.2\% & 0.45 & 28.3\% \\
 & \whitespaces & 0.69 & 5.6\% & 0.62 & 20.7\% & \textbf{0.80} & \cellcolor[HTML]{FFFFFF}1.8\% \\
 & \tokenCount & 0.67 & 5.3\% & 0.59 & 24.8\% & 0.70 & 3.9\% \\
 & \nloc & 0.64 & 4.2\% & 0.57 & 20.7\% & 0.70 & 0.1\% \\
\multirow{1}{*}{\textit{\textbf{Effect Modifiers}}}  & \complexity & 0.43 & 4.3\% & 0.40 & 16.8\% & 0.47 & 0.9\% \\
 & \ASTnodes & 0.72 & 7.8\% & 0.62 & 29.4\% & 0.77 & 4.3\% \\
 & \ASTerrors & 0.02 & -2.4\% & 0.05 & 3.7\% & 0.18 & 2.3\% \\
 & \ASTlevels & 0.40 & 9.9\% & 0.31 & 30.4\% & 0.44 & 8.1\% \\ \hline
\multicolumn{8}{l}{\textbf{Causal Effects $(T\rightarrow Y)$}} \\ \hline
 \textit{\textbf{Score Matching}} & ATE & - & - & 104.02 & -3.7\% & -314.36 & 6.9\% \\
 & Placebo & - & - & -0.21 &   \textbf{298\%} & 0.02 & 0.1\% \\
 & RCC & - & - & 112.14 & -5.2\% & -102.71 & 3.3\% \\
 & Subset & - & - & 110.85 & -5.1\% & -101.6 & 3.3\% \\
\textit{\textbf{Stratification}} & ATE & - & - & 111.05 & -5.1\% & -101.73 & 3.3\% \\
 & Placebo & - & - & -0.17 & 0.04\% & 0.01 & {\ul0.05\%} \\
 & RCC & - & - & 111.17 & -5.1\% & -101.7 & 3.3\% \\
 & Subset & - & - & 110.95 & -5.2\% & -101.49 & 3.3\% \\
 \textit{\textbf{IPW}} & ATE & - & - & 111.05 & -5.1\% & -101.73 & 3.3\% \\
 & Placebo & - & - & -0.54 & -0.02\% & -1.30 & {\ul-0.07\%} \\
 & RCC & - & - & 111.04 & -5.1\% & -101.74 & 3.3\% \\
 & Subset & - & - & 111.12 & -5.1\% & -101.47 & 3.3\% \\ \hline
\end{tabular}
\vspace{4cm}

\end{adjustbox}
{
 bold: highest correlation, underline: null effect. 
}
\end{table}

%% file: chapters/part_02_chap_03/sec_04_discussion.tex
\section{Discussion}
\label{ch-06:sec-04-wm}

Considerable research attention has been devoted to data collection and benchmarking for \llmc. \tabref{tab:benchmark-comparison} showcases eight qualitative properties that we use to compare three state-of-art benchmarks (\ie \textit{CodeXGLUE}, \textit{IdBench}, and \textit{MultiPL-E}) with \galeras. Firstly, Husain \etal introduced \textit{CodeSearchNet} for code retrieval automation \citep{husain2019codesearchnet}. Their datasets have been mostly employed to pre-train \llms rather than benchmarking software tasks. Later, researchers at Microsoft extended \textit{CodeSearchNet} and amalgamated 12 SE-related datasets for other relevant downstream tasks (\eg clone detection, refinement, translation) \citep{lu2021codexglue}. These datasets and benchmarks are known as \textit{CodeXGLUE}, which partially support some accuracy and distance metrics. Secondly, Wainakh \etal proposed \textit{IdBench} to evaluate generated identifiers by measuring similarity distances of semantic representations \citep{wainakh_evaluating_2019}. Finally, Chen \etal notably posed \textit{HumanEval} to validate the functional correctness of generated code \citep{Chen2021}. Cassano \etal  amplified \textit{HumanEval} to create \textit{MultiPL-E} for code translation \citep{cassano_multipl-e_2022}. Although these three benchmarks have been successfully employed for evaluating \llmc, these benchmarking strategies were not conceived to address the \textit{interpretation} of models' outputs. 

\input{tables/chap_6_tab4_benchmarks}

As \llmc are quickly evolving due to data and hyperparameter augmentation, current models (\eg \chatgpt, AlfaCode, Copilot) could have been trained on samples already used for evaluation (\aka data snooping) and datasets such as \textit{BigQuery} \citep{GoogleBigQuery}, \textit{BigPython}\citep{Nijkamp2022CodeGenAO}, and the \textit{Pile} \citep{gao2020pile} have omitted the importance of interpreting \llmc' performance. \galeras, however, offers curated testbeds for enabling prompt engineering evaluation. This evaluation includes an interpretability analysis based on \textit{causal inference} in the form of Structural Causal Models (SCM). What is more, \galeras provides a pipeline to collect and access confounders and treatment data. Such data is plugged into the SCM to estimate the causal effects between treatments and outcomes. Estimating these casual effects promotes statistical rigor in evaluating SE-based generative tasks. 

In summary, this study used a qualitative technique to analyze the causal effect of SE-oriented treatments on the performance of \llmc. Such a technique is embedded into a benchmarking strategy named \galeras. Our benchmarking enables researchers to interpret \textit{why} a given \llmc is reporting a particular accuracy metric. We curated two raw Python testbeds: \textit{RawData} with only mined code and \textit{RawDataDocstring} with the corresponding documentation from GitHub. We also provide five SE Python testbeds for three SE tasks (\ie code completion, code summarization, and commit generation), we proposed a pipeline for collecting testbeds from git repositories. Finally, we conducted a rigorous evaluation of code completion with \chatgpt. Our causal study suggests that \chatgpt's performance is not only affected by the prompt size but also by the prompt semantics.  Future research will focus on determining whether other unmeasured confounders are affecting \llmc's prediction by augmenting the number of testbeds.

%% file: tables/chap_6_tab4_benchmarks.tex
\begin{table}[ht]
\centering
\caption{SOTA Benchmark qualitative properties comparison.}
\label{tab:benchmark-comparison}

\begin{adjustbox}{width=\textwidth}
\setlength{\tabcolsep}{4pt} 

\begin{tabular}{ccccccc}
\hline
\multicolumn{2}{c}{\textbf{}} & \multicolumn{1}{l}{} & \multicolumn{4}{c}{\textit{\textbf{Benchmarks}}} \\ \cline{4-7} 
\multicolumn{2}{c}{\textbf{Qualitative Properties}} & \multicolumn{1}{l}{} & \textbf{CodeXGLUE} & \textbf{IdBench} & \textbf{MultiPL-E} & \textbf{Galeras} \\ \hline
 & Clone detection &  & \checkmark & \xmark & \xmark & \xmark \\
 & Defect detection &  & \checkmark & \xmark & \xmark & \xmark \\
 & Type Inferring &  & \xmark & \checkmark & \xmark & \xmark \\
 & Summarization &  & \xmark & \xmark & \xmark & \checkmark \\
 & Code generation &  & \xmark & \xmark & \xmark & \checkmark \\
 & Commit generation &  & \xmark & \xmark & \xmark & \checkmark \\
 & Repair &  & \checkmark & \xmark & \xmark & \xmark \\
 & Translation &  & \checkmark & \xmark & \checkmark & \xmark \\
\multirow{-9}{*}{\textit{\begin{tabular}[c]{@{}c@{}}Software\\ Tasks\end{tabular}}} & Search &  & \checkmark & \xmark & \xmark & \xmark \\ \hline
 & code-code &  & \checkmark & \xmark & \xmark & \checkmark \\
 & code-text &  & \checkmark & \checkmark & \xmark & \checkmark \\
\multirow{-3}{*}{\textit{I/O}} & text-code &  & \checkmark & \xmark & \checkmark & \xmark \\ \hline
 & Identifiers &  & \xmark & \checkmark & \xmark & \xmark \\
 & Code line &  & \checkmark & \xmark & \xmark & \xmark \\
 & Method &  & \checkmark & \xmark & \checkmark & \checkmark \\
\multirow{-4}{*}{\textit{\begin{tabular}[c]{@{}c@{}}Output\\ Granularity\end{tabular}}} & Files &  & \checkmark & \xmark & \xmark & \xmark \\ \hline
 & Words &  & \xmark & \checkmark & \xmark & \xmark \\
 & Tokens &  & \checkmark & \xmark & \xmark & \checkmark \\
 & Snippets &  & \checkmark & \xmark & \checkmark & \checkmark \\
\multirow{-4}{*}{\textit{\begin{tabular}[c]{@{}c@{}}Type of\\ Datum\end{tabular}}} & Prompts &  & \xmark & \xmark & \xmark & \cellcolor[HTML]{EFEFEF}\checkmark \\ \hline
 & Size &  & 416K & 500 answers & 164 problems & 227K \\
\multirow{-2}{*}{\textit{Dimension}} & Languages &  & $\approx 12$ & 3 & 19 & 1 \\ \hline
 & BLEU &  & \checkmark & \checkmark & \xmark & \checkmark \\
 & CodeBLEU &  & \checkmark & \xmark & \xmark & \checkmark \\
 & Cloze testing &  & \checkmark & \xmark & \xmark & \xmark \\
 & Levenshtein &  & \xmark & \checkmark & \xmark & \checkmark \\
 & Accuracy &  & \checkmark & \xmark & \xmark & \xmark \\
\multirow{-6}{*}{\textit{\begin{tabular}[c]{@{}c@{}}Supported\\ Metrics\end{tabular}}} & Causal Effect &  & \xmark & \xmark & \xmark & \cellcolor[HTML]{EFEFEF}\checkmark \\ \hline
\textit{Prompt}\citep{liu2023improving} & Single-step &  & \xmark & \xmark & \checkmark & \checkmark \\
\textit{Engineering} & Multiple-step &  & \xmark & \xmark & \xmark & \cellcolor[HTML]{EFEFEF}\checkmark \\ \hline
 & Confounders &  & \xmark & \xmark & \xmark & \cellcolor[HTML]{EFEFEF}\checkmark tab.\ref{tab:statistical-description}\\
\multirow{-2}{*}{\textit{\begin{tabular}[c]{@{}c@{}}Causal\\ Evaluation\end{tabular}}} & Inference &  & \xmark & \xmark & \xmark & \cellcolor[HTML]{EFEFEF}\checkmark \\ \hline
\end{tabular}
\vspace{-3.90cm}

\end{adjustbox}
{
Shadowed cells indicate \galeras only.
}
\end{table}

%% file: chapters/part_03_chap_01/autopoietic.tex
\chapter{On Enabling An Artificial Self-Construction Software Life-cycle via Autopoietic Architectures}
\label{ch:autopoietic}

Software engineering research has focused on automating maintenance and evolution processes to reduce cost and improve reliability. However, despite advancements in areas such as automated testing, program repair, and deployment, these efforts remain dependent on external interventions and lack intrinsic self-adaptive capabilities. Drawing inspiration from Artificial Life (\textit{ALife}), we propose a fundamental shift in the Software Development Life-Cycle (\textit{SDLC}) by introducing self-construction mechanisms that enable software to evolve and maintain autonomously. This chapter explores the potential of Autopoietic Architectures, specifically $\Psi$-\textit{Arch}, as a foundational framework for self-constructing software. We first analyze the limitations of traditional maintenance approaches and identify gaps in current SDLC automation. Subsequently, we outline the core challenges in achieving self-construction, including the development of intelligent reasoning units and the establishment of novel architectural paradigms.  To realize this vision, we focused on integrating self-construction capabilities into modern software engineering practices and redefining the role of developers as Artificial Software Engineers who design and oversee maintenance software products. Although this chapter does not present an ultimate solution, it seeks to catalyze discourse and inspire research toward a new paradigm in software engineering --one where self-constructed software represents the next frontier in SDLC automation.

\input{chapters/part_03_chap_01/sec_01_intro}
\input{chapters/part_03_chap_01/sec_02_background}
\input{chapters/part_03_chap_01/sec_03_position}
\input{chapters/part_03_chap_01/sec_04_discussion}
\input{chapters/part_03_chap_01/sec_05_conclusion}

%% file: chapters/part_03_chap_01/sec_01_intro.tex
\section{Introduction}
\label{ch-autopoietic:sec-01-wm}

Software Development Life-Cycle (\sdlc) outlines five phases to produce a software artifact \ie designing, development, testing, deployment, and maintenance \citep{meza2024,Lehman1980ProgramsEvolution}. Peculiarly, the software maintenance phase embodies the most critical and costly in the \sdlc\citep{10.1007/978-981-16-0739-4_28,895984}. Software maintenance entails \textit{any modifications done to a system after deployment}. Those modifications, in fact, involve practitioners' knowledge\citep{meza2024} to design, construct, and improve the source code. Recent software maintenance (\sm) research has focused on automating repetitive tasks using Machine Learning~ (ML).

Automating software maintenance ensures that software systems remain efficient, reliable, and relevant over time \citep{9568959,meza2024}. Several studies have investigated software maintenance automation on downstream tasks such as testing \citep{gruber2023automatictestgenerationtools,10765036}, and program repair \citep{10172693,10.1145/3597503.3623310}. Moreover, code generation has been studied in the field of software maintenance~\citep{Dit2013b,Tufano2018,Wang2017Semantics-awareCode,BinghamBrown2017TheMutants,White2016} to assist practitioners. However, as software products grow in size and complexity, \textit{canonical} ML-assisted maintenance becomes increasingly challenging and error-prone \citep{10.1145/3597503.3639095}. 

\begin{marginfigure}
 \centering
    \includegraphics[width=\linewidth]{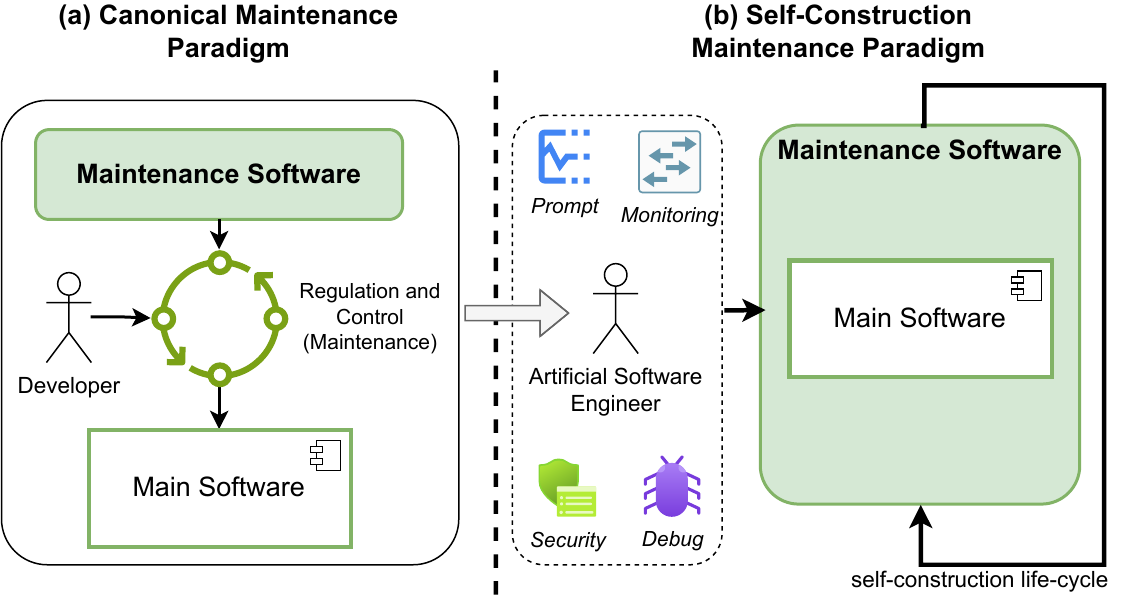}
     \caption{{Maintenance Paradigm Shift: a) Software Maintenance (\sm) is independent from the main software, and b) \sm is wrapping the main software.}} 
     \label{fig:newD}
\end{marginfigure}
This canonical vision of software maintenance has permeated research methods across the software community, leaving no room for alternative approaches due to the remarkably successful use of statistical learning technologies such as transformers \citep{vaswani2017transformers} or reinforcement learning \citep{wei2025swerladvancingllmreasoning}. Irrespective of the exponentially growing canonical research on maintenance, a significant gap remains as this \sm automation has not yet been addressed from an Artificial Life (\alife) vision: the \textit{self-construction} perspective. 

\alife is an interdisciplinary field that studies life-like behaviors- such as self-organization, adaptation, and evolution- in both natural and artificial systems \citep{Dorin2024Alife}. By integrating \alife principles, we can develop maintenance automation frameworks that not only execute predefined tasks but also autonomously adapt and evolve in response to changing conditions. The self-construction maintenance replaces the canonical maintenance paradigm at the \sdlc. This chapter challenges the canonical definition of \sm and introduces a new concept of an autonomous mechanism that supports the software construction life-cycle. This autonomous mechanism is achieved by incorporating \alife elements to enable \textbf{autopoietic} features (\eg self-organization, self-assembly, or self-replication) for \textit{self-construction}~\citep{Sayama_2024,Dorin2024Alife,arcas2024computationallifewellformedselfreplicating}.

This chapter introduces an \textbf{autopoietic architecture} called \autopo for the artificial self-construction software life-cycle. \autopo is an extended and adapted version of von Neumann's self-replicating model\citep{VonNeumann1966}. We present a first draft for the formal notation of \autopo and a complete description of an additional self-construction reasoning unit $\Gamma$. \autopo establishes the foundation for the next-generation \sdlc through \textit{artificial software engineering}.

We envision a formal and nature-inspired approach that artificially enables the \textit{self-construction} of the Software Engineering Life-Cycle using \autopo. \autoref{fig:newD} illustrates the paradigm shift from a canonical ML-driven to a self-construction maintenance technique. This self-construction paradigm allows software systems to create better replicas of themselves at run-time. 

We assert the software maintenance research done for automatic techniques up to this day will be easily integrated within the autopoietic architecture (\autopo). Consider the following scenario, a software engineer wants to eliminate a useless functionality of a software system $S$. The useless functionality can be mapped into sections of the source code that are executed, but its results are never required. This is a typical example of \textit{dead code} elimination. A software engineer develops a solution based on Machine Learning (ML). This engineer wants the system $S$ to self-identify dead code and eliminate it with his ML-based solution. Using an autopoietic architecture (\autopo), the ML-based solution is controlled by a reasoning unit $\Gamma$ that decides how to react in the presence of dead code and create a refactoring. The autopoietic architecture automatically generates a new refactored replica to replace the actual system after guaranteeing that this replica is running stably. Indeed, we can argue that the replicated software system $S$ exhibits \textit{complex behavior} \citep{MaturanaHumberto1980AutopoesisCognition}.

%% file: chapters/part_03_chap_01/sec_02_background.tex
\section{Background \& Context}
\label{ch-autopoietic:sec-02-wm}

The Software Development Life-Cycle (\sdlc) enables a software product by involving technological resources and stakeholders into a well-designed pipeline. This pipeline consists of several phases, including planning, analysis, design, development, testing, deployment, and \textit{maintenance}. Each phase has its own set of activities and deliverables that feed into the next phase. Software Maintenance (SM), the last phase, is the longest and, therefore, most costly as it stays active for as long as the software remains in production\citep{10.1007/978-981-16-0739-4_28,895984}.

Software Maintenance (\sm) research mainly focuses on automating activities (\eg performance optimization, security patching, updates, and enhancements) through the maintenance phase. Considerable research attention has been devoted to \sm on the automation of specific \textit{tasks}. These tasks assist maintenance activities such as clone detection\citep{White2016,White:ICSE15}, bug-fixing\citep{Tufano2018, Tufano:MSR18}, impact analysis, traceability\citep{moran_improving_2020}, refactoring\citep{Nader-palacio2018AssessingDetection,Mkaouer2016AOpportunities}, bad smell detection \citep{Romano2018ACode,Tufano2017WhenAway}, and vulnerability detection\citep{10.1145/3597503.3623345, 10.1145/3597503.3639173, 10764854}. 

It is generally accepted that \sm automation operates using machine learning (ML) approaches or evolutionary computation (EC) solutions (\ie Search-Based Software Engineering \citep{10.1145/2379776.2379787}). Consequently, several maintenance tasks, within the Software Development Life-Cycle \sdlc, still require the active intervention of practitioners. For example, automating \textit{bug-fixing} with a statistical learning approach depends upon collecting, filtering, deduplicating, and transforming code-based datasets, tasks generally performed by practitioners. Despite the breadth and depth of existing research for automating maintenance tasks, none of those ML and EC approaches have been addressed from the point of view of a \textit{self-adaptive technology}. 

Self-adaptivity comprises the ability of a software system to assess and regulate its behavior\citep{Macias-Escriva2013Self-adaptiveApplications}. Recent progress in self-adaptivity investigates and develops sophisticated systems such as multi-agent software \citep{marl-book}, component-based software engineering \citep{10481576}, configurable systems \citep{ye2025distilledlifelongselfadaptationconfigurable}, and model-driven architectures \citep{10608368}. These systems can adapt their behavior in response to environmental changes (\eg new requirements, limited resources, or code changes). Remarkably, this adaptive capability has also been further studied in the context of Artificial Life (\alife) -particularly- with the introduction, definition, and formalization of \textbf{autopoietic systems}~\citep{Dorin2024Alife}.

The term \textit{autopoiesis} refers to a process by which systems continuously regenerate and sustain their components and connections autonomously \citep{MaturanaHumberto1980AutopoesisCognition,Bedau2007ArtificialLife}. We can extend the concept of autopoiesis to formulate new procedures and practices that guide the automation of some maintenance tasks. Under this autopoietic automation vision, software systems are perceived as dynamic entities with inherent self-regulation, adaptability, and, in particular, \textit{self-construction} behaviors. 

%% file: chapters/part_03_chap_01/sec_03_position.tex
\section{From Canonical to Self-Construction Life-Cycle}
\label{ch-autopoietic:sec-03-wm}

\begin{figure*}[ht]
\centering
    \includegraphics[width=\textwidth]{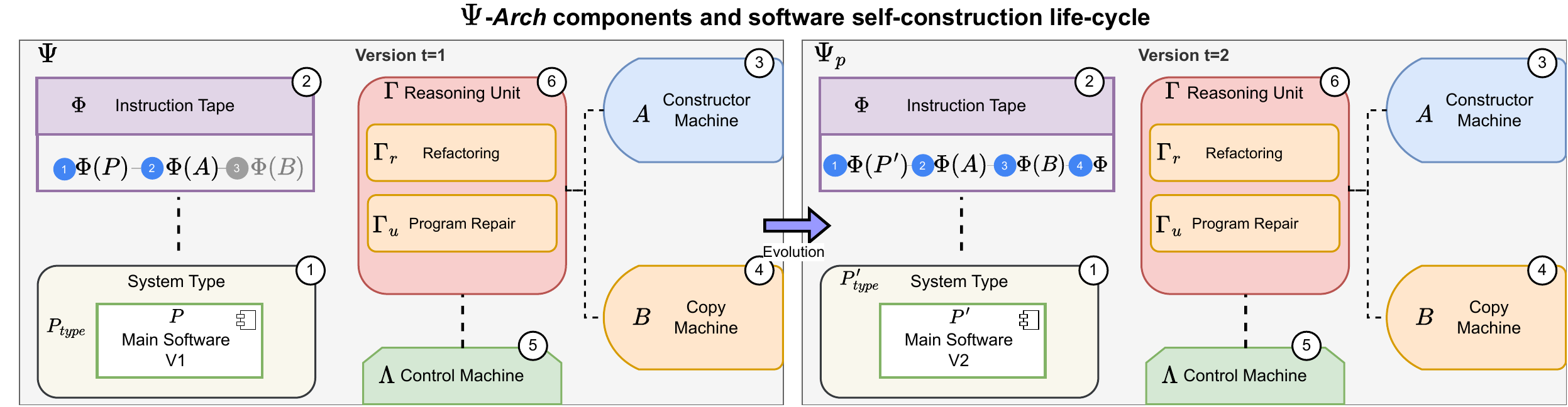}
     \caption{Autopoietic Architecture \autopo to enable artificial self-construction software life-cycle}
     \label{fig:arch}
\end{figure*}

In his tale \textit{-the last question-}, Isaac Asimov presented a self-replicating machine with the ability to answer any question of the universe. Following Asimov's vision, this chapter examines the relationship between the premise \textit{software systems can self-replicate into a ``better, decentralized, and intelligent'' version of themselves} and the Software Development Life-Cycle (\sdlc). This self-replication process embodies the \textit{self-construction} of adapted internal mechanisms, which compose the original software, to attend an \textit{environment} of functional (\ie what the system should do) and non-functional (\ie how the system should perform) requirements.

Assuming the autopoietic automation vision, we must then recognize and acknowledge the feasibility of a new type of autonomous architecture leveraging an artificial self-construction life-cycle. We refer to this autonomous mechanism as \textit{Autopoietic Architectures} (\autopo). To attain an artificial self-construction life-cycle paradigm, beyond the canonical perspective (see \autoref{fig:newD}-a), we argue that \autopo can be used as a self-construction standard to implement systems whose \textit{self-replicating components} are calibrated and well communicated. This self-construction paradigm draws inspiration from the Von Neumann kinetic beast\citep{VonNeumann1951,VonNeumann1966}. Von Neumann's approach comprises a foundational model simulating a cellular automaton's self-replication process \footnote{Langton later enhanced this foundational model by introducing a simpler self-replicating automaton to address the problem of evolutionary capabilities \citep{Langton1989ALife}}. 

The \autopo embodies six self-replicating components: the main (or original) software, an instruction tape, three machines (\eg constructor, copy, and control machines), and a reasoning unit. The self-replicating process starts when the control machine $\Lambda$ causes the copy machine $B$ to replicate the instruction tape $\Phi$.

The \textit{first} component encapsulates the original software to replicate $P$ (\autoref{fig:arch}-\circled{1}). From a software engineering perspective, $P_{type}$ refers to Lehman's categorization of software systems~\citep{Lehman1980ProgramsEvolution} encompassing specified systems (S-Systems), problem-solving systems (P-Systems), and evolving systems (E-systems). The \textit{second} component (\autoref{fig:arch}-\circled{2}) consists of an \textit{instruction tape}, $\Phi$, which stores the meta-data of \autopo. This meta-data contains detailed instructions on constructing the entire architecture from its source code. The instruction tape $\Phi$ could incorporate advanced, goal-oriented techniques for system maintenance (\ie automated refactoring, patching, repairing, and re-engineering processes). The \textit{third} component (\autoref{fig:arch}-\circled{3}) reflects a \textit{constructor machine}, denoted as $A$, which is responsible for reading any instruction $\phi$ from the instruction tape $\Phi$ to construct components. The \textit{fourth} component is a \textit{copy machine} $B$ (\autoref{fig:arch}-\circled{4}). $B$ can replicate $\Phi$ instruction, \ie $B$ can copy $\Phi$ meta-data -including $\Phi(B)$. The \textit{fifth} component is the control machine $\Lambda$ (\autoref{fig:arch}-\circled{5}). $\Lambda$ produces an isolated copy of $\Psi$ and communicates with the \textit{reasoning unit} $\Gamma$ (\autoref{fig:arch}-\circled{6}). $\Gamma$ is the \textit{sixth} -and last- component oracle that formulates causal questions to make decisions~\citep{Tucci2013IntroductionDo-Calculus}.

\begin{defn}
The copy process generates a $\Phi$ final instruction tape  
\begin{equation} 
B + \Phi(\Psi) \rightarrow \Phi(\Psi)
\end{equation}
where the operand `$+$' is the use of two components ($B$ and $\Phi$) and the operand `$\rightarrow$' is a realization or copy. In other words, $B$ takes the meta-data instructions and generates a copy of  $\Phi(\Psi),$ which is the architecture ``DNA''. Define $\Delta = A+B$. Thus, $\Psi$ would be the $\Delta + P$ composition. Replacing in the previous formula:

\begin{equation} 
B + \Phi(\Delta + P) \rightarrow \Phi(\Delta + P)
\end{equation}
\end{defn}

\begin{defn}
Now, $\Lambda$ causes A to construct the components described by the instruction in $\Phi(\Psi)$.  
The construction process generates a copy of $\Psi$ 
\begin{equation} 
A + \Phi(\Psi) \rightarrow \Psi
\end{equation}
\end{defn}

\begin{defn}
The control machine $\Lambda$ synchronizes the machines correctly to produce the tape $\Phi$ and the $\Psi$-Architecture by adding the copy of $\Phi(\Psi)$ to the new $\Psi$
\begin{equation} 
A + B + \Gamma + \Phi(\Psi) \rightarrow \Psi + \Phi(\Psi)
\end{equation}
Now, let's include in the tape machine the information of the main software or $P$, the reasoning unit $\Gamma$, and rename the entire architecture as $\Psi_P$, therefore $\Delta= A + B + \Gamma$: 
\begin{equation} 
 \Psi_P = \Delta + \Phi(\Delta+P) = A + B + \Gamma + \Phi(A + B + \Gamma + P)
\end{equation}
so we can observe that 
\begin{equation} 
\Psi_P \rightarrow A + B + \Gamma + P + \Phi(A + B + \Gamma + P)
\end{equation}
which means:
\begin{equation} 
\Psi_P \rightarrow \Psi_P + P
\end{equation}
\end{defn}

Finally, after following previous definitions, we obtain the original software $P$, with the respective optimizations $P'$, and its instruction tape. $P$ is embedded in the autopoietic architecture \autopo. Note that $P'$ optimizations originate from the reasoning unit $\Gamma$. $\Gamma$ introduces small instructions to the instruction tape so $\Phi(P)$ evolves to $P'$. $\Gamma$ can include other complementary maintainability tasks \eg performing control tasks, such as software traceability.

Although autopoietic architectures \autopo pose a promising self-construction framework, the vast majority of work in the \sdlc research field has focused on traditional (\ie canonical) maintenance tasks. Studies indicate that over 70\% of software lifecycle costs are attributed to maintenance and evolution efforts~\citep{895984,Lehman1980ProgramsEvolution}. Conversely, our autopoietic architectures \autopo can leverage an artificial self-construction life-cycle that continuously monitors, modifies, and optimizes targeted software. These architectures encapsulate the original software system within an intelligent maintenance layer, operating as an autonomous entity.  

%% file: chapters/part_03_chap_01/sec_04_discussion.tex
\section{Discussion \& Implications}
\label{ch-autopoietic:sec-04-wm}

\input{tables/chap_autopoietic/tab1_reasoning_unit}

The autopoietic architecture (\autopo) incorporates a reasoning unit $\Gamma$, responsible for making maintenance-related decisions. This reasoning component is the most challenging to design as it comprises unique and necessary knowledge to address self-construction tasks. To properly define this reasoning unit, we decompose it, as detailed in \autoref{table:reasoning}, into six key dimensions (\eg component, representation, purpose, source of information, algorithm employed, and agent question). 

In addition to those dimensions, we can classify reasoning tasks into three levels according to Pearl's causality approach: observation, intervention, and retrospection \citep{pearl2009overview}. Firstly, the goal of the \textit{observation} level is to predict a maintenance task $M$ based on given evidence $\epsilon$. The prediction is a conditional distribution $p(M|\epsilon)$ where $M$ can be corrective $M_c$, adaptive $M_a$, perfective $M_p$, or preventive $M_v$ activity. Secondly, the \textit{intervention} level aims to measure the causal effect when the reasoning unit $\Gamma$ modifies a sub-component in $P$ to accomplish a maintenance task $M$. Finally, the \textit{retrospection} level aims to estimate the expected value of hypothetical conditions. For example, consider the reasoning unit $\Gamma$ performed an adaptive maintenance task $M_a$ with an observed result $Y=y$, but now the reasoning unit wants to assess what would have happened ($Y=?$) if, instead of $M_a$, the unit had performed $M_v$. The previous situation formally defines a counterfactual prediction. The reasoning unit can measure this hypothetical condition using the expected value $E(Y_{M=M_v} | M=M_a, Y=y)$.

Because the reasoning unit automates some maintenance tasks, a replica of the autopoietic architecture \autopo serves as an enhanced copy of the original (\ie main) software. This enhancement formally represents what software researchers refer to as \textit{software evolution}\citep{10.1145/2593882.2593893}. In light of our autopoietic architecture \autopo, the proposed work seeks to explore the following research questions (RQs):

\begin{itemize}[leftmargin=*, labelwidth=0pt, labelindent=0pt]
	\item $RQ_1$ \textit{How effective is using \alife elements to create software systems that exhibit self-construction behavior?} The effectiveness of \alife to create software systems that exhibit living features (\ie self-replication) has been widely studied \citep{Bedau2007ArtificialLife}. However, we require researching methodologies compatible with the software development life-cycle similar to the software transplantaton \citep{10.1145/3695987}.
	\item $RQ_2$ \textit{To what extent is software maintenance automated under the autopoietic \autopo scheme?} We can measure maintenance levels in a software system by considering corrective, adaptive, perfective, and preventive tasks.  
	\item $RQ_3$ \textit{Should software designs have a dedicated component to self-construction?} This question explores if and when incorporating \alife elements in software design. 
	\item $RQ_4$ \textit{How can we migrate from canonical maintenance to self-construction maintenance paradigm?} We can leverage software engineering research to migrate any potential system to an autopoietic architecture \autopo.
\end{itemize}

To address the RQs, developing and deploying \autopo will require extensive expertise in training and programming \textit{AI-agents}. An increased interest in AI-agents has emerged in recent years; this technology has enhanced the performance of challenging tasks\citep{yang2024sweagentagentcomputerinterfacesenable,wang2024executablecodeactionselicit,jimenez2024swebenchlanguagemodelsresolve}. We believe AI-agents theory and practice can contribute to the implementation of the proposed self-replicating components. For example, the control mechanism could be an AI agent trained on architectural patterns data. This control agent can also formulate questions to or communicate with the reasoning unit, another independent agent.

While the exact nature of inputs and outputs within autopoietic architectures \autopo remains uncertain, the overarching goal is clear: to enable self-construction and maintenance through replication and adaptability processes inspired by \alife. It may be observed that the proposed architecture does not account for possible security issues or any other type of non-functional constraints. Moreover, the environmental variables are not studied in-depth.

%% file: tables/chap_autopoietic/tab1_reasoning_unit.tex
\begin{table}[ht]
\caption{The Reasoning Unit Depends Upon The Maintenance Task}
\label{table:reasoning}
\centering

\begin{adjustbox}{width=\textwidth}

\def\arraystretch{1}\tabcolsep=10pt
\begin{tabular}{p{0.2\linewidth}|p{0.32\linewidth}p{0.32\linewidth}p{0.32\linewidth}}
\toprule
\multicolumn{1}{c|}{\textbf{Feature}} & \multicolumn{3}{c}{\textbf{Causal Reasoning Unit $\Gamma$}} \\ \hline
\multicolumn{1}{l|}{\textbf{Level}} & \multicolumn{1}{c}{\textit{Observation}} & \multicolumn{1}{c}{\textit{Intervention}} & \multicolumn{1}{c}{\textit{Retrospection}} \\ \hline
\textbf{Representation} & Probability $p(M|\epsilon)$ & Causal Effect $do\{p(M|\epsilon)\}$ & Counterfactuals $E{[}Y_M|M,Y{]}$ \\
\textbf{Purpose} & Take decisions about maintenance tasks based on observations & To intervene in the architecture to assess the causal effect of a new replica & To imagine possible scenarios where a modification  on the architecture takes place \\
\textbf{Source of Inf.} & Quality SE Metrics $\Pi$ & Traceability Links $L$, SE Metrics $\Pi$ & Depends upon the maintenance task $M$ \\
\textbf{Algorithm} & Variational Inference & Front/Back door adjustment & Front/Back door adjustment \\
\textbf{Agent Question} & What if the unit sees a metric $M$ under desired thresholds? & If the unit modifies a sub-component $P$ in the main system to fulfill a maintenance task $M$, will enhance a $\Pi$ metric? & What if the unit had assessed one possible maintenance solution $M$ that was not being considered before? \\ \hline
\end{tabular}%

\end{adjustbox}

\end{table}

%% file: chapters/part_03_chap_01/sec_05_conclusion.tex
\section{Conclusion}
\label{ch-autopoietic:sec-05-wm}

Software maintenance and evolution are challenging activities on \sdlc that require the application of Artificial Life (\alife) techniques to advance next-generation artificial software engineering. This chapter highlights the benefits of closing the gap between \alife and Software Engineering research by formulating \textit{autopoietic architectures} to enable an artificial self-construction software life-cycle. The autopoietic architecture, namely \autopo, aims to follow Lehman laws for software evolution on a \sdlc (\ie continuing change, self-regulation, stability). \autopo elicits the next generation of artificial software engineers for new \sdlc activities (\ie prompt-engineering, agentAI monitoring, and LLMs for SE debugging). Furthermore, \autopo requires architectural data and benchmarking to assess the extent of self-construction of the original software system.

%% file: chapters/part_04_chap_01/conclusions.tex
\chapter{Conclusions}
\label{ch:conclusion}

The ultimate goal of interpretability in Deep Learning for Software Engineering is to generate explanations of \nlms' predictions. We achieved this goal by formulating causal interpretations of such predictions, quantified through performance values (\eg Cross-Entropy, Next Token Prediction, or Distance Metrics). We demonstrate that regardless of conventional evaluations of \nlms (\eg BLEU, Accuracy, Perplexity), producing causal interpretations is vital for model understanding. For example, even if a model demonstrates a deficient BLEU score when its outputs are evaluated against the ground truth for a downstream task, we can still generate causal explanations of those outputs by defining an appropriate SCM.

Our interpretability method, \docode, generates causal explanations (or interpretations) based on the idea of estimating \textit{the effect of \textbf{interventions}}. These interventions can be applied to the data inputs or the configuration parameters of the \nlms. Furthermore, estimating the causal effects of such interventions stems from  Pearl's theory of causation~\citep{Pearl2009Causality, Pearl2016Causality, Pearl2018Causality}. 

Below, we pose three aspects of the discussion: 1) some general insights from the case study, 2) a brief explanation of the guidelines to use \docode in practice, and 3) a list of challenges and opportunities practitioners might face when adapting \docode for their analyses.

\input{chapters/part_04_chap_01/sec_01_insights}

\input{chapters/part_04_chap_01/sec_02_inpractice}
\input{chapters/part_04_chap_01/sec_03_actionability} 
\input{chapters/part_04_chap_01/sec_03_challenges}

%% file: chapters/part_04_chap_01/sec_01_insights.tex
\section{Insights From \texorpdfstring{\docode}{docode} Case Study}
\label{ch:conclusion:sec_01}

\begin{figure*}[ht]
  \includegraphics[width=\textwidth]{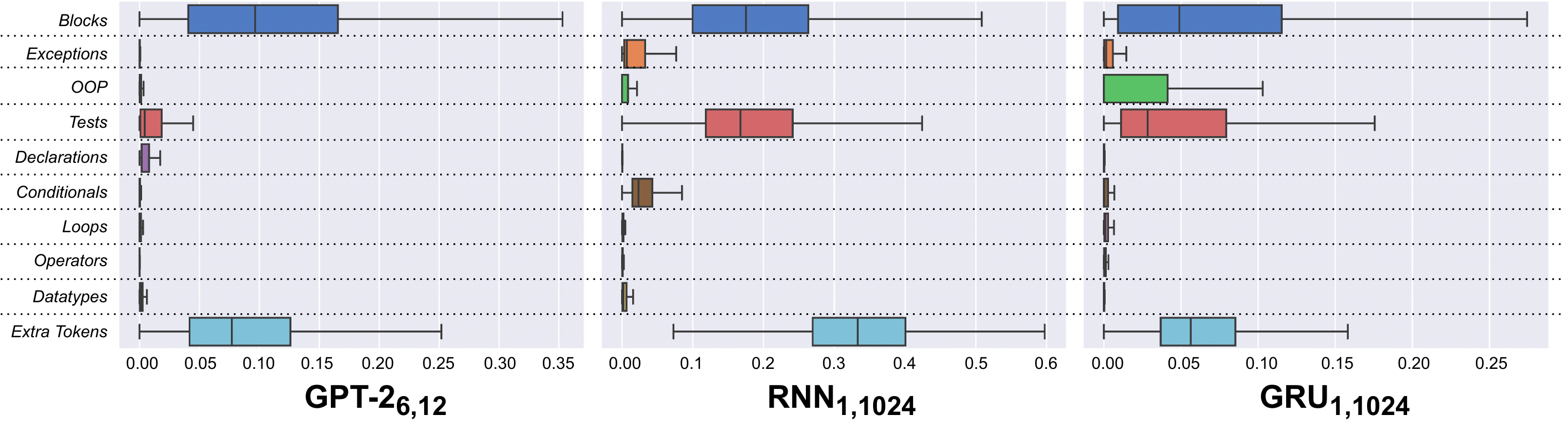}
  \caption{{Descriptive Error Analysis. Keyword-Based Categories probabilities (Normalized $Y_l$ - \ntp) for \tf \rnn and \gru on the \BuggyTB dataset. Higher values indicate less erroneous predictions.}}
  \label{fig:bug_rnn_tax_ntp}
\end{figure*}

In \docode, we examine model semantics by categorizing code tokens into different conceptual categories and evaluating raw model performance according to these groups, along with the causal relationships of these categories across treatments. Syntax can be measured by how often the model predicts a given token that is syntactically correct. Although we did not directly report syntactic correctness, in our observations, our studied models rarely made syntactic errors. However, the local prediction performance of tokens in different semantic categories tends to vary quite a bit. This is partially explained by the prevalence of different token categories in the training set, but further work on model architectures that improve the performance of underperforming token groups could help advance research on \nlms.

For each \nlm, a strong correlation across testbeds for a given code token category means that we observe the prediction performance has \textit{changed} significantly across SE-based treatments (\eg commented/uncommented code). This change could correspond to the fluctuation in prediction performance within the token categories. However, because our approach uses a causal graph (\ie Structural Causal Model), we can determine the amount of causal effect between the correlated variables based on covariates, and decide whether or not they are \textbf{spurious correlations}, due to \textit{confounding bias} \figref{fig:covariate}, or true causal relationships caused by the change in treatment. Based on our experiments, NCMs seem to have a stronger \textit{statistical understanding} of syntax and a more limited understanding of semantics, according to our definition of semantics. However, once we started exploring Masking Language Models such as BERT-like code models, we found the opposite: AST node tokens are not better predicted than a random and, therefore, unstructured set of tokens.

As the results suggest, our \nlms under study learn to predict tokens related to code blocks (\eg brackets, parentheses, semicolons) more effectively than most other code token types (\eg loops, conditionals, datatypes), we found that our \nlms are \textit{sensitive to seemingly subtle changes in code syntax}, reinforcing previous studies concluding the same~\citep{rabin2021generalizability}, and our models are only \textit{marginally impacted} by the presence of comments and bugs, which challenges findings from previous work~\citep{Baishakhi2016buggy}. Consequently, practitioners can identify which \textit{categories} of code tokens were important in the model’s decision-making process. In other words, \docode supports the identification of which 1) tokens, 2) layers, or 3) hyperparameters are impacting code predictions, which we describe below:

\textbf{Tokens.} In \secref{ch:case:sec_04}, we propose a \textit{syntax clustering} function for grouping the \nlms' code predictions into more understandable syntax categories (\ie keyword-based and AST/Grammar-based). Although the syntax clustering formalism was originally omitted from the \docode pipeline, we realized during the experimentation phase that a clustering strategy is vital as practitioners must make sense of generated subwords or fine-grained levels of code.

Our syntax clustering formalism stems from the need to make fine-grained code predictions more understandable to practitioners. As such, our approach enables us to determine which token categories a model can learn (or not) in different application settings (\eg SE-based interventions). For example, given the normalized \ntp values computed for scenario \textit{A}, as depicted in \figref{fig:bug_rnn_tax_ntp}, one could conclude that block-category tokens are more effectively predicted by \tf, \rnn, and \gru. Yet such \nlms typically struggled to predict operator-category tokens. More importantly, we can observe causal relationships related to the importance of code tokens in decision-making \textit{changes} across treatments (\ie data interventions or model parameters alterations).

\textbf{Layers.} Our method supports interventions on the \textit{number of layers} for a given architecture. Therefore, \docode can identify if the layers, as a hyperparameter, influence the performance. Nonetheless, identifying specific layers to influence the performance would require going beyond the causal interpretability scope and exploring the inner mechanisms of the neural network \citep{molnar2025}.

\textbf{Hyperparemeters.} $do_{code}$ supports the identification of hyperparameters that influence neural network performance by extending SE-based intervention definitions at the model level (see \secref{sec:seintervention}). Our case study supports two types of parameters: \textit{NumberOfLayers} and \textit{NumberOfUnits}.

\textbf{About Causal Transportability.} Although we concentrated on datasets and keyword-based categories for Java (\ie scenarios $A-F$) in our exploratory analyses, the assumptions we made to build the SCMs still hold for scenario $G$. This \textit{transportability} of assumptions across similar domains have been researched in Causal Inference \citep{pearl_transportability}. As such, the causal information learned from our experiments can be reused in analogous settings if there is homogeneity in the effect modifiers. Transportability allows us to maintain our assumptions across scenarios, as long as the underlying structure of the causal graph remains consistent, keeping the same type of treatments, potential outcomes, and confounding variables.

%% file: chapters/part_04_chap_01/sec_02_inpractice.tex
\section{\texorpdfstring{\docode}{docode} in practice}
\label{ch:conclusion:sec_02}

While it may appear that \nlms have begun to achieve promising performance, it is insufficient to \textit{simply} generate code predictions (\ie the \textit{\textbf{what}} of \nlms' decision). This current status quo, at best, provides an incomplete picture of the limitations and caveats of \nlms for code. Given the potential impact and consequences of these models and their resulting applications, there is a clear need to strive for a complete understanding of how they function in practice. As such, we must push to understand how \nlms arrive at their predictions (\ie the \textit{\textbf{why}} of \nlms' decision) as shown in our proposed Structural Causal Models for each scenario. With the design and results of interpretability scenarios, we demonstrate that \docode comprises a causal explanation method that aims to make Deep Learning for Software Engineering \nlms, and their decision-making process, understandable to practitioners. 

\begin{figure*}[ht]
		\centering
		\includegraphics[width=0.99\textwidth]{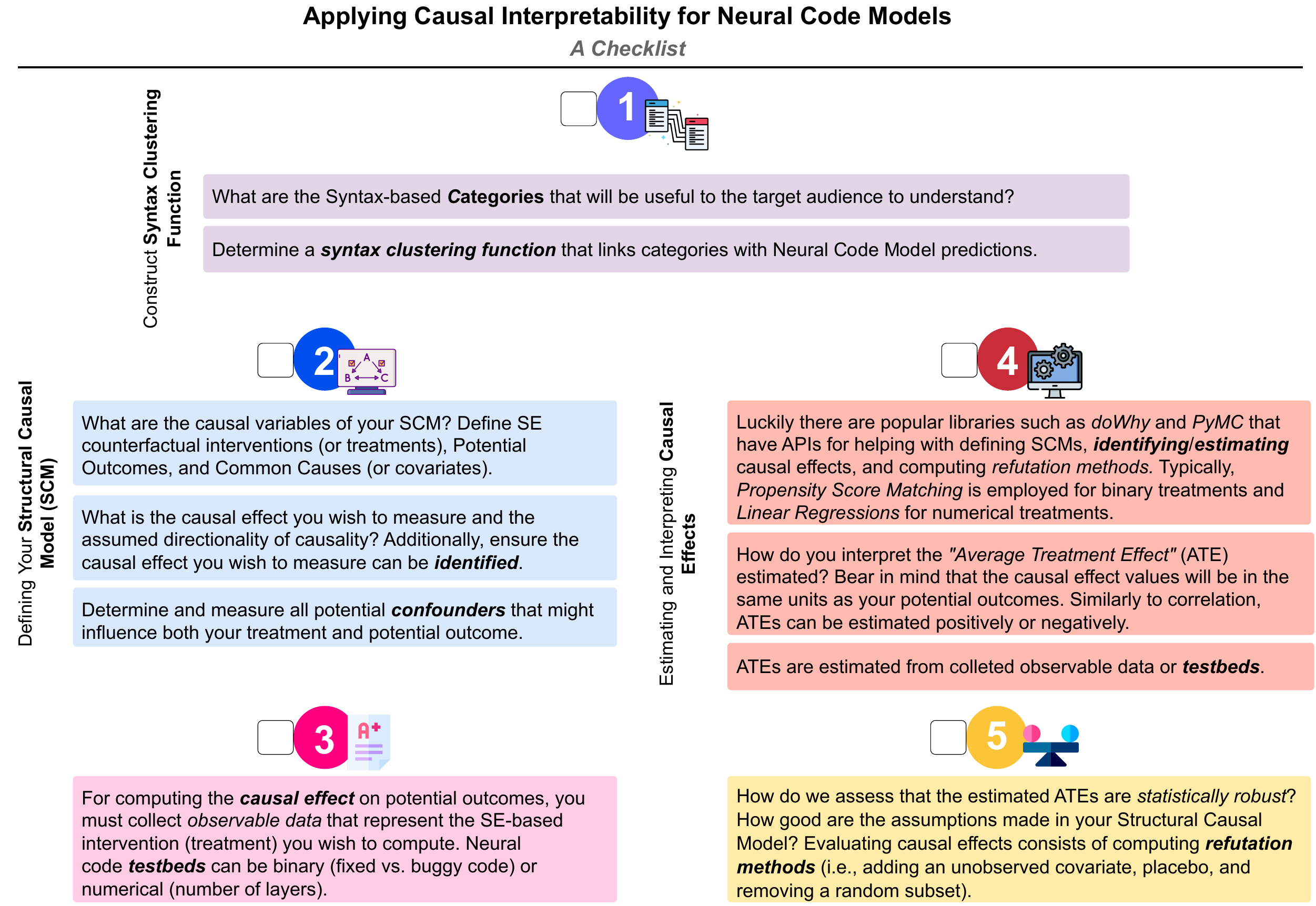}
		\caption{ Five Guidelines for Applying Causal Interpretability to Neural Code Models. } 
        \vspace{0.5cm}
        \label{fig:checklist}
\end{figure*}

To that end, we have created a checklist that outlines the general process researchers can use to apply causal interpretability to Neural Code Models \ref{fig:checklist}. {For instance, one of the practical applications of \docode is facilitating the debugging process of \nlms. By debugging a \nlm we refer to the tasks of reducing the amount of \textit{confounding bias} between SE-interventions and code predictions. Our proposed guidelines below help to design a pipeline in which \docode facilitates the detection of this bias for a given setting.}

As shown in \figref{fig:checklist}, the proposed checklist comprises five guidelines, which correspond to both the approach pipeline (see \chapref{ch:docode}) and a definition of a syntax clustering function (see \secref{ch:case:sec_04}), starting with the construction of the clustering function. In the first guideline, a researcher determines what would be understandable to their target audience and constructs a syntax clustering function that translates the model under study's fine-grained code predictions to the target audience. Once you have this clustering, you can move on to the second guideline that revolves around defining a Structural Causal Model (SCM). It is important to have this step after your clustering function so that you can properly model the interventions, outcomes, and confounders. With your SCM in hand, in the third guideline, you can now collect your data that contains the observable SE-based interventions (\eg buggy vs. fixed code) that will be used to estimate the causal effect of the treatment. Then, in the fourth guideline, you can use existing popular libraries (\eg doWhy) to estimate your causal effect. And lastly, in the fifth guideline, you must check your assumptions (\ie your confounders and SCM) using refutation methods. With this checklist, we hope to ease the complexity around causal analysis for researchers.

%% file: chapters/part_04_chap_01/sec_03_actionability.tex
\section{Who benefits from a scientific view of causal interpretability in \dlse? and How?}
\label{ch:conclusion:sec_03}

This dissertation suggests the need to integrate causal interpretability into post hoc analysis of Neural Code Models (\nlms). We hope to provide researchers with an approach that integrates causal analysis into their research. To that end, we made the following contributions. 

\underline{$Product_1$:} \textit{A Syntax (De)Composition Statistical Technique To Enable Code-based Explanations.} This document explores how syntactic (AST-based) and semantic (keyword-based) clustering of code predictions can be used within the \docode framework to provide more interpretable explanations of model behavior. More details in \chapref{ch:decomposition}.

\underline{$Product_2$:} \textit{A Hierarchi\textbf{C}al Pr\textbf{O}babilistic \textbf{M}odel for Softwar\textbf{E} \textbf{T}raceability.} \Comet is a tool to improve the effectiveness of software traceability automation. Furthermore, this technique demonstrates that software artifacts and their relationships can be represented as statistical distributions. More details in \chapref{ch:bayes}. 

\underline{$Product_3$:} \textit{A Method That Leverages Information Theory for Interpretability.} Information theory metrics can provide valuable formal analyses on the information flow and relationships between software artifacts, helping to interpret the effectiveness of traceability models. See \infotheory in \chapref{ch:nature}.

\underline{$Product_4$:} \textit{A Patent for a Debugging Tool and an Intepretability Technique for Code-based Explanations.} We introduce \textit{code rationales} (\codeRational), a technique with rigorous mathematical and statistical foundation, to identify subsets of tokens that can explain individual code predictions. This technique was patented as a debugging tool \citep{clement2024debugging}. More details in \chapref{ch:rationales}. 

\underline{$Product_5$:} \textit{A Methodology for Causal Interpretability (\docode).} This dissertation emphasizes that simply identifying correlations between software artifacts or code features and model predictions is insufficient for truly understanding and leveraging \dlse models for intervention. To effectively use \dlse for tasks requiring action and understanding of consequences, a shift towards causal interpretability is necessary. This involves identifying and quantifying cause-and-effect relationships. The \docode approach provides a structured methodology to achieve causal interpretability in \dlse, involving the formulation of \scms, defining interventions, estimating causal effects, and performing refutation tests. More details in \chapref{ch:docode}.

\underline{$Product_6$:} \textit{Several Empirical Intepretability Scenarios in Software Engineering.} We illustrate how \docode is actionable in Software Engineering. Seven scenarios were proposed within a permutation framework that can be extended as practitioners or researchers need it. More details in \chapref{ch:case}.  

\underline{$Product_7$:} \textit{A Benchmarking Strategy to Enable Causal Evaluations.} This work stresses the need to consider interventions that are meaningful within the software engineering domain (\eg un-commenting code, changing the number of layers) and to account for potential confounders (\ie common causes) that might lead to spurious correlations. The \Galeras benchmark is introduced as a tool for causal evaluation in code generation, emphasizing the need to go beyond traditional performance metrics and assess the causal effects of code features on model predictions. More details in \chapref{ch:benchmarking}. 

\underline{$Product_8$:} \textit{A Prospective Analysis on Counterfactual Applications in Software Engineering.} Drawing inspiration from Artificial Life (\textit{ALife}), we propose a fundamental shift in the Software Development Life-Cycle (\textit{SDLC}) by introducing self-construction mechanisms that enable software to evolve and maintain autonomously. Autopoietic Architectures, specifically $\Psi$-\textit{Arch}, as a foundational framework for self-constructing software are proposed in \chapref{ch:autopoietic}.  

\underline{$Artifacts$:} \textit{Replication Packages and Repositories.} To encourage the use of our methodology in future work on developing and evaluating code generation models and for replicating our experiments: \docode~\citep{icodegen},\codeRational~\citep{CodeQ2025}, \astrust~\citep{RepoASTrust24}.

%% file: chapters/part_04_chap_01/sec_03_challenges.tex
\section{Challenges \& Future Work}
\label{ch:conclusion:sec_04}

In this section, we list some challenges ($CHs$) that practitioners might face when adapting our method to their interpretability analyses. 

\textbf{$CH_1$: Proposing a new syntax clustering criterion.} {Our proposed category system to group fine-grained code predictions into more understandable categories is based entirely on current Programming Languages definitions and guidelines to facilitate automatic categorization. This might represent a limitation if a practitioner requires a more in-depth analysis of the syntax interactions. \docode introduces a clustering formalism that can be extended and reformulated to specific interpretability needs. Different clustering functions give rise to different challenges. For instance, one BPE token can be part of different categories or overlap different AST nodes. How do practitioners deal with this overlapping? Are they forcing the BPE token as we did in our keyword-based clustering? or Do they allow for some statistically controlled overlaps as we did in AST/Grammar-based clustering?}

\textbf{$CH_2$: Collecting data for formulating SE-based interventions.} \docode was not implemented to perform \textit{causal discovery}, that is unveiling the causes of code predictions directly from observations. In fact, \docode is restricted to estimate causal inference measures (\eg correlations, ATEs, CATEs) based on hypotheses and assumptions, from domain knowledge, embedded in Structural Causal Models. Therefore, users of \docode must define and formulate specific SE-based interventions depending on the available data and, also, provide the possible confounding factors that can alter the effect on code predictions. Future work can be oriented to facilitate the definition of the Structural Causal Model into a more automatic approach. Such an automatic approach should include recent techniques on causal discovery.

\textbf{$CH_3$: Integrating \docode in Deep Learning for Software Engineering life-cycle.} {In order to use \docode in practical settings, we must propose some strategies for how to integrate interpretability approaches on the production pipeline of Software Engineering solutions based on Deep Learning models. For instance, researchers are encouraged to explore Deep Learning Continuous Integration solutions that incorporate not only \docode but also any future interpretability method in a production or monitoring pipeline. 

\textbf{$CH_4$: Creating the Structural Causal Model.} The quality of the causal explanations generated by \docode heavily relies on the structure of the causal graph. Consequently, employing \docode in practical scenarios would require statistical evidence supporting each \scm's component. Although we provide evidence supporting these components for the proposed scenarios, it may not hold for every new setting. For instance, the confounders and code syntax clustering introduced in our study might differ for functional programming languages. Building a sound \scm would entail automatically identifying potential confounders and formulating assumptions from a rigorous causal discovery process not covered in our pipeline.

In summary, our interpretability method \docode is based on the idea of outlining causal queries given a defined SCM from domain knowledge. These causal queries are obtained by estimating an interventional distribution where the potential outcome is generally a prediction performance value of the Neural Code Model under study and the interventions are a set of software-based properties that help us construct explanations about the generative model. We have implemented \docode in an open-source library~\citep{icodegen}.}

%% file: appendixA.tex
\chapter{Glossary of Terms and Concepts}
\label{appendixA}

\begin{itemize}
\item \textbf{Abstract Syntax Tree (AST)}: A tree representation of the abstract syntactic structure of source code. ASTs are used to decompose code into terminal and non-terminal nodes, providing the basis for syntax-grounded explanations and enabling clustering of code predictions into syntax categories.

\item \textbf{Alignment Function ($\delta$):} A function mapping code tokens to terminal nodes in an AST. Formally, $\delta: w_{\leq i} \rightarrow \vec{\lambda}$, where $w_{\leq i}$ is a code sub-sequence whose tokens are associated with the corresponding terminal node vector $\vec{\lambda}$ of syntax subcategories.

\item \textbf{ASTrust:} An interpretability method for Large Language Models (LLMs) of code that generates explanations grounded in the relationship between model confidence and syntactic structures. ASTrust aligns next-token predictions with AST nodes and clusters them into syntax categories, supporting both local and global explanations.

\item \textbf{Associational Interpretability:} The first level in Pearl's Ladder of Causation, focused on identifying correlations between inputs and outputs (\ie answering ``what is?'' questions) by estimating $p(Y|T)$ using statistical association methods.

\item \textbf{Average Treatment Effect (ATE):} A statistical measure in causal inference quantifying the causal effect of a treatment on an outcome. Used in \docode to estimate the causal effect of interventions on model predictions.

\item \textbf{Causal Inference:} A formal framework for reasoning about cause and effect, introduced by Judea Pearl. It uses Structural Causal Models (SCMs) and do-calculus to distinguish between association, intervention, and counterfactual reasoning.

\item \textbf{Causal Interpretability:} A global post hoc approach to explaining Neural Code Models (NCMs) based on causal assumptions encoded in SCMs. It uses the Ladder of Causation to estimate quantifiable causal effects of interventions on model predictions.

\item \textbf{Causal Interpretability Hypothesis:} The hypothesis that any NCM prediction can be causally explained using the Ladder of Causation. The \docode method operationalizes this hypothesis.

\item \textbf{Clustering Function ($\theta$):} A function that aggregates token-level predictions from AST nodes into syntax categories. Formally, $\theta: \vec{\lambda} \rightarrow \operatorname{median}(\vec{n})$, where $\vec{\lambda}$ is a vector of predictions for a sub-sequence and $\vec{n}$ is the vector of associated non-terminal nodes.

\item \textbf{COMET (Hierarchical Probabilistic Model for Software Traceability):} A tool that models software artifacts and their relationships as statistical distributions, improving software traceability using IR/ML techniques, developer feedback, and transitive links.

\item \textbf{Confounding Bias:} The distortion in the estimated effect of a treatment $T$ on outcome $Y$ due to confounding variables $Z$ that influence both $T$ and $Y$. Confounding bias is present when $p(y|t) \neq p(y|do(t))$.

\item \textbf{Counterfactual Reasoning:} The third level in Pearl's Ladder of Causation, dealing with ``why?'' questions and reasoning about what would have happened under different circumstances, expressed as $p(y_t|t', y')$.

\item \textbf{Deep Learning for Software Engineering (DL4SE):} The application of deep learning techniques to automate or enhance software engineering tasks, leveraging automatic feature extraction from large-scale software data.

\item \textbf{do-calculus:} A symbolic framework for predicting the effect of interventions and controlling for confounding bias, using the do-operator ($do(t)$) to distinguish between observational and interventional distributions.

\item \textbf{\docode:} A post hoc interpretability method for NCMs, leveraging causal inference to provide programming language-oriented explanations of model predictions. It involves modeling a causal problem, identifying the causal estimand, estimating causal effects, and refuting effect estimates.

\item \textbf{Explainability vs. Interpretability:} Explainability refers to understanding how a model operates internally (\eg via layer analysis), while interpretability refers to mapping model predictions to understandable concepts (\eg syntax categories).

\item \textbf{Information Theory:} A mathematical framework used to analyze information flow and relationships between software artifacts. In the dissertation, it is applied via \infotheory to interpret the effectiveness of traceability models using metrics like mutual information and information loss.

\item \textbf{Interventional Interpretability:} The second level in the Ladder of Causation, focused on understanding the effect of interventions (``what if?'' questions), by estimating $p(y|do(t))$.

\item \textbf{Jensen-Shannon Distance:} A statistical measure of similarity between probability distributions, used to assess associations $p(Y|T)$ for binary treatments in code interventions.

\item \textbf{Ladder of Causation:} A hierarchy introduced by Pearl, with three levels: association (seeing), intervention (doing), and counterfactual (imagining). Each level answers increasingly complex causal queries.

\item \textbf{Large Language Models (LLMs):} Neural network models trained on vast text data, capable of generating human-like text. In the context of code, LLMs are evaluated for their trustworthiness and interpretability.

\item \textbf{Neural Code Models (NCMs):} Deep learning models specialized for code-related tasks, such as code completion and generation. Their predictions are often opaque, motivating the need for interpretability.

\item \textbf{Next Token Predictions (NTP):} The probabilities assigned by a language model to each token in its vocabulary for the next token in a sequence. Used in ASTrust to align predictions with AST nodes.

\item \textbf{Post Hoc Interpretability:} Interpretability methods applied after model training to explain predictions, as opposed to intrinsic methods that build interpretability into the model architecture.

\item \textbf{Potential Outcomes:} In causal inference, the possible values an outcome variable would take under different treatment conditions (\eg cross-entropy, next token predictions).

\item \textbf{SE-based Interventions:} Changes in input data distributions representing meaningful software engineering concepts (\eg buggy vs. fixed code), used to attribute causal effects in model performance.

\item \textbf{Spurious Correlation:} A correlation between variables that does not reflect a causal relationship, often due to confounding variables.

\item \textbf{Structural Causal Models (SCMs):} Mathematical models representing causal assumptions and observable data, consisting of graphical models (DAGs), structural equations, and logic for counterfactuals/interventions.

\item \textbf{Syntax Categories (SCs):} Human-understandable groupings of syntax elements in code. ASTrust defines ten such categories (\eg Decisions, Data Structures, Operators) for interpretability.

\item \textbf{\infotheory:} A method leveraging information theory to interpret software traceability, analyzing metrics such as self-information, mutual information, and information loss.

\item \textbf{Trustworthiness:} The intrinsic quality of a system ensuring reliable, transparent, and ethical performance. In LLMs for code, trustworthiness is linked to the interpretability of model predictions.

\end{itemize}

%% file: appendixB.tex
\chapter{Information Transmission Exploratory Results}
\label{appendixB}

Probability distribution of similarity measures and information metrics.
\begin{figure}[hb]
		\centering
		\includegraphics[width=0.5\textwidth]{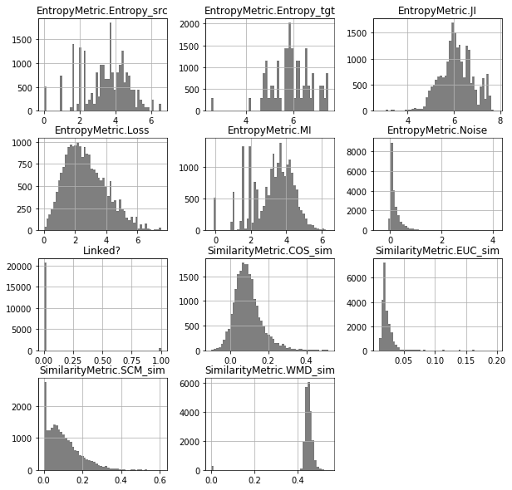}
		\caption{Probability distribution of measures of similarity and information metrics.}
        \vspace{0.3cm}
        \label{fig:manifold_info}
\end{figure}

Distribution of similarity measures and information metrics grouped by Ground Truth.
\begin{figure}[hb]
		\centering
		\includegraphics[width=0.50\textwidth]{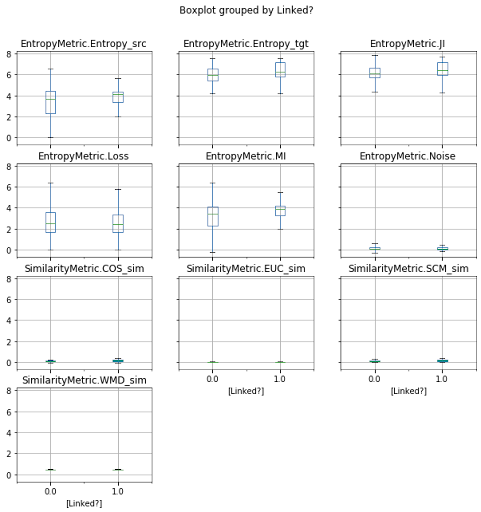}
		\caption{Distribution of measures of similarity and information metrics grouped by Ground Truth.}
        \vspace{0.3cm}
        \label{fig:manifold_gt}
\end{figure}

Scatter plot for measures of information vs. distances. 
\begin{figure}[hb]
		\centering
		\includegraphics[width=0.7\textwidth]{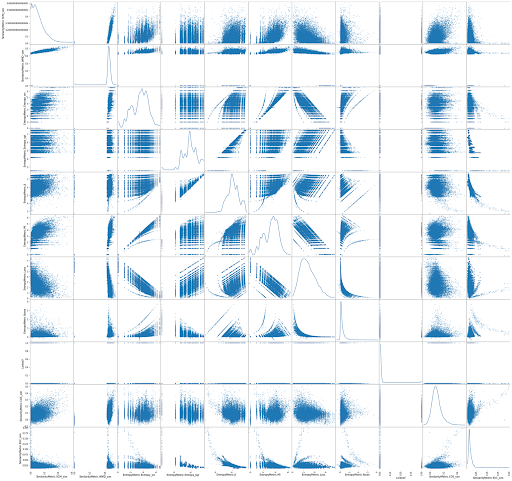}
		\caption{Correlation Analysis of Similarity and Information Measures.}
        \vspace{0.3cm}
        \label{fig:scatter}
\end{figure}

%% file: main.bbl
\newcommand{\etalchar}[1]{$^{#1}$}
\begin{thebibliography}{OdSSdASN{\etalchar{+}}25}

\bibitem[AAT10]{Asuncion:ICSE'10}
Hazeline~U. Asuncion, Arthur~U. Asuncion, and Richard~N. Taylor.
\newblock Software traceability with topic modeling.
\newblock In {\em Proceedings of the 32nd ACM/IEEE International Conference on Software Engineering - Volume 1}, ICSE '10, pages 95--104, 2010.

\bibitem[ABC{\etalchar{+}}16]{tensorflow2015-whitepaper}
Mart\'{\i}n Abadi, Paul Barham, Jianmin Chen, Zhifeng Chen, Andy Davis, Jeffrey Dean, Matthieu Devin, Sanjay Ghemawat, Geoffrey Irving, Michael Isard, Manjunath Kudlur, Josh Levenberg, Rajat Monga, Sherry Moore, Derek~G. Murray, Benoit Steiner, Paul Tucker, Vijay Vasudevan, Pete Warden, Martin Wicke, Yuan Yu, and Xiaoqiang Zheng.
\newblock Tensorflow: a system for large-scale machine learning.
\newblock In {\em Proceedings of the 12th USENIX Conference on Operating Systems Design and Implementation}, OSDI'16, page 265–283, USA, 2016. USENIX Association.

\bibitem[ABK18]{allamanis2018learning}
Miltiadis Allamanis, Marc Brockschmidt, and Mahmoud Khademi.
\newblock Learning to represent programs with graphs.
\newblock In {\em International Conference on Learning Representations}, 2018.

\bibitem[ACCL00]{Antoniol:ICSE'00}
Antoniol, Canfora, Casazza, and De~Lucia.
\newblock Information retrieval models for recovering traceability links between code and documentation.
\newblock In {\em Proceedings of the International Conference on Software Maintenance}, ICSE'00, pages 40--49, Oct 2000.

\bibitem[ACRC]{ahmad_unified_2021}
Wasi~Uddin Ahmad, Saikat Chakraborty, Baishakhi Ray, and Kai-Wei Chang.
\newblock Unified pre-training for program understanding and generation.

\bibitem[ACS24]{marl-book}
Stefano~V. Albrecht, Filippos Christianos, and Lukas Sch\"afer.
\newblock {\em Multi-Agent Reinforcement Learning: Foundations and Modern Approaches}.
\newblock MIT Press, 2024.

\bibitem[AHS20]{Aung2020ALR}
Thazin Win~Win Aung, Huan Huo, and Yulei Sui.
\newblock A literature review of automatic traceability links recovery for software change impact analysis.
\newblock {\em 2020 IEEE/ACM 28th International Conference on Program Comprehension (ICPC)}, pages 14--24, 2020.

\bibitem[All19]{Allamanis19}
Miltiadis Allamanis.
\newblock The adverse effects of code duplication in machine learning models of code.
\newblock In {\em Onward! OOPLSA 2019}, pages 143--153, 2019.

\bibitem[AMMIL12]{abu-mastafa}
Yaser~S. Abu-Mostafa, Malik Magdon-Ismail, and Hsuan-Tien Lin.
\newblock {\em Learning from Data: A Short Course}.
\newblock AMLBook, United States, 2012.

\bibitem[Ant24]{TheC3}
Anthropic.
\newblock The claude 3 model family: Opus, sonnet, haiku, 2024.
\newblock Preprint.

\bibitem[AON{\etalchar{+}}21]{austin2021program}
Jacob Austin, Augustus Odena, Maxwell Nye, Maarten Bosma, Henryk Michalewski, David Dohan, Ellen Jiang, Carrie Cai, Michael Terry, Quoc Le, and Charles Sutton.
\newblock Program synthesis with large language models, 2021.

\bibitem[BB24]{bishop_deep_2024}
Christopher~M. Bishop and Hugh Bishop.
\newblock {\em Deep Learning: Foundations and Concepts}.
\newblock Springer International Publishing, 2024.

\bibitem[BBH{\etalchar{+}}22]{GPTNeoX}
Sid Black, Stella Biderman, Eric Hallahan, Quentin Anthony, Leo Gao, Laurence Golding, Horace He, Connor Leahy, Kyle McDonell, Jason Phang, Michael Pieler, USVSN~Sai Prashanth, Shivanshu Purohit, Laria Reynolds, Jonathan Tow, Ben Wang, and Samuel Weinbach.
\newblock Gpt-neox-20b: An open-source autoregressive language model, 2022.

\bibitem[BD18]{Bassett2018MaximumEstimators}
Robert Bassett and Julio Deride.
\newblock {Maximum a posteriori estimators as a limit of Bayes estimators}, 2018.

\bibitem[BDV03]{Bengio2003AModel}
Yoshua Bengio, Réjean Ducharme, and Pascal Vincent.
\newblock {A neural probabilistic language model}.
\newblock {\em Advances in Neural Information Processing Systems}, 3:1137--1155, 2003.

\bibitem[Bed07]{Bedau2007ArtificialLife}
M~A Bedau.
\newblock {Artificial Life}.
\newblock {\em Philosophy of Biology}, pages 585--603, 2007.

\bibitem[BF20]{bastings_elephant_2020}
Jasmijn Bastings and Katja Filippova.
\newblock The elephant in the interpretability room: Why use attention as explanation when we have saliency methods?
\newblock In Afra Alishahi, Yonatan Belinkov, Grzegorz Chrupa{\l}a, Dieuwke Hupkes, Yuval Pinter, and Hassan Sajjad, editors, {\em Proceedings of the Third BlackboxNLP Workshop on Analyzing and Interpreting Neural Networks for NLP}, pages 149--155, Online, November 2020. Association for Computational Linguistics.

\bibitem[BGMMM21]{bender2021parrots}
Emily~M. Bender, Timnit Gebru, Angelina McMillan-Major, and Margaret Mitchell.
\newblock On the dangers of stochastic parrots: Can language models be too big?
\newblock In {\em Proceedings of the 2021 ACM Conference on Fairness, Accountability, and Transparency}, FAccT '21, page 610–623, New York, NY, USA, 2021. Association for Computing Machinery.

\bibitem[Bis06]{Bishop:2006}
Christopher~M. Bishop.
\newblock {\em Pattern Recognition and Machine Learning (Information Science and Statistics)}.
\newblock Springer-Verlag New York, Inc., 2006.

\bibitem[BKL09]{nltk}
Steven Bird, Ewan Klein, and Edward Loper.
\newblock {\em Natural Language Processing with Python}.
\newblock O'Reilly Media Inc., 2009.

\bibitem[BL04]{bird-loper-2004-nltk}
Steven Bird and Edward Loper.
\newblock {NLTK}: The natural language toolkit.
\newblock In {\em Proceedings of the {ACL} Interactive Poster and Demonstration Sessions}, pages 214--217, Barcelona, Spain, July 2004. Association for Computational Linguistics.

\bibitem[BLB{\etalchar{+}}13]{sklearn_api}
Lars Buitinck, Gilles Louppe, Mathieu Blondel, Fabian Pedregosa, Andreas Mueller, Olivier Grisel, Vlad Niculae, Peter Prettenhofer, Alexandre Gramfort, Jaques Grobler, Robert Layton, Jake VanderPlas, Arnaud Joly, Brian Holt, and Ga{\"{e}}l Varoquaux.
\newblock {API} design for machine learning software: experiences from the scikit-learn project.
\newblock In {\em ECML PKDD Workshop: Languages for Data Mining and Machine Learning}, pages 108--122, 2013.

\bibitem[BLKE24]{10608368}
Mayuri Bhadra, Daniela~Sanchez Lopera, Robert Kunzelmann, and Wolfgang Ecker.
\newblock A model-driven architecture approach to accelerate software code generation.
\newblock In {\em 2024 7th International Conference on Software and System Engineering (ICoSSE)}, pages 23--30, 2024.

\bibitem[BLP{\etalchar{+}}20]{brunner2020identifiability}
Gino Brunner, Yang Liu, Damián Pascual, Oliver Richter, Massimiliano Ciaramita, and Roger Wattenhofer.
\newblock On identifiability in transformers, 2020.

\bibitem[BMR{\etalchar{+}}20]{brown2020languagemodelsfewshotlearners}
Tom~B. Brown, Benjamin Mann, Nick Ryder, Melanie Subbiah, Jared Kaplan, Prafulla Dhariwal, Arvind Neelakantan, Pranav Shyam, Girish Sastry, Amanda Askell, Sandhini Agarwal, Ariel Herbert-Voss, Gretchen Krueger, Tom Henighan, Rewon Child, Aditya Ramesh, Daniel~M. Ziegler, Jeffrey Wu, Clemens Winter, Christopher Hesse, Mark Chen, Eric Sigler, Mateusz Litwin, Scott Gray, Benjamin Chess, Jack Clark, Christopher Berner, Sam McCandlish, Alec Radford, Ilya Sutskever, and Dario Amodei.
\newblock Language models are few-shot learners.
\newblock In {\em Proceedings of the 34th International Conference on Neural Information Processing Systems}, NIPS '20, Red Hook, NY, USA, 2020. Curran Associates Inc.

\bibitem[BMW94]{Biggerstaff:ACM'94}
Ted~J. Biggerstaff, Bharat~G. Mitbander, and Dallas~E. Webster.
\newblock Program understanding and the concept assignment problem.
\newblock {\em Commun. ACM}, 37(5):72--82, May 1994.

\bibitem[BQH{\etalchar{+}}25]{tree-sitter}
Max Brunsfeld, Amaan Qureshi, Andrew Hlynskyi, Patrick Thomson, ObserverOfTime, Will Lillis, Josh Vera, dundargoc, Phil Turnbull, Timothy Clem, Douglas Creager, Andrew Helwer, Rob Rix, Daumantas Kavolis, Hendrik van Antwerpen, Michael Davis, Christian Clason, Ika, Amin Ya, Riley Bruins, Tuan-Anh Nguyen, Stafford Brunk, Matt Massicotte, bfredl, Niranjan Hasabnis, Mingkai Dong, Samuel Moelius, Steven Kalt, and Kolja.
\newblock tree-sitter/tree-sitter: v0.25.3, March 2025.

\bibitem[BR21]{baltes_sampling_2021}
Sebastian Baltes and Paul Ralph.
\newblock Sampling in {Software} {Engineering} {Research}: {A} {Critical} {Review} and {Guidelines}, October 2021.
\newblock arXiv:2002.07764 [cs].

\bibitem[Bro96]{Brooke:96}
J.~Brooke.
\newblock {{SUS}}: {{A}} quick and dirty usability scale.
\newblock In P.~W. Jordan, B.~Weerdmeester, A.~Thomas, and I.~L. Mclelland, editors, {\em Usability Evaluation in Industry}. {Taylor and Francis}, London, 1996.

\bibitem[BSB{\etalchar{+}}23]{burnell_rethink_2023}
Ryan Burnell, Wout Schellaert, John Burden, Tomer~D. Ullman, Fernando Martinez-Plumed, Joshua~B. Tenenbaum, Danaja Rutar, Lucy~G. Cheke, Jascha Sohl-Dickstein, Melanie Mitchell, Douwe Kiela, Murray Shanahan, Ellen~M. Voorhees, Anthony~G. Cohn, Joel~Z. Leibo, and Jose Hernandez-Orallo.
\newblock Rethink reporting of evaluation results in ai.
\newblock {\em Science}, 380(6641):136--138, 2023.

\bibitem[Bun59]{bunge1959causality}
Mario Bunge.
\newblock {\em Causality: The Place of the Causal Principle in Modern Science}.
\newblock Harvard University Press, 1959.

\bibitem[Bun03]{bunge2003emergence}
Mario Bunge.
\newblock {\em Emergence and Convergence: Qualitative Novelty and the Unity of Knowledge}.
\newblock University of Toronto Press, 2003.

\bibitem[Bun11]{bunge2011philosophy}
Mario Bunge.
\newblock {\em Philosophy of Science: Volume 1 and 2}.
\newblock Routledge, 2011.

\bibitem[BVLR17]{BinghamBrown2017TheMutants}
David~Bingham Brown, Michael Vaughn, Ben Liblit, and Thomas Reps.
\newblock The care and feeding of wild-caught mutants.
\newblock In {\em Proceedings of the 2017 11th Joint Meeting on Foundations of Software Engineering}, ESEC/FSE 2017, page 511–522, New York, NY, USA, 2017. Association for Computing Machinery.

\bibitem[BYR{\etalchar{+}}24]{balayn2024empiricalexplorationtrustdynamics}
Agathe Balayn, Mireia Yurrita, Fanny Rancourt, Fabio Casati, and Ujwal Gadiraju.
\newblock An empirical exploration of trust dynamics in llm supply chains, 2024.

\bibitem[Car89]{cartwright1989capacities}
Nancy Cartwright.
\newblock {\em Nature’s Capacities and Their Measurement}.
\newblock Oxford University Press, 1989.

\bibitem[Car99]{cartwright1999dappled}
Nancy Cartwright.
\newblock {\em The Dappled World: A Study of the Boundaries of Science}.
\newblock Cambridge University Press, 1999.

\bibitem[Car07]{cartwright2007hunting}
Nancy Cartwright.
\newblock {\em Hunting Causes and Using Them: Approaches in Philosophy and Economics}.
\newblock Cambridge University Press, 2007.

\bibitem[CCP{\etalchar{+}}21a]{Ciniselli.MSR}
Matteo Ciniselli, Nathan Cooper, Luca Pascarella, Denys Poshyvanyk, Massimiliano Di~Penta, and Gabriele Bavota.
\newblock An empirical study on the usage of bert models for code completion.
\newblock In {\em 2021 IEEE/ACM 18th International Conference on Mining Software Repositories (MSR)}, pages 108--119, 2021.

\bibitem[CCP{\etalchar{+}}21b]{MSR-Completion}
Matteo Ciniselli, Nathan Cooper, Luca Pascarella, Denys Poshyvanyk, Massimiliano~Di Penta, and Gabriele Bavota.
\newblock An empirical study on the usage of {BERT} models for code completion.
\newblock {\em CoRR}, abs/2103.07115, 2021.

\bibitem[CCP{\etalchar{+}}22]{Ciniselli.TSE}
Matteo Ciniselli, Nathan Cooper, Luca Pascarella, Antonio Mastropaolo, Emad Aghajani, Denys Poshyvanyk, Massimiliano Di~Penta, and Gabriele Bavota.
\newblock An empirical study on the usage of transformer models for code completion.
\newblock {\em IEEE Transactions on Software Engineering}, 48(12):4818--4837, 2022.

\bibitem[CDK15]{Corley2015}
C.~S. Corley, K.~Damevski, and N.~A. Kraft.
\newblock Exploring the use of deep learning for feature location.
\newblock In {\em 2015 {{IEEE International Conference}} on {{Software Maintenance}} and {{Evolution}} ({{ICSME}})}, ICSME'15, pages 556--560, September 2015.
\newblock ISSN:.

\bibitem[CDMC22]{cito_counterfactual_2022}
Jürgen Cito, Isil Dillig, Vijayaraghavan Murali, and Satish Chandra.
\newblock Counterfactual explanations for models of code.
\newblock In {\em Proceedings of the 44th {International} {Conference} on {Software} {Engineering}: {Software} {Engineering} in {Practice}}, pages 125--134, Pittsburgh Pennsylvania, May 2022. ACM.

\bibitem[CGN{\etalchar{+}}22]{cassano_multipl-e_2022}
Federico Cassano, John Gouwar, Daniel Nguyen, Sydney Nguyen, Luna Phipps-Costin, Donald Pinckney, Ming-Ho Yee, Yangtian Zi, Carolyn~Jane Anderson, Molly~Q. Feldman, Arjun Guha, Michael Greenberg, and Abhinav Jangda.
\newblock {MultiPL}-{E}: {A} {Scalable} and {Extensible} {Approach} to {Benchmarking} {Neural} {Code} {Generation}, December 2022.
\newblock arXiv:2208.08227 [cs].

\bibitem[Cha06]{kathy2006}
Kathy Charmaz.
\newblock {\em Constructing Grounded Theory: A Practical Guide through Qualitative Analysis}.
\newblock SAGE Publications Inc., 2006.

\bibitem[CHCC03]{Cleland-Huang:TSE'03}
Jane Cleland-Huang, Carl~K. Chang, and Mark Christensen.
\newblock Event-based traceability for managing evolutionary change.
\newblock {\em IEEE Trans. Softw. Eng.}, 29(9), September 2003.

\bibitem[CHCGE10]{Cleland-Huang:ICSE'10}
Jane Cleland-Huang, Adam Czauderna, Marek Gibiec, and John Emenecker.
\newblock A machine learning approach for tracing regulatory codes to product specific requirements.
\newblock In {\em Proceedings of the 32nd ACM/IEEE International Conference on Software Engineering}, ICSE'10, pages 155--164. ACM, 2010.

\bibitem[CHCP22]{CanWeFix}
A.~Connor, A.~Harris, N.~Cooper, and D.~Poshyvanyk.
\newblock Can we automatically fix bugs by learning edit operations?
\newblock In {\em 2022 IEEE International Conference on Software Analysis, Evolution and Reengineering (SANER)}, pages 782--792, Los Alamitos, CA, USA, mar 2022. IEEE Computer Society.

\bibitem[CHGHH{\etalchar{+}}14]{Cleland-Huang:FOSE'14}
Jane Cleland-Huang, Orlena C.~Z. Gotel, Jane Huffman~Hayes, Patrick M\"{a}der, and Andrea Zisman.
\newblock Software traceability: Trends and future directions.
\newblock In {\em Proceedings of the on Future of Software Engineering}, FOSE'14, pages 55--69. ACM, 2014.

\bibitem[CHGZ12]{Cleland-Huang:Springer'12}
Jane Cleland-Huang, Orlena Gotel, and Andrea Zisman.
\newblock {\em Software and Systems Traceability}.
\newblock Springer Publishing Company, Incorporated, 2012.

\bibitem[CHRM14]{Cleland-Huang:FSE'14}
Jane Cleland-Huang, Mona Rahimi, and Patrick M\"{a}der.
\newblock Achieving lightweight trustworthy traceability.
\newblock In {\em Proceedings of the 22nd ACM SIGSOFT International Symposium on Foundations of Software Engineering}, FSE'14, pages 849--852, 2014.

\bibitem[CKT{\etalchar{+}}21]{Chen2019}
Zimin Chen, Steve Kommrusch, Michele Tufano, Louis-Noël Pouchet, Denys Poshyvanyk, and Martin Monperrus.
\newblock Sequencer: Sequence-to-sequence learning for end-to-end program repair.
\newblock {\em IEEE Transactions on Software Engineering}, 47(9):1943--1959, 2021.

\bibitem[CLL{\etalchar{+}}]{clement_long-range_2021}
Colin~B. Clement, Shuai Lu, Xiaoyu Liu, Michele Tufano, Dawn Drain, Nan Duan, Neel Sundaresan, and Alexey Svyatkovskiy.
\newblock Long-range modeling of source code files with {eWASH}: Extended window access by syntax hierarchy.

\bibitem[Clo25]{GoogleBigQuery}
Google Cloud.
\newblock Bigquery: Serverless, highly scalable, and cost-effective multi-cloud data warehouse, 2025.
\newblock Accessed: 2025-04-07.

\bibitem[CPS{\etalchar{+}}24]{clement2024debugging}
Colin~Bruce Clement, David Alberto~Nader Palacio, Neelakantan Sundaresan, Alexey Svyatkovskiy, and Michele Tufano.
\newblock Debugging tool for code generation neural language models, March 2024.
\newblock Application No. 18/082366, filed on December 15, 2022.

\bibitem[CS13]{libest}
Inc. Cisco~Systems.
\newblock Libest: Enrollment over secure transport (est) library, 2013.
\newblock Accessed: 2025-04-07.

\bibitem[CSGH24]{10.1145/3597503.3639173}
Yizhou Chen, Zeyu Sun, Zhihao Gong, and Dan Hao.
\newblock Improving smart contract security with contrastive learning-based vulnerability detection.
\newblock In {\em Proceedings of the IEEE/ACM 46th International Conference on Software Engineering}, ICSE '24, New York, NY, USA, 2024. Association for Computing Machinery.

\bibitem[CSH{\etalchar{+}}22]{chen_what_2022}
Zixi Chen, Varshini Subhash, Marton Havasi, Weiwei Pan, and Finale Doshi-Velez.
\newblock What {Makes} a {Good} {Explanation}?: {A} {Harmonized} {View} of {Properties} of {Explanations}, December 2022.
\newblock arXiv:2211.05667 [cs].

\bibitem[CSW{\etalchar{+}}24]{10764854}
Sicong Cao, Xiaobing Sun, Xiaoxue Wu, David Lo, Lili Bo, Bin Li, Xiaolei Liu, Xingwei Lin, and Wei Liu.
\newblock Snopy: Bridging sample denoising with causal graph learning for effective vulnerability detection.
\newblock In {\em 2024 39th IEEE/ACM International Conference on Automated Software Engineering (ASE)}, pages 606--618, 2024.

\bibitem[CTJ{\etalchar{+}}21]{Chen2021}
Mark Chen, Jerry Tworek, Heewoo Jun, Qiming Yuan, Henrique~Ponde de~Oliveira~Pinto, Jared Kaplan, Harri Edwards, Yuri Burda, Nicholas Joseph, Greg Brockman, Alex Ray, Raul Puri, Gretchen Krueger, Michael Petrov, Heidy Khlaaf, Girish Sastry, Pamela Mishkin, Brooke Chan, Scott Gray, Nick Ryder, Mikhail Pavlov, Alethea Power, Lukasz Kaiser, Mohammad Bavarian, Clemens Winter, Philippe Tillet, Felipe~Petroski Such, Dave Cummings, Matthias Plappert, Fotios Chantzis, Elizabeth Barnes, Ariel Herbert-Voss, William~Hebgen Guss, Alex Nichol, Alex Paino, Nikolas Tezak, Jie Tang, Igor Babuschkin, Suchir Balaji, Shantanu Jain, William Saunders, Christopher Hesse, Andrew~N. Carr, Jan Leike, Josh Achiam, Vedant Misra, Evan Morikawa, Alec Radford, Matthew Knight, Miles Brundage, Mira Murati, Katie Mayer, Peter Welinder, Bob McGrew, Dario Amodei, Sam McCandlish, Ilya Sutskever, and Wojciech Zaremba.
\newblock Evaluating large language models trained on code, 2021.

\bibitem[CvMBB14]{cho_properties_2014}
Kyunghyun Cho, Bart van Merrienboer, Dzmitry Bahdanau, and Yoshua Bengio.
\newblock On the {Properties} of {Neural} {Machine} {Translation}: {Encoder}-{Decoder} {Approaches}, October 2014.
\newblock arXiv:1409.1259 [cs, stat].

\bibitem[dBLMT11]{softtrust}
Vieri del Bianco, Luigi Lavazza, Sandro Morasca, and Davide Taibi.
\newblock A survey on open source software trustworthiness.
\newblock {\em IEEE Software}, 28(5):67--75, 2011.

\bibitem[DDH{\etalchar{+}}21]{dai2021knowledge}
Damai Dai, Li~Dong, Yaru Hao, Zhifang Sui, and Furu Wei.
\newblock Knowledge neurons in pretrained transformers.
\newblock {\em CoRR}, abs/2104.08696, 2021.

\bibitem[DGM{\etalchar{+}}13]{Dasgupta2013}
Tathagata Dasgupta, Mark Grechanik, Evan Moritz, Bogdan Dit, and Denys Poshyvanyk.
\newblock {Enhancing software traceability by automatically expanding corpora with relevant documentation}.
\newblock {\em IEEE International Conference on Software Maintenance, ICSM}, pages 320--329, 2013.

\bibitem[DHS{\etalchar{+}}07]{Dekhtyar:RE'07}
Alex Dekhtyar, Jane~Huffman Hayes, Senthil~Karthikeyan Sundaram, Elizabeth~Ashlee Holbrook, and Olga Dekhtyar.
\newblock Technique integration for requirements assessment.
\newblock {\em Proceedings of the 15th IEEE International Requirements Engineering Conference}, pages 141--150, 2007.

\bibitem[DhSqH{\etalchar{+}}20]{Du2020AutomaticTL}
Tianlong Du, Guo hua Shen, Zhi qiu Huang, Yaoliang Yu, and De~xiang Wu.
\newblock Automatic traceability link recovery via active learning.
\newblock {\em Frontiers of Information Technology \& Electronic Engineering}, 21:1217 -- 1225, 2020.

\bibitem[DJP{\etalchar{+}}24]{dubey2024llama3herdmodels}
Abhimanyu Dubey, Abhinav Jauhri, Abhinav Pandey, Abhishek Kadian, Ahmad Al-Dahle, Aiesha Letman, Akhil Mathur, Alan Schelten, Amy Yang, Angela Fan, Anirudh Goyal, Anthony Hartshorn, Aobo Yang, Archi Mitra, Archie Sravankumar, Artem Korenev, Arthur Hinsvark, Arun Rao, Aston Zhang, Aurelien Rodriguez, Austen Gregerson, Ava Spataru, Baptiste Roziere, Bethany Biron, Binh Tang, Bobbie Chern, Charlotte Caucheteux, Chaya Nayak, Chloe Bi, Chris Marra, Chris McConnell, Christian Keller, Christophe Touret, Chunyang Wu, Corinne Wong, Cristian~Canton Ferrer, Cyrus Nikolaidis, Damien Allonsius, Daniel Song, Danielle Pintz, Danny Livshits, David Esiobu, Dhruv Choudhary, Dhruv Mahajan, Diego Garcia-Olano, Diego Perino, Dieuwke Hupkes, Egor Lakomkin, Ehab AlBadawy, Elina Lobanova, Emily Dinan, Eric~Michael Smith, Filip Radenovic, Frank Zhang, Gabriel Synnaeve, Gabrielle Lee, Georgia~Lewis Anderson, Graeme Nail, Gregoire Mialon, Guan Pang, Guillem Cucurell, Hailey Nguyen, Hannah Korevaar, Hu~Xu, Hugo Touvron, Iliyan Zarov,
  Imanol~Arrieta Ibarra, Isabel Kloumann, Ishan Misra, Ivan Evtimov, Jade Copet, Jaewon Lee, Jan Geffert, Jana Vranes, Jason Park, Jay Mahadeokar, Jeet Shah, Jelmer van~der Linde, Jennifer Billock, Jenny Hong, Jenya Lee, Jeremy Fu, Jianfeng Chi, Jianyu Huang, Jiawen Liu, Jie Wang, Jiecao Yu, Joanna Bitton, Joe Spisak, Jongsoo Park, Joseph Rocca, Joshua Johnstun, Joshua Saxe, Junteng Jia, Kalyan~Vasuden Alwala, Kartikeya Upasani, Kate Plawiak, Ke~Li, Kenneth Heafield, Kevin Stone, Khalid El-Arini, Krithika Iyer, Kshitiz Malik, Kuenley Chiu, Kunal Bhalla, Lauren Rantala-Yeary, Laurens van~der Maaten, Lawrence Chen, Liang Tan, Liz Jenkins, Louis Martin, Lovish Madaan, Lubo Malo, Lukas Blecher, Lukas Landzaat, Luke de~Oliveira, Madeline Muzzi, Mahesh Pasupuleti, Mannat Singh, Manohar Paluri, Marcin Kardas, Mathew Oldham, Mathieu Rita, Maya Pavlova, Melanie Kambadur, Mike Lewis, Min Si, Mitesh~Kumar Singh, Mona Hassan, Naman Goyal, Narjes Torabi, Nikolay Bashlykov, Nikolay Bogoychev, Niladri Chatterji, Olivier
  Duchenne, Onur Çelebi, Patrick Alrassy, Pengchuan Zhang, Pengwei Li, Petar Vasic, Peter Weng, Prajjwal Bhargava, Pratik Dubal, Praveen Krishnan, Punit~Singh Koura, Puxin Xu, Qing He, Qingxiao Dong, Ragavan Srinivasan, Raj Ganapathy, Ramon Calderer, Ricardo~Silveira Cabral, Robert Stojnic, Roberta Raileanu, Rohit Girdhar, Rohit Patel, Romain Sauvestre, Ronnie Polidoro, Roshan Sumbaly, Ross Taylor, Ruan Silva, Rui Hou, Rui Wang, Saghar Hosseini, Sahana Chennabasappa, Sanjay Singh, Sean Bell, Seohyun~Sonia Kim, Sergey Edunov, Shaoliang Nie, Sharan Narang, Sharath Raparthy, Sheng Shen, Shengye Wan, Shruti Bhosale, Shun Zhang, Simon Vandenhende, Soumya Batra, Spencer Whitman, Sten Sootla, Stephane Collot, Suchin Gururangan, Sydney Borodinsky, Tamar Herman, Tara Fowler, Tarek Sheasha, Thomas Georgiou, Thomas Scialom, Tobias Speckbacher, Todor Mihaylov, Tong Xiao, Ujjwal Karn, Vedanuj Goswami, Vibhor Gupta, Vignesh Ramanathan, Viktor Kerkez, Vincent Gonguet, Virginie Do, Vish Vogeti, Vladan Petrovic, Weiwei Chu,
  Wenhan Xiong, Wenyin Fu, Whitney Meers, Xavier Martinet, Xiaodong Wang, Xiaoqing~Ellen Tan, Xinfeng Xie, Xuchao Jia, Xuewei Wang, Yaelle Goldschlag, Yashesh Gaur, Yasmine Babaei, Yi~Wen, Yiwen Song, Yuchen Zhang, Yue Li, Yuning Mao, Zacharie~Delpierre Coudert, Zheng Yan, Zhengxing Chen, Zoe Papakipos, Aaditya Singh, Aaron Grattafiori, Abha Jain, Adam Kelsey, Adam Shajnfeld, Adithya Gangidi, Adolfo Victoria, Ahuva Goldstand, Ajay Menon, Ajay Sharma, Alex Boesenberg, Alex Vaughan, Alexei Baevski, Allie Feinstein, Amanda Kallet, Amit Sangani, Anam Yunus, Andrei Lupu, Andres Alvarado, Andrew Caples, Andrew Gu, Andrew Ho, Andrew Poulton, Andrew Ryan, Ankit Ramchandani, Annie Franco, Aparajita Saraf, Arkabandhu Chowdhury, Ashley Gabriel, Ashwin Bharambe, Assaf Eisenman, Azadeh Yazdan, Beau James, Ben Maurer, Benjamin Leonhardi, Bernie Huang, Beth Loyd, Beto~De Paola, Bhargavi Paranjape, Bing Liu, Bo~Wu, Boyu Ni, Braden Hancock, Bram Wasti, Brandon Spence, Brani Stojkovic, Brian Gamido, Britt Montalvo, Carl
  Parker, Carly Burton, Catalina Mejia, Changhan Wang, Changkyu Kim, Chao Zhou, Chester Hu, Ching-Hsiang Chu, Chris Cai, Chris Tindal, Christoph Feichtenhofer, Damon Civin, Dana Beaty, Daniel Kreymer, Daniel Li, Danny Wyatt, David Adkins, David Xu, Davide Testuggine, Delia David, Devi Parikh, Diana Liskovich, Didem Foss, Dingkang Wang, Duc Le, Dustin Holland, Edward Dowling, Eissa Jamil, Elaine Montgomery, Eleonora Presani, Emily Hahn, Emily Wood, Erik Brinkman, Esteban Arcaute, Evan Dunbar, Evan Smothers, Fei Sun, Felix Kreuk, Feng Tian, Firat Ozgenel, Francesco Caggioni, Francisco Guzmán, Frank Kanayet, Frank Seide, Gabriela~Medina Florez, Gabriella Schwarz, Gada Badeer, Georgia Swee, Gil Halpern, Govind Thattai, Grant Herman, Grigory Sizov, Guangyi, Zhang, Guna Lakshminarayanan, Hamid Shojanazeri, Han Zou, Hannah Wang, Hanwen Zha, Haroun Habeeb, Harrison Rudolph, Helen Suk, Henry Aspegren, Hunter Goldman, Igor Molybog, Igor Tufanov, Irina-Elena Veliche, Itai Gat, Jake Weissman, James Geboski, James Kohli,
  Japhet Asher, Jean-Baptiste Gaya, Jeff Marcus, Jeff Tang, Jennifer Chan, Jenny Zhen, Jeremy Reizenstein, Jeremy Teboul, Jessica Zhong, Jian Jin, Jingyi Yang, Joe Cummings, Jon Carvill, Jon Shepard, Jonathan McPhie, Jonathan Torres, Josh Ginsburg, Junjie Wang, Kai Wu, Kam~Hou U, Karan Saxena, Karthik Prasad, Kartikay Khandelwal, Katayoun Zand, Kathy Matosich, Kaushik Veeraraghavan, Kelly Michelena, Keqian Li, Kun Huang, Kunal Chawla, Kushal Lakhotia, Kyle Huang, Lailin Chen, Lakshya Garg, Lavender A, Leandro Silva, Lee Bell, Lei Zhang, Liangpeng Guo, Licheng Yu, Liron Moshkovich, Luca Wehrstedt, Madian Khabsa, Manav Avalani, Manish Bhatt, Maria Tsimpoukelli, Martynas Mankus, Matan Hasson, Matthew Lennie, Matthias Reso, Maxim Groshev, Maxim Naumov, Maya Lathi, Meghan Keneally, Michael~L. Seltzer, Michal Valko, Michelle Restrepo, Mihir Patel, Mik Vyatskov, Mikayel Samvelyan, Mike Clark, Mike Macey, Mike Wang, Miquel~Jubert Hermoso, Mo~Metanat, Mohammad Rastegari, Munish Bansal, Nandhini Santhanam, Natascha
  Parks, Natasha White, Navyata Bawa, Nayan Singhal, Nick Egebo, Nicolas Usunier, Nikolay~Pavlovich Laptev, Ning Dong, Ning Zhang, Norman Cheng, Oleg Chernoguz, Olivia Hart, Omkar Salpekar, Ozlem Kalinli, Parkin Kent, Parth Parekh, Paul Saab, Pavan Balaji, Pedro Rittner, Philip Bontrager, Pierre Roux, Piotr Dollar, Polina Zvyagina, Prashant Ratanchandani, Pritish Yuvraj, Qian Liang, Rachad Alao, Rachel Rodriguez, Rafi Ayub, Raghotham Murthy, Raghu Nayani, Rahul Mitra, Raymond Li, Rebekkah Hogan, Robin Battey, Rocky Wang, Rohan Maheswari, Russ Howes, Ruty Rinott, Sai~Jayesh Bondu, Samyak Datta, Sara Chugh, Sara Hunt, Sargun Dhillon, Sasha Sidorov, Satadru Pan, Saurabh Verma, Seiji Yamamoto, Sharadh Ramaswamy, Shaun Lindsay, Shaun Lindsay, Sheng Feng, Shenghao Lin, Shengxin~Cindy Zha, Shiva Shankar, Shuqiang Zhang, Shuqiang Zhang, Sinong Wang, Sneha Agarwal, Soji Sajuyigbe, Soumith Chintala, Stephanie Max, Stephen Chen, Steve Kehoe, Steve Satterfield, Sudarshan Govindaprasad, Sumit Gupta, Sungmin Cho, Sunny
  Virk, Suraj Subramanian, Sy~Choudhury, Sydney Goldman, Tal Remez, Tamar Glaser, Tamara Best, Thilo Kohler, Thomas Robinson, Tianhe Li, Tianjun Zhang, Tim Matthews, Timothy Chou, Tzook Shaked, Varun Vontimitta, Victoria Ajayi, Victoria Montanez, Vijai Mohan, Vinay~Satish Kumar, Vishal Mangla, Vlad Ionescu, Vlad Poenaru, Vlad~Tiberiu Mihailescu, Vladimir Ivanov, Wei Li, Wenchen Wang, Wenwen Jiang, Wes Bouaziz, Will Constable, Xiaocheng Tang, Xiaofang Wang, Xiaojian Wu, Xiaolan Wang, Xide Xia, Xilun Wu, Xinbo Gao, Yanjun Chen, Ye~Hu, Ye~Jia, Ye~Qi, Yenda Li, Yilin Zhang, Ying Zhang, Yossi Adi, Youngjin Nam, Yu, Wang, Yuchen Hao, Yundi Qian, Yuzi He, Zach Rait, Zachary DeVito, Zef Rosnbrick, Zhaoduo Wen, Zhenyu Yang, and Zhiwei Zhao.
\newblock The llama 3 herd of models, 2024.

\bibitem[DLOS06]{DeLucia:ICSM'06}
Andrea De~Lucia, Rocco Oliveto, and Paola Sgueglia.
\newblock Incremental approach and user feedbacks: a silver bullet for traceability recovery.
\newblock In {\em Proceedings of the International Conference on Software Maintenance}, ICSM'06, pages 299--309, 2006.

\bibitem[DLOT08]{DeLucia:ASE'08}
Andrea De~Lucia, Rocco Oliveto, and Genoveffa Tortora.
\newblock Ir-based traceability recovery processes: An empirical comparison of one-shot and incremental processes.
\newblock In {\em Proceedings of the 2008 23rd IEEE/ACM International Conference on Automated Software Engineering}, pages 39--48. IEEE Computer Society, 2008.

\bibitem[DLOT09]{DeLucia:EMSE'09}
Andrea De~Lucia, Rocco Oliveto, and Genoveffa Tortora.
\newblock Assessing ir-based traceability recovery tools through controlled experiments.
\newblock {\em Empirical Software Engineering}, 14(1):57--92, 2009.

\bibitem[DMLVP13]{Dit2013}
Bogdan Dit, Evan Moritz, Mario Linares-V{\'{a}}squez, and Denys Poshyvanyk.
\newblock {Supporting and accelerating reproducible research in software maintenance using tracelab component library}.
\newblock {\em IEEE International Conference on Software Maintenance, ICSM}, pages 330--339, 2013.

\bibitem[DPM{\etalchar{+}}13]{Dit2013b}
Bogdan Dit, Annibale Panichella, Evan Moritz, Rocco Oliveto, Massimilano Di~Penta, Denys Poshyvanyk, and Andrea De~Lucia.
\newblock Configuring topic models for software engineering tasks in tracelab.
\newblock In {\em 2013 7th International Workshop on Traceability in Emerging Forms of Software Engineering (TEFSE)}, pages 105--109, 2013.

\bibitem[DRP13]{Dit2013a}
Bogdan Dit, Meghan Revelle, and Denys Poshyvanyk.
\newblock {Integrating information retrieval, execution and link analysis algorithms to improve feature location in software}.
\newblock {\em Empirical Software Engineering}, 18(2):277--309, 2013.

\bibitem[DS24]{Dorin2024Alife}
Alan Dorin and Susan Stepney.
\newblock What is artificial life today, and where should it go?
\newblock {\em Artificial Life}, 30(1):1--15, 02 2024.

\bibitem[DTG{\etalchar{+}}21]{dehghani2021benchmark}
Mostafa Dehghani, Yi~Tay, Alexey~A. Gritsenko, Zhe Zhao, Neil Houlsby, Fernando Diaz, Donald Metzler, and Oriol Vinyals.
\newblock The benchmark lottery.
\newblock {\em CoRR}, abs/2107.07002, 2021.

\bibitem[DVK17]{doshivelez2017rigorousscienceinterpretablemachine}
Finale Doshi-Velez and Been Kim.
\newblock Towards a rigorous science of interpretable machine learning, 2017.

\bibitem[DVK18]{Doshi-Velez2018ConsiderationsLearning}
Finale Doshi-Velez and Been Kim.
\newblock {\em Considerations for Evaluation and Generalization in Interpretable Machine Learning}, pages 3--17.
\newblock Springer International Publishing, Cham, 2018.

\bibitem[EGK{\etalchar{+}}01]{895984}
S.G. Eick, T.L. Graves, A.F. Karr, J.S. Marron, and A.~Mockus.
\newblock Does code decay? assessing the evidence from change management data.
\newblock {\em IEEE Transactions on Software Engineering}, 27(1):1--12, 2001.

\bibitem[ELGB24]{10.1145/3597503.3639095}
Hadeel Eladawy, Claire Le~Goues, and Yuriy Brun.
\newblock Automated program repair, what is it good for? not absolutely nothing!
\newblock In {\em Proceedings of the IEEE/ACM 46th International Conference on Software Engineering}, ICSE '24, New York, NY, USA, 2024. Association for Computing Machinery.

\bibitem[FDPCC17]{Falessi:EMSE17}
Davide Falessi, Massimiliano Di~Penta, Gerardo Canfora, and Giovanni Cantone.
\newblock Estimating the number of remaining links in traceability recovery.
\newblock {\em Empirical Software Engineering}, 22(3):996--1027, 2017.

\bibitem[FFT19]{Furia:TSE'19}
Carlo~A. Furia, Robert Feldt, and Richard Torkar.
\newblock Bayesian data analysis in empirical software engineering research.
\newblock {\em IEEE Transactions on Software Engineering}, abs/1811.05422, 2019.

\bibitem[FGM{\etalchar{+}}23]{synerror}
Zhiyu Fan, Xiang Gao, Martin Mirchev, Abhik Roychoudhury, and Shin~Hwei Tan.
\newblock Automated repair of programs from large language models.
\newblock In {\em 2023 IEEE/ACM 45th International Conference on Software Engineering (ICSE)}, pages 1469--1481, 2023.

\bibitem[FGT{\etalchar{+}}20]{feng2020codebertpretrainedmodelprogramming}
Zhangyin Feng, Daya Guo, Duyu Tang, Nan Duan, Xiaocheng Feng, Ming Gong, Linjun Shou, Bing Qin, Ting Liu, Daxin Jiang, and Ming Zhou.
\newblock {C}ode{BERT}: A pre-trained model for programming and natural languages.
\newblock In Trevor Cohn, Yulan He, and Yang Liu, editors, {\em Findings of the Association for Computational Linguistics: EMNLP 2020}, pages 1536--1547, Online, November 2020. Association for Computational Linguistics.

\bibitem[FNT{\etalchar{+}}23]{fu2023vul-explain}
Michael Fu, Van Nguyen, Chakkrit~Kla Tantithamthavorn, Trung Le, and Dinh Phung.
\newblock Vulexplainer: A transformer-based hierarchical distillation for explaining vulnerability types.
\newblock {\em IEEE Transactions on Software Engineering}, 49(10):4550--4565, 2023.

\bibitem[FPMH]{flora_comparing_2022}
Montgomery Flora, Corey Potvin, Amy {McGovern}, and Shawn Handler.
\newblock Comparing explanation methods for traditional machine learning models part 1: An overview of current methods and quantifying their disagreement.

\bibitem[FWF{\etalchar{+}}24]{10765036}
Yujia Fan, Sinan Wang, Zebang Fei, Yao Qin, Huaxuan Li, and Yepang Liu.
\newblock Can cooperative multi-agent reinforcement learning boost automatic web testing? an exploratory study.
\newblock In {\em 2024 39th IEEE/ACM International Conference on Automated Software Engineering (ASE)}, pages 14--26, 2024.

\bibitem[FZ16]{Furtado:RE'16}
Felipe Furtado and Andrea Zisman.
\newblock Trace++: A traceability approach to support transitioning to agile software engineering.
\newblock In {\em Requirements Engineering Conference (RE), 2016 IEEE 24th International}, pages 66--75. IEEE, 2016.

\bibitem[GBB{\etalchar{+}}20]{gao2020pile}
Leo Gao, Stella Biderman, Sid Black, Laurence Golding, Travis Hoppe, Charles Foster, Jason Phang, Horace He, Anish Thite, Noa Nabeshima, Shawn Presser, and Connor Leahy.
\newblock The pile: An 800gb dataset of diverse text for language modeling, 2020.

\bibitem[GBB{\etalchar{+}}21]{codeparrot}
Leo Gao, Stella Biderman, Sid Black, Laurence Golding, Travis Hoppe, Charles Foster, Jason Phang, Horace He, Anish Thite, Noa Nabeshima, et~al.
\newblock Codeparrot, 2021.

\bibitem[GCCH17]{Guo:ICSE'17}
Jin Guo, Jinghui Cheng, and Jane Cleland-Huang.
\newblock Semantically enhanced software traceability using deep learning techniques.
\newblock In {\em Proceedings of the 39th International Conference on Software Engineering}, ICSE'17, pages 3--14. IEEE Press, 2017.

\bibitem[GCHB13]{Guo2013FoundationsTraceability}
Jin Guo, Jane Cleland-Huang, and Brian Berenbach.
\newblock Foundations for an expert system in domain-specific traceability.
\newblock In {\em 2013 21st IEEE International Requirements Engineering Conference (RE)}, pages 42--51, 2013.

\bibitem[GDMH12]{Ge2012}
Xi~Ge, Quinton~L Dubose, and Emerson Murphy-Hill.
\newblock {Reconciling Manual and Automatic Refactoring}.
\newblock In {\em 2012 34th International Conference on Software Engineering (ICSE)}, pages 211 -- 221, Zurich, 2012. IEEE.

\bibitem[git20]{github}
github.
\newblock Github, 2020.

\bibitem[{Git}25]{github_copilot}
{GitHub}.
\newblock Github copilot, 2025.
\newblock Accessed: 2025-04-07.

\bibitem[GOPL11a]{Gethers:ICSM'11}
M.~Gethers, R.~Oliveto, D.~Poshyvanyk, and A.~D. Lucia.
\newblock On integrating orthogonal information retrieval methods to improve traceability recovery.
\newblock In {\em Proceedings of the International Conference on Software Maintenance}, ICSM'11, pages 133--142, 2011.

\bibitem[GOPL11b]{6080780}
Malcom Gethers, Rocco Oliveto, Denys Poshyvanyk, and Andrea~De Lucia.
\newblock On integrating orthogonal information retrieval methods to improve traceability recovery.
\newblock In {\em 2011 27th IEEE International Conference on Software Maintenance (ICSM)}, pages 133--142, 2011.

\bibitem[GRCH{\etalchar{+}}16]{Guo:MSR'16}
Jin Guo, Mona Rahimi, Jane Cleland-Huang, Alexander Rasin, Jane~Huffman Hayes, and Michael Vierhauser.
\newblock Cold-start {{Software Analytics}}.
\newblock In {\em Proceedings of the 13th {{International Conference}} on {{Mining Software Repositories}}}, MSR '16, pages 142--153, Austin, Texas, 2016. {ACM}.

\bibitem[GRL{\etalchar{+}}21]{guo2021graphcodebert}
Daya Guo, Shuo Ren, Shuai Lu, Zhangyin Feng, Duyu Tang, Shujie LIU, Long Zhou, Nan Duan, Alexey Svyatkovskiy, Shengyu Fu, Michele Tufano, Shao~Kun Deng, Colin Clement, Dawn Drain, Neel Sundaresan, Jian Yin, Daxin Jiang, and Ming Zhou.
\newblock Graphcode{\{}bert{\}}: Pre-training code representations with data flow.
\newblock In {\em International Conference on Learning Representations}, 2021.

\bibitem[GRM21]{Gadelha2021TraceabilityRB}
Guilherme Gadelha, Franklin Ramalho, and Tiago~Lima Massoni.
\newblock Traceability recovery between bug reports and test cases-a mozilla firefox case study.
\newblock {\em Automated Software Engineering}, 28, 2021.

\bibitem[GRP{\etalchar{+}}23]{gruber2023automatictestgenerationtools}
Martin Gruber, Muhammad~Firhard Roslan, Owain Parry, Fabian Scharnböck, Phil McMinn, and Gordon Fraser.
\newblock Do automatic test generation tools generate flaky tests?, 2023.

\bibitem[GWZK19]{Ghorbani19}
Amirata Ghorbani, James Wexler, James Zou, and Been Kim.
\newblock {\em Towards automatic concept-based explanations}.
\newblock Curran Associates Inc., Red Hook, NY, USA, 2019.

\bibitem[GZK18]{Gu2018}
Xiaodong Gu, Hongyu Zhang, and Sunghun Kim.
\newblock Deep code search.
\newblock In {\em Proceedings of the 40th International Conference on Software Engineering}, ICSE '18, pages 933--944, New York, NY, USA, 2018. ACM.

\bibitem[HBK{\etalchar{+}}21]{hendrycksapps2021}
Dan Hendrycks, Steven Basart, Saurav Kadavath, Mantas Mazeika, Akul Arora, Ethan Guo, Collin Burns, Samir Puranik, Horace He, Dawn Song, and Jacob Steinhardt.
\newblock Measuring coding challenge competence with apps.
\newblock In J.~Vanschoren and S.~Yeung, editors, {\em Proceedings of the Neural Information Processing Systems Track on Datasets and Benchmarks}, volume~1, 2021.

\bibitem[HBS{\etalchar{+}}12]{hindle2012natural}
Abram Hindle, Earl~T. Barr, Zhendong Su, Mark Gabel, and Premkumar Devanbu.
\newblock On the naturalness of software.
\newblock In {\em Proceedings of the 34th International Conference on Software Engineering}, ICSE '12, page 837–847. IEEE Press, 2012.

\bibitem[HD17]{Hellendoorn2017AreCode}
Vincent~J. Hellendoorn and Premkumar Devanbu.
\newblock {Are deep neural networks the best choice for modeling source code?}
\newblock {\em ESEC/FSE 2017: Proceedings of the 2017 11th Joint Meeting on Foundations of Software Engineering}, pages 763--773, 2017.

\bibitem[HDS06]{Hayes:TSE'06}
Jane~Huffman Hayes, Alex Dekhtyar, and Senthil~Karthikeyan Sundaram.
\newblock Advancing candidate link generation for requirements tracing: The study of methods.
\newblock {\em IEEE Transactions on Software Engineering}, 32(1):4, 2006.

\bibitem[HG14]{Hoffman2011TheCarlo}
Matthew~D. Homan and Andrew Gelman.
\newblock The no-u-turn sampler: adaptively setting path lengths in hamiltonian monte carlo.
\newblock {\em J. Mach. Learn. Res.}, 15(1):1593–1623, January 2014.

\bibitem[HHZW20]{Hussain2020DeepTL}
Yasir Hussain, Zhiqiu Huang, Yu~Zhou, and Senzhang Wang.
\newblock Deep transfer learning for source code modeling.
\newblock {\em Int. J. Softw. Eng. Knowl. Eng.}, 30:649--668, 2020.

\bibitem[HLX{\etalchar{+}}18]{Hu2018comment}
Xing Hu, Ge~Li, Xin Xia, David Lo, and Zhi Jin.
\newblock Deep code comment generation.
\newblock In {\em Proceedings of the 26th Conference on Program Comprehension}, ICPC '18, page 200–210, New York, NY, USA, 2018. Association for Computing Machinery.

\bibitem[HMU06]{10.5555/1196416}
John~E. Hopcroft, Rajeev Motwani, and Jeffrey~D. Ullman.
\newblock {\em Introduction to Automata Theory, Languages, and Computation (3rd Edition)}.
\newblock Addison-Wesley Longman Publishing Co., Inc., USA, 2006.

\bibitem[HMZ12]{10.1145/2379776.2379787}
Mark Harman, S.~Afshin Mansouri, and Yuanyuan Zhang.
\newblock Search-based software engineering: Trends, techniques and applications.
\newblock {\em ACM Comput. Surv.}, 45(1), December 2012.

\bibitem[HSL]{han_which_2022}
Tessa Han, Suraj Srinivas, and Himabindu Lakkaraju.
\newblock Which explanation should i choose? a function approximation perspective to characterizing post hoc explanations.

\bibitem[HSW{\etalchar{+}}24]{trustLLM}
Yue Huang, Lichao Sun, Haoran Wang, Siyuan Wu, Qihui Zhang, Yuan Li, Chujie Gao, Yixin Huang, Wenhan Lyu, Yixuan Zhang, Xiner Li, Hanchi Sun, Zhengliang Liu, Yixin Liu, Yijue Wang, Zhikun Zhang, Bertie Vidgen, Bhavya Kailkhura, Caiming Xiong, Chaowei Xiao, Chunyuan Li, Eric~P. Xing, Furong Huang, Hao Liu, Heng Ji, Hongyi Wang, Huan Zhang, Huaxiu Yao, Manolis Kellis, Marinka Zitnik, Meng Jiang, Mohit Bansal, James Zou, Jian Pei, Jian Liu, Jianfeng Gao, Jiawei Han, Jieyu Zhao, Jiliang Tang, Jindong Wang, Joaquin Vanschoren, John Mitchell, Kai Shu, Kaidi Xu, Kai-Wei Chang, Lifang He, Lifu Huang, Michael Backes, Neil~Zhenqiang Gong, Philip~S. Yu, Pin-Yu Chen, Quanquan Gu, Ran Xu, Rex Ying, Shuiwang Ji, Suman Jana, Tianlong Chen, Tianming Liu, Tianyi Zhou, William~Yang Wang, Xiang Li, Xiangliang Zhang, Xiao Wang, Xing Xie, Xun Chen, Xuyu Wang, Yan Liu, Yanfang Ye, Yinzhi Cao, Yong Chen, and Yue Zhao.
\newblock Position: {T}rust{LLM}: Trustworthiness in large language models.
\newblock In Ruslan Salakhutdinov, Zico Kolter, Katherine Heller, Adrian Weller, Nuria Oliver, Jonathan Scarlett, and Felix Berkenkamp, editors, {\em Proceedings of the 41st International Conference on Machine Learning}, volume 235 of {\em Proceedings of Machine Learning Research}, pages 20166--20270. PMLR, 21--27 Jul 2024.

\bibitem[HT22]{pmlr-v162-hu22b}
Yaojie Hu and Jin Tian.
\newblock Neuron dependency graphs: A causal abstraction of neural networks.
\newblock In Kamalika Chaudhuri, Stefanie Jegelka, Le~Song, Csaba Szepesvari, Gang Niu, and Sivan Sabato, editors, {\em Proceedings of the 39th International Conference on Machine Learning}, volume 162 of {\em Proceedings of Machine Learning Research}, pages 9020--9040. PMLR, 17--23 Jul 2022.

\bibitem[Hug]{codeparrotDataset}
Hugginface.
\newblock Codeparrot(codeparrot).
\newblock Accessed: 23 July 2024.

\bibitem[HWG{\etalchar{+}}19]{husain2019codesearchnet}
Hamel Husain, Ho-Hsiang Wu, Tiferet Gazit, Miltiadis Allamanis, and Marc Brockschmidt.
\newblock {CodeSearchNet} challenge: Evaluating the state of semantic code search.
\newblock {\em arXiv preprint arXiv:1909.09436}, 2019.

\bibitem[JEC18]{dit}
R.~G. James, C.~J. Ellison, and J.~P. Crutchfield.
\newblock {dit}: a {P}ython package for discrete information theory.
\newblock {\em The Journal of Open Source Software}, 3(25):738, 2018.

\bibitem[JG20]{jacovi2020faithfully}
Alon Jacovi and Yoav Goldberg.
\newblock Towards faithfully interpretable nlp systems: How should we define and evaluate faithfulness?, 2020.

\bibitem[JNC{\etalchar{+}}08]{Jiang:ASE'08}
Hsin-Yi Jiang, Tien~N Nguyen, Xiang Chen, Hojun Jaygarl, and Carl~K Chang.
\newblock Incremental latent semantic indexing for automatic traceability link evolution management.
\newblock In {\em Proceedings of the 2008 23rd IEEE/ACM International Conference on Automated Software Engineering}, ASE'08, pages 59--68. IEEE Computer Society, 2008.

\bibitem[JQC{\etalchar{+}}25]{ji2024ai}
Jiaming Ji, Tianyi Qiu, Boyuan Chen, Borong Zhang, Hantao Lou, Kaile Wang, Yawen Duan, Zhonghao He, Lukas Vierling, Donghai Hong, Jiayi Zhou, Zhaowei Zhang, Fanzhi Zeng, Juntao Dai, Xuehai Pan, Kwan~Yee Ng, Aidan O'Gara, Hua Xu, Brian Tse, Jie Fu, Stephen McAleer, Yaodong Yang, Yizhou Wang, Song-Chun Zhu, Yike Guo, and Wen Gao.
\newblock Ai alignment: A comprehensive survey, 2025.

\bibitem[JW19]{jain_attention_2019}
Sarthak Jain and Byron~C. Wallace.
\newblock Attention is not {Explanation}, May 2019.
\newblock arXiv:1902.10186 [cs].

\bibitem[JWY18]{tabnine}
Jacob Jackson, Dror Weiss, and Eran Yahav.
\newblock Tabnine: Ai code assistant, 2018.
\newblock AI code assistant that accelerates and simplifies software development while keeping code private, secure, and compliant.

\bibitem[JYW{\etalchar{+}}24]{jimenez2024swebenchlanguagemodelsresolve}
Carlos~E. Jimenez, John Yang, Alexander Wettig, Shunyu Yao, Kexin Pei, Ofir Press, and Karthik Narasimhan.
\newblock Swe-bench: Can language models resolve real-world github issues?, 2024.

\bibitem[KB15]{Kingma2015AdamAM}
Diederik~P. Kingma and Jimmy Ba.
\newblock Adam: A method for stochastic optimization.
\newblock {\em CoRR}, abs/1412.6980, 2015.

\bibitem[KBR{\etalchar{+}}20a]{Karampatsis2020BigCode}
Rafael~Michael Karampatsis, Hlib Babii, Romain Robbes, Charles Sutton, and Andrea Janes.
\newblock {Big code != big vocabulary: Open-vocabulary models for source code}.
\newblock {\em Proceedings - International Conference on Software Engineering}, pages 1073--1085, 2020.

\bibitem[KBR{\etalchar{+}}20b]{Karampatsis2020Open-VocabularyAbstract}
Rafael~Michael Karampatsis, Hlib Babii, Romain Robbes, Charles Sutton, and Andrea Janes.
\newblock {Open-Vocabulary Models for Source Code (Extended Abstract)}.
\newblock {\em Proceedings - 2020 ACM/IEEE 42nd International Conference on Software Engineering: Companion, ICSE-Companion 2020}, pages 294--295, 2020.

\bibitem[KC25]{jenkins}
Kohsuke Kawaguchi and Jenkins Community.
\newblock Jenkins: The leading open source automation server, 2025.
\newblock Accessed: 2025-04-07.

\bibitem[KCM20]{Kolchinsky2020DecomposingTransformation}
Artemy Kolchinsky and Bernat Corominas-Murtra.
\newblock {Decomposing information into copying versus transformation}.
\newblock {\em Journal of the Royal Society Interface}, 17(162):1--17, 2020.

\bibitem[KFG{\etalchar{+}}21]{kocmi2021ship}
Tom Kocmi, Christian Federmann, Roman Grundkiewicz, Marcin Junczys{-}Dowmunt, Hitokazu Matsushita, and Arul Menezes.
\newblock To ship or not to ship: An extensive evaluation of automatic metrics for machine translation.
\newblock {\em CoRR}, abs/2107.10821, 2021.

\bibitem[KGDP11]{Poshyvanyk:TEFSE'11}
Samuel Klock, Malcom Gethers, Bogdan Dit, and Denys Poshyvanyk.
\newblock Traceclipse: an eclipse plug-in for traceability link recovery and management.
\newblock In {\em Proceedings of the international workshop on traceability in emerging forms of software engineering}, TEFSE'11, pages 24--30, 2011.

\bibitem[KHQJ18]{khandelwal2018sharp}
Urvashi Khandelwal, He~He, Peng Qi, and Dan Jurafsky.
\newblock Sharp nearby, fuzzy far away: How neural language models use context.
\newblock In {\em Proceedings of the 56th Annual Meeting of the Association for Computational Linguistics (Volume 1: Long Papers)}, pages 284--294, Melbourne, Australia, July 2018. Association for Computational Linguistics.

\bibitem[KJL15]{karpathy2015understand}
Andrej Karpathy, Justin Johnson, and Fei{-}Fei Li.
\newblock Visualizing and understanding recurrent networks.
\newblock {\em CoRR}, abs/1506.02078, 2015.

\bibitem[KLP{\etalchar{+}}25]{khati2025mappingtrustterrainllms}
Dipin Khati, Yijin Liu, David~N. Palacio, Yixuan Zhang, and Denys Poshyvanyk.
\newblock Mapping the trust terrain: Llms in software engineering -- insights and perspectives, 2025.

\bibitem[KMB{\etalchar{+}}17]{Kitchenham2017RobustEngineering}
Barbara Kitchenham, Lech Madeyski, David Budgen, Jacky Keung, Pearl Brereton, Stuart Charters, Shirley Gibbs, and Amnart Pohthong.
\newblock {Robust Statistical Methods for Empirical Software Engineering}.
\newblock {\em Empirical Software Engineering}, 22(2):579--630, 2017.

\bibitem[KMK{\etalchar{+}}]{karimi_relationship_2023}
Amir-Hossein Karimi, Krikamol Muandet, Simon Kornblith, Bernhard Schölkopf, and Been Kim.
\newblock On the relationship between explanation and prediction: A causal view.

\bibitem[KNH{\etalchar{+}}17]{Kuang:SANER'17}
Hongyu Kuang, Jia Nie, Hao Hu, Patrick Rempel, Jian L{\"u}, Alexander Egyed, and Patrick M{\"a}der.
\newblock Analyzing closeness of code dependencies for improving ir-based traceability recovery.
\newblock In {\em Proceedings of the 24th International Conference on Software Analysis, Evolution and Reengineering}, SANER'17, pages 68--78, 2017.

\bibitem[KR24]{sentencepiece}
Taku Kudo and John Richardson.
\newblock Sentencepiece: A simple and language-independent subword tokenizer and detokenizer for neural text processing.
\newblock \url{https://github.com/google/sentencepiece}, 2024.
\newblock Accessed: 2024-11-21.

\bibitem[KS83]{Kalai1983OnWS}
Ehud Kalai and Dov Samet.
\newblock On weighted shapley values.
\newblock {\em International Journal of Game Theory}, 16:205--222, 1983.

\bibitem[KS19]{Karampatsis2019}
Rafael-Michael Karampatsis and Charles Sutton.
\newblock Maybe deep neural networks are the best choice for modeling source code, 2019.

\bibitem[KWG{\etalchar{+}}18]{kim_interpretability_2018}
Been Kim, Martin Wattenberg, Justin Gilmer, Carrie Cai, James Wexler, Fernanda Viegas, and Rory Sayres.
\newblock Interpretability {Beyond} {Feature} {Attribution}: {Quantitative} {Testing} with {Concept} {Activation} {Vectors} ({TCAV}), June 2018.
\newblock arXiv:1711.11279 [stat].

\bibitem[Lab25]{daniel23}
SEMERU Lab.
\newblock Galeras benchmark: Benchmarking causal analysis for interpreting llms for source code.
\newblock \url{https://github.com/WM-SEMERU/galeras-benchmark}, 2025.
\newblock Accessed: 2025-04-07.

\bibitem[Lan89]{Langton1989ALife}
Christopher~G. Langton.
\newblock Artificial life: The proceedings of an interdisciplinary workshop on the synthesis and simulation of living systems.
\newblock In {\em Artificial Life}, volume~VI of {\em SFI Studies in the Sciences of Complexity}, Redwood City, CA, 1989. Addison-Wesley.

\bibitem[LAZ{\etalchar{+}}23]{li2023starcodersourceyou}
Raymond Li, Loubna~Ben Allal, Yangtian Zi, Niklas Muennighoff, Denis Kocetkov, Chenghao Mou, Marc Marone, Christopher Akiki, Jia Li, Jenny Chim, Qian Liu, Evgenii Zheltonozhskii, Terry~Yue Zhuo, Thomas Wang, Olivier Dehaene, Mishig Davaadorj, Joel Lamy-Poirier, João Monteiro, Oleh Shliazhko, Nicolas Gontier, Nicholas Meade, Armel Zebaze, Ming-Ho Yee, Logesh~Kumar Umapathi, Jian Zhu, Benjamin Lipkin, Muhtasham Oblokulov, Zhiruo Wang, Rudra Murthy, Jason Stillerman, Siva~Sankalp Patel, Dmitry Abulkhanov, Marco Zocca, Manan Dey, Zhihan Zhang, Nour Fahmy, Urvashi Bhattacharyya, Wenhao Yu, Swayam Singh, Sasha Luccioni, Paulo Villegas, Maxim Kunakov, Fedor Zhdanov, Manuel Romero, Tony Lee, Nadav Timor, Jennifer Ding, Claire Schlesinger, Hailey Schoelkopf, Jan Ebert, Tri Dao, Mayank Mishra, Alex Gu, Jennifer Robinson, Carolyn~Jane Anderson, Brendan Dolan-Gavitt, Danish Contractor, Siva Reddy, Daniel Fried, Dzmitry Bahdanau, Yacine Jernite, Carlos~Muñoz Ferrandis, Sean Hughes, Thomas Wolf, Arjun Guha, Leandro von
  Werra, and Harm de~Vries.
\newblock Starcoder: may the source be with you!, 2023.

\bibitem[LAZCH13]{Lohar:FSE'13}
Sugandha Lohar, Sorawit Amornborvornwong, Andrea Zisman, and Jane Cleland-Huang.
\newblock Improving trace accuracy through data-driven configuration and composition of tracing features.
\newblock In {\em Proceedings of the 2013 9th Joint Meeting on Foundations of Software Engineering}, ESEC/FSE 2013, pages 378--388. ACM, 2013.

\bibitem[LBL{\etalchar{+}}]{liang_holistic_2022}
Percy Liang, Rishi Bommasani, Tony Lee, Dimitris Tsipras, Dilara Soylu, Michihiro Yasunaga, Yian Zhang, Deepak Narayanan, Yuhuai Wu, Ananya Kumar, Benjamin Newman, Binhang Yuan, Bobby Yan, Ce~Zhang, Christian Cosgrove, Christopher~D. Manning, Christopher Ré, Diana Acosta-Navas, Drew~A. Hudson, Eric Zelikman, Esin Durmus, Faisal Ladhak, Frieda Rong, Hongyu Ren, Huaxiu Yao, Jue Wang, Keshav Santhanam, Laurel Orr, Lucia Zheng, Mert Yuksekgonul, Mirac Suzgun, Nathan Kim, Neel Guha, Niladri Chatterji, Omar Khattab, Peter Henderson, Qian Huang, Ryan Chi, Sang~Michael Xie, Shibani Santurkar, Surya Ganguli, Tatsunori Hashimoto, Thomas Icard, Tianyi Zhang, Vishrav Chaudhary, William Wang, Xuechen Li, Yifan Mai, Yuhui Zhang, and Yuta Koreeda.
\newblock Holistic evaluation of language models.

\bibitem[LBM21]{leclair_ensemble_2021}
Alexander LeClair, Aakash Bansal, and Collin McMillan.
\newblock Ensemble {Models} for {Neural} {Source} {Code} {Summarization} of {Subroutines}, July 2021.
\newblock arXiv:2107.11423 [cs].

\bibitem[LBZ{\etalchar{+}}23]{liu2023improving}
Chao Liu, Xuanlin Bao, Hongyu Zhang, Neng Zhang, Haibo Hu, Xiaohong Zhang, and Meng Yan.
\newblock Improving chatgpt prompt for code generation, 2023.

\bibitem[LCH{\etalchar{+}}19]{lage_human_2019}
Isaac Lage, Emily Chen, Jeffrey He, Menaka Narayanan, Been Kim, Samuel~J. Gershman, and Finale Doshi-Velez.
\newblock Human {Evaluation} of {Models} {Built} for {Interpretability}.
\newblock {\em Proceedings of the AAAI Conference on Human Computation and Crowdsourcing}, 7:59--67, October 2019.

\bibitem[Leh80]{Lehman1980ProgramsEvolution}
Meir Lehman.
\newblock {Programs, Life Cycles, and Laws of Software Evolution}.
\newblock {\em Proceedings of the IEEE}, 1980.

\bibitem[LFOT04]{Lucia:ICSM'04}
A.~De Lucia, F.~Fasano, R.~Oliveto, and G.~Tortora.
\newblock Enhancing an artefact management system with traceability recovery features.
\newblock In {\em Proceedings of the 20th IEEE International Conference on Software Maintenance, 2004. Proceedings.}, ICSM'04, pages 306--315, Sept 2004.

\bibitem[LGM20]{Liao_2020}
Q.~Vera Liao, Daniel Gruen, and Sarah Miller.
\newblock Questioning the ai: Informing design practices for explainable ai user experiences.
\newblock In {\em Proceedings of the 2020 CHI Conference on Human Factors in Computing Systems}, CHI '20, page 1–15, New York, NY, USA, 2020. Association for Computing Machinery.

\bibitem[LGR{\etalchar{+}}21]{lu2021codexglue}
Shuai Lu, Daya Guo, Shuo Ren, Junjie Huang, Alexey Svyatkovskiy, Ambrosio Blanco, Colin~B. Clement, Dawn Drain, Daxin Jiang, Duyu Tang, Ge~Li, Lidong Zhou, Linjun Shou, Long Zhou, Michele Tufano, Ming Gong, Ming Zhou, Nan Duan, Neel Sundaresan, Shao~Kun Deng, Shengyu Fu, and Shujie Liu.
\newblock Codexglue: {A} machine learning benchmark dataset for code understanding and generation.
\newblock {\em CoRR}, abs/2102.04664, 2021.

\bibitem[Lip]{lipton_mythos_2017}
Zachary~C. Lipton.
\newblock The mythos of model interpretability.

\bibitem[LL17]{lundberg2017unified}
Scott~M. Lundberg and Su-In Lee.
\newblock A unified approach to interpreting model predictions.
\newblock In {\em Proceedings of the 31st International Conference on Neural Information Processing Systems}, NIPS'17, page 4768–4777, Red Hook, NY, USA, 2017. Curran Associates Inc.

\bibitem[LLG{\etalchar{+}}19]{lewis2019bartdenoisingsequencetosequencepretraining}
Mike Lewis, Yinhan Liu, Naman Goyal, Marjan Ghazvininejad, Abdelrahman Mohamed, Omer Levy, Ves Stoyanov, and Luke Zettlemoyer.
\newblock Bart: Denoising sequence-to-sequence pre-training for natural language generation, translation, and comprehension, 2019.

\bibitem[LLZ{\etalchar{+}}21]{transformer_trace_21}
Jinfeng Lin, Yalin Liu, Qingkai Zeng, Meng Jiang, and Jane Cleland-Huang.
\newblock Traceability transformed: Generating more accurate links with pre-trained bert models.
\newblock In {\em Proceedings of the 43rd International Conference on Software Engineering}, ICSE '21, page 324–335. IEEE Press, 2021.

\bibitem[Lo]{lo_trustworthy_2023}
David Lo.
\newblock Trustworthy and synergistic artificial intelligence for software engineering: Vision and roadmaps.

\bibitem[Lon23]{distrust2023}
Luca Longo, editor.
\newblock {\em Explainable Artificial Intelligence: First World Conference, xAI 2023, Lisbon, Portugal, July 26–28, 2023, Proceedings, Part I}, volume 1901 of {\em Communications in Computer and Information Science}. Springer, 2023.

\bibitem[LPK21]{linardatosExplainable}
Pantelis Linardatos, Vasilis Papastefanopoulos, and Sotiris Kotsiantis.
\newblock Explainable ai: A review of machine learning interpretability methods.
\newblock {\em Entropy}, 23(1), 2021.

\bibitem[LPY{\etalchar{+}}22]{Lin2022EnhancingAS}
Jinfeng Lin, A.~Poudel, Wenhao Yu, Qingkai Zeng, Meng Jiang, and Jane Cleland-Huang.
\newblock Enhancing automated software traceability by transfer learning from open-world data.
\newblock {\em ArXiv}, abs/2207.01084, 2022.

\bibitem[LS04]{Lee_See_2004}
John~D. Lee and Katrina~A. See.
\newblock Trust in automation: Designing for appropriate reliance.
\newblock {\em Human Factors}, 46(1):50--80, 2004.
\newblock PMID: 15151155.

\bibitem[LSW{\etalchar{+}}21]{synerror1}
Zheng Li, Fuxiang Sun, Haifeng Wang, Yifan Ding, Yong Liu, and Xiang Chen.
\newblock Clacer: A deep learning-based compilation error classification method for novice students’ programs.
\newblock In {\em 2021 IEEE 45th Annual Computers, Software, and Applications Conference (COMPSAC)}, pages 74--83, 2021.

\bibitem[LTL{\etalchar{+}}23]{synerror2}
Zhijie Liu, Yutian Tang, Xiapu Luo, Yuming Zhou, and Liang~Feng Zhang.
\newblock No need to lift a finger anymore? assessing the quality of code generation by chatgpt, 2023.

\bibitem[LTLL22]{liu2022explainable}
Yue Liu, Chakkrit Tantithamthavorn, Li~Li, and Yepang Liu.
\newblock Explainable ai for android malware detection: Towards understanding why the models perform so well?
\newblock In {\em 2022 IEEE 33rd International Symposium on Software Reliability Engineering (ISSRE)}, pages 169--180, 2022.

\bibitem[LTLL24]{liu2024reliability}
Yue Liu, Chakkrit Tantithamthavorn, Yonghui Liu, and Li~Li.
\newblock On the reliability and explainability of language models for program generation.
\newblock {\em ACM Trans. Softw. Eng. Methodol.}, 33(5), June 2024.

\bibitem[LWCS22]{lopez_ast-probe_2022}
José Antonio~Hernández López, Martin Weyssow, Jesús~Sánchez Cuadrado, and Houari Sahraoui.
\newblock {AST}-{Probe}: {Recovering} abstract syntax trees from hidden representations of pre-trained language models, September 2022.
\newblock arXiv:2206.11719 [cs].

\bibitem[LWX18]{Li2018}
D.~{Li}, Z.~{Wang}, and Y.~{Xue}.
\newblock Fine-grained android malware detection based on deep learning.
\newblock In {\em 2018 IEEE Conference on Communications and Network Security (CNS)}, pages 1--2, May 2018.

\bibitem[LXWZ23]{liu2023code}
Jiawei Liu, Chunqiu~Steven Xia, Yuyao Wang, and Lingming Zhang.
\newblock Is your code generated by chatgpt really correct? rigorous evaluation of large language models for code generation, 2023.

\bibitem[LXZ18]{Liu2018}
Hui Liu, Zhifeng Xu, and Yanzhen Zou.
\newblock Deep learning based feature envy detection.
\newblock In {\em Proceedings of the 33rd ACM/IEEE International Conference on Automated Software Engineering}, ASE 2018, pages 385--396, New York, NY, USA, 2018. ACM.

\bibitem[LYT{\etalchar{+}}24]{liu2024trustworthyllmssurveyguideline}
Yang Liu, Yuanshun Yao, Jean-Francois Ton, Xiaoying Zhang, Ruocheng Guo, Hao Cheng, Yegor Klochkov, Muhammad~Faaiz Taufiq, and Hang Li.
\newblock Trustworthy llms: a survey and guideline for evaluating large language models' alignment, 2024.

\bibitem[LZL{\etalchar{+}}23]{li2023}
Yao Li, Tao Zhang, Xiapu Luo, Haipeng Cai, Sen Fang, and Dawei Yuan.
\newblock Do pretrained language models indeed understand software engineering tasks?
\newblock {\em IEEE Transactions on Software Engineering}, 49(10):4639--4655, 2023.

\bibitem[LZP{\etalchar{+}}17]{Liu2017}
P.~{Liu}, X.~{Zhang}, M.~{Pistoia}, Y.~{Zheng}, M.~{Marques}, and L.~{Zeng}.
\newblock Automatic text input generation for mobile testing.
\newblock In {\em 2017 IEEE/ACM 39th International Conference on Software Engineering (ICSE)}, pages 643--653, May 2017.

\bibitem[Mac03]{MacKay2003InformationAlgorithms}
David J.~C. MacKay.
\newblock {\em {Information theory, inference, and learning algorithms}}.
\newblock Cambridge University Press, 2003.

\bibitem[Mat80]{MaturanaHumberto1980AutopoesisCognition}
Varela~Francisco Maturana, Humberto.
\newblock {\em {Autopoesis and Cognition}}.
\newblock Springer, 1980.

\bibitem[MB23]{10172693}
Manish Motwani and Yuriy Brun.
\newblock Better automatic program repair by using bug reports and tests together.
\newblock In {\em 2023 IEEE/ACM 45th International Conference on Software Engineering (ICSE)}, pages 1225--1237, May 2023.

\bibitem[MCB21]{Molnar2020InterpretableChallenges}
Christoph Molnar, Giuseppe Casalicchio, and Bernd Bischl.
\newblock Interpretable machine learning – a brief history, state-of-the-art and challenges.
\newblock In Mohammed~J. Zaki, Panagiotis Papapetrou, Victor~S. Sheng, Aristides Gionis, Lars Schmidt-Thieme, and Wolfgang~J. Krause, editors, {\em ECML PKDD 2020 Workshops}, pages 417--431. Springer International Publishing, 2021.

\bibitem[MEAH18]{Mills:ICSME18}
Chris Mills, Javier Escobar-Avila, and Sonia Haiduc.
\newblock {Automatic} {Traceability} {Maintenance} via {Machine} {Learning} {Classification}.
\newblock In {\em Proceedings of the 34th {IEEE} {International} {Conference} on {Software} {Maintenance} and {Evolution}}, ICSME'18, pages 369--380, Madrid, Spain, September 2018. ACM.

\bibitem[MEHDTH13]{Macias-Escriva2013Self-adaptiveApplications}
Frank~D. Mac{\'{i}}as-Escriv{\'{a}}, Rodolfo Haber, Raul Del~Toro, and Vicente Hernandez.
\newblock {Self-adaptive systems: A survey of current approaches, research challenges and applications}.
\newblock {\em Expert Systems with Applications}, 40(18):7267--7279, 2013.

\bibitem[MGF07]{4027145}
Tim Menzies, Jeremy Greenwald, and Art Frank.
\newblock Data mining static code attributes to learn defect predictors.
\newblock {\em IEEE Transactions on Software Engineering}, 33(1):2--13, 2007.

\bibitem[{Mic}25]{intellicode}
{Microsoft}.
\newblock Visual studio intellicode, 2025.
\newblock Accessed: 2025-04-07.

\bibitem[Mil19]{miller2018explanation}
Tim Miller.
\newblock Explanation in artificial intelligence: Insights from the social sciences.
\newblock {\em Artificial Intelligence}, 267:1--38, 2019.

\bibitem[MJ23]{molak_causal_2023}
Aleksander Molak and Ajit Jaokar.
\newblock {\em Causal inference and discovery in Python: unlock the secrets of modern causal machine learning with {DoWhy}, {EconML}, {PyTorch} and more}.
\newblock Packt Publishing Limited, 2023.

\bibitem[MJXZ{\etalchar{+}}18]{Ma2018}
Lei Ma, Felix Juefei-Xu, Fuyuan Zhang, Jiyuan Sun, Minhui Xue, Bo~Li, Chunyang Chen, Ting Su, Li~Li, Yang Liu, Jianjun Zhao, and Yadong Wang.
\newblock Deepgauge: Multi-granularity testing criteria for deep learning systems.
\newblock In {\em Proceedings of the 33rd ACM/IEEE International Conference on Automated Software Engineering}, ASE 2018, pages 120--131, New York, NY, USA, 2018. ACM.

\bibitem[MJZCH13]{Mader:Soft'13}
P.~M\"{a}der, P.~L. Jones, Y.~Zhang, and J.~Cleland-Huang.
\newblock Strategic traceability for safety-critical projects.
\newblock {\em IEEE Software}, 30(3):58--66, May 2013.

\bibitem[MKC{\etalchar{+}}16]{Mkaouer2016AOpportunities}
Mohamed~Wiem Mkaouer, Marouane Kessentini, Mel Cinn{\'{e}}ide, Shinpei Hayashi, and Kalyanmoy Deb.
\newblock {A robust multi-objective approach to balance severity and importance of refactoring opportunities}, 2016.

\bibitem[MKF06]{murphy2006ide}
Gail~C. Murphy, Mik Kersten, and Leah Findlater.
\newblock How are java software developers using the eclipse ide?
\newblock {\em IEEE Softw.}, 23(4):76–83, July 2006.

\bibitem[MKG{\etalchar{+}}20]{moraffah2020causalinterpretabilitymachinelearning}
Raha Moraffah, Mansooreh Karami, Ruocheng Guo, Adrienne Raglin, and Huan Liu.
\newblock Causal interpretability for machine learning -- problems, methods and evaluation, 2020.

\bibitem[MLY18]{murdoch2018beyond}
W.~James Murdoch, Peter~J. Liu, and Bin Yu.
\newblock Beyond word importance: Contextual decomposition to extract interactions from {LSTM}s.
\newblock In {\em International Conference on Learning Representations}, 2018.

\bibitem[MM03]{Marcus:ICSE'03}
Andrian Marcus and Jonathan~I. Maletic.
\newblock Recovering documentation-to-source-code traceability links using latent semantic indexing.
\newblock In {\em Proceedings of the 25th International Conference on Software Engineering}, ICSE '03, pages 125--135, Washington, DC, USA, 2003. IEEE Computer Society.

\bibitem[MNN{\etalchar{+}}20]{mohankumar_towards_2020}
Akash~Kumar Mohankumar, Preksha Nema, Sharan Narasimhan, Mitesh~M. Khapra, Balaji~Vasan Srinivasan, and Balaraman Ravindran.
\newblock Towards {Transparent} and {Explainable} {Attention} {Models}.
\newblock In {\em Proceedings of the 58th {Annual} {Meeting} of the {Association} for {Computational} {Linguistics}}, pages 4206--4216, Online, July 2020. Association for Computational Linguistics.

\bibitem[MNX12]{Mahmoud:ICPC'12}
A.~Mahmoud, N.~Niu, and S.~Xu.
\newblock A semantic relatedness approach for traceability link recovery.
\newblock In {\em Proceedings of the 20th IEEE International Conference on Program Comprehension}, ICPC'12, pages 183--192, June 2012.

\bibitem[Mol25]{molnar2025}
Christoph Molnar.
\newblock {\em Interpretable Machine Learning}.
\newblock 3 edition, 2025.

\bibitem[Mor04]{Morville:04}
Peter Morville.
\newblock User experience design, 2004.
\newblock Accessed: 2025-04-07.

\bibitem[MP24]{10481576}
Pradeep~Kumar Mahapatro and Neelamadhab Padhy.
\newblock Reviewing the landscape: Component-based software engineering practices and challenges.
\newblock In {\em 2024 International Conference on Emerging Systems and Intelligent Computing (ESIC)}, pages 360--365, 2024.

\bibitem[MPBC{\etalchar{+}}20a]{moran_improving_2020}
Kevin Moran, David~N. Palacio, Carlos Bernal-C\'{a}rdenas, Daniel McCrystal, Denys Poshyvanyk, Chris Shenefiel, and Jeff Johnson.
\newblock Improving the effectiveness of traceability link recovery using hierarchical bayesian networks.
\newblock In {\em Proceedings of the ACM/IEEE 42nd International Conference on Software Engineering}, ICSE '20, page 873–885, New York, NY, USA, 2020. Association for Computing Machinery.

\bibitem[MPBC{\etalchar{+}}20b]{comet2020}
Kevin Moran, David~N. Palacio, Carlos Bernal-Cardenas, Daniel McCrystal, Denys Poshyvanyk, Chris Shenefiel, and Jeff Johnson.
\newblock Online appendix for "improving the effectiveness of traceability link recovery using hierarchical bayesian networks", 2020.
\newblock Accessed: 2025-04-07.

\bibitem[MPR09]{McMillan:TEFSE'09}
Collin McMillan, Denys Poshyvanyk, and Meghan Revelle.
\newblock Combining textual and structural analysis of software artifacts for traceability link recovery.
\newblock In {\em Proceedings of the ICSE Workshop on Traceability in Emerging Forms of Software Engineering}, TEFSE '09, pages 41--48. IEEE Computer Society, 2009.

\bibitem[MSC{\etalchar{+}}21]{Mastropaolo2021StudyingTasks}
Antonio Mastropaolo, Simone Scalabrino, Nathan Cooper, David~Nader Palacio, Denys Poshyvanyk, Rocco Oliveto, and Gabriele Bavota.
\newblock Studying the usage of text-to-text transfer transformer to support code-related tasks.
\newblock In {\em Proceedings of the 43rd International Conference on Software Engineering}, ICSE '21, page 336–347. IEEE Press, 2021.

\bibitem[MTH23]{mohammadkhani2023explain}
Ahmad~Haji Mohammadkhani, Chakkrit Tantithamthavorn, and Hadi Hemmatif.
\newblock Explaining transformer-based code models: What do they learn? when they do not work?
\newblock In {\em 2023 IEEE 23rd International Working Conference on Source Code Analysis and Manipulation (SCAM)}, pages 96--106, 2023.

\bibitem[Mur12]{Murphy:2012}
Kevin~P. Murphy.
\newblock {\em Machine Learning: A Probabilistic Perspective}.
\newblock The MIT Press, 2012.

\bibitem[MW16]{Mahmoud:RE'16}
Anas Mahmoud and Grant Williams.
\newblock Detecting, classifying, and tracing non-functional software requirements.
\newblock {\em Requirements Engineering}, 21(3):357--381, Sep 2016.

\bibitem[MYP{\etalchar{+}}22]{Moran.SANER.2022}
Kevin Moran, Ali Yachnes, George Purnell, Junayed Mahmud, Michele Tufano, Carlos~Bernal Cardenas, Denys Poshyvanyk, and Zach H'Doubler.
\newblock An empirical investigation into the use of image captioning for automated software documentation.
\newblock In {\em 2022 IEEE International Conference on Software Analysis, Evolution and Reengineering (SANER)}, pages 514--525, 2022.

\bibitem[Neu51]{VonNeumann1951}
John~Von Neumann.
\newblock The general and logical theory of automata.
\newblock In L.A. Jeffress, editor, {\em Cerebral Mechanisms in Behaviour—The Hixon Symposium}, pages 1--41, New York, NY, USA, 1951. Wiley.

\bibitem[Neu66]{VonNeumann1966}
John~Von Neumann.
\newblock {\em Theory of Self-Reproducing Automata}.
\newblock University of Illinois Press, Champaign, IL, USA, 1966.

\bibitem[Ney37]{Neyman:37}
J.~Neyman.
\newblock {\em Outline of a Theory of Statistical Estimation Based on the Classical Theory of Probability}.
\newblock Royal Society, 1937.

\bibitem[NN15]{Nguyen:ICSE15}
Anh~Tuan Nguyen and Tien~N. Nguyen.
\newblock Graph-based statistical language model for code.
\newblock In {\em Proceedings of the 37th International Conference on Software Engineering - Volume 1}, ICSE '15, page 858–868. IEEE Press, 2015.

\bibitem[NNN13]{Nguyen2013ASS}
T.~Nguyen, A.~Nguyen, and H.~Nguyen.
\newblock A statistical semantic language model for source code.
\newblock In {\em ESEC/FSE 2013}, 2013.

\bibitem[NNNN13]{nguyen_statistical_2013}
Tung~Thanh Nguyen, Anh~Tuan Nguyen, Hoan~Anh Nguyen, and Tien~N. Nguyen.
\newblock A statistical semantic language model for source code.
\newblock In {\em Proceedings of the 2013 9th Joint Meeting on Foundations of Software Engineering}, ESEC/FSE 2013, page 532–542, New York, NY, USA, 2013. Association for Computing Machinery.

\bibitem[NPH{\etalchar{+}}23]{Nijkamp2022CodeGenAO}
Erik Nijkamp, Bo~Pang, Hiroaki Hayashi, Lifu Tu, Huan Wang, Yingbo Zhou, Silvio Savarese, and Caiming Xiong.
\newblock Codegen: An open large language model for code with multi-turn program synthesis.
\newblock In {\em The Eleventh International Conference on Learning Representations}, 2023.

\bibitem[NpRcG18]{Nader-palacio2018AssessingDetection}
David Nader-palacio, Daniel Rodr{\'{i}}guez-c{\'{a}}rdenas, and Jonatan Gomez.
\newblock {Assessing Single-Objective Performance Convergence and Time Complexity for Refactoring Detection}.
\newblock In {\em Proceedings of the Genetic and Evolutionary Computation Conference Companion on - GECCO '18}, pages 1606--1613, 2018.

\bibitem[NSF{\etalchar{+}}12]{Nejati:IST'12}
Shiva Nejati, Mehrdad Sabetzadeh, Davide Falessi, Lionel Briand, and Thierry Coq.
\newblock A sysml-based approach to traceability management and design slicing in support of safety certification: Framework, tool support, and case studies.
\newblock {\em Inf. Softw. Technol.}, 54(6):569--590, June 2012.

\bibitem[NWF{\etalchar{+}}15]{Nishikawa:ICSME'15}
Kazuki Nishikawa, Hironori Washizaki, Yoshiaki Fukazawa, Keishi Oshima, and Ryota Mibe.
\newblock Recovering transitive traceability links among software artifacts.
\newblock In {\em Proceedings of the IEEE International Conference on Software Maintenance and Evolution}, ICSME'15, pages 576--580. IEEE, 2015.

\bibitem[NYZ{\etalchar{+}}15]{Nhlabatsi:SST'15}
Armstrong Nhlabatsi, Yijun Yu, Andrea Zisman, Thein Tun, Niamul Khan, Arosha Bandara, Khaled~M. Khan, and Bashar Nuseibeh.
\newblock Managing security control assumptions using causal traceability.
\newblock In {\em Proceedings of the 8th International Symposium on Software and Systems Traceability}, SST '15, pages 43--49. IEEE Press, 2015.

\bibitem[OdSSdASN{\etalchar{+}}25]{10.1145/3695987}
Leandro Oliveria~de Souza, Eduardo Santana~de Almeida, Paulo Anselmo da~Mota Silveira~Neto, Earl~T. Barr, and Justyna Petke.
\newblock Software product line engineering via software transplantation.
\newblock {\em ACM Trans. Softw. Eng. Methodol.}, 34(2), January 2025.

\bibitem[oEfSSTC25]{coest-datasets}
Center of~Excellence~for Software \& Systems Traceability~(CoEST).
\newblock Coest traceability datasets, 2025.
\newblock Accessed: 2025-04-07.

\bibitem[OGPDL10]{Oliveto:ICPC'10}
Rocco Oliveto, Malcom Gethers, Denys Poshyvanyk, and Andrea De~Lucia.
\newblock On the equivalence of information retrieval methods for automated traceability link recovery.
\newblock In {\em Proceedings of the 18th International Conference on Program Comprehension}, ICPC '10, pages 68--71, 2010.

\bibitem[OO22]{10.1007/978-981-16-0739-4_28}
Oluwaseyi~Ezekiel Olorunshola and Francisca~Nonyelum Ogwueleka.
\newblock Review of system development life cycle (sdlc) models for effective application delivery.
\newblock In Amit Joshi, Mufti Mahmud, Roshan~G. Ragel, and Nileshsingh~V. Thakur, editors, {\em Information and Communication Technology for Competitive Strategies (ICTCS 2020)}, pages 281--289, Singapore, 2022. Springer Singapore.

\bibitem[Ope23]{openai2023gpt4}
OpenAI.
\newblock Gpt-4 technical report, 2023.

\bibitem[Ora24]{OracleJavaKeywords}
Oracle.
\newblock Java language keywords, 2024.
\newblock Accessed: 2025-04-07.

\bibitem[Pal23]{palacio2023tracexplainer}
David~N. Palacio.
\newblock Information theory for interpreting software traceability.
\newblock \url{https://danaderp.github.io/danaderp/projects/1_project/}, 2023.

\bibitem[PB11]{pearl_transportability}
Judea Pearl and Elias Bareinboim.
\newblock Transportability of causal and statistical relations: A formal approach.
\newblock In {\em 2011 IEEE 11th International Conference on Data Mining Workshops}, pages 540--547, 2011.

\bibitem[PCR{\etalchar{+}}23]{palacio_toward_2023}
David~N. Palacio, Nathan Cooper, Alvaro Rodriguez, Kevin Moran, and Denys Poshyvanyk.
\newblock Toward a {Theory} of {Causation} for {Interpreting} {Neural} {Code} {Models}, February 2023.
\newblock arXiv:2302.03788 [cs, stat].

\bibitem[PDE{\etalchar{+}}19]{dlse19-report}
Devanbu Prem, Matthew Dwyer, Sebastian Elbaum, Michael Lowry, Kevin Moran, Denys Poshyvanyk, Baishakhi Ray, Rishabh Singh, and Xiangyu Zhang.
\newblock Deep learning \& software engineering: State of research and future directions.
\newblock In {\em Proceedings of the 2019 NSF Workshop on Deep Learning and Software Engineering}, 2019.

\bibitem[Pea09a]{pearl2009overview}
Judea Pearl.
\newblock Causal inference in statistics: An overview.
\newblock {\em Statistics Surveys}, 3:96--146, 2009.

\bibitem[Pea09b]{Pearl2009Causality}
Judea Pearl.
\newblock {\em Causality: Models, Reasoning and Inference}.
\newblock Cambridge University Press, USA, 2nd edition, 2009.

\bibitem[Pea18]{Pearl_WSDM_2018}
Judea Pearl.
\newblock Theoretical impediments to machine learning with seven sparks from the causal revolution.
\newblock In {\em Proceedings of the Eleventh ACM International Conference on Web Search and Data Mining}, WSDM '18, page~3, New York, NY, USA, 2018. Association for Computing Machinery.

\bibitem[PGM{\etalchar{+}}19]{pytorch}
Adam Paszke, Sam Gross, Francisco Massa, Adam Lerer, James Bradbury, Gregory Chanan, Trevor Killeen, Zeming Lin, Natalia Gimelshein, Luca Antiga, Alban Desmaison, Andreas Kopf, Edward Yang, Zachary DeVito, Martin Raison, Alykhan Tejani, Sasank Chilamkurthy, Benoit Steiner, Lu~Fang, Junjie Bai, and Soumith Chintala.
\newblock Pytorch: An imperative style, high-performance deep learning library.
\newblock In H.~Wallach, H.~Larochelle, A.~Beygelzimer, F.~d\textquotesingle Alch\'{e}-Buc, E.~Fox, and R.~Garnett, editors, {\em Advances in Neural Information Processing Systems 32}, pages 8024--8035. Curran Associates, Inc., 2019.

\bibitem[PGP16]{Pearl2016Causality}
Judea Pearl, Madelyn Glymour, and Nicholas P.Jewell.
\newblock {\em {Causal Inference in Statistics, A Primer}}.
\newblock Wiley, 2016.

\bibitem[PHM19]{prabhakaran2019perturbation}
Vinodkumar Prabhakaran, Ben Hutchinson, and Margaret Mitchell.
\newblock Perturbation sensitivity analysis to detect unintended model biases.
\newblock In {\em Proceedings of the 2019 Conference on Empirical Methods in Natural Language Processing and the 9th International Joint Conference on Natural Language Processing (EMNLP-IJCNLP)}, pages 5740--5745, Hong Kong, China, November 2019. Association for Computational Linguistics.

\bibitem[PKRC{\etalchar{+}}25]{CodeQ2025}
David~N. Palacio, Dipin Khati, Daniel Rodriguez-Cardenas, Alejandro Velasco, and Denys Poshyvanyk.
\newblock Codeq: On explaining (large) language models for code using global code-based explanations, 2025.

\bibitem[PM18]{Pearl2018Causality}
Judea Pearl and Dana Mackenzie.
\newblock {\em The Book of Why: The New Science of Cause and Effect}.
\newblock Basic Books, Inc., USA, 1st edition, 2018.

\bibitem[PMM{\etalchar{+}}19]{palacio_security}
David~N. Palacio, Daniel McCrystal, Kevin Moran, Carlos Bernal-Cárdenas, Denys Poshyvanyk, and Chris Shenefiel.
\newblock Learning to identify security-related issues using convolutional neural networks.
\newblock In {\em 2019 IEEE International Conference on Software Maintenance and Evolution (ICSME)}, pages 140--144, 2019.

\bibitem[PRCV{\etalchar{+}}24a]{RepoASTrust24}
David~N. Palacio, Daniel Rodriguez-Cardenas, Alejandro Velasco, Dipin Khati, Kevin Moran, and Denys Poshyvanyk.
\newblock Astrust: Github repository, 2024.

\bibitem[PRCV{\etalchar{+}}24b]{palacio2024trustworthyinterpretablellmscode}
David~N. Palacio, Daniel Rodriguez-Cardenas, Alejandro Velasco, Dipin Khati, Kevin Moran, and Denys Poshyvanyk.
\newblock Towards more trustworthy and interpretable llms for code through syntax-grounded explanations, 2024.

\bibitem[PRWZ02]{papineni_bleu_2002}
Kishore Papineni, Salim Roukos, Todd Ward, and Wei-Jing Zhu.
\newblock {BLEU}: a method for automatic evaluation of machine translation.
\newblock In {\em Proceedings of the 40th {Annual} {Meeting} on {Association} for {Computational} {Linguistics}}, {ACL} '02, pages 311--318, USA, July 2002. Association for Computational Linguistics.

\bibitem[PSGH{\etalchar{+}}21]{poursabzi-sangdeh_manipulating_2021}
Forough Poursabzi-Sangdeh, Daniel~G. Goldstein, Jake~M. Hofman, Jennifer~Wortman Vaughan, and Hanna Wallach.
\newblock Manipulating and {Measuring} {Model} {Interpretability}, August 2021.
\newblock arXiv:1802.07810 [cs].

\bibitem[PTJ{\etalchar{+}}21]{pornprasit2021explain}
Chanathip Pornprasit, Chakkrit Tantithamthavorn, Jirayus Jiarpakdee, Michael Fu, and Patanamon Thongtanunam.
\newblock Pyexplainer: Explaining the predictions of just-in-time defect models.
\newblock In {\em 2021 36th IEEE/ACM International Conference on Automated Software Engineering (ASE)}, pages 407--418, 2021.

\bibitem[Qua25]{noauthor_qualtrics_nodate}
Qualtrics.
\newblock Qualtrics experience management platform, 2025.
\newblock Accessed: 2025-04-10.

\bibitem[Raj14]{10.1145/2593882.2593893}
V\'{a}clav Rajlich.
\newblock Software evolution and maintenance.
\newblock In {\em Future of Software Engineering Proceedings}, FOSE 2014, page 133–144, New York, NY, USA, 2014. Association for Computing Machinery.

\bibitem[RBW{\etalchar{+}}21]{rabin2021generalizability}
Md~Rafiqul~Islam Rabin, Nghi~DQ Bui, Ke~Wang, Yijun Yu, Lingxiao Jiang, and Mohammad~Amin Alipour.
\newblock On the generalizability of neural program models with respect to semantic-preserving program transformations.
\newblock {\em Information and Software Technology}, 135:106552, 2021.

\bibitem[Rep24]{RepoTraceXplainer24}
RepoTrace.
\newblock tracexplainer, 2024.
\newblock [Accessed 11-24-2024].

\bibitem[RGL{\etalchar{+}}20]{Ren2020codebleu}
Shuo Ren, Daya Guo, Shuai Lu, Long Zhou, Shujie Liu, Duyu Tang, Neel Sundaresan, Ming Zhou, Ambrosio Blanco, and Shuai Ma.
\newblock Codebleu: a method for automatic evaluation of code synthesis.
\newblock {\em CoRR}, abs/2009.10297, 2020.

\bibitem[RHG{\etalchar{+}}16]{Baishakhi2016buggy}
Baishakhi Ray, Vincent Hellendoorn, Saheel Godhane, Zhaopeng Tu, Alberto Bacchelli, and Premkumar Devanbu.
\newblock On the "naturalness" of buggy code.
\newblock In {\em Proceedings of the 38th International Conference on Software Engineering}, ICSE '16, page 428–439, New York, NY, USA, 2016. Association for Computing Machinery.

\bibitem[RJL18]{rajpurkar2018squad}
Pranav Rajpurkar, Robin Jia, and Percy Liang.
\newblock Know what you don{'}t know: Unanswerable questions for {SQ}u{AD}.
\newblock In {\em Proceedings of the 56th Annual Meeting of the Association for Computational Linguistics (Volume 2: Short Papers)}, pages 784--789, Melbourne, Australia, July 2018. Association for Computational Linguistics.

\bibitem[RLCL20]{Roziere2020transcoder}
Baptiste Roziere, Marie-Anne Lachaux, Lowik Chanussot, and Guillaume Lample.
\newblock Unsupervised translation of programming languages.
\newblock In H.~Larochelle, M.~Ranzato, R.~Hadsell, M.~F. Balcan, and H.~Lin, editors, {\em Advances in Neural Information Processing Systems}, volume~33, pages 20601--20611. Curran Associates, Inc., 2020.

\bibitem[RMKCH14]{Rempel:ICSE'14}
Patrick Rempel, Patrick Mader, Tobias Kuschke, and Jane Cleland-Huang.
\newblock Mind the {{Gap}}: {{Assessing}} the {{Conformance}} of {{Software Traceability}} to {{Relevant Guidelines}}.
\newblock In {\em Proceedings of the 36th {{International Conference}} on {{Software Engineering}}}, ICSE'14, pages 943--954, Hyderabad, India, 2014. {ACM}.

\bibitem[RRG{\etalchar{+}}18]{Rath:ICSE'18}
Michael Rath, Jacob Rendall, Jin~LC Guo, Jane Cleland-Huang, and Patrick M{\"a}der.
\newblock Traceability in the wild: automatically augmenting incomplete trace links.
\newblock In {\em Proceedings of the 40th International Conference on Software Engineering}, ICSE'18, pages 834--845. ACM, 2018.

\bibitem[RS61]{Raiffa:61}
H.~Ra{\"\i}ffa and R.~Schlaifer.
\newblock {\em Applied statistical decision theory}.
\newblock Studies in managerial economics. Division of Research, Graduate School of Business Adminitration, Harvard University, 1961.

\bibitem[RSFL20]{rei2020comet}
Ricardo Rei, Craig Stewart, Ana~C Farinha, and Alon Lavie.
\newblock {COMET}: A neural framework for {MT} evaluation.
\newblock In {\em Proceedings of the 2020 Conference on Empirical Methods in Natural Language Processing (EMNLP)}, pages 2685--2702, Online, November 2020. Association for Computational Linguistics.

\bibitem[RSG16]{ribeiro2016why}
Marco~Tulio Ribeiro, Sameer Singh, and Carlos Guestrin.
\newblock "why should i trust you?": Explaining the predictions of any classifier.
\newblock In {\em Proceedings of the 22nd ACM SIGKDD International Conference on Knowledge Discovery and Data Mining}, KDD '16, page 1135–1144, New York, NY, USA, 2016. Association for Computing Machinery.

\bibitem[RVSP18]{Romano2018ACode}
Simone Romano, Christopher Vendome, Giuseppe Scanniello, and Denys Poshyvanyk.
\newblock {A Multi-study Investigation Into Dead Code}.
\newblock {\em IEEE Transactions on Software Engineering}, 2018.

\bibitem[RVY14]{Raychev2014CodeCW}
Veselin Raychev, Martin~T. Vechev, and Eran Yahav.
\newblock Code completion with statistical language models.
\newblock {\em Proceedings of the 35th ACM SIGPLAN Conference on Programming Language Design and Implementation}, 2014.

\bibitem[RWGS20]{ribeiro2020checklist}
Marco~Tulio Ribeiro, Tongshuang Wu, Carlos Guestrin, and Sameer Singh.
\newblock Beyond accuracy: Behavioral testing of {NLP} models with {C}heck{L}ist.
\newblock In {\em Proceedings of the 58th Annual Meeting of the Association for Computational Linguistics}, pages 4902--4912, Online, July 2020. Association for Computational Linguistics.

\bibitem[SF19]{DesiderataSokol}
Kacper Sokol and Peter Flach.
\newblock Desiderata for interpretability: Explaining decision tree predictions with counterfactuals.
\newblock {\em Proceedings of the AAAI Conference on Artificial Intelligence}, 33:10035--10036, 07 2019.

\bibitem[SGK19]{shrikumar_learning_2019}
Avanti Shrikumar, Peyton Greenside, and Anshul Kundaje.
\newblock Learning {Important} {Features} {Through} {Propagating} {Activation} {Differences}, October 2019.
\newblock arXiv:1704.02685 [cs].

\bibitem[SGL24]{10.1145/3597503.3623345}
Benjamin Steenhoek, Hongyang Gao, and Wei Le.
\newblock Dataflow analysis-inspired deep learning for efficient vulnerability detection.
\newblock In {\em Proceedings of the IEEE/ACM 46th International Conference on Software Engineering}, ICSE '24, New York, NY, USA, 2024. Association for Computing Machinery.

\bibitem[SGP{\etalchar{+}}24]{spiess2024quality}
Claudio Spiess, David Gros, Kunal~Suresh Pai, Michael Pradel, Md~Rafiqul~Islam Rabin, Amin Alipour, Susmit Jha, Prem Devanbu, and Toufique Ahmed.
\newblock Calibration and correctness of language models for code, 2024.

\bibitem[SGZ03]{Spanoudakis:SEKE'03}
George Spanoudakis, Artur S~d'Avila Garcez, and Andrea Zisman.
\newblock Revising rules to capture requirements traceability relations: A machine learning approach.
\newblock In {\em SEKE}, SEKE'03, pages 570--577, 2003.

\bibitem[SHB15]{sennrich2015neural}
Rico Sennrich, Barry Haddow, and Alexandra Birch.
\newblock Neural machine translation of rare words with subword units.
\newblock {\em arXiv preprint arXiv:1508.07909}, 2015.

\bibitem[SHW{\etalchar{+}}24]{sun2024trustllmtrustworthinesslargelanguage}
Lichao Sun, Yue Huang, Haoran Wang, Siyuan Wu, Qihui Zhang, Yuan Li, Chujie Gao, Yixin Huang, Wenhan Lyu, Yixuan Zhang, Xiner Li, Zhengliang Liu, Yixin Liu, Yijue Wang, Zhikun Zhang, Bertie Vidgen, Bhavya Kailkhura, Caiming Xiong, Chaowei Xiao, Chunyuan Li, Eric Xing, Furong Huang, Hao Liu, Heng Ji, Hongyi Wang, Huan Zhang, Huaxiu Yao, Manolis Kellis, Marinka Zitnik, Meng Jiang, Mohit Bansal, James Zou, Jian Pei, Jian Liu, Jianfeng Gao, Jiawei Han, Jieyu Zhao, Jiliang Tang, Jindong Wang, Joaquin Vanschoren, John Mitchell, Kai Shu, Kaidi Xu, Kai-Wei Chang, Lifang He, Lifu Huang, Michael Backes, Neil~Zhenqiang Gong, Philip~S. Yu, Pin-Yu Chen, Quanquan Gu, Ran Xu, Rex Ying, Shuiwang Ji, Suman Jana, Tianlong Chen, Tianming Liu, Tianyi Zhou, William Wang, Xiang Li, Xiangliang Zhang, Xiao Wang, Xing Xie, Xun Chen, Xuyu Wang, Yan Liu, Yanfang Ye, Yinzhi Cao, Yong Chen, and Yue Zhao.
\newblock Trustllm: Trustworthiness in large language models, 2024.

\bibitem[SK{\etalchar{+}}19]{dowhy}
Amit Sharma, Emre Kiciman, et~al.
\newblock Do{W}hy: {A Python package for causal inference}.
\newblock https://github.com/microsoft/dowhy, 2019.

\bibitem[SLS{\etalchar{+}}24]{meza2024}
Adriana~Meza Soria, Taylor Lopez, Elizabeth Seero, Negin Mashhadi, Emily Evans, Janet Burge, and Andr\'{e} Van~der Hoek.
\newblock Characterizing software maintenance meetings: Information shared, discussion outcomes, and information captured.
\newblock In {\em Proceedings of the IEEE/ACM 46th International Conference on Software Engineering}, ICSE '24, New York, NY, USA, 2024. Association for Computing Machinery.

\bibitem[SMMDE21]{9568959}
Saad Shafiq, Atif Mashkoor, Christoph Mayr-Dorn, and Alexander Egyed.
\newblock A literature review of using machine learning in software development life cycle stages.
\newblock {\em IEEE Access}, 9:140896--140920, 2021.

\bibitem[SN24]{Sayama_2024}
Hiroki Sayama and Chrystopher~L. Nehaniv.
\newblock Self-reproduction and evolution in cellular automata: 25 years after evoloops.
\newblock {\em Artificial Life}, 31(1):81–95, February 2024.

\bibitem[Sol22]{management_solutions_xai}
Management Solutions.
\newblock Explainable artificial intelligence (xai). challenges of model interpretability.
\newblock \url{https://www.managementsolutions.com/sites/default/files/minisite/static/22959b0f-b3da-47c8-9d5c-80ec3216552b/iax/pdf/explainable-artificial-intelligence-en-04.pdf}, 2022.
\newblock Accessed: 18 June 2024.

\bibitem[Spe22]{speicher2022usability}
Maximilian Speicher.
\newblock What is usability? a characterization based on iso 9241-11 and iso/iec 25010, 2022.

\bibitem[SR15]{Svajlenko2015EvaluatingBigCloneBench}
Jeffrey Svajlenko and Chanchal~K. Roy.
\newblock {Evaluating clone detection tools with BigCloneBench}.
\newblock {\em 2015 IEEE 31st International Conference on Software Maintenance and Evolution, ICSME 2015 - Proceedings}, pages 131--140, 2015.

\bibitem[SS19]{serrano_is_2019}
Sofia Serrano and Noah~A. Smith.
\newblock Is {Attention} {Interpretable}?, June 2019.
\newblock arXiv:1906.03731 [cs].

\bibitem[SSZK21]{Sharma2021DoWhyAssumptions}
Amit Sharma, Vasilis Syrgkanis, Cheng Zhang, and Emre Kıcıman.
\newblock Dowhy: Addressing challenges in expressing and validating causal assumptions, 2021.

\bibitem[STY17]{sundararajan2017axiomatic}
Mukund Sundararajan, Ankur Taly, and Qiqi Yan.
\newblock Axiomatic attribution for deep networks.
\newblock In {\em Proceedings of the 34th International Conference on Machine Learning - Volume 70}, ICML'17, page 3319–3328. JMLR.org, 2017.

\bibitem[SvK22]{Scholkopf2022}
Bernhard Schölkopf and Julius von Kügelgen.
\newblock From statistical to causal learning, 2022.

\bibitem[TC22]{troshin_probing_2022}
Sergey Troshin and Nadezhda Chirkova.
\newblock Probing {Pretrained} {Models} of {Source} {Code}, November 2022.
\newblock arXiv:2202.08975 [cs].

\bibitem[TCHC23]{tantithamthavorn2023explain}
Chakkrit Tantithamthavorn, Jürgen Cito, Hadi Hemmati, and Satish Chandra.
\newblock Explainable ai for se: Challenges and future directions.
\newblock {\em IEEE Software}, 40(3):29--33, 2023.

\bibitem[TDP19]{tenney2019bert}
Ian Tenney, Dipanjan Das, and Ellie Pavlick.
\newblock {BERT} rediscovers the classical {NLP} pipeline.
\newblock In {\em Proceedings of the 57th Annual Meeting of the Association for Computational Linguistics}, pages 4593--4601, Florence, Italy, July 2019. Association for Computational Linguistics.

\bibitem[{Ten}25]{tensorflowWikipediaDataset}
{TensorFlow Datasets}.
\newblock Wikipedia dataset, 2025.
\newblock Accessed: 2025-04-07.

\bibitem[{The}16]{theano}
{Theano Development Team}.
\newblock Theano: A {Python} framework for fast computation of mathematical expressions.
\newblock {\em arXiv e-prints}, abs/1605.02688, 2016.

\bibitem[Tho24]{thorneInterplay}
Simon Thorne.
\newblock Understanding the interplay between trust, reliability, and human factors in the age of generative ai.
\newblock {\em International Journal of Simulation: Systems, Science \& technology}, May 2024.

\bibitem[TMM{\etalchar{+}}22]{Tufano.ICSE.2022}
Rosalia Tufano, Simone Masiero, Antonio Mastropaolo, Luca Pascarella, Denys Poshyvanyk, and Gabriele Bavota.
\newblock Using pre-trained models to boost code review automation.
\newblock In {\em 2022 IEEE/ACM 44th International Conference on Software Engineering (ICSE)}, pages 2291--2302, 2022.

\bibitem[TMP18]{Tufano:MSR18}
Watson C Bavota G Di Penta M White~M Tufano~M. and D~Poshyvanyk.
\newblock {Deep Learning Similarities from Different Representations of Source Code}.
\newblock In {\em Proceedings of the 15th IEEE/ACM Conference on Mining Software Repositories (MSR’18)}, MSR '18, Gothenburg, Sweden, 2018.

\bibitem[TPB{\etalchar{+}}17]{Tufano2017WhenAway}
Michele Tufano, Fabio Palomba, Gabriele Bavota, Rocco Oliveto, Massimiliano~Di Penta, Andrea De~Lucia, and Denys Poshyvanyk.
\newblock {When and Why Your Code Starts to Smell Bad (and Whether the Smells Go Away)}.
\newblock {\em IEEE Transactions on Software Engineering}, 43(11):1063--1088, 2017.

\bibitem[TPJR18]{Tian2018a}
Yuchi Tian, Kexin Pei, Suman Jana, and Baishakhi Ray.
\newblock Deeptest: Automated testing of deep-neural-network-driven autonomous cars.
\newblock In {\em Proceedings of the 40th International Conference on Software Engineering}, ICSE '18, pages 303--314, New York, NY, USA, 2018. ACM.

\bibitem[TPT{\etalchar{+}}21]{Tufano:icse2021}
Rosalia Tufano, Luca Pascarella, Michele Tufano, Denys Poshyvanyk, and Gabriele Bavota.
\newblock Towards automating code review activities.
\newblock In {\em 43rd International Conference on Software Engineering, {ICSE}'21}, 2021.

\bibitem[TPW{\etalchar{+}}19]{Tufano:icse2019}
Michele Tufano, Jevgenija Pantiuchina, Cody Watson, Gabriele Bavota, and Denys Poshyvanyk.
\newblock On learning meaningful code changes via neural machine translation.
\newblock In {\em Proceedings of the 41st International Conference on Software Engineering, {ICSE} 2019, Montreal, QC, Canada, May 25-31, 2019}, pages 25--36, 2019.

\bibitem[TSD14]{tu2014local}
Zhaopeng Tu, Zhendong Su, and Premkumar Devanbu.
\newblock On the localness of software.
\newblock In {\em Proceedings of the 22nd ACM SIGSOFT International Symposium on Foundations of Software Engineering}, FSE 2014, page 269–280, New York, NY, USA, 2014. Association for Computing Machinery.

\bibitem[Tuc13]{Tucci2013IntroductionDo-Calculus}
Robert~R. Tucci.
\newblock Introduction to judea pearl's do-calculus, 2013.

\bibitem[TWB{\etalchar{+}}18a]{Tufano.MSR.2018}
Michele Tufano, Cody Watson, Gabriele Bavota, Massimiliano Di~Penta, Martin White, and Denys Poshyvanyk.
\newblock Deep learning similarities from different representations of source code.
\newblock In {\em 2018 IEEE/ACM 15th International Conference on Mining Software Repositories (MSR)}, pages 542--553, 2018.

\bibitem[TWB{\etalchar{+}}18b]{Tufano2018}
Michele Tufano, Cody Watson, Gabriele Bavota, Massimiliano Di~Penta, Martin White, and Denys Poshyvanyk.
\newblock An empirical investigation into learning bug-fixing patches in the wild via neural machine translation.
\newblock In {\em Proceedings of the 33rd ACM/IEEE International Conference on Automated Software Engineering}, ASE 2018, pages 832--837, New York, NY, USA, 2018. ACM.

\bibitem[TWB{\etalchar{+}}19a]{Tufano2019LearningBugFixes}
Michele Tufano, Cody Watson, Gabriele Bavota, Massimiliano Di~Penta, Martin White, and Denys Poshyvanyk.
\newblock {Learning How to Mutate Source Code from Bug-Fixes}.
\newblock {\em ICSME 2019}, pages 301--312, 2019.

\bibitem[TWB{\etalchar{+}}19b]{Tufano:tosem2019}
Michele Tufano, Cody Watson, Gabriele Bavota, Massimiliano~Di Penta, Martin White, and Denys Poshyvanyk.
\newblock An empirical study on learning bug-fixing patches in the wild via neural machine translation.
\newblock {\em {ACM} Trans. Softw. Eng. Methodol.}, 28(4):19:1--19:29, 2019.

\bibitem[VBF{\etalchar{+}}24]{chen_generation_2021}
Helena Vasconcelos, Gagan Bansal, Adam Fourney, Q.~Vera Liao, and Jennifer~Wortman Vaughan.
\newblock Generation probabilities are not enough: Uncertainty highlighting in ai code completions.
\newblock {\em ACM Trans. Comput.-Hum. Interact.}, October 2024.
\newblock Just Accepted.

\bibitem[VDBR]{vafa_rationales_2021}
Keyon Vafa, Yuntian Deng, David~M. Blei, and Alexander~M. Rush.
\newblock Rationales for sequential predictions.

\bibitem[VSP{\etalchar{+}}17]{vaswani2017transformers}
Ashish Vaswani, Noam Shazeer, Niki Parmar, Jakob Uszkoreit, Llion Jones, Aidan~N. Gomez, {\L}ukasz Kaiser, and Illia Polosukhin.
\newblock Attention is all you need.
\newblock In {\em Proceedings of the 31st International Conference on Neural Information Processing Systems}, NeurIPs'17, page 6000–6010, Red Hook, NY, USA, 2017. Curran Associates Inc.

\bibitem[WCG19]{wang_neural_2019}
Changhan Wang, Kyunghyun Cho, and Jiatao Gu.
\newblock Neural {Machine} {Translation} with {Byte}-{Level} {Subwords}, December 2019.
\newblock arXiv:1909.03341.

\bibitem[WCN{\etalchar{+}}20]{watson2020dl4se}
Cody Watson, Nathan Cooper, David Nader{-}Palacio, Kevin Moran, and Denys Poshyvanyk.
\newblock A systematic literature review on the use of deep learning in software engineering research.
\newblock {\em CoRR}, abs/2009.06520, 2020.

\bibitem[WCY{\etalchar{+}}24]{wang2024executablecodeactionselicit}
Xingyao Wang, Yangyi Chen, Lifan Yuan, Yizhe Zhang, Yunzhu Li, Hao Peng, and Heng Ji.
\newblock Executable code actions elicit better llm agents, 2024.

\bibitem[WDC{\etalchar{+}}25]{wei2025swerladvancingllmreasoning}
Yuxiang Wei, Olivier Duchenne, Jade Copet, Quentin Carbonneaux, Lingming Zhang, Daniel Fried, Gabriel Synnaeve, Rishabh Singh, and Sida~I. Wang.
\newblock Swe-rl: Advancing llm reasoning via reinforcement learning on open software evolution, 2025.

\bibitem[WDH{\etalchar{+}}21]{trustincollab}
David~Gray Widder, Laura Dabbish, James~D. Herbsleb, Alexandra Holloway, and Scott Davidoff.
\newblock Trust in collaborative automation in high stakes software engineering work: A case study at nasa.
\newblock In {\em Proceedings of the 2021 CHI Conference on Human Factors in Computing Systems}, CHI '21, New York, NY, USA, 2021. Association for Computing Machinery.

\bibitem[WDS{\etalchar{+}}20]{wolf2020transformers}
Thomas Wolf, Lysandre Debut, Victor Sanh, Julien Chaumond, Clement Delangue, Anthony Moi, Pierric Cistac, Tim Rault, Rémi Louf, Morgan Funtowicz, Joe Davison, Sam Shleifer, Patrick von Platen, Clara Ma, Yacine Jernite, Julien Plu, Canwen Xu, Teven~Le Scao, Sylvain Gugger, Mariama Drame, Quentin Lhoest, and Alexander~M. Rush.
\newblock Transformers: State-of-the-art natural language processing.
\newblock In {\em Proceedings of the 2020 Conference on Empirical Methods in Natural Language Processing: System Demonstrations}, pages 38--45, Online, October 2020. Association for Computational Linguistics.

\bibitem[Wel22]{weller2019transparency}
Adrian Weller.
\newblock {\em Transparency: Motivations and Challenges}, page 23–40.
\newblock Springer-Verlag, Berlin, Heidelberg, 2022.

\bibitem[Whi15]{White:ICSE15}
M~White.
\newblock {Deep Representations for Software Engineering}.
\newblock In {\em Proceedings of the 37th IEEE/ACM International Conference on Software Engineering (ICSE'15)}, volume~2 of {\em ICSE '15}, pages 781--783, 5 2015.

\bibitem[WK20]{white_reassert_2020}
Robert White and Jens Krinke.
\newblock {ReAssert}: {Deep} {Learning} for {Assert} {Generation}, November 2020.
\newblock arXiv:2011.09784 [cs].

\bibitem[WLT16]{Wang2016}
S.~{Wang}, T.~{Liu}, and L.~{Tan}.
\newblock Automatically learning semantic features for defect prediction.
\newblock In {\em 2016 IEEE/ACM 38th International Conference on Software Engineering (ICSE)}, pages 297--308, May 2016.

\bibitem[WPR19]{wainakh_evaluating_2019}
Yaza Wainakh, Michael Pradel, and Moiz Rauf.
\newblock Evaluating semantic representations of source code.
\newblock {\em arXiv}, 2019.
\newblock arXiv: 1910.05177.

\bibitem[WRHW19]{wu2019errudite}
Tongshuang Wu, Marco~Tulio Ribeiro, Jeffrey Heer, and Daniel Weld.
\newblock {E}rrudite: Scalable, reproducible, and testable error analysis.
\newblock In {\em Proceedings of the 57th Annual Meeting of the Association for Computational Linguistics}, pages 747--763, Florence, Italy, July 2019. Association for Computational Linguistics.

\bibitem[WRHW21]{wu-etal-2021-polyjuice}
Tongshuang Wu, Marco~Tulio Ribeiro, Jeffrey Heer, and Daniel Weld.
\newblock Polyjuice: Generating counterfactuals for explaining, evaluating, and improving models.
\newblock In {\em Proceedings of the 59th Annual Meeting of the Association for Computational Linguistics and the 11th International Joint Conference on Natural Language Processing (Volume 1: Long Papers)}, pages 6707--6723, Online, August 2021. Association for Computational Linguistics.

\bibitem[WS25]{icodegen}
WM-SEMERU.
\newblock Causalse: Causal interpretability for software engineering.
\newblock \url{https://github.com/WM-SEMERU/CausalSE}, 2025.
\newblock Accessed: 2025-04-07.

\bibitem[WSM{\etalchar{+}}19]{wang2018glue}
Alex Wang, Amanpreet Singh, Julian Michael, Felix Hill, Omer Levy, and Samuel~R. Bowman.
\newblock {GLUE}: A multi-task benchmark and analysis platform for natural language understanding.
\newblock In {\em International Conference on Learning Representations}, 2019.

\bibitem[WTM{\etalchar{+}}19]{SANER.2019}
Martin White, Michele Tufano, Matías Martínez, Martin Monperrus, and Denys Poshyvanyk.
\newblock Sorting and transforming program repair ingredients via deep learning code similarities.
\newblock In {\em 2019 IEEE 26th International Conference on Software Analysis, Evolution and Reengineering (SANER)}, pages 479--490, 2019.

\bibitem[WTM{\etalchar{+}}20]{Watson:ICSE20}
Cody Watson, Michele Tufano, Kevin Moran, Gabriele Bavota, and Denys Poshyvanyk.
\newblock On learning meaningful assert statements for unit test cases.
\newblock In {\em Proceedings of the ACM/IEEE 42nd International Conference on Software Engineering}, ICSE '20, page 1398–1409, New York, NY, USA, 2020. Association for Computing Machinery.

\bibitem[WTVP16]{White2016}
M.~White, M.~Tufano, C.~Vendome, and D.~Poshyvanyk.
\newblock Deep learning code fragments for code clone detection.
\newblock In {\em 2016 31st {{IEEE}}/{{ACM International Conference}} on {{Automated Software Engineering}} ({{ASE}})}, ASE'16, pages 87--98, September 2016.
\newblock ISSN:.

\bibitem[WVLVP15]{White:MSR15}
Martin White, Christopher Vendome, Mario Linares-V{\'{a}}squez, and Denys Poshyvanyk.
\newblock {Toward Deep Learning Software Repositories}.
\newblock In {\em Proceedings of the 12th IEEE Working Conference on Mining Software Repositories (MSR'15)}, MSR '15, pages 334--345, Piscataway, NJ, USA, 2015. IEEE Press.

\bibitem[WWJH21]{wang-etal-2021-codet5}
Yue Wang, Weishi Wang, Shafiq Joty, and Steven~C.H. Hoi.
\newblock {C}ode{T}5: Identifier-aware unified pre-trained encoder-decoder models for code understanding and generation.
\newblock In Marie-Francine Moens, Xuanjing Huang, Lucia Specia, and Scott Wen-tau Yih, editors, {\em Proceedings of the 2021 Conference on Empirical Methods in Natural Language Processing}, pages 8696--8708, Online and Punta Cana, Dominican Republic, November 2021. Association for Computational Linguistics.

\bibitem[WWW17]{Wang2017Semantics-awareCode}
Shuai Wang, Pei Wang, and Dinghao Wu.
\newblock {Semantics-aware machine learning for function recognition in binary code}.
\newblock In {\em Proceedings - 2017 IEEE International Conference on Software Maintenance and Evolution, ICSME 2017}, 2017.

\bibitem[WZY{\etalchar{+}}18]{Wan2018}
Yao Wan, Zhou Zhao, Min Yang, Guandong Xu, Haochao Ying, Jian Wu, and Philip~S. Yu.
\newblock Improving automatic source code summarization via deep reinforcement learning.
\newblock In {\em Proceedings of the 33rd ACM/IEEE International Conference on Automated Software Engineering}, ASE 2018, pages 397--407, New York, NY, USA, 2018. ACM.

\bibitem[WZZ{\etalchar{+}}22]{wan_what_2022}
Yao Wan, Wei Zhao, Hongyu Zhang, Yulei Sui, Guandong Xu, and Hai Jin.
\newblock What {Do} {They} {Capture}? -- {A} {Structural} {Analysis} of {Pre}-{Trained} {Language} {Models} for {Source} {Code}, February 2022.
\newblock arXiv:2202.06840 [cs].

\bibitem[XANH22]{xu_systematic_2022}
Frank~F. Xu, Uri Alon, Graham Neubig, and Vincent~J. Hellendoorn.
\newblock A {Systematic} {Evaluation} of {Large} {Language} {Models} of {Code}, May 2022.
\newblock arXiv:2202.13169 [cs].

\bibitem[XYX{\etalchar{+}}16]{Xu2016}
B.~Xu, D.~Ye, Z.~Xing, X.~Xia, G.~Chen, and S.~Li.
\newblock Predicting semantically linkable knowledge in developer online forums via convolutional neural network.
\newblock In {\em 2016 31st {{IEEE}}/{{ACM International Conference}} on {{Automated Software Engineering}} ({{ASE}})}, ASE'16, pages 51--62, September 2016.
\newblock ISSN:.

\bibitem[yAAE{\etalchar{+}}24]{arcas2024computationallifewellformedselfreplicating}
Blaise~Agüera y~Arcas, Jyrki Alakuijala, James Evans, Ben Laurie, Alexander Mordvintsev, Eyvind Niklasson, Ettore Randazzo, and Luca Versari.
\newblock Computational life: How well-formed, self-replicating programs emerge from simple interaction, 2024.

\bibitem[YCL25]{ye2025distilledlifelongselfadaptationconfigurable}
Yulong Ye, Tao Chen, and Miqing Li.
\newblock Distilled lifelong self-adaptation for configurable systems, 2025.

\bibitem[YJW{\etalchar{+}}24]{yang2024sweagentagentcomputerinterfacesenable}
John Yang, Carlos~E. Jimenez, Alexander Wettig, Kilian Lieret, Shunyu Yao, Karthik Narasimhan, and Ofir Press.
\newblock Swe-agent: Agent-computer interfaces enable automated software engineering, 2024.

\bibitem[ZBO21]{openai_codex}
Wojciech Zaremba, Greg Brockman, and OpenAI.
\newblock Openai codex, Aug 2021.

\bibitem[ZCZ{\etalchar{+}}23]{zan_large_2023}
Daoguang Zan, Bei Chen, Fengji Zhang, Dianjie Lu, Bingchao Wu, Bei Guan, Yongji Wang, and Jian-Guang Lou.
\newblock Large {Language} {Models} {Meet} {NL2Code}: {A} {Survey}, May 2023.
\newblock arXiv:2212.09420 [cs].

\bibitem[ZLL{\etalchar{+}}24]{10.1145/3597503.3623310}
Wenkang Zhong, Chuanyi Li, Kui Liu, Tongtong Xu, Jidong Ge, Tegawende~F. Bissyande, Bin Luo, and Vincent Ng.
\newblock Practical program repair via preference-based ensemble strategy.
\newblock In {\em Proceedings of the IEEE/ACM 46th International Conference on Software Engineering}, ICSE '24, New York, NY, USA, 2024. Association for Computing Machinery.

\bibitem[ZLS{\etalchar{+}}19]{zhou_devign_nodate}
Yaqin Zhou, Shangqing Liu, Jingkai Siow, Xiaoning Du, and Yang Liu.
\newblock {\em Devign: effective vulnerability identification by learning comprehensive program semantics via graph neural networks}.
\newblock Curran Associates Inc., Red Hook, NY, USA, 2019.

\bibitem[ZZL{\etalchar{+}}23]{zhao_survey_2023}
Wayne~Xin Zhao, Kun Zhou, Junyi Li, Tianyi Tang, Xiaolei Wang, Yupeng Hou, Yingqian Min, Beichen Zhang, Junjie Zhang, Zican Dong, Yifan Du, Chen Yang, Yushuo Chen, Zhipeng Chen, Jinhao Jiang, Ruiyang Ren, Yifan Li, Xinyu Tang, Zikang Liu, Peiyu Liu, Jian-Yun Nie, and Ji-Rong Wen.
\newblock A {Survey} of {Large} {Language} {Models}, April 2023.
\newblock arXiv:2303.18223 [cs].

\end{thebibliography}
